\documentclass[%
 reprint,
superscriptaddress,
 amsmath,amssymb,
 aps,
]{revtex4-2}

\usepackage{graphicx}% Include figure files
\usepackage[caption=false]{subfig}
\usepackage{dcolumn}% Align table columns on decimal point
\usepackage{bm}% bold math
\usepackage{hyperref}% add hypertext capabilities
\hypersetup{
  colorlinks   = true, %Colours links instead of ugly boxes
  urlcolor     = blue, %Colour for external hyperlinks
  linkcolor    = blue, %Colour of internal links
  citecolor   = blue %Colour of citations
}
\usepackage{siunitx}
\newcommand{\mmicro}{\si\micro} %- consistent text micro symbol
\usepackage{float}
\usepackage{longtable}
\newcolumntype{P}[1]{>{\centering\arraybackslash}p{#1}}
\newcolumntype{L}[1]{>{\raggedright\arraybackslash}p{#1}}
\usepackage{multirow}
\usepackage{boldline}
\begin{document}

%\preprint{APS/123-QED}

%\title{Nuclear Excitation Functions from 35 to 200 MeV Proton Irradiations of Arsenic, Titanium, and Copper
%Nuclear Excitation Functions of Proton-Induced Reactions from 35 to 200 MeV on Arsenic, Titanium, and Copper
%with Analysis of Novel Production Pathways to the $^{72}$Se/$^{72}$As and $^{68}$Ge/$^{68}$Ga Generators
%Including High-Energy Reaction Modeling Analysis}
%\title{Nuclear Excitation Functions of Proton-Induced Reactions from 35 to 200 MeV on Arsenic, Titanium, and Copper
%with Analysis of the $^{72}$Se/$^{72}$As and $^{68}$Ge/$^{68}$Ga Generators and Charged-Particle Modeling}
%\title{Proton-Induced Reactions on Arsenic from 35 to 200 MeV with Analysis of the $^{72}$Se/$^{72}$As and $^{68}$Ge/$^{68}$Ga Generators and Reaction Modeling}
%\title{Analysis of the $^{72}$Se/$^{72}$As and $^{68}$Ge/$^{68}$Ga Generators and Reaction Modeling}
%\title{Measurement of Proton-Induced Reactions on Arsenic from 35 to 200 MeV and Residual Production Modeling with Study of Off-Stability Nuclear Level Densities}
\title{Measurement and Modeling of Proton-Induced Reactions on\\ Arsenic from 35 to 200 MeV}

\author{Morgan B. Fox}
\email{morganbfox@berkeley.edu}
\affiliation{Department of Nuclear Engineering, University of California, Berkeley, Berkeley, California 94720, USA}
\author{Andrew S. Voyles}
\email{asvoyles@lbl.gov}
\affiliation{Department of Nuclear Engineering, University of California, Berkeley, Berkeley, California 94720, USA}
\affiliation{Lawrence Berkeley National Laboratory, Berkeley, California 94720, USA}
\author{Jonathan T. Morrell}
\affiliation{Department of Nuclear Engineering, University of California, Berkeley, Berkeley, California 94720, USA}
\author{Lee A. Bernstein}
\affiliation{Department of Nuclear Engineering, University of California, Berkeley, Berkeley, California 94720, USA}
\affiliation{Lawrence Berkeley National Laboratory, Berkeley, California 94720, USA}
\author{Jon C. Batchelder}
\affiliation{Department of Nuclear Engineering, University of California, Berkeley, Berkeley, California 94720, USA}
\author{Eva R. Birnbaum}
\affiliation{Los Alamos National Laboratory, Los Alamos, New Mexico 87544, USA}
\author{Cathy S. Cutler}
\affiliation{Brookhaven National Laboratory, Upton, New York 11973, USA}
\author{Arjan J. Koning}
\affiliation{International Atomic Energy Agency, P.O. Box 100, A-1400 Vienna, Austria}
\author{Amanda M. Lewis}
\affiliation{Department of Nuclear Engineering, University of California, Berkeley, Berkeley, California 94720, USA}
\author{Dmitri G. Medvedev}
\affiliation{Brookhaven National Laboratory, Upton, New York 11973, USA}
\author{Francois M. Nortier}
\affiliation{Los Alamos National Laboratory, Los Alamos, New Mexico 87544, USA}
\author{Ellen M. O'Brien}
\affiliation{Los Alamos National Laboratory, Los Alamos, New Mexico 87544, USA}
\author{Christiaan Vermeulen}
\affiliation{Los Alamos National Laboratory, Los Alamos, New Mexico 87544, USA}

\date{\today}

\begin{abstract}
$^{72}$As is a promising positron emitter for diagnostic imaging that can be employed locally using a $^{72}$Se generator. However, current reaction pathways to $^{72}$Se have insufficient nuclear data for efficient production using regional 100--200\,MeV high-intensity proton accelerators. In order to address this deficiency, stacked-target irradiations were performed at LBNL, LANL, and BNL to measure the production of the $^{72}$Se/$^{72}$As PET generator system via $^{75}$As(p,x) between 35 and 200\,MeV. This work provides the most well-characterized excitation function for $^{75}$As(p,4n)$^{72}$Se starting from threshold. Additional focus was given to report the first measurements of $^{75}$As(p,x)$^{68}$Ge and bolster an already robust production capability for the highly valuable $^{68}$Ge/$^{68}$Ga PET generator. Thick target yield comparisons with prior established formation routes to both generators are made. In total, high-energy proton-induced cross sections are reported for 55 measured residual products from $^{75}$As, $^{\textnormal{nat}}$Cu, and $^{\textnormal{nat}}$Ti targets, where the latter two materials were present as monitor foils. These results were compared with literature data as well as the default theoretical calculations of the nuclear model codes TALYS, CoH, EMPIRE, and ALICE. Reaction modeling at these energies is typically unsatisfactory due to few prior published data and many interacting physics models. Therefore, a detailed assessment of the TALYS code was performed with simultaneous parameter adjustments applied according to a standardized procedure. Particular attention was paid to the formulation of the two-component exciton model in the transition between the compound and pre-equilibrium regions, with a linked investigation of level density models for nuclei off of stability and their impact on modeling predictive power. This paper merges experimental work and evaluation techniques for high-energy charged-particle isotope production in an extension to an earlier study of this kind.

\end{abstract}

\maketitle

\section{\label{Introduction}Introduction}
%XXX The continued rise of nuclear medicine to study physiological processes and diagnose, stage, and treat diseases, prompts not only the increased creation of currently used radionuclides but also the development of improved production routes and entirely novel radioisotopes altogether \cite{NucMedGeneral}. The implementation of new methodologies or products in nuclear medicine is a multifaceted problem but fundamentally depends on accurate and precise nuclear reaction cross section data. Well-characterized excitation functions for radionuclides are needed to properly inform and optimize their large scale production required for clinical use.
%In the continuation of the YATC effort to improve the state of charged-particle induced nuclear reaction data for the formation of radionuclides with properties of interest to the medical community, proton irradiations of arsenic have been performed. Desired knowledge of $^{75}$As(p,x)$^{72}$Se, $^{68}$Ge is motivated by the production ``cow" properties of the radionuclides for their decay daughters, $^{72}$As and $^{68}$Ga, respectively.
%. and use with monoclonal antibodies of long biological half-lives applicable to oncological 
Multi-hundred MeV proton accelerators are promising sites for the large scale production of medical radionuclides due to the high production rates enabled by their high-intensity beam capabilities and the long range of high-energy protons. However, the ability to reliably conduct isotope production at these accelerators and model relevant (p,x) reactions in the 100--200\,MeV range is hampered by a lack of measured data.  

In the effort to improve this state of proton-induced nuclear reaction data, irradiations of arsenic have been performed. The formation of $^{72}$Se and $^{68}$Ge from $^{75}$As(p,x) is of particular interest for their application in diagnostic imaging as generators or ``cows" for their decay daughters, $^{72}$As and $^{68}$Ga, respectively. The present general production data for $^{72}$Se at incident proton energies in the 35--200\,MeV range are scarce to non-existent. Low-energy $^{68}$Ge production data have been thoroughly assessed and already contribute to a robust production capability set over the past decade, but extending knowledge for $^{68}$Ge formation at higher-energies too should benefit its overall application. The 35--200\,MeV range is especially relevant because it is characteristic of the Los Alamos Isotope Production Facility (IPF) and the Brookhaven LINAC Isotope Producer (BLIP), where medical isotopes are created for widespread use.
%In the effort to improve this state of proton-induced nuclear reaction data, irradiations of arsenic have been performed. The formation of $^{72}$Se and $^{68}$Ge from $^{75}$As(p,x) are of particular interest for their application in diagnostic imaging as generators or ``cows" for their decay daughters, $^{72}$As and $^{68}$Ga, respectively. The present production data for these generator radionuclides for incident proton energies in the 35--200\,MeV range are scarce to non-existent. This energy range is especially relevant because it is characteristic of the Los Alamos Isotope Production Facility (IPF) and the Brookhaven LINAC Isotope Producer (BLIP), where medical isotopes are created for widespread use.
%under the U.S. Department of Energy's Office of Science.

$^{72}$As ($t_{1/2}=26.0\ (1) $\,h, 87.8 (22)\%  $\beta^+$ \cite{DataSheetsA72}) is a favourable positron emitting radionuclide for the imaging of slower biological processes. Its half-life makes $^{72}$As-labelled radiopharmaceuticals useful for the observation of long-term metabolic processes, such as the enrichment and distribution of antibodies in tumour tissue, by positron emission tomography (PET) \cite{Jennewein2005:SeAsGenerator,Tarkanyi2019:MedicalIsotopesDataNeeds}. $^{72}$As offers the similar slow kinetic behaviour as the PET isotope $^{124}$I ($t_{1/2}=4.1760\ (3)$\,d, 22.7 (13)\%  $\beta^+$ \cite{DataSheetsA124}) but with a higher positron emission decay branch \cite{Cascini2014:124I}. Furthermore, $^{72}$As can form a promising pair with $^{77}$As ($t_{1/2}=38.83\ (5)$\,h, 100.0 (4)\%  $\beta^-$, 683.2 (17)\,keV $E_{\beta^-,max}$ \cite{DataSheetsA77}) for combined imaging and radiotherapy \cite{Degraffenreid2019:BNLAs,Jennewein2008:MedRadioarsenic,Sanders2020:Radioarsenic}. The high sulfur affinity of arsenic, promoting its covalent binding to thiol groups, along with the toxicity of the $^{77}$As decay spectrum, make $^{72}$As/$^{77}$As an unique theranostic candidate \cite{Jennewein2008:MedRadioarsenic,Ellison2017:72AsRadioarsenic}.

Current production methods for $^{72}$As require a charged-particle beam in an accelerator setting. Existing accelerator pathways rely on $^{\mathrm{nat}}$Ge targets via the $^{\mathrm{nat}}$Ge(p/d,xn)$^{72}$As mechanisms in the 10--50\,MeV incident particle energy range \cite{Tarkanyi2019:MedicalIsotopesDataNeeds,Ballard2012:SeAsGenerator}. However, these direct routes to $^{72}$As constrain its use to medical centres nearby the production facility due to a half-life not appropriate for shipping or dispensing from a storage inventory. Additionally, direct production from $^{\mathrm{nat}}$Ge suffers from low thick target yields at these low incident energies and from co-production of the longer-lived radioisotopic impurities $^{74,73,71}$As \cite{Tarkanyi2019:MedicalIsotopesDataNeeds,Ballard2012:SeAsGenerator}. Instead, recognition of the longer-lived $^{72}$Se ($t_{1/2}=8.40\ (8)$\,d \cite{DataSheetsA72}) as the parent precursor to $^{72}$As creates the possibility for a $^{72}$Se/$^{72}$As generator system  \cite{Jennewein2005:SeAsGenerator,Ballard2012:SeAsGenerator,Sanders2020:Radioarsenic}. Production of a generator results in $^{72}$As free from other radioarsenic contaminants, on account of advantageous lifetime differences between $^{72}$Se and neighboring Se nuclei, and availability restrictions at medical facilities across the globe. Measurements of a $^{\mathrm{nat}}$Br(p,x)$^{72}$Se production route have been undertaken but the thick target yields, even approaching 200 MeV incident protons, are relatively low \cite{Tarkanyi2019:MedicalIsotopesDataNeeds,Degraffenreid2019:BNLAs,Ballard2012:Br,Fassbender2001:Br}. Bromine targets subjected to high power may also pose heating and/or reactivity problems \cite{Ballard2012:Br,Fassbender2001:Br}. The alternatively explored formation mechanism of $^{\mathrm{nat}/70}$Ge($\alpha$,xn)$^{72}$Se also suffers from low yields due to the short range of lower energy $\alpha$-particles combined with a relatively small ($<100$\,mb) production peak \cite{Takacs2016:GetoSe}.

In contrast, proton-induced reactions on arsenic offer a potentially improved production pathway to the $^{72}$Se/$^{72}$As generator system. The combination of an expected sufficient cross section over a wide energy range with a naturally monoisotopic ($^{75}$As), stable material that can be appropriately formed into production targets makes high-intensity, high-energy proton irradiations an enticing approach.
%($t_{1/2}=65.30$\,h \cite{DataSheetsA71}) and $^{74}$As ($t_{1/2}=17.77$\,d \cite{DataSheetsA74})

$^{68}$Ga ($t_{1/2}=67.71\ (8)$\,min, 88.91 (9)\%  $\beta^+$ \cite{DataSheetsA68}) has emerged as a significant short-lived positron emitter alongside the ubiquitous $^{18}$F for PET imaging in cases of general cancer, glioma, hypoxia, neuroendocrine tumours, and more \cite{Martiniova2016:68GaInfo,IAEA2010:68Ge}. $^{68}$Ga readily forms stable complexes with DOTA (a synthetically flexible metal chelating agent) and HBED, allowing peptides and other small molecules to be radiolabeled at high specific activities \cite{Burke2020:68Ga,LANLResearchQuarterly}. NETSpot, using $^{68}$Ga-DOTA, is an FDA approved PET imaging agent for neuroendocrine cancers \cite{Burke2020:68Ga}. Further, the compatibility of $^{68}$Ga with a prostate-specific membrane antigen targeting ligand (PSMA-11 with HBED chelator) has led to a sought-after, highly successful PET tracer for the diagnosis of prostate cancer \cite{Martiniova2016:68GaInfo,Kratochwil2018:225Ac,Burke2020:68Ga}. However, in a similar fashion to $^{72}$As, direct production by typical $^{65}$Cu($\alpha$,n)$^{68}$Ga and $^{68}$Zn(p,n)$^{68}$Ga routes suffer from the same local accelerator production and shipping time constraints that inhibit widespread use \cite{Tarkanyi2019:MedicalIsotopesDataNeeds}. Conversely, an indirect pathway to $^{68}$Ga, through its long-lived $^{68}$Ge ($t_{1/2}=270.93\ (13)$ d \cite{DataSheetsA68}) parent, constitutes an effective generator system more applicable for societal application. 

While the elution and separation chemistry of the $^{68}$Ge/$^{68}$Ga system has been extensively developed, nuclear data for $^{68}$Ge production remains partially incomplete \cite{IAEA2010:68Ge}. The $^{\mathrm{nat}}$Ga(p,xn)$^{68}$Ge route is the heavily studied, successful favourite of accelerator sites globally -- particularly the prominent facilities of IPF, BLIP, and iThemba labs -- but data only reaches up to 100\,MeV. Other $^{69}$Ga(p,xn)$^{68}$Ge, $^{\mathrm{nat}}$Ge(p,pxn)$^{68}$Ge, and $^{66}$Zn($\alpha$,2n)$^{68}$Ge low-energy pathways have been explored but are less ideal due to excitation functions that peak in the 15--35\,MeV range, which may be suboptimal for thick target yields, and present target manufacturing and purity concerns \cite{IAEA2010:68Ge,LANLResearchQuarterly}. Studying proton-induced reactions on arsenic gives a chance to strengthen the community’s total understanding of $^{68}$Ge/$^{68}$Ga formation.
%While the elution and separation chemistry of the $^{68}$Ge/$^{68}$Ga system has been extensively developed, nuclear data for $^{68}$Ge production remains incomplete \cite{IAEA2010:68Ge}. The $^{\mathrm{nat}/69}$Ga(p,xn)$^{68}$Ge, $^{\mathrm{nat}}$Ge(p,pxn)$^{68}$Ge, and $^{66}$Zn($\alpha$,2n)$^{68}$Ge pathways are the current favourites of accelerator sites globally. These excitation functions peak in the 15--35\,MeV range, which may be suboptimal for both thick target yields and regional facility beam operations \cite{IAEA2010:68Ge,LANLResearchQuarterly}. Additionally, gallium metal (melting point: 39$^{\circ}$C) is a difficult target to manipulate and presents thermal stability issues that may limit $^{68}$Ge production \cite{IAEA2010:68Ge}.
%
%In contrast, proton-induced reactions on arsenic offer potentially improved production pathways to these $^{72}$Se/$^{72}$As and $^{68}$Ge/$^{68}$Ga generator systems. Arsenic is naturally monoisotopic ($^{75}$As), stable, and can be appropriately formed into thick targets.

In this work, proton-induced nuclear reaction data for $^{75}$As were measured for energies 35--200\,MeV using the stacked-target method as part of the DOE Isotope Program's Tri-laboratory Effort in Nuclear Data (TREND) between Lawrence Berkeley National Laboratory (LBNL), Los Alamos National Laboratory (LANL), and Brookhaven National Laboratory (BNL) \cite{Fox2020:NbLa}. We report the first cross section measurements for $^{75}$As(p,x)$^{68}$Ge and the most well-characterized excitation function of $^{75}$As(p,4n)$^{72}$Se to-date. Thick target yields are additionally calculated from the measured excitation functions and compared to established formation routes for the generator radionuclides to better inform accelerator facilities of optimal production parameters.

This stacked-target work has further provided 53 other high-energy (p,x) production cross section datasets for residual nuclei stemming from $^{75}$As, $^{\mathrm{nat}}$Cu, and $^{\mathrm{nat}}$Ti targets.
%These additional products find applications as other medically useful therapeutic and diagnostic isotopes or help to improve nuclear data.

These extensive measurements were also used to assess the predictions of multiple nuclear reaction codes. The standardized fitting procedure for reaction model parameters and pre-equilibrium adjustments developed in \textcite{Fox2020:NbLa} was applied to the arsenic data, with an investigative focus to check if the proposed exciton model trends are seen.

In addition to studying pre-equilibrium, the fitting procedure provided insight into the appropriate level density models for a swath of nuclei. A discussion of the impact of level density knowledge on modeling predictive power is presented with a reflection of the limitations imposed on creating recommended high-energy charged-particle data.
%A necessary study of level density models for nuclei off of stability developed from applying the fitting procedure. A discussion of the impact of level density knowledge on modeling predictive power is presented with a reflection of the consequent limitations imposed on creating recommended high-energy charged-particle data.

The combination of experimental measurement and evaluation study presented in this work creates data with immediate application while contributing to an increasingly prioritized future need for high-energy modeling in the nuclear data community \cite{WANDA2021}.

\section{\label{2}Experimental Methods and Materials}
This work is an outcome of the same set of experimental irradiations and activations performed for \textcite{Fox2020:NbLa} but gives a focus to the analysis and interpretation of arsenic, titanium, and copper target foils not previously discussed. Charged-particle stacked-target irradiations were carried out at the 88-Inch Cyclotron at LBNL for proton energies of $E_p<55$\,MeV, at IPF at LANL for $50<E_p<100$\,MeV, and at BLIP at BNL for $100<E_p<200$\,MeV.

The stacked-target technique is a typical methodology for charged-particle irradiations to simultaneously measure multiple high-fidelity energy-separated cross section values per reaction channel. A stacked-target includes thin foils of a target of interest in combination with thick degraders and monitor foils. The degraders selectively reduce the primary beam energy throughout the stack while the monitor foils can be used to characterize the evolving beam properties as it propagates through the targets. Detailed explanations of the technique can be read in \cite{Fox2020:NbLa,Voyles2018:Nb,Morrell2020,Graves2016:StackTarget,Engle2013:ProtonsThorium,Engle2016:p_Tb}.

\subsection{\label{2.}Stacked-Target Design}
Individual stacks were created for each irradiation at each experimental site. The three stacks differed slightly in composition according to the physical constraints of each site's irradiation geometry and as a function of expected residual radionuclide production based on beam current and energy parameters.

\subsubsection{\label{2.1}LBNL Stack and Irradiation}
The 88-Inch Cyclotron stack consisted of 25 \mmicro m $^{\mathrm{nat}}$Cu foils (99.95\%, CU000420,  Goodfellow Metals, Coraopolis, PA 15108-9302, USA) and thin metallic $^{75}$As layers electroplated onto 10 \mmicro m or 25 \mmicro m $^{\mathrm{nat}}$Ti foil backings (99.6\%, TI000213/TI000290, Goodfellow Metals).

\begin{figure}[H]
{\includegraphics[width=1.0\columnwidth]{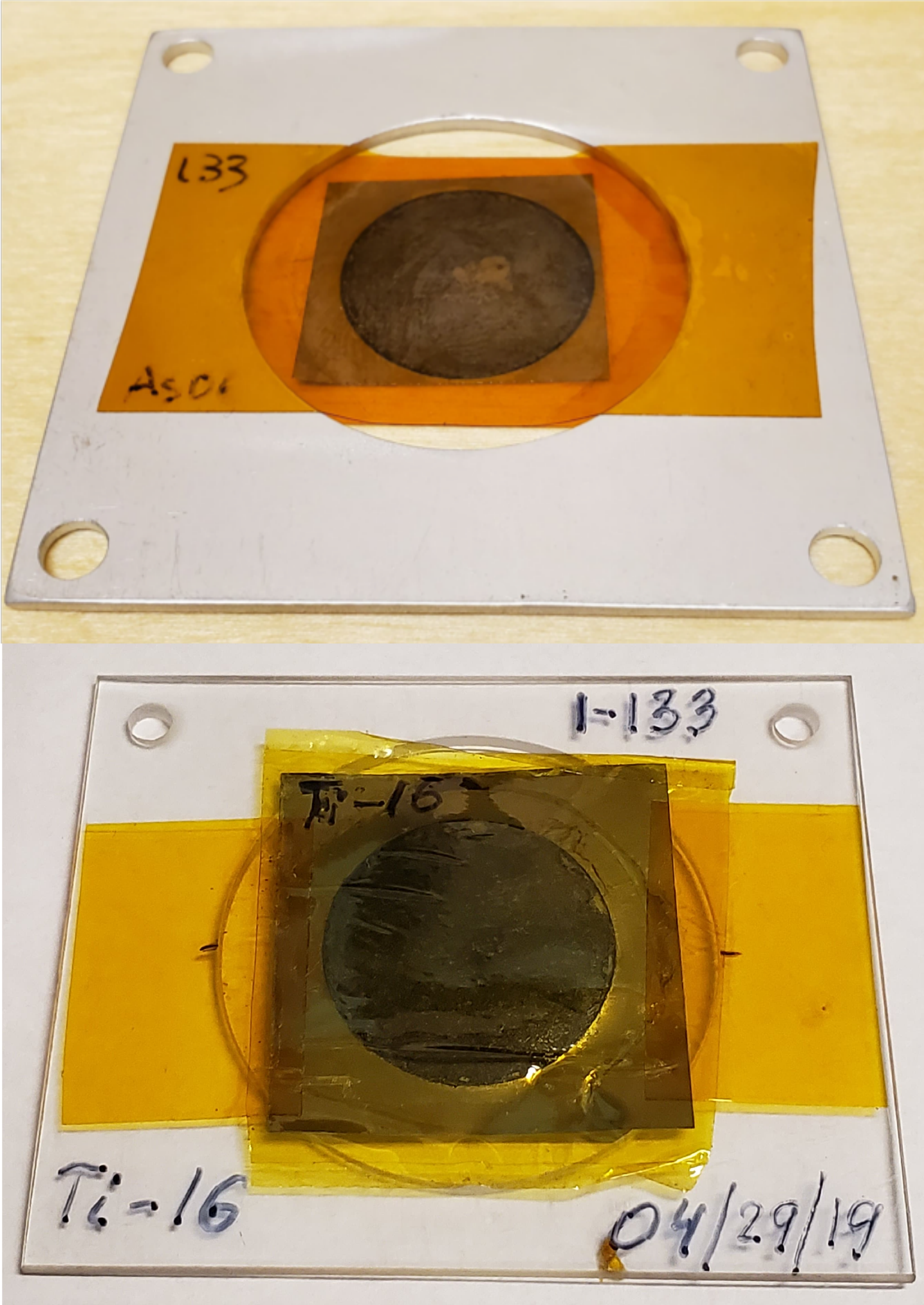}}
\caption{View of individual electroplated arsenic depositions on titanium backings within Kapton seals. The top target is sampled from the LBNL stack and is pictured after proton irradiation, where slight bubbling in the Kapton seal exists as a result of beam heating. The bottom target is part of the BNL stack prior to proton irradiation.}
\label{AsFoilElectroplated}
\end{figure}

Nine copper and titanium foils each were cut into 2.5\,cm $\times$ 2.5\,cm squares and characterized by taking four length and width measurements using a digital caliper (Mitutoyo America Corp.) and four thickness measurements taken at different locations using a digital micrometer (Mitutoyo America Corp.). Each foil was also massed multiple times using an analytical balance at 0.1\,mg precision after being cleaned with isopropyl alcohol. The characterization of the approximately 2.25 cm diameter arsenic depositions onto titanium, pictured in Figure \ref{AsFoilElectroplated}, was a more intensive process involving particle transmission and neutron activation analysis. These details and the description of the associated electroplating creation process are given in \textcite{Fox2021:Targets}, while the resulting thickness and areal density values can be seen in Table \ref{LBNLStack}.
%The creation and characterization of the approximately 2.25 cm diameter arsenic depositions onto titanium, pictured in Figure \ref{AsFoilElectroplated}, is described in detail by [VOYLES TARGETRY PAPER XXXX]. 

\begin{figure}[H]
\centering{{\includegraphics[width=0.75\columnwidth]{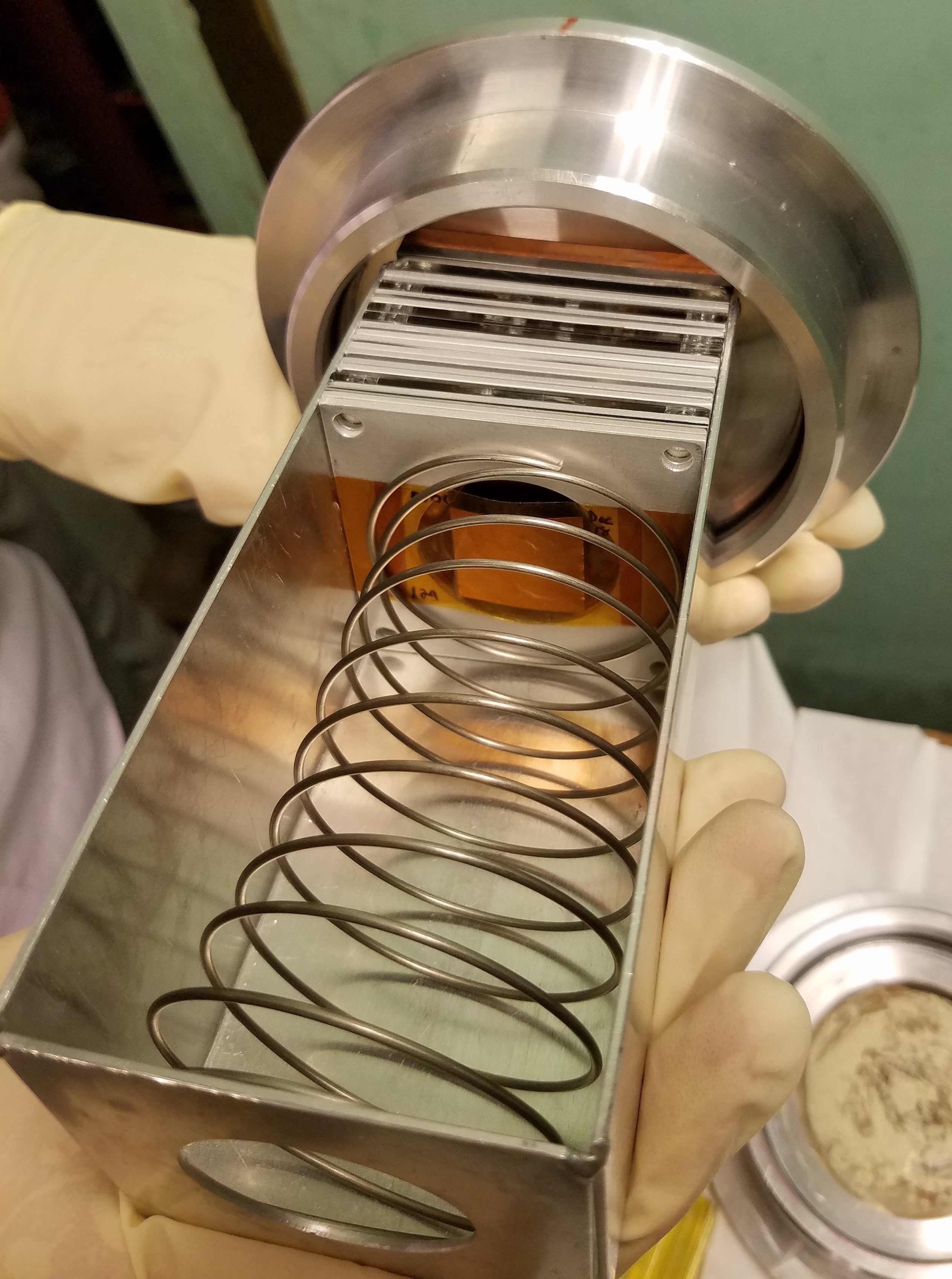}}}
\caption{A top view of the assembled LBNL target stack prior to loading into the cyclotron beam pipe. The beam is first incident on the front facing copper target shown in the photo, as described in Table \ref{LBNLStack}.}%showing the ten target ``compartments" separated by aluminum degraders.
\label{LBNLTargetBox}
\end{figure}

All targets were then sealed using DuPont Kapton polyimide film tape of either 43.2\,\mmicro m of silicone adhesive on 25.4\,\mmicro m of polyimide backing (total nominal 7.77\,mg/cm$^2$) or 43.2\,\mmicro m of silicone adhesive on 50.8\,\mmicro m of polyimide backing (total nominal 11.89\,mg/cm$^2$). The encapsulated foils were mounted to the center of hollow 5.7\,cm $\times$ 5.7\,cm aluminum frames. The frames protected the foils during handling and centered them in the beam pipe after the stack was fully arranged in the target box seen in Figure \ref{LBNLTargetBox}.
%5.715\,cm $\times$ 5.715\,cm

Multiple aluminum degraders were characterized in the same manner as the copper foils and included in the stack to yield nine different beam energy ``compartments" for cross section measurements. One copper foil and one electroplated arsenic foil were placed into each of the nine compartments in the target box. Stainless steel plates (approximately 100 mg/cm$^2$) were placed near the front and back of the stack for post-irradiation dose mapping using radiochromic film (Gafchromic EBT3) in order to examine the spatial profile of the beam entering and exiting the stack. The full detailed target stack ordering and properties for the LBNL irradiation are given in Table \ref{LBNLStack}. 

The stack was irradiated at the 88-Inch Cyclotron for 3884 seconds with a nominal 192\,nA H$^+$ beam. The total collected charge of the beam was measured using a current integrator connected to the electrically-isolated target holder, which was used to determine that the beam current was stable over the duration of the experiment. The mean beam energy extracted was 55.4 MeV with an approximately 1\% uncertainty.

\begin{table}[!t]
\caption{\footnotesize{Target stack design for irradiation at the 88-Inch Cyclotron. The proton beam initially hits the Cu-SN1 target and is subsequently transported through the rest of the shown stack order. The thickness and areal density measurements are prior to any application of the variance minimization techniques described in this work.}}\label{LBNLStack}
\begin{ruledtabular}
\begin{tabular}{lP{1.3cm}P{1.8cm}P{1.8cm}}
\multirow{4}{*}{Target Layer}&\multirow{4}{=}{\centering Thickness [$\mmicro$m]}&\multirow{4}{=}{\centering Areal Density [mg/cm$^2$]}&\multirow{4}{=}{\centering Areal Density Uncertainty [\%]}\\ \\ \\ \\[0.1cm]
\hline\\[-0.22cm]
Cu-SN1 &24.81 &22.23  &0.33  \\[0.1cm]

As-SN1 &3.24  &1.85 &9.8  \\[0.1cm]

Ti-SN1 &25.00 &11.265 &1.0 \\[0.1cm]

SS Profile Monitor &130.0 &100.12&0.07 \\[0.1cm]

Al Degrader E1 &253.0 &68.31 &0.10 \\[0.1cm]

Al Degrader E2 &252.7&68.24 &0.10  \\[0.1cm]

Cu-SN2 &24.88 &22.29  &0.08  \\[0.1cm]

As-SN2 &1.69   &0.97 &9.9 \\[0.1cm]

Ti-SN2 &25.00 &11.265 &1.0  \\[0.1cm]

Al Degrader D1 &674.2 &174.44 &0.05 \\[0.1cm]

Cu-SN3 &24.88  &22.29 &0.06 \\[0.1cm]

As-SN3 &1.81   &1.04 &9.9 \\[0.1cm]

Ti-SN3 &25.00  &11.265 &1.0  \\[0.1cm]

Al Degrader D2 &664.5  &174.87 &0.06 \\[0.1cm]

Cu-SN4 &24.87  &22.28 &0.04 \\[0.1cm]

As-SN4 &2.22   &1.27 &10 \\[0.1cm]

Ti-SN4 &25.00  &11.265 &1.0  \\[0.1cm]

Al Degrader E3 &253.1  &68.35 &0.10 \\[0.1cm]

Cu-SN5 &24.97  &22.37 &0.06 \\[0.1cm]

As-SN5 &1.95  &1.12 &9.9 \\[0.1cm]

Ti-SN5 &25.00  &11.265 &1.0  \\[0.1cm]

Al Degrader F1 &181.5 &46.91 &0.12 \\[0.1cm]

Al Degrader F2 &192.2 &48.97 &0.14  \\[0.1cm]

Cu-SN6 &24.85  &22.27 &0.09 \\[0.1cm]

As-SN6 &1.30   &0.74 &11 \\[0.1cm]

Ti-SN6 &25.00  &11.265 &1.0  \\[0.1cm]

Al Degrader E4 &252.9  &68.29 &0.10 \\[0.1cm]

Cu-SN7 &24.67  &22.11 &0.39 \\[0.1cm]

As-SN7 &2.36   &1.35 &8.9 \\[0.1cm]

Ti-SN7 &10.00  &4.506 &1.0  \\[0.1cm]

Al Degrader C1 &970.0  &261.48 &0.03 \\[0.1cm]

Cu-SN8 &24.80 &22.22 &0.06 \\[0.1cm]

As-SN8&0.94  &0.54 &9.7\\[0.1cm]

Ti-SN8 &25.00  &11.265 &1.0  \\[0.1cm]

Al Degrader E5 &252.7  &68.24 &0.10 \\[0.1cm]

Cu-SN9 &24.90  &22.31 &0.10 \\[0.1cm]

As-SN9 &0.57  &0.32 &10 \\[0.1cm]

Ti-SN9 &25.00 &11.265 &1.0  \\[0.1cm]

SS Profile Monitor &130.0 &100.48&0.07 \\[0.1cm]
\end{tabular}
\end{ruledtabular}
\end{table}

\subsubsection{\label{2.2}LANL Stack and Irradiation}
The LANL stack included copper, niobium, aluminum, and electroplated arsenic targets. The stack composition is described in detail in \textcite{Fox2020:NbLa}, where characterization procedures were very similar to the LBNL setup. A summary of the stack is provided in this paper in Table \ref{LANLStack} (see Appendix \ref{Appendix_OtherStacks}). The stack was irradiated for 7203 seconds with an H$^+$ beam of 100\,nA nominal current. The beam current, measured using an inductive pickup, remained stable under these conditions for the duration of the irradiation. The mean beam energy extracted was 100.16 MeV with an approximately 0.1\% uncertainty.

\subsubsection{\label{2.3}BNL Stack and Irradiation}
The BNL stack was composed of copper, niobium, and electroplated arsenic targets. The exact specifications of the stack are given in \textcite{Fox2020:NbLa} and a summary can be seen in Table \ref{BNLStack} (see Appendix \ref{Appendix_OtherStacks}). The stack was irradiated for 3609 seconds with an H$^+$ beam of 200 nA nominal current. The beam current during operation was recorded using toroidal beam transformers and shown to remain stable under these conditions for the duration of the irradiation. The mean beam energy extracted was 200 MeV with an approximately 0.2\% uncertainty \cite{Degraffenreid2019:BNLAs}.

\subsection{\label{}Gamma Spectroscopy and Measurement of Foil Activities}

\subsubsection{\label{}LBNL}
The gamma spectroscopy at the 88-Inch Cyclotron utilized an ORTEC GMX series (model GMX-50220-S) High-Purity Germanium (HPGe) detector and seven ORTEC IDM-200-VTM HPGe detectors. The GMX is a nitrogen-cooled coaxial n-type HPGe with a 0.5\,mm beryllium window, and a 64.9\,mm diameter, 57.8\,mm long crystal. The IDMs are mechanically-cooled coaxial p-type HPGes with single, large-area 85\,mm diameter $\times$ 30\,mm length crystals and built-in spectroscopy electronics. The energy and absolute photopeak efficiency of the GMX and IDMs were calibrated using standard $^{133}$Ba, $^{137}$Cs, and $^{152}$Eu sources. The efficiency model used in this work is the physical model presented by \textcite{Gallagher1974:EffModel}.

Foil activity data was first collected from counts beginning approximately 45 minutes after the end-of-bombardment (EoB) and removal of the target stack from the beamline. The copper and electroplated arsenic foils were initially cycled through multiple 5--30 minute counts on the GMX during the 24 hours immediately following the irradiation. The counting distances from the GMX detector face were varied from 80 cm to 15 cm during this period subject to dead time constraints. Each electroplated arsenic foil was then transferred to an individual IDM detector where counts were collected in 1 hour intervals at a 10 cm distance from the IDM face over the next three weeks. The repeated counts of each foil helped to establish consistent decay curves for residual nuclides and reduce uncertainty in the spectroscopy analysis, particularly aiding in the determination of longer-lived products. Final 12--24 hour counts for the copper foils were captured on the GMX near the end of the three week period to record appropriate statistics for long-lived monitor channels.

The radiochromic film, developed by the stainless steel plates, showed that an $\approx$1\,cm diameter proton beam was centered on the stack foils and properly inscribed within the size-limiting borders of the arsenic deposits throughout the stack.

\subsubsection{\label{}LANL}
The LANL experiment used a series of GEM and IDM HPGe detectors. The foil counting at LANL followed a similar cycling routine to LBNL, with counting times ranging from 10 minutes during the first hours after EoB to upwards of 8 hours over the course of 6 weeks after the irradiation for the stack's 40 total targets. The LANL counting scheme is given explicitly in \textcite{Fox2020:NbLa}. Notably, the electroplated arsenic targets of the LANL stack were shipped to LBNL in order to perform multi-week long counts with the LBNL GMX to better capture the $^{68}$Ge signal, which remained weak in the longest of the LANL counts.

\subsubsection{\label{}BNL}
The BNL gamma spectroscopy setup incorporated two EURISYS MESURES 2 Fold Segmented ``Clover" detectors in addition to one GMX and two GEM detectors. Foils were cycled in front of the many detectors for repeated short counts of 30 minutes or less during the first 24 hours after EoB. Data collection at BNL continued with multi-hour target counts for an additional day before the targets were shipped back to LBNL, arriving within two weeks after EoB. The LBNL GMX was used for multi-day to week-long counts of the copper, electroplated arsenic, and niobium foils over the course of the next 2+ months.

Further details of the BLIP activation and spectroscopy is provided in \textcite{Fox2020:NbLa}.

\subsubsection{\label{ActivationAnalysis}Activation Analysis}
\begin{figure*}[t]
{\includegraphics[width=1.0\textwidth]{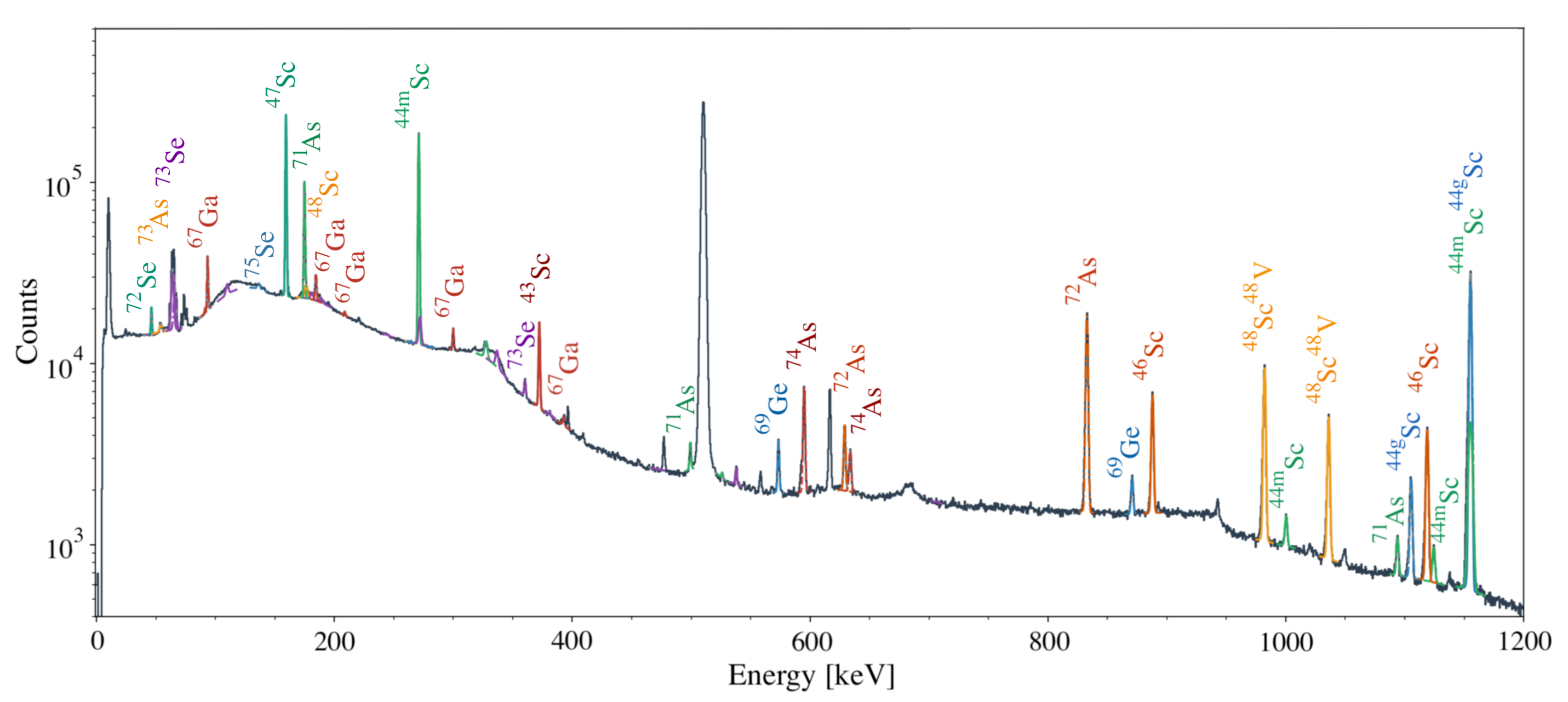}}
%\vspace{-0.55cm}
\caption{Example gamma-ray spectrum from the induced activation of an electroplated arsenic target in the LANL stack at approximately $E_p=91$\,MeV. The spectrum was taken slightly beyond 2 days after EoB and the smooth fits to the peaks of interest shown are produced by the Curie package \cite{Morrell2019}.}
\label{As_spectrum_LANL_Countroom}
\end{figure*}

The UC Berkeley code package Curie \cite{MorrellCurie} was used to analyze the collected gamma spectra from each irradiation. Decay curves for observed residual products were constructed from the count data with appropriate timing, efficiency, and attenuation corrections. EoB activities $A_0$ were then determined by fitting decay curves to the applicable Bateman equations \cite{Morrell2020,Fox2020:NbLa,Voyles2018:Nb}. A sample gamma-ray spectrum from an electroplated arsenic target is given in Figure \ref{As_spectrum_LANL_Countroom}.

Independent, $(i)$, $A_0$ results were determined from decay curve fits where decay contributions from any precursors of a residual product could be distinguished or where no parent decay in-feeding existed. In cases where precursor contributions could not be distinguished, either due to timing or decay property limitations, cumulative, $(c)$, $A_0$ values for a residual product within a decay chain were instead calculated.

The total uncertainties in the determined EoB activities had contributions from fitted peak areas, evaluated half-lives and gamma intensities, regression parameters, and detector efficiency calibrations. Each contribution to the total uncertainty was assumed to be independent and was added in quadrature. The impact of calculated $A_0$ uncertainties on final cross section results is detailed in Section \ref{xscalc}.

\subsection{\label{}Stack Current and Energy Properties}
The proton beam energy and current at each target in a given stack was determined by monitor foil activation data, Curie's Andersen \& Ziegler-based Monte Carlo particle transport code, and a ``variance minimization" approach, following the established methodology presented in \textcite{Voyles2018:Nb,Morrell2020,Graves2016:StackTarget}.

\begin{figure}[H]
{\includegraphics[width=1.0\columnwidth]{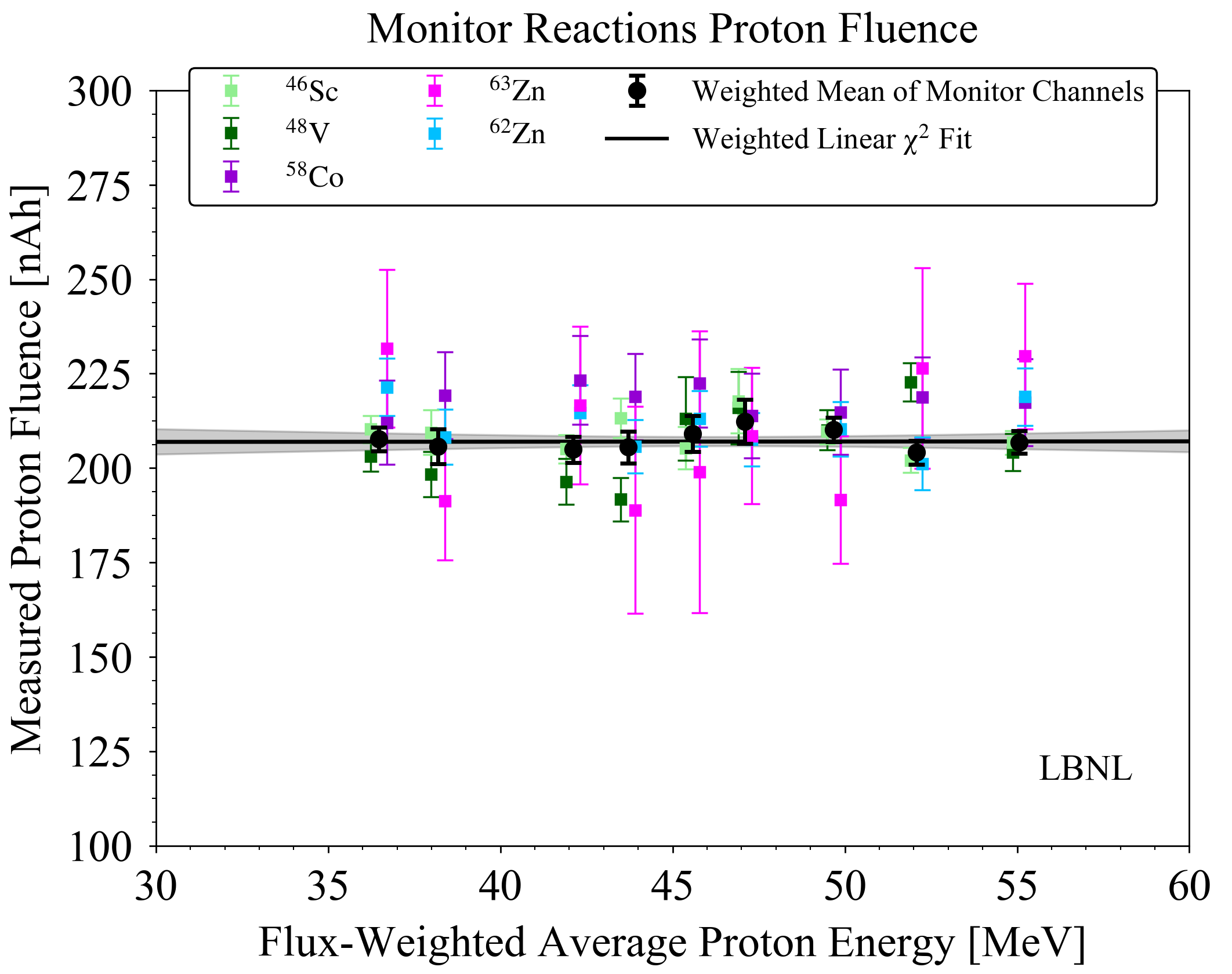}}
\caption{Plot of the proton beam fluence measured by monitor reactions in the LBNL stack following adjustments made by the variance minimization technique.}
\label{CovFluenceLBNL}
\end{figure}

The $^{\mathrm{nat}}$Ti(p,x)$^{48}$V, $^{46}$Sc and $^{\mathrm{nat}}$Cu(p,x)$^{63,62}$Zn, $^{58}$Co monitor reactions, taken from the IAEA-recommended data reference for charged-particle reactions \cite{IAEARefCrossSections}, were used for the LBNL beam characterization. The results after variance minimization are shown in Figure \ref{CovFluenceLBNL} with plotted weighted averages of all the monitor reaction fluence predictions in each stack compartment. The weighted averages account for data and measurement correlations between the monitor reaction channels at each position in the stack and were used to create the uncertainty-weighted linear fit, also included in Figure \ref{CovFluenceLBNL} \cite{Voyles2021:Fe}. The fit is a global model applied due to the observed flat fluence depletion and provides an interpolation for the fluence and energy of each individual target of interest in the stack. This optimized linear model after variance minimization shows an approximately constant 207\,nAh fluence throughout the LBNL stack.%%%%%dp=1.0423

Further details of the monitor foil calculations, variance minimization approach, and energy determinations for the LBNL experiment can be reviewed in Appendix \ref{Appendix_VarMin}. An in-depth discussion of this same beam characterization procedure for the LANL and BNL stacks is provided in \textcite{Fox2020:NbLa}. Recall that this work and \textcite{Fox2020:NbLa} are outcomes of the same set of target stacks and irradiations meaning that the LANL and BNL fluence results and energy assignments from \textcite{Fox2020:NbLa} are identically applied here.

The final deduced energy assignments, with associated uncertainties, for targets in all three stacks are provided in Tables \ref{As_xsValues}, \ref{Cu_xsValues}, and \ref{Ti_xsValues}.

\subsection{\label{xscalc}Cross Section Determination}
Cross sections for observed products in this work were calculated from the typical activation formula,
\begin{gather}\label{CrossSectionCalc}
\sigma=\frac{A_0}{I_p(\rho_N\Delta r)(1-e^{-\lambda t_{irr}})},
\end{gather}
where $I_p$ is the beam current in protons per second at a given foil in a stack, $\rho_N\Delta r$ is the relevant foil's areal number density, $\lambda$ is the decay constant for the observed residual product of interest, and $t_{irr}$ is the beam-on irradiation time.

Measured $^{75}$As(p,x) cross sections are reported in Table \ref{As_xsValues} for $^{75,73,72}$Se, $^{74-70}$As, $^{72,68-66}$Ga, $^{69,68,66}$Ge, $^{69\textnormal{m},65}$Zn, and $^{60,58-56}$Co.
% $^{75}$Se, $^{74}$As, $^{73}$Se, $^{73}$As, $^{72}$Se, $^{72}$As, $^{72}$Ga, $^{71}$As, $^{70}$As, $^{69}$Ge, $^{69\textnormal{m}}$Zn, $^{68}$Ge, $^{68}$Ga, $^{67}$Ga, $^{66}$Ge, $^{66}$Ga, $^{65}$Zn, $^{60}$Co, $^{58}$Co, $^{57}$Co, and $^{56}$Co.

$^{\mathrm{nat}}$Cu(p,x) production cross sections for $^{65,63,62}$Zn, $^{64,61,60}$Cu, $^{60,57-55}$Co, $^{59}$Fe, $^{57,56}$Ni, $^{56,54,52}$Mn, $^{51,49,48}$Cr, $^{48}$V, and $^{47,46,44\textnormal{m}}$Sc are given in Table~\ref{Cu_xsValues}.
%$^{65}$Zn, $^{64}$Cu, $^{63}$Zn, $^{62}$Zn, $^{61}$Cu, $^{60}$Cu, $^{60}$Co, $^{59}$Fe, $^{57}$Ni, $^{57}$Co, $^{56}$Ni, $^{56}$Mn, $^{56}$Co, $^{55}$Co, $^{54}$Mn, $^{52}$Mn, $^{51}$Cr, $^{49}$Cr, $^{48}$Cr, $^{48}$V, $^{47}$Sc, $^{46}$Sc, and $^{44\textnormal{m}}$Sc

$^{\mathrm{nat}}$Ti(p,x) experimental cross section results for $^{48}$V, $^{48-46,44\textnormal{m},44\textnormal{g},43}$Sc, $^{47}$Ca, $^{44}$Ti, and $^{43,42}$K are listed in Table \ref{Ti_xsValues}.
%$^{48}$V, $^{48}$Sc, $^{47}$Sc, $^{47}$Ca, $^{46}$Sc, $^{44\textnormal{m}}$Sc, $^{44\textnormal{g}}$Sc, $^{44}$Ti, $^{43}$Sc, $^{43}$K, and $^{42}$K

In Tables \ref{As_xsValues} \ref{Cu_xsValues}, and \ref{Ti_xsValues}, the cross sections for residual products are marked as either independent, $(i)$, or cumulative, $(c)$, referencing the distinction discussed in Section \ref{ActivationAnalysis} surrounding decay chains.

The final uncertainty contributions to the cross section measurements include uncertainties in evaluated decay constants (0.02--1.0\%), foil areal density measurements (0.05--11\%), proton current determination calculated from monitor fluence measurements and variance minimization (1.1--3.4\%), and $A_0$ quantification that accounts for efficiency uncertainty in addition to other factors listed in Section \ref{ActivationAnalysis} (1.5--14\%). These contributions were added in quadrature to give uncertainty in the final cross section results at the 3.5--17\% level.

\section{\label{Results}Results and Discussion}
The measured data from select reactions of particular interest to the medical applications community or for nuclear reaction modeling purposes are discussed in detail below. Plots of all other reported cross sections are given in Appendix \ref{Appendix_Plots} (Figures \ref{Ti_42K}--\ref{As_75SE}).

The experimentally extracted cross sections are compared with the predictions of nuclear reaction modeling codes TALYS-1.95 \cite{TalysManual}, CoH-3.5.3 \cite{Kawano2019:CoHoverview}, EMPIRE-3.2.3 \cite{EmpireOverview}, and ALICE-20 \cite{Blann1982}, each using default settings and parameters. A discussion of these default conditions and assumptions is provided in \textcite{Fox2020:NbLa}. Comparisons with the TENDL-2019 library \cite{KoningRochman2012:MethodTALYSEval} are also made. Additionally, the cross section measurements in this work are compared to the existing body of literature data, retrieved from EXFOR \cite{Degraffenreid2019:BNLAs,Mushtaq1988:ProtonsAs,Qaim1988:73SeIsomer,Brodovitch1976:ProtonsPEM,Levkovski1991:MiddleMass,
Cohen1953,Neumann1999:Thesis,Garrido2016:ProtonsTiNiCu,Brodzinki1971:Ti,Michel1985:Ti,Fink1990:Ti,Michel1978:ProtonsTi,
Michel1997:ProtonsTiCuNb,Khandaker2009:ProtonsTi,Zarie2006:Ti,Cervenak2020:Ti,Kopecky1993:Ti,Bringas2005:Ti,Hermanne2014:Ti,
Dittrich1988,Tarkanyi1991,Walton1976,Takacs2002:TiCu,Mills1992:ProtonsCu,Graves2016:StackTarget,Voyles2018:Nb,
Grutter1982:ProtonsCuAl,Heydegger1972,Orth1978,Yashima2003:Cu,Greenwood1984,Williams1967:ProtonsAlFeCu,Shahid2015:ProtonsCu,
Morrell2020,Uddin2004,Hermanne1999,Kopecky1985,Buthelezi2006,Khandaker2007,Yoshizawa1976,AlSaleh2006,Titarenko2003,
Aleksandrov1987,Kuhnhenn2001,Jost2013}.

%, to initially explore variations between the codes and their sensitivity to pre-equilibrium reaction dy- namics. Where measured cumulative cross sections are plotted, the corresponding code calculations shown also include the necessary parent production to estimate cu- mulative yields. Note that ALICE-20 is not suited to calculate independent isomer or ground state production due to a lack of detailed angular momentum modeling. Comparisons with the TENDL-2019 library [21] are also made.

\begin{table*}
\footnotesize
\caption{Summary of arsenic cross sections measured in this work. Subscripts $(i)$ and $(c)$ indicate independent and cumulative cross sections, respectively. Uncertainties are listed in the least significant digit, that is, 49.5 (14)\,MeV means 49.5 $\pm$ 1.4\,MeV. Stack ID specifies which irradiation each measurement belongs to - Stack ``BR" designates the Brookhaven irradiation, Stack ``LA" designates the Los Alamos irradiation, and Stack ``LB" designates the Lawrence Berkeley irradiation.} 
\label{As_xsValues}
\begin{ruledtabular}
\begin{tabular}{L{1.8cm}P{1.61cm}P{1.61cm}P{1.61cm}P{1.61cm}P{1.61cm}P{1.61cm}P{1.61cm}P{1.61cm}P{1.61cm}}
%\hlineB{2.5}\\[-0.22cm]
\multicolumn{10}{c}{\bfseries $^{75}$As(p,x) Production Cross Sections [mb]}\\[+0.1cm]
%\hlineB{2.5}\\[-0.22cm]
\hline\\[-0.22cm]
$E\mathrm{_p}$ [MeV] & 192.28 (49) & 177.01 (51) & 163.21 (54) & 148.55 (58) & 133.75 (62) & 119.66 (67) & 104.09 (73) & 91.09 (51) & 79.19 (56)\\[+0.1cm]
Stack ID & BR & BR & BR & BR & BR & BR & BR & LA & LA\\[+0.1cm]
\hline\\[-0.22cm]

$^{56}$Co$_{(c)}$ & 0.823 (98) & 0.337 (34) & 0.581 (64) & 0.436 (48) & 0.169 (28) & - & - & - & - \\[0.1cm]

$^{57}$Co$_{(c)}$ & 3.04 (46) & 1.36 (18) & 2.03 (28) & 1.68 (25) & 0.51 (17) & - & - & - & - \\[0.1cm]

$^{58}$Co$_{(i)}$  & 3.62 (80) & - & 2.81 (33) & 2.25 (26) & 0.84 (11) & 0.32 (24) & 0.07 (8) & - & - \\[0.1cm]

$^{60}$Co$_{(i)}$  & 8.8 (11) & 1.89 (20) & 1.70 (20) & 1.06 (14) & - & - & - & - & - \\[0.1cm]

$^{65}$Zn$_{(c)}$ & 45.8 (77) & 47.6 (58) & 47.4 (63) & 35.1 (43) & 29.2 (38) & 31.4 (38) & 10.8 (15) & - & - \\[0.1cm]

$^{66}$Ga$_{(c)}$ & 11.1 (66) & 24.9 (66) & 31 (17) & 24.1 (98) & 16.0 (42) & - & 14.6 (39) & 5.43 (89) & 5.33 (95) \\[0.1cm]

$^{66}$Ge$_{(c)}$ & - & - & - & 1.15 (49) & 1.18 (22) & - & - & - & - \\[0.1cm]

$^{67}$Ga$_{(c)}$ & 39.1 (46) & 44.8 (45) & 43.2 (47) & 42.1 (43) & 38.6 (49) & 36.7 (36) & 35.0 (39) & 20.6 (19) & 25.5 (24) \\[0.1cm]

$^{68}$Ga$_{(i)}$ & 41.7 (83) & 39.2 (62) & 41.3 (58) & 40.7 (69) & 35.5 (53) & 42.8 (55) & 39.5 (52) & - & - \\[0.1cm]

$^{68}$Ge$_{(c)}$ & 30.7 (46) & 26.9 (30) & 26.4 (32) & 22.8 (27) & 21.9 (30) & 20.3 (23) & 13.0 (16) & 11.1 (22) & 24.1 (41) \\[0.1cm]

$^{69\textnormal{m}}$Zn$_{(i)}$ & 1.24 (19) & 1.38 (22) & 1.38 (17) & 1.26 (14) & 1.02 (24) & 1.29 (13) & 0.75 (13) & - & - \\[0.1cm]

$^{69}$Ge$_{(c)}$ & 36.9 (43) & 40.5 (43) & 41.6 (50) & 37.0 (42) & 36.9 (49) & 42.5 (44) & 35.0 (39) & 19.8 (20) & 16.2 (16) \\[0.1cm]

$^{70}$As$_{(c)}$ & 15.9 (18) & 16.4 (17) & 17.7 (19) & 16.4 (17) & 17.2 (21) & 23.2 (23) & 27.1 (28) & 36.9 (39) & 43.7 (45) \\[0.1cm]

$^{71}$As$_{(c)}$ & 40.0 (45) & 49.2 (51) & 55.2 (64) & 55.8 (60) & 64.3 (79) & 76.2 (76) & 73.4 (75) & - & 91.8 (85) \\[0.1cm]

$^{72}$Ga$_{(c)}$ & - & - & - & 1.39 (57) & 3.07 (95) & 1.89 (68) & 3.38 (82) & 2.20 (29) & 2.31 (49) \\[0.1cm]

$^{72}$As$_{(i)}$ & 70.3 (77) & 82.6 (82) & 80.3 (90) & 89.2 (94) & 97 (12) & 122 (12) & 116 (12) & - & 108.8 (99) \\[0.1cm]

$^{72}$Se$_{(i)}$ & 6.12 (72) & 6.90 (75) & 8.12 (94) & 8.09 (89) & 8.4 (11) & 11.2 (12) & 11.6 (13) & - & 15.2 (16) \\[0.1cm]

$^{73}$As$_{(i)}$ & 95 (17) & 125 (19) & 138 (24) & 128 (24) & 138 (26) & 166 (28) & 172 (31) & 180 (42) & 174 (24)\\[0.1cm]

$^{73}$Se$_{(c)}$ & 11.9 (15) & 14.0 (16) & 14.8 (17) & 15.6 (18) & 18.0 (24) & 23.0 (25) & 23.5 (27) & 22.8 (29) & 25.7 (35) \\[0.1cm]

$^{74}$As$_{(i)}$ & 98 (11) & 112 (12) & 113 (16) & 118 (14) & 124 (18) & 138 (14) & 148 (18) & - & 123 (12) \\[0.1cm]

$^{75}$Se$_{(i)}$ & 5.55 (59) & 6.65 (63) & 7.47 (79) & 6.80 (69) & 7.44 (89) & 9.23 (88) & 9.48 (95) & 6.08 (52) & 10.10 (87) \\[0.1cm]

\hline\\[-0.22cm]
$E\mathrm{_p}$ [MeV] & 72.39 (60) & 67.00 (64) & 62.92 (67) & 59.93 (69) & 57.31 (72) & 55.42 (74) & 54.9 (13) & 53.46 (76) & 52.0 (14)\\[+0.1cm]
Stack ID & LA & LA & LA & LA & LA & LA & LB & LA & LB\\[+0.1cm]
\hline\\[-0.22cm]

$^{66}$Ga$_{(c)}$ & 2.88 (64) & - & - & - & - & - & - & - & - \\[0.1cm]

$^{67}$Ga$_{(c)}$ & 16.4 (18) & 6.2 (10) & 2.32 (77) & 1.00 (78) & 0.91 (74) & - & - & - & - \\[0.1cm]

$^{68}$Ge$_{(c)}$ & 41.4 (72) & 39.2 (69) & 31.1 (54) & 14.1 (20) & - & - & - & - & - \\[0.1cm]

$^{69}$Ge$_{(c)}$ & 17.6 (19) & 20.6 (22) & 25.6 (26) & 34.5 (40) & 39.4 (42) & 37.4 (40) & 41.5 (44) & 39.6 (45) & 35.8 (39) \\[0.1cm]

$^{70}$As$_{(c)}$ & 33.1 (40) & - & 2.3 (10) & - & - & - & - & 2.3 (16) & - \\[0.1cm]

$^{71}$As$_{(c)}$ & 131 (13) & 143 (14) & 130 (12) & 128 (14) & 103 (11) & 74.9 (77) & 63.9 (65) & 53.6 (61) & 32.3 (34) \\[0.1cm]

$^{72}$Ga$_{(c)}$ & 1.72 (51) & 2.25 (47) & - & 1.26 (47) & - & 1.31 (48) & - & 1.03 (34) & - \\[0.1cm]

$^{72}$As$_{(i)}$ & 146 (14) & 169 (17) & 188 (18) & 238 (26) & 262 (26) & 249 (24) & 277 (28) & 266 (28) & 246 (25) \\[0.1cm]

$^{72}$Se$_{(i)}$ & 23.0 (25) & 28.5 (31) & 34.2 (36) & 49.3 (58) & 57.1 (62) & 57.9 (62) & 59.8 (63) & 62.7 (71) & 80 (12)  \\[0.1cm]

$^{73}$As$_{(i)}$ & 229 (32) & 244 (35) & 252 (35) & 323 (47) & 325 (47) & 282 (40) & - & 346 (60) & 320 (53) \\[0.1cm]

$^{73}$Se$_{(c)}$ & 37.4 (48)  & 39.0 (55) & 45.2 (55) & 54.2 (81) & 62.1 (82) & 57.0 (80) & 60.1 (69) & 65.4 (89)  & 65.4 (76)  \\[0.1cm]

$^{74}$As$_{(i)}$ & 153 (16) & 158 (17) & 157 (16) & 186 (21) & 185 (19) & 169 (17) & 188 (20) & 170 (19) & 182 (19) \\[0.1cm]

$^{75}$Se$_{(i)}$ & 13.2 (12) & 14.2 (13) & 14.4 (13) & 16.9 (18) & 17.7 (17) & 16.2 (15) & 15.2 (16) & 16.9 (18) &  16.1 (18) \\[0.1cm]

\hline\\[-0.22cm]
$E\mathrm{_p}$ [MeV] & 51.44 (78) & 49.5 (14) & 47.0 (15) & 45.4 (15) & 43.6 (16) & 41.9 (16) & 38.0 (17) & 36.3 (18) & \\[+0.1cm]
Stack ID & LA & LB & LB & LB & LB & LB & LB & LB & \\[+0.1cm]
\hline\\[-0.22cm]

$^{69}$Ge$_{(c)}$ & 40.6 (48) & 31.5 (34) & 27.6 (31) & 17.6 (20) & 13.0 (16) & 12.0 (12) & - & - & \\[0.1cm]

$^{71}$As$_{(c)}$ & 39.6 (48) & 17.4 (19) & 9.7 (11) & 6.44 (77) & 8.2 (11) & 3.46 (39) & - & - & \\[0.1cm]

$^{72}$Ga$_{(c)}$ & - & - & - & - & - & - & 0.21 (13) & - \\[0.1cm]

$^{72}$As$_{(i)}$ & 280 (30) & 226 (23) & 219 (22) & 207 (22) & - &  131 (12) & 73.8 (85) & 41.9 (55) & \\[0.1cm]

$^{72}$Se$_{(i)}$ & 79.3 (92) & 85 (14) & 87 (13) & 93 (10) & 72.0 (85) & 58.3 (73) & 25.4 (40) & 9.3 (14)\\[0.1cm]

$^{73}$As$_{(i)}$ & 345 (52) & 359 (65) & 469 (79) & 460 (69) & 570 (100) & 587 (85) & 680 (110) & 600 (94)\\[0.1cm]

$^{73}$Se$_{(c)}$ & 80 (12) & 69.6 (79) & 91 (10) & 92 (11) & 114 (14) & 205 (21) & 235 (26) & 307 (37)\\[0.1cm]

$^{74}$As$_{(i)}$ & 186 (22) & 181 (19) & 194 (21) & 193 (21) & - & 234 (23) & 218 (24) & 239 (27)\\[0.1cm]

$^{75}$Se$_{(i)}$ & 18.0 (19) & 17.8 (20) & 17.0 (18) & 17.2 (20) & 21.8 (35) & 23.8 (23) & 25.0 (31) & 26.5 (39)\\[0.1cm]
\end{tabular}
\end{ruledtabular}
\end{table*}

\begin{table*}
\scriptsize
\caption{Summary of copper cross sections measured in this work. Subscripts $(i)$ and $(c)$ indicate independent and cumulative cross sections, respectively. Uncertainties are listed in the least significant digit, that is, 90.94 (52)\,MeV means 90.94 $\pm$ 0.52\,MeV. Stack ID specifies which irradiation each measurement belongs to - Stack ``BR" designates the Brookhaven irradiation, Stack ``LA" designates the Los Alamos irradiation, and Stack ``LB" designates the Lawrence Berkeley irradiation.}
\label{Cu_xsValues}
\begin{ruledtabular}
\begin{tabular}{L{1.35cm}P{1.39cm}P{1.39cm}P{1.39cm}P{1.39cm}P{1.39cm}P{1.39cm}P{1.39cm}P{1.39cm}P{1.39cm}P{1.39cm}}
\multicolumn{11}{c}{\bfseries $^{\mathrm{nat}}$Cu(p,x) Production Cross Sections [mb]}\\[+0.1cm]
\hline\\[-0.20cm]
$E\mathrm{_p}$ [MeV] & 192.54 (49) & 177.28 (52) & 163.49 (54) & 148.86 (58) & 134.08 (62) & 120.02 (67) & 104.49 (74) & 90.94 (52) & 79.03 (57) & 72.22 (61) \\[+0.1cm]
Stack ID & BR & BR & BR & BR & BR & BR & BR & LA & LA & LA\\[+0.1cm]
\hline\\[-0.24cm]

$^{44m}$Sc$_{(i)}$ & 0.289 (12) & 0.1338 (63) & 0.0784 (85) & 0.0444 (40) & - & - & - & - & - & - \\[0.1cm]

$^{46}$Sc$_{(i)}$ & 0.572 (21) & 0.335 (11) & 0.2381 (65) & 0.1065 (59) & 0.0616 (30) & 0.0375 (24) & - & - & - & -\\[0.1cm]

$^{47}$Sc$_{(c)}$ & 0.261 (46) & 0.182 (31) & 0.218 (26) & - & - & - & - & - & - & -\\[0.1cm]

$^{48}$V$_{(c)}$ & 2.346 (84) & 1.560 (47) & 1.162 (30) & 0.689 (29) & 0.499 (15) & 0.298 (45) & - & - & - & -\\[0.1cm]

$^{48}$Cr$_{(c)}$ & 0.0707 (35) & 0.0437 (19) & 0.0263 (27) & 0.0207 (11) & - & - & - & - & - & -\\[0.1cm]

$^{49}$Cr$_{(c)}$ & 0.943 (67) & 0.624 (60) & 0.411 (46) & - & - & - & - & - & - & -\\[0.1cm]

$^{51}$Cr$_{(c)}$ & 11.59 (42) & 9.79 (29) & 8.44 (21) & 6.46 (26) & 5.33 (13) & 4.35 (13) & 1.676 (68) & 1.220 (61) & 0.427 (49) & 0.469 (43) \\[0.1cm]

$^{52}$Mn$_{(c)}$ & 5.34 (19) & 4.72 (14) & 4.22 (11) & 3.34 (12) & 2.733 (70) & 1.934 (59) & 1.727 (70) & 1.759 (67) & 0.509 (22) & 0.1008 (63) \\[0.1cm]

$^{54}$Mn$_{(i)}$  & 16.26 (59) & 15.72 (48) & 14.88 (38) & 13.4 (12) & 12.48 (31) & 11.05 (32) & 7.30 (27) & 6.63 (23) & 3.87 (15) & 3.86 (17) \\[0.1cm]

$^{55}$Co$_{(c)}$  & 2.04 (11) & 2.12 (11) & 1.995 (97) & 2.06 (10) & 1.813 (91) & 1.679 (90) & 1.77 (10) & 2.50 (18) & 1.43 (11) & 0.647 (60) \\[0.1cm]

$^{56}$Mn$_{(c)}$ & 2.52 (15) & 2.54 (15) & 2.46 (14) & 2.18 (13) & 2.07 (13) & 1.85 (11) & 1.40 (10) & 1.186 (57) & 1.106 (54) & 0.927 (43) \\[0.1cm]

$^{56}$Co$_{(i)}$ & 12.50 (43) & 12.65 (35) & 12.57 (29) & 13.18 (34) & 12.29 (27) & 11.55 (31) & 10.51 (37) & 10.31 (44) & 12.12 (49) & 12.68 (56) \\[0.1cm]

$^{56}$Ni$_{(c)}$  & 0.072 (59) & 0.089 (12) & 0.116 (12) & 0.105 (13) & 0.131 (15) & 0.093 (15) & - & 0.0884 (75) & 0.1103 (82) & 0.1070 (81) \\[0.1cm]

$^{57}$Co$_{(c)}$ & 43.0 (35) & 42.3 (14) & 43.1 (12) & 43.6 (11) & 44.5 (12) & 44.7 (14) & 42.2 (16) & 44.7 (14) & 37.7 (11) & 36.9 (11) \\[0.1cm]

$^{57}$Ni$_{(c)}$ & 1.687 (85) & 1.787 (66) & 1.820 (61) & 1.776 (57) & - & - & - & 1.76 (11) & 1.286 (83) & 1.391 (89) \\[0.1cm]

$^{59}$Fe$_{(c)}$  & 1.180 (51) & 1.209 (45) & 1.189 (40) & 1.100 (50) & 1.097 (36) & 1.045 (38) & 0.923 (40) & 0.931 (33) & 0.867 (29) & 0.817 (29) \\[0.1cm]

$^{60}$Co$_{(i)}$  & 11.72 (47) & 13.66 (61) & 13.73 (48) & 11.28 (55) & 12.41 (35) & 12.24 (38) & 12.01 (48) & 14.21 (42) & 12.50 (37) & 11.48 (36) \\[0.1cm]

$^{60}$Cu$_{(c)}$  & 8.01 (42) & 9.37 (48) & 10.75 (57) & 13.77 (77) & 11.4 (10) & 15.1 (14) & 16.5 (19) & 16.87 (75) & 16.0 (10) & 17.38 (90) \\[0.1cm]

$^{61}$Cu$_{(c)}$  & 29.9 (16) & 33.2 (16) & 36.4 (17) & 39.0 (17) & 42.9 (19) & 46.6 (22) & 55.7 (29) & 60.6 (30) & 54.3 (29) & 72.5 (35) \\[0.1cm]

$^{62}$Zn$_{(i)}$ & 1.71 (11) & 2.16 (13) & 1.86 (12) & 2.44 (14) & 2.39 (15) & 3.42 (19) & 3.26 (21) & - & - & -\\[0.1cm]

$^{63}$Zn$_{(i)}$ & 3.52 (34) & 4.32 (45) & 5.25 (63) & 6.05 (87) & 5.52 (97) & 5.73 (93) & - & 8.40 (52) & 10.90 (71) & 12.98 (78) \\[0.1cm]

$^{64}$Cu$_{(i)}$ & 26.3 (15) & 31.7 (18) & 30.8 (34) & 35.1 (18) & 36.6 (35) & 40.7 (22) & 44.7 (39) & 52.0 (57) & 40.4 (55) & 50.3 (51) \\[0.1cm]

$^{65}$Zn$_{(i)}$ & 1.13 (26) & 1.52 (20) & 1.61 (16) & 1.53 (11) & 1.938 (83) & 2.200 (78) & 2.69 (11) & 2.868 (95) & 3.257 (95) & 3.68 (11) \\[0.1cm]

\hline\\[-0.20cm]
$E\mathrm{_p}$ [MeV] & 66.81 (65) & 62.73 (68) & 59.73 (71) & 57.11 (73) & 55.21 (75) & 55.2 (13) & 53.24 (77) & 52.2 (14) & 51.22 (80) & 49.9 (14) \\[+0.1cm]
Stack ID & LA & LA & LA & LA & LA & LB & LA & LB & LA & LB\\[+0.1cm]
\hline\\[-0.24cm]

$^{51}$Cr$_{(c)}$ & 0.512 (37) & 0.409 (38) & 0.328 (33) & 0.278 (29) & - & - & - & - & - & -\\[0.1cm]

$^{54}$Mn$_{(i)}$ & 4.70 (17) & 4.95 (33) & 4.70 (27) & 4.10 (20)  & 3.41 (15) & 3.58 (14) & 2.65 (11) & 2.31 (13) & 1.848 (74) & 1.25 (10) \\[0.1cm]

$^{55}$Co$_{(c)}$ & 0.169 (22) & 0.077 (15) & 0.060 (20) & 0.043 (12)  & 0.0394 (92) & 0.0127 (40) & - & - & 0.0162 (69) & - \\[0.1cm]

$^{56}$Mn$_{(c)}$ & 0.644 (33) & 0.460 (25) & 0.243 (18) & 0.171 (15) & 0.161 (14)  & 0.101 (13) & 0.089 (11) & - & 0.0541 (91) & - \\[0.1cm]

$^{56}$Co$_{(i)}$ & 10.95 (46) & 7.66 (32) & 4.47 (18) & 2.405 (99) & 1.272 (62) & -  & 0.713 (39) & - & 0.373 (57) & - \\[0.1cm]

$^{56}$Ni$_{(c)}$ & 0.0837 (61) & 0.0518 (37) & 0.0330 (28) & 0.0144 (28) & 0.0082 (26) & - & 0.0076 (22) & - & 0.0043 (13) & -  \\[0.1cm]

$^{57}$Co$_{(c)}$ & 42.4 (13) & 50.0 (21) & 55.9 (23) & 59.5 (26) & 58.7 (26) & 64.6 (50) & 58.0 (25) & 55.6 (12) & 54.8 (24) & 49.9 (10) \\[0.1cm]

$^{57}$Ni$_{(c)}$ & 1.78 (11) & 2.32 (10) & 2.61 (12) & 2.73 (12) & 2.60 (12) & 2.608 (99) & 2.38 (11) & 1.942 (62) & 1.985 (90) & 1.502 (47) \\[0.1cm]

$^{59}$Fe$_{(c)}$ & 0.775 (27) & 0.690 (29) & 0.618 (26) & 0.516 (22) & 0.419 (19) & - & 0.322 (14) & - & 0.227 (10) & - \\[0.1cm]

$^{60}$Co$_{(i)}$ & 11.68 (36) & 12.22 (49) & 12.15 (47) & 11.60 (46) & 10.88 (51) & 10.34 (41) & 10.77 (49) & 10.04 (39) & 10.28 (40) & 9.53 (36)  \\[0.1cm]

$^{60}$Cu$_{(c)}$ & 18.6 (15) & 27.2 (23) & - & 26.1 (38) & 26.4 (29) & 30.1 (27) & - & 33.6 (25) & - & 29.5 (25) \\[0.1cm]

$^{61}$Cu$_{(c)}$ & 82.8 (39) & 89.7 (42) & - & 91.9 (44) & 94.2 (45) & 91.1 (42) & 93.6 (45) & 94.0 (42) & 97.5 (47) & 103.7 (45)  \\[0.1cm]

$^{63}$Zn$_{(i)}$ & 12.29 (88) & 14.0 (11) & 16.3 (13) & 17.5 (16) & 17.9 (20) & -  & - & - & - & -\\[0.1cm]

$^{64}$Cu$_{(i)}$ & 61.7 (60) & 51.4 (56) & 63.0 (62) & 66.6 (66) & 59.7 (56) & 60.7 (30) & 55.4 (59) & 56.1 (28) & 62.7 (62) & 57.3 (32) \\[0.1cm]

$^{65}$Zn$_{(i)}$ & 4.05 (11) & 4.21 (20) & 4.39 (19) & 4.66 (21) & 4.79 (24) & 4.53 (23) & 5.32 (28) & 4.65 (25) & 5.30 (26) & 5.51 (28) \\[0.1cm]

\hline\\[-0.20cm]
$E\mathrm{_p}$ [MeV] & 47.3 (15) & 45.8 (15) & 43.9 (16) & 42.3 (16) & 38.4 (17) & 36.7 (18)\\[+0.1cm]
Stack ID & LB & LB & LB & LB & LB & LB & & & & \\[+0.1cm]
\hline\\[-0.24cm]

$^{54}$Mn$_{(i)}$ & 0.533 (15) & 0.160 (43) & 0.091 (29) & 0.020 (18) & 0.076 (30) & 0.092 (28) \\[0.1cm]

$^{57}$Co$_{(c)}$ & 36.36 (68) & 29.27 (61) & 17.91 (41) & 11.09 (29) & 1.446 (96) & 0.398 (34) \\[0.1cm]

$^{57}$Ni$_{(c)}$  & 0.909 (32) & 0.634 (26) & 0.309 (19) & 0.1257 (93) & - & -  \\[0.1cm]

$^{60}$Co$_{(i)}$ & 8.78 (16) & 7.72 (31) & 7.12 (31) & 5.95 (32) & 4.95 (27) & 4.35 (24) \\[0.1cm]

$^{60}$Cu$_{(c)}$ & 19.3 (22) & 9.3 (21) & 5.5 (17) & 4.7 (18) & - & - \\[0.1cm]

$^{61}$Cu$_{(c)}$ & 112.6 (48) & 125.9 (54) & 137.7 (59) & 156.9 (67) & 179.8 (77) & 187.4 (82) \\[0.1cm]

$^{64}$Cu$_{(i)}$ & 58.1 (31) & 66.5 (33) & 59.7 (30) & 64.9 (31) & 63.1 (33) & 74.4 (36) \\[0.1cm]

$^{65}$Zn$_{(i)}$ & 5.57 (12) & 5.50 (26) & 6.19 (27) & 6.32 (29) & 6.97 (30) & 7.33 (34) \\[0.1cm]
\end{tabular}
\end{ruledtabular}
\end{table*}

\begin{table*}
\footnotesize
\caption{Summary of titanium cross sections measured in this work. Subscripts $(i)$ and $(c)$ indicate independent and cumulative cross sections, respectively. Uncertainties are listed in the least significant digit, that is, 72.34 (61)\,MeV means 72.34 $\pm$ 0.61\,MeV. Stack ID specifies which irradiation each measurement belongs to - Stack ``BR" designates the Brookhaven irradiation, Stack ``LA" designates the Los Alamos irradiation, and Stack ``LB" designates the Lawrence Berkeley irradiation.}
\label{Ti_xsValues}
\begin{ruledtabular}
\begin{tabular}{L{1.8cm}P{1.61cm}P{1.61cm}P{1.61cm}P{1.61cm}P{1.61cm}P{1.61cm}P{1.61cm}P{1.61cm}P{1.61cm}}
\multicolumn{10}{c}{\bfseries $^{\mathrm{nat}}$Ti(p,x) Production Cross Sections [mb]}\\[+0.1cm]
\hline\\[-0.22cm]
$E\mathrm{_p}$ [MeV] & 192.26 (49) & 176.99 (51) & 163.18 (54) & 148.52 (58) & 133.72 (62) & 119.63 (67) & 104.05 (74) & 91.05 (51) & 79.15 (57)\\[+0.1cm]
Stack ID & BR & BR & BR & BR & BR & BR & BR & LA & LA\\[+0.1cm]
\hline\\[-0.22cm]

$^{42}$K$_{(i)}$ & 7.54 (78) & 6.45 (70) & 6.83 (66) & 6.34 (67) & 6.92 (64) & 5.56 (62) & 6.10 (88) & 6.73 (47) & 6.48 (43) \\[0.1cm]

$^{43}$K$_{(c)}$ & 2.62 (10) & 2.493 (90) & 2.84 (11) & 2.34 (10) & 2.23 (10) & 2.116 (83) & 1.95 (13) & 1.830 (58) & 1.349 (45) \\[0.1cm]

$^{43}$Sc$_{(c)}$ & 16.5 (11) & 15.9 (11) & 12.8 (22) & 15.17 (95) & 17.1 (11) & 20.0 (14) & - & 22.8 (19) & 15.0 (16) \\[0.1cm]

$^{44\textnormal{g}}$Sc$_{(i)}$ & 25.1 (13) & 26.5 (16) & 27.9 (13) & 28.49 (97) & 28.5 (10) & 31.5 (17) & 31.7 (15) & 32.2 (19) & 39.3 (22) \\[0.1cm]

$^{44\textnormal{m}}$Sc$_{(i)}$ & 11.46 (44) & 11.88 (40) & 12.71 (37) & 13.43 (39) & 14.47 (80) & 14.82 (85) & 19.1 (16) & 21.34 (72) & 22.29 (73) \\[0.1cm]

$^{44}$Ti$_{(c)}$ & 2.7 (18) & 2.8 (11) & 3.3 (10) & 4.37 (42) & 3.3 (17) & 4.55 (49) & - & - & - \\[0.1cm]

$^{46}$Sc$_{(i)}$ & 34.0 (13) & 36.1 (12) & 38.2 (11) & 39.3 (10) & 39.3 (11) & 40.9 (13) & 41.5 (16) & 42.1 (15) & 42.3 (13) \\[0.1cm]

$^{47}$Ca$_{(c)}$ & 0.167 (22) & 0.187 (27) & 0.168 (30) & 0.158 (39) & - & - & - & - & - \\[0.1cm]

$^{47}$Sc$_{(i)}$ & 25.7 (21) & 25.84 (98) & 26.53 (87) & 26.82 (84) & 26.2 (13) & 26.70 (97) & 26.0 (28) & 23.5 (12) & 22.4 (11) \\[0.1cm]

$^{48}$Sc$_{(i)}$ & 2.31 (15) & 2.35 (16) & 1.85 (44) & 1.88 (13) & 2.53 (31) & - & 2.65 (42) & 2.45 (13) & 2.35 (13) \\[0.1cm]

$^{48}$V$_{(i)}$ & 3.62 (13) & 4.11 (13) & 4.16 (12) & 4.86 (12) & 5.60 (17) & 6.24 (20) & 7.06 (28) & - & - \\[0.1cm]

\hline\\[-0.22cm]
$E\mathrm{_p}$ [MeV] & 72.34 (61) & 66.95 (64) & 62.87 (67) & 59.88 (70) & 57.26 (72) & 55.36 (74) & 54.9 (13) & 53.40 (76) & 51.9 (14)\\[+0.1cm]
Stack ID & LA & LA & LA & LA & LA & LA & LB & LA & LB\\[+0.1cm]
\hline\\[-0.22cm]

$^{42}$K$_{(i)}$ & 6.94 (49) & 7.32 (51) & 6.57 (43) & 5.62 (37) & 4.30 (31) & 3.23 (23) & 2.86 (20) & 2.77 (22) & 1.72 (11) \\[0.1cm]

$^{43}$K$_{(c)}$ & 1.295 (46) & 1.358 (44) & 1.339 (45) & 1.425 (48) & 1.408 (48) & 1.532 (51) & 1.400 (34) & 1.439 (54) & 1.333 (28) \\[0.1cm]

$^{43}$Sc$_{(c)}$ & 15.4 (14) & 13.9 (15) & 15.2 (14) & 15.7 (17) & 17.9 (17) & 18.6 (20) & 14.22 (84) & 19.0 (17) & 15.83 (88) \\[0.1cm]

$^{44\textnormal{g}}$Sc$_{(i)}$ & 35.4 (23) & - & 30.4 (17) & 27.7 (17) & 21.3 (27) & 24.65 (78) & 21.3 (12) & 22.22 (71) & 22.2 (12) \\[0.1cm]

$^{44\textnormal{m}}$Sc$_{(i)}$ & 23.03 (78) & 21.13 (69) & 18.18 (61) & 15.97 (53) & 14.23 (47) & 13.52 (45) & 12.02 (24) & 12.79 (42) & 10.48 (22) \\[0.1cm]

$^{46}$Sc$_{(i)}$ & 44.8 (16) & 48.0 (16) & 50.0 (17) & 51.8 (21) & 53.2 (19) & 55.5 (21) & - & 55.3 (18) & - \\[0.1cm]

$^{47}$Sc$_{(i)}$ & 23.2 (11) & 23.7 (11) & 23.8 (11) & 23.9 (11) & 23.6 (11) & 23.5 (11) & 20.82 (65) & 22.7 (11) & 19.08 (64) \\[0.1cm]

$^{48}$Sc$_{(i)}$ & 2.33 (12) & 2.30 (12) & 2.28 (13) & 2.18 (15) & 2.131 (87) & 2.02 (13) & 1.649 (85) & 2.01 (12) & 1.596 (44) \\[0.1cm]

\hline\\[-0.22cm]
$E\mathrm{_p}$ [MeV] & 51.39 (79) & 49.5 (14) & 46.9 (15) & 45.4 (15) & 43.5 (16) & 41.9 (16) & 38.0 (17) & 36.2 (18) & \\[+0.1cm]
Stack ID & LA & LB & LB & LB & LB & LB & LB & LB & \\[+0.1cm]
\hline\\[-0.22cm]

$^{42}$K$_{(i)}$ & 1.67 (16) & 1.151 (90) & 0.786 (67) & 0.670 (80) & 0.571 (55) & - & 0.378 (45) & - & \\[0.1cm]

$^{43}$K$_{(c)}$ & 1.394 (52) & 1.169 (25) & 0.863 (19) & 0.645 (17) & 0.473 (12) & - & 0.1122 (65) & - & \\[0.1cm]

$^{43}$Sc$_{(c)}$ & 20.6 (22) & 16.12 (90) & 16.32 (91) & 15.80 (92) & - & 13.18 (85) & 9.38 (58) & 6.54 (42) & \\[0.1cm]

$^{44\textnormal{g}}$Sc$_{(i)}$ & 23.24 (72) & 19.4 (12) & 22.7 (15) & 22.3 (16) & 23.8 (17) & 25.1 (11) & 29.8 (11) & 33.73 (97) & \\[0.1cm]

$^{44\textnormal{m}}$Sc$_{(i)}$ & 12.86 (42) & 11.54 (26) & 12.00 (28) & 11.61 (22) & 12.16 (28) & 12.24 (26) & 15.03 (39) & 13.45 (32) & \\[0.1cm]

$^{46}$Sc$_{(i)}$ & 59.7 (21) & - & - & - & - & - & - & - & \\[0.1cm]

$^{47}$Sc$_{(i)}$ & 23.0 (11) & 20.41 (87) & 20.89 (92) & 19.87 (50) & 20.37 (72) & 19.16 (57) & 23.62 (93) & 22.35 (70) & \\[0.1cm]

$^{48}$Sc$_{(i)}$ & 2.01 (12) & 1.836 (70) & 1.809 (51) & 1.684 (91) & 1.627 (49) & 1.370 (52) & 1.296 (62) & 1.003 (70) & \\[0.1cm]

\end{tabular}
\end{ruledtabular}
\end{table*}

\subsection{\label{}{$^{75}$As(p,4n)$^{72}$Se Cross Section}}
\begin{figure}[H]
{\includegraphics[width=1.0\columnwidth]{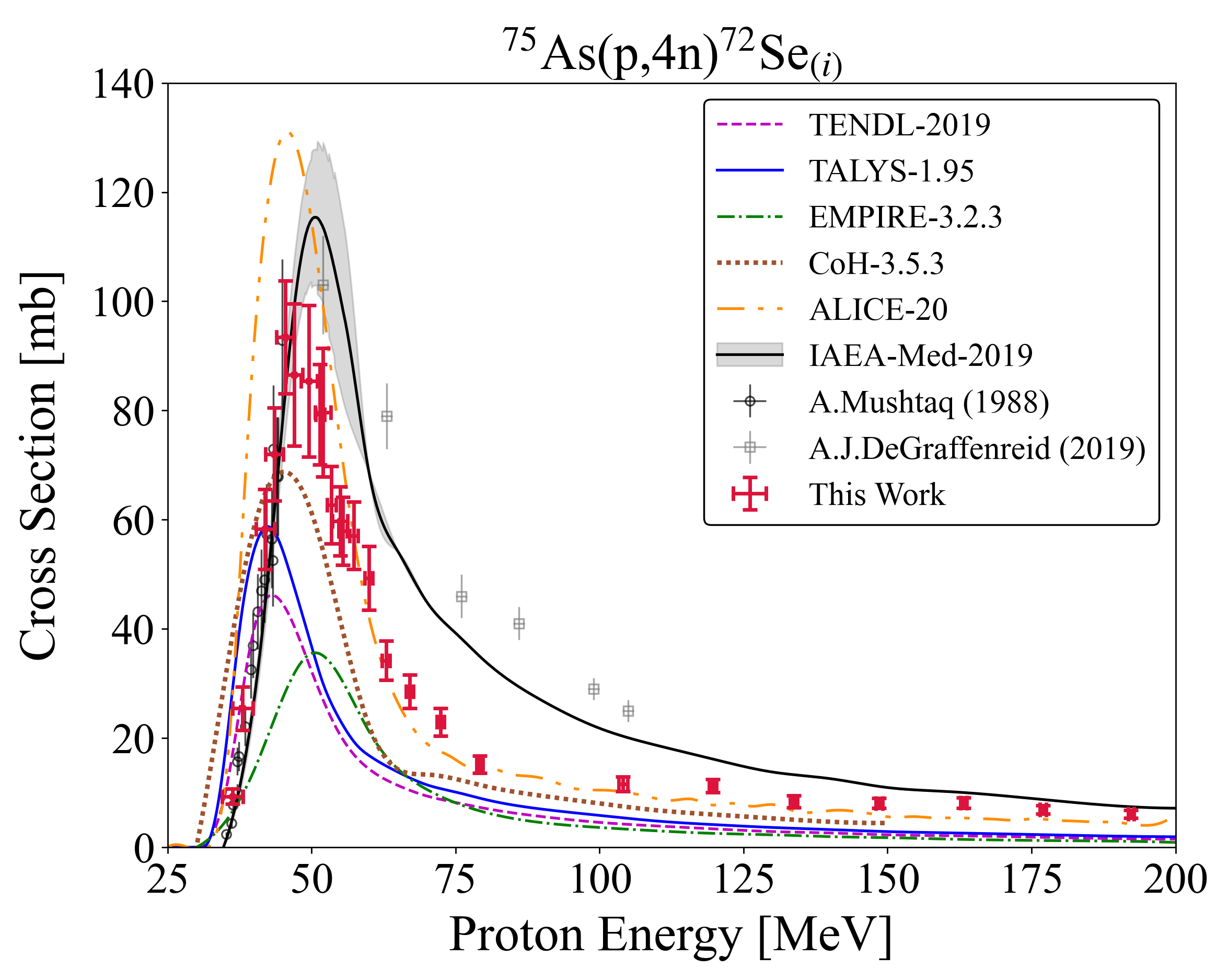}}
\vspace{-0.65cm}
\caption{Experimental and theoretical cross sections for $^{72}$Se production, peaking near 90 mb around 50\,MeV.}
\label{As_72SE}
\end{figure}

$^{72}$Se decays 100\% by electron capture to the first 1$^+$ excited state in $^{72}$As. This leaves a 45.89\,keV ($I_\gamma=57.2\ (4)$\%) $\gamma$-ray as the only direct detectable signature of $^{72}$Se formation from the HPGe equipment used in this work. However, $^{72}$Se production could additionally be quantified using the $^{72}$As decay gamma-rays after $^{72}$Se/$^{72}$As were in secular equilibrium at least 11 days after EoB. The results from each measurement method were seen to be very comparable but only the secular equilibrium values were recorded, and plotted in Figure \ref{As_72SE}, due to comparatively reduced uncertainties.

\vspace{-0.15cm}
\vspace{-0.3cm}

Only two prior experimental datasets partially measured this excitation function. The \textcite{Mushtaq1988:ProtonsAs} results cover the low energy production from threshold towards the maximum of the compound peak near 50\,MeV and agree well with the measurements of this work. The second prior experimental dataset from \textcite{Degraffenreid2019:BNLAs} covers a broader higher-energy portion of the excitation function between 52--105\,MeV. A large discrepancy exists between the \textcite{Degraffenreid2019:BNLAs} data and the values reported here. This difference is most evident for the cross section above 60\,MeV where our measurements demonstrate a much more constrained ``bell-shape" for the compound peak with a pre-equilibrium ``tail" that decreases in magnitude quicker than expressed by \textcite{Degraffenreid2019:BNLAs}. These differences are possibly partly a function of the contrasting experimental methodologies between this work and \textcite{Degraffenreid2019:BNLAs}. \textcite{Degraffenreid2019:BNLAs} did not use a stacked-target technique, but instead used multiple irradiations with thicker GaAs wafer targets, a much larger beam current, and analysis by chemical dissolution of the targets with subsequent radioassays on an HPGe using solution aliquots.

The TALYS, CoH, and ALICE reaction codes, along with the TENDL evaluation, demonstrate a similar shape though all but ALICE underpredict the compound peak cross section magnitude. Incorrect compound peak energy centroids are a pervasive error among all the calculations for this channel, generally as a function of the codes' poor threshold predictions. TENDL perhaps best matches the experimental threshold and rising edge behaviour of the excitation function but its incorrect magnitude, on account of misestimated competition with adjacent channels, muddles some of the comparison of the evaluation to the data.

In general, the variation in peak centroid location between the codes is typical and is a function of the differing pre-equilibrium calculations. Small differences between pre-equilibrium models in the codes can amplify the impact caused by particles emitted in pre-equilibrium that carry a significant amount of energy, which ultimately alter which compound nucleus is formed at a given incident energy \cite{Morrell2020}. Consequently, the improper pre-equilibrium tail modeling among TALYS, CoH, EMPIRE, and TENDL is noteworthy because it is an error that will propagate to the thresholding and rising edge behaviour in residual products that are energetically downstream of this (p,4n) channel.

Moreover, EMPIRE performs worst among the codes likely on account of these incorrect pre-equilibrium results for residual products closer in mass to the target nucleus. In this $^{72}$Se channel, the errors in EMPIRE manifest as an estimated rising edge with a much too small of a slope and the largest magnitude underprediction.

The production cross section of $^{72}$Se has also been evaluated as part of an IAEA coordinated research project (IAEA-Med-2019) focused on the recommendation of data for medical radionuclides, and in specific, diagnostic positron emitters \cite{Tarkanyi2019:MedicalIsotopesDataNeeds}. The \textcite{Degraffenreid2019:BNLAs} data were not avaialble at the time of the IAEA evaluation and though the IAEA prediction reaches a similar peak to \textcite{Degraffenreid2019:BNLAs}, which is above the peak predicted in this work, the IAEA recommendation does not support the very broad compound peak.

It is worth reflecting that these $^{72}$Se production results, i.e., the proper characterization of an excitation function from threshold to 200 MeV where little prior data existed, are emblematic of the overall TREND endeavour.

\subsection{\label{}{$^{75}$As(p,x)$^{68}$Ge Cross Section}}
The results reported here represent the first measurement of this channel. The $^{68}$Ge production cross section proved difficult to quantify in this work due to its long half-life ($t_{1/2}=270.93\ (13)$\,d \cite{DataSheetsA68}) and the lack of gamma-ray emissions. $^{68}$Ge decays 100\% by electron capture directly to the ground state of $^{68}$Ga. As a result, it was necessary to rely on the still weak, but strongest available, 1077.34\,keV ($I_\gamma=3.22\ (3)$\%) $\gamma$-ray from the decay of $^{68}$Ga to measure the $^{68}$Ge formation cross section \cite{DataSheetsA67}. $^{68}$Ga is short-lived with a 67.71 (8) min half-life and it quickly falls into secular equilibrium with $^{68}$Ge \cite{DataSheetsA68}. Therefore, all 1077.34\,keV emissions measured in the arsenic target spectra taken months after the irradiation dates were solely attributable to the decay of the initial cumulative $^{68}$Ge population. Multi-week-long counts were required to achieve reasonable statistics for the 1077.34\,keV signal.

\begin{figure}[H]
{\includegraphics[width=1.0\columnwidth]{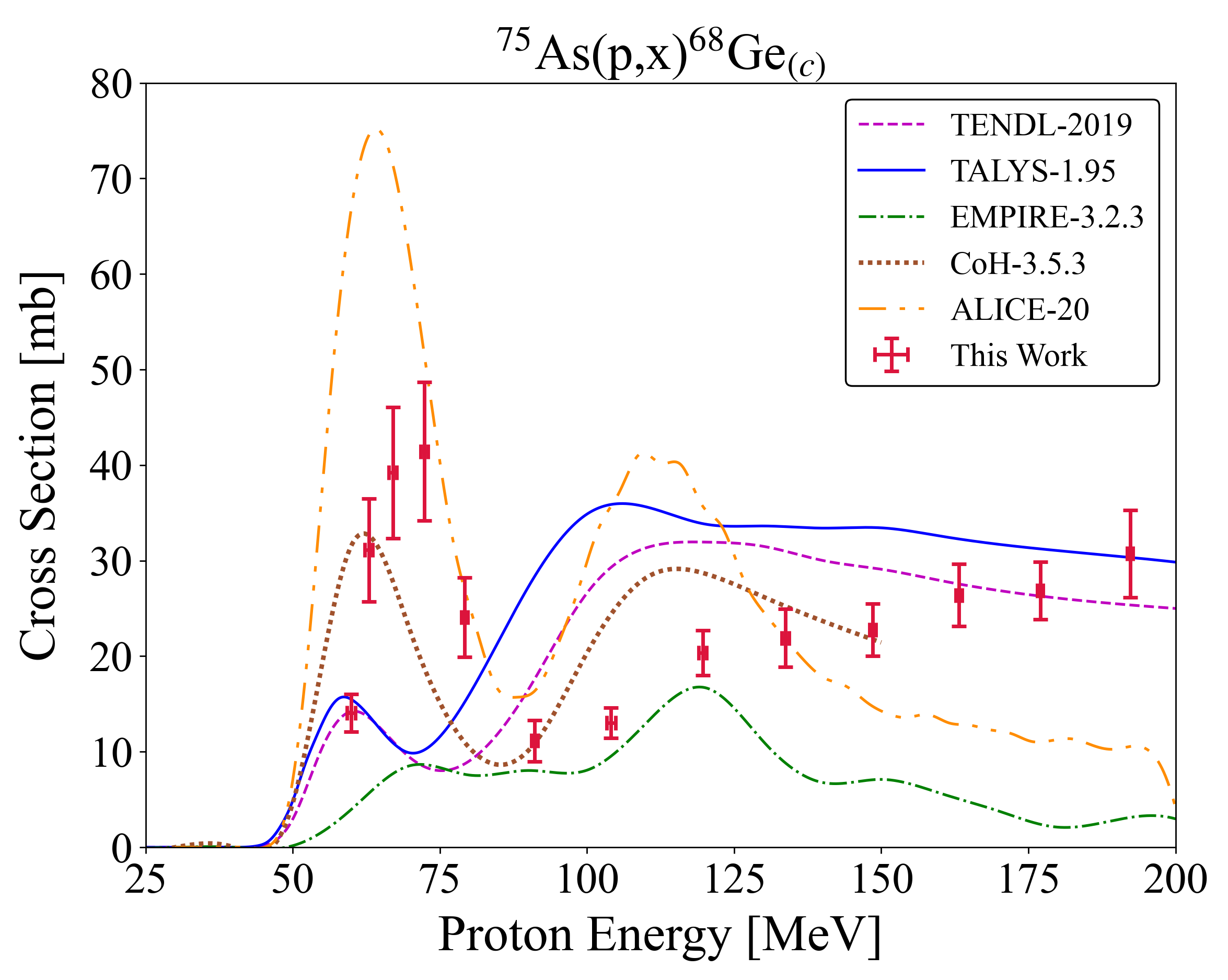}}
\vspace{-0.65cm}
\caption{Experimental and theoretical cross sections for $^{68}$Ge production, peaking near 42 mb around 72\,MeV.}
\label{As_68GE}
\end{figure}

The ensuing measured $^{75}$As(p,x)$^{68}$Ge excitation function is given in Figure \ref{As_68GE}. No cross sections were extracted from the LBNL irradiation or the rear-end of the LANL stack as the incident proton energies were below or too near threshold for measurable $^{68}$Ge production. The given excitation function in Figure \ref{As_68GE} is the first measurement of $^{68}$Ge formation from arsenic up to 200\,MeV. The excitation function shows a peak of approximately 42\,mb at 72\,MeV due to the $^{75}$As(p,$\alpha$4n)$^{68}$Ge pathway and a high-energy increasing pre-equilibrium tail from formation mechanisms where $\alpha$-particle emission is replaced by 2p2n. The cross section is additionally impacted by the shape of the $^{68}$As excitation function since the given result is cumulative.

\vspace{-0.15cm}
\vspace{-0.3cm}

Interestingly, EMPIRE's overprediction of the compound peak energy centroid for $^{72}$Se production versus all other codes (Figure \ref{As_72SE}) is also seen for the $^{68}$Ge excitation function except it is a fairly accurate representation of reality in Figure \ref{As_68GE}. However, this energy comparison is the endpoint of EMPIRE's accuracy as its excitation function shape and magnitude are markedly incorrect.

ALICE continues to overestimate the compound peak magnitude and it even incorrectly predicts a higher-energy second compound peak rather than a pre-equilibrium tail.  CoH performs similarly to ALICE but at a more correct magnitude albeit at a shifted centroid energy of near 10 MeV below the experimental data. Both TALYS and TENDL correctly demonstrate a significant pre-equilibrium tail with an approximately correct shape, similar to CoH, but the relative magnitudes between their peaks and tails are erroneous. 

It is important to temper expectations for the predictive power of these codes in calculating the $^{68}$Ge production seen here since this is a cumulative result. Note that in the cumulative cases of this work, the code calculations shown include necessary summing of decay precursor contributions. $^{68}$Ge therefore requires calculation contributions from three residual products and ultimately only makes up a minor $\approx$5\% of the total non-elastic cross section, which creates a difficult predictive case.

\subsection{\label{}{$^{75}$As(p,3n)$^{73}$Se Cross Section}}
\begin{figure}[H]
{\includegraphics[width=1.0\columnwidth]{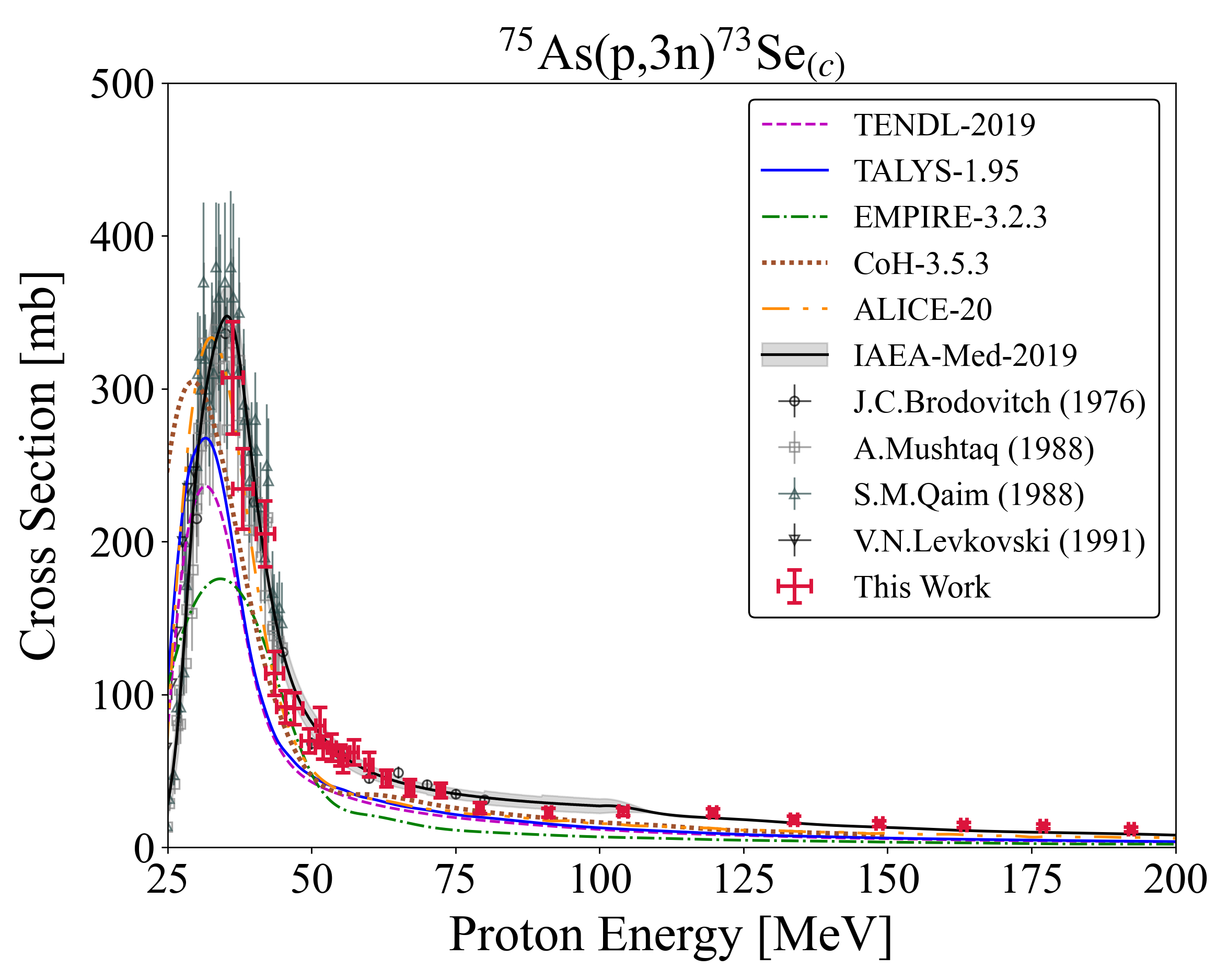}}
\vspace{-0.65cm}
\caption{Experimental and theoretical cross sections for $^{73}$Se production, peaking near 330 mb around 35\,MeV.}
\label{As_73SE}
\end{figure}

The $^{75}$As(p,3n)$^{73}$Se excitation function is the most well-characterized residual product channel in the prior literature data. The measured cross sections extracted from the LBNL and LANL irradiations are shown in Figure \ref{As_73SE} to agree very well with these existing results. Note that the reported cross sections are cumulative and include the formation contribution from the short-lived parent isomer $^{\textnormal{73m}}$Se ($t_{1/2}=39.8\ (13)$\,min) in addition to the longer-lived ($t_{1/2}=7.15\ (8)$\,h) ground state \cite{DataSheetsA73}. The results of the BNL irradiation help to extend the excitation function and characterize its tail behaviour up to 200\,MeV. The consistency between our results and the literature data compiled in EXFOR builds confidence in the energy and current assignments determined in this work as well as the overall measurement and data analysis methodology.

\vspace{-0.15cm}
\vspace{-0.3cm}

The default TALYS and EMPIRE predictions both underestimate the compound peak magnitude, EMPIRE decidedly more so than TALYS, while TALYS also shifts the peak energy lower than experimentally observed. The ALICE calculation performs best here with an appropriate peak magnitude and nearly proper tail shape, which is just incorrectly shifted similar to TALYS. TENDL replicates TALYS very closely other than a slightly reduced peak magnitude. CoH significantly mispredicts the channel's rising edge resulting in a more severe energy shift than both TALYS and ALICE.

The measured falling edge of the compound peak is additionally relevant to the medical community as $^{75}$As(p,3n) has been shown as the most advantageous route to the nonstandard positron emitter $^{73}$Se \cite{Qaim2018}. In this vein, the production of $^{73}$Se has also been evaluated by the IAEA and this recommended fit is given in Figure \ref{As_73SE} \cite{Tarkanyi2019:MedicalIsotopesDataNeeds}. The IAEA fit is seen to agree very well with the measured data in this paper.

It is worth noting that although the cross section averages only $\approx$40\,mb from 50--200\,MeV, the greater range of incident protons at 200\,MeV as compared to 50\,MeV would lead to a more than doubling in the overall $^{73}$Se production yield. This brief consideration is representative of the value inherent to high-current, high-energy proton accelerator facilities and rationalizes the effort to measure high-energy reaction data for potential production targets such as arsenic.
%about 2.4 times increase

\subsection{\label{}{$^{75}$As(p,p3n)$^{72}$As Cross Section}}
\begin{figure}[H]
{\includegraphics[width=1.0\columnwidth]{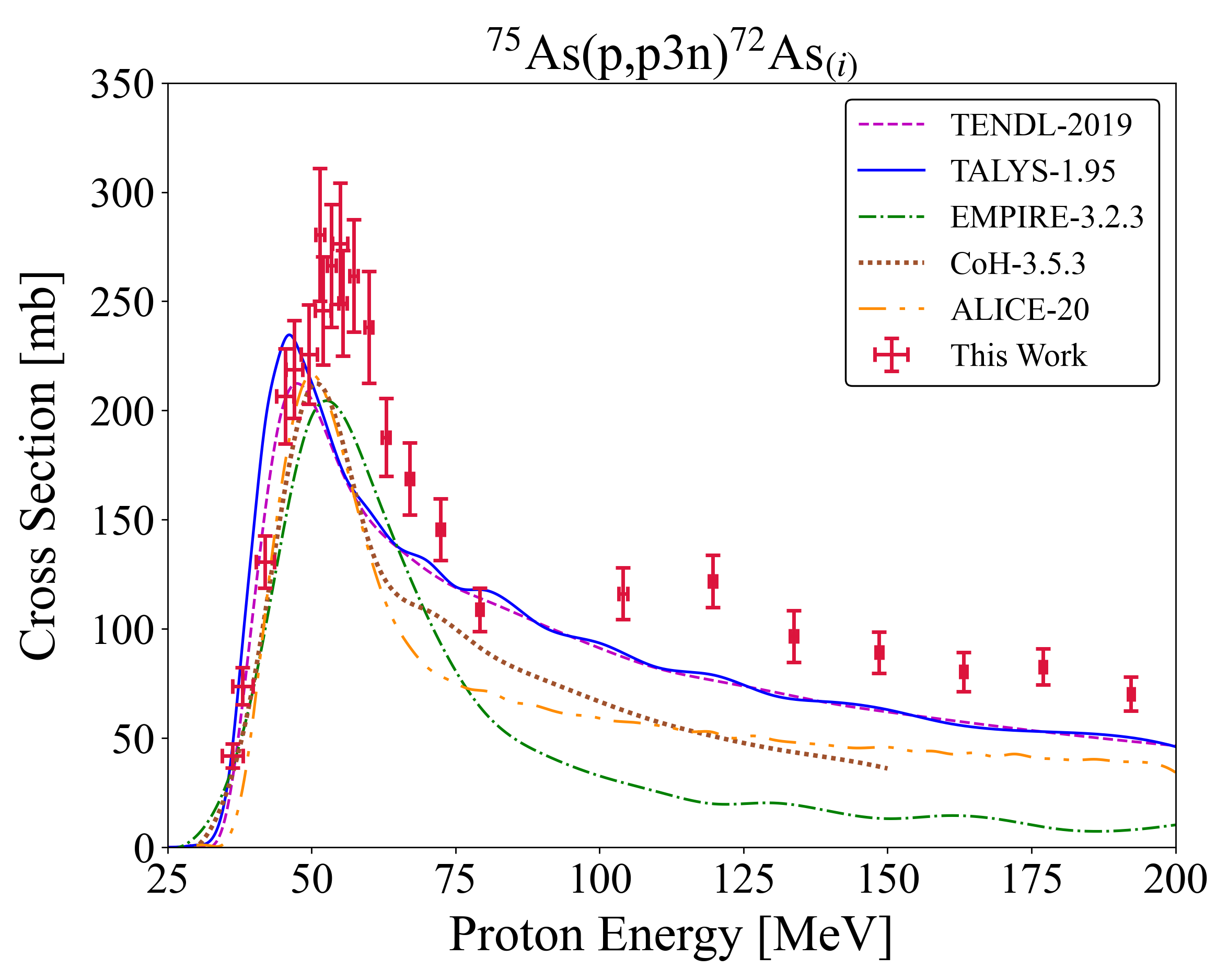}}
\vspace{-0.65cm}
\caption{Experimental and theoretical cross sections for $^{72}$As production, peaking near 275 mb around 55\,MeV.}
\label{As_72AS}
\end{figure}

The direct measurement of $^{72}$Se decay allowed for the subsequent independent cross section quantification of $^{72}$As. The cross section results are presented in Figure \ref{As_72AS} and are the measured first data for this reaction channel.

The modeling predictions all perform similarly in this channel, in contrast to the large variations seen for nearby $^{72}$Se and $^{73}$Se production. EMPIRE, CoH, and ALICE underpredict the high-energy pre-equilibrium tail for $^{72}$As relative to TALYS and TENDL, though the former trio of codes have the better energy placement of the compound peak centroid.

\vspace{-0.15cm}
\vspace{-0.3cm}

%\subsection{\label{}{$^{75}$As(p,pn)$^{74}$As Cross Section}}
%The arsenic isotope that has been used in nuclearmedicine the lon-
%gest is arsenic-74 (74As). It has a half-life of 17.8 d and decays through two pathways, either via positron emission (66\%) to stable germanium-74 (74Ge) with a maximum beta energy of 1.54 MeV and an associated gamma emission of 595.8 keV (59\%) or by beta emission (34\%) to stable selenium-74. The beta minus emission has amaximum beta energy of 1.352 MeV and an emitted gamma ray of 634.8 keV (15.4\%). The modes of decay for 74As allow for this isotope to be theragnostic, where the isotope could be used for both imaging and treatment \cite{Sanders2020:Radioarsenic}.

\subsection{\label{}{$^{\mathrm{nat}}$Ti(p,x)$^{\mathrm{44m/g}}$Sc Cross Section}}
\begin{figure}[H]
{\includegraphics[width=1.0\columnwidth]{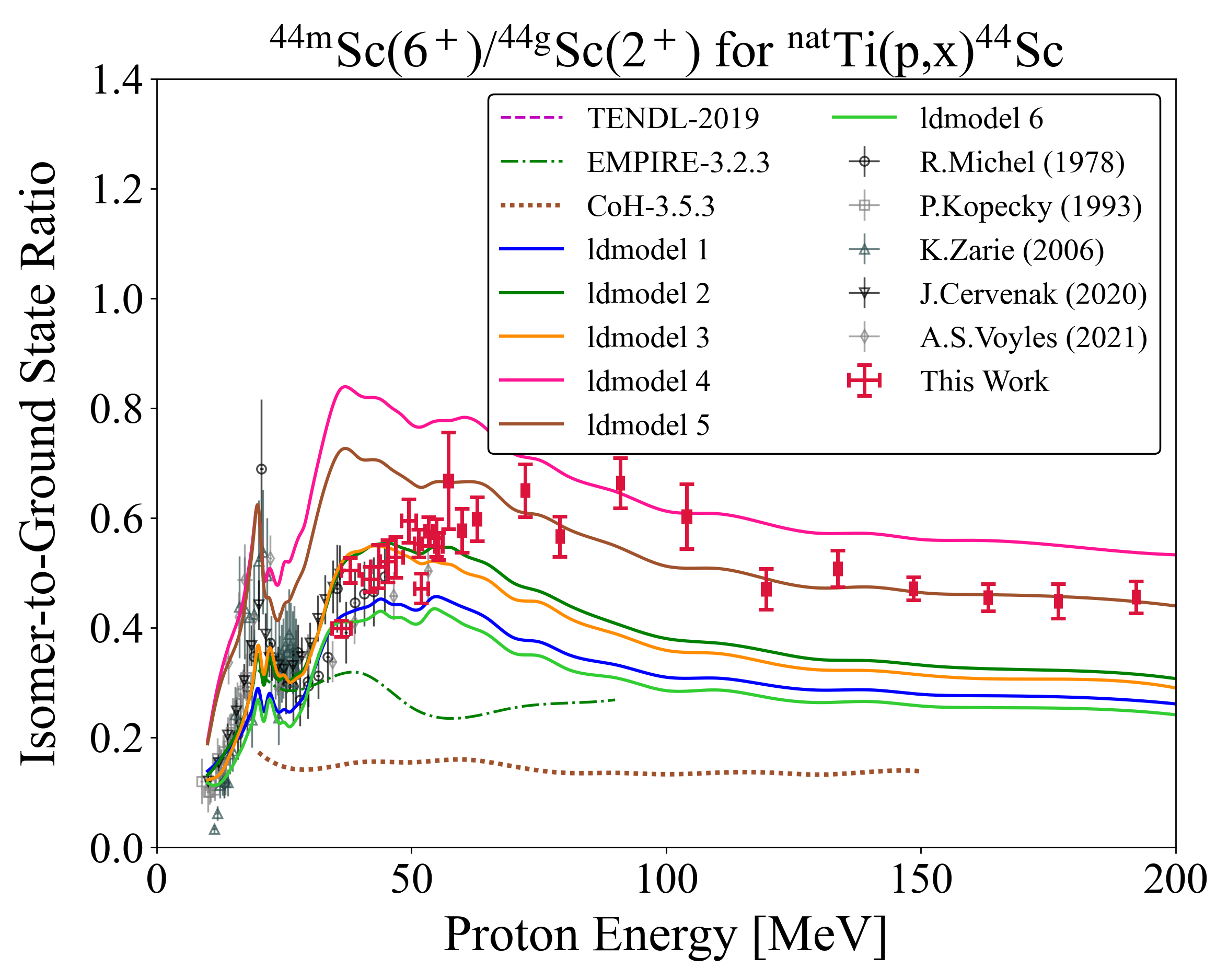}}
\vspace{-0.65cm}
\caption{Experimental and theoretical results for the isomer-to-ground state production ratio for $^{\mathrm{nat}}$Ti(p,x)$^{\mathrm{44m/g}}$Sc. The predictions from all 6 TALYS level density models are shown, where \texttt{ldmodel} 1 is equivalent to the TALYS default.}
\label{Ti_44Sc_mgRatio}
\end{figure}

The production of $^{\mathrm{44g}}$Sc  ($t_{1/2}=3.97\ (4)$\,h \cite{DataSheetsA44}) is of general interest as an emerging radiometal for nuclear imaging and theranostic purposes \cite{Tarkanyi2019:MedicalIsotopesDataNeeds,Chaple2018,Ferguson2019,Qaim2018}. While the measurements of the $^{\mathrm{nat}}$Ti(p,x)$^{\mathrm{44m/g}}$Sc excitation functions extracted from the titanium monitor foils included in the target stacks may not give an ideal production route for this medical application, these cross section results do give the only observable isomer and ground state pair from the three irradiations. As a result, this work provides a large update to the $^{\mathrm{44m}}$Sc ($t_{1/2}=58.61\ (10)$ h, $J^\pi=6^+$) to $^{\mathrm{44g}}$Sc ($t_{1/2}=3.97\ (4)$ h, $J^\pi=2^+$) \cite{DataSheetsA44} isomer-to-ground state ratio via $^{\mathrm{nat}}$Ti(p,x), as seen in Figure \ref{Ti_44Sc_mgRatio} and recorded in Table \ref{44Sc_mgRatio_table}.

\vspace{-0.2cm}
\begin{table}[H]
\caption{Isomer-to-ground state production ratio for $^{\mathrm{nat}}$Ti(p,x)$^{\mathrm{44m/g}}$Sc covering incident proton energies from 36 to 192\,MeV.}
\label{44Sc_mgRatio_table}
\begin{ruledtabular}
\begin{tabular}{lc}
$E\mathrm{_p}$ [MeV] &$\sigma$($^{\mathrm{44m}}$Sc)/$\sigma$($^{\mathrm{44g}}$Sc)\\[0.1cm]
\hline\\[-0.25cm]
192.26 (49) & 0.456 (29) \\[0.1cm]
176.99 (51) &  0.449 (32) \\[0.1cm]
163.18 (54) & 0.455 (25)\\[0.1cm]
148.52 (58) & 0.471 (21)\\[0.1cm]
133.72 (62) & 0.508 (34)\\[0.1cm]
119.63 (67) & 0.470 (37)\\[0.1cm]
104.05 (74) & 0.603 (59)\\[0.1cm]
91.05 (51) & 0.664 (45)\\[0.1cm]
79.15 (57) & 0.566 (37)\\[0.1cm]
72.34 (61) & 0.650 (48)\\[0.1cm]
62.87 (67) & 0.598 (40)\\[0.1cm]
59.88 (70) & 0.577 (40)\\[0.1cm]
57.26 (72) & 0.668 (88)\\[0.1cm]
55.36 (74) & 0.548 (25)\\[0.1cm]
54.9 (13) & 0.563 (34)\\[0.1cm]
53.40 (76) & 0.576 (26)\\[0.1cm]
51.9 (14) & 0.472 (27)\\[0.1cm]
51.39 (79) & 0.554 (25)\\[0.1cm]
49.5 (14) & 0.595 (40)\\[0.1cm]
46.9 (15) & 0.529 (37)\\[0.1cm]
45.4 (15) & 0.521 (39)\\[0.1cm]
43.5 (16) & 0.512 (39)\\[0.1cm]
41.9 (16) & 0.488 (23)\\[0.1cm]
38.0 (17) & 0.505 (23)\\[0.1cm]
36.2 (18) & 0.399 (15)\\
\end{tabular}
\end{ruledtabular}
\end{table}

\vspace{-0.15cm}
\vspace{-0.3cm}

Multiple experiments have measured this ratio previously for less than 50\,MeV and there is agreement between the high-energy end of those measurements and the lowest-energy results of this work \cite{Zarie2006:Ti,Michel1978:ProtonsTi,Kopecky1993:Ti,Cervenak2020:Ti,Voyles2021:Fe}. This new data extension could be used by the reaction modeling community to gain insight into angular momentum deposition over a broad range of incident particle energies.

The EMPIRE, CoH, and TENDL predictions for the isomer-to-ground state ratio are also shown in Figure \ref{Ti_44Sc_mgRatio} for comparison. The EMPIRE and CoH predictions markedly underestimate the ratio, however this result is a function of varying errors. In EMPIRE's case, the ratio is incorrect due to an overestimation of $^{\mathrm{nat}}$Ti(p,x)$^{\mathrm{44g}}$Sc production (see Figure \ref{Ti_44gSC} in Appendix \ref{Appendix_Plots}) while the CoH misprediction is instead a function of underestimation for $^{\mathrm{nat}}$Ti(p,x)$^{\mathrm{44m}}$Sc production (Figure \ref{Ti_44mSC} in Appendix \ref{Appendix_Plots}).

In the compound peak energy region of the $^{\mathrm{44m/g}}$Sc excitation functions (25--45\,MeV), competition with other exit residual product channels is minimized. Hence the optical model impact and transmission coefficient effects are minimized and the isomer-to-ground state data in Figure \ref{Ti_44Sc_mgRatio} is largely a function of the level density of $^{44}$Sc. Consequently, comparing the isomer-to-ground state predictions from TALYS's numerous nuclear level density models is a conventional brief investigation of this data. These TALYS predictions are the remaining comparisons shown in Figure \ref{Ti_44Sc_mgRatio}.

The \texttt{ldmodel} 1 in TALYS is the default Gilbert-Cameron constant temperature and Fermi gas model, but \texttt{ldmodel} 2, the Back-shifted Fermi gas model, appears to perform best in Figure \ref{Ti_44Sc_mgRatio} over the largest energy range. Though, it is perhaps noteworthy that the high-energy portion of the data is best reproduced by two of TALYS's microscopic level density models - \texttt{ldmodel} 4 and \texttt{ldmodel} 5. The exact nature of these microscopic models, and all six models in total, can be reviewed in the TALYS-1.95 manual \cite{TalysManual}.

\begin{figure}[H]
{\includegraphics[width=1.0\columnwidth]{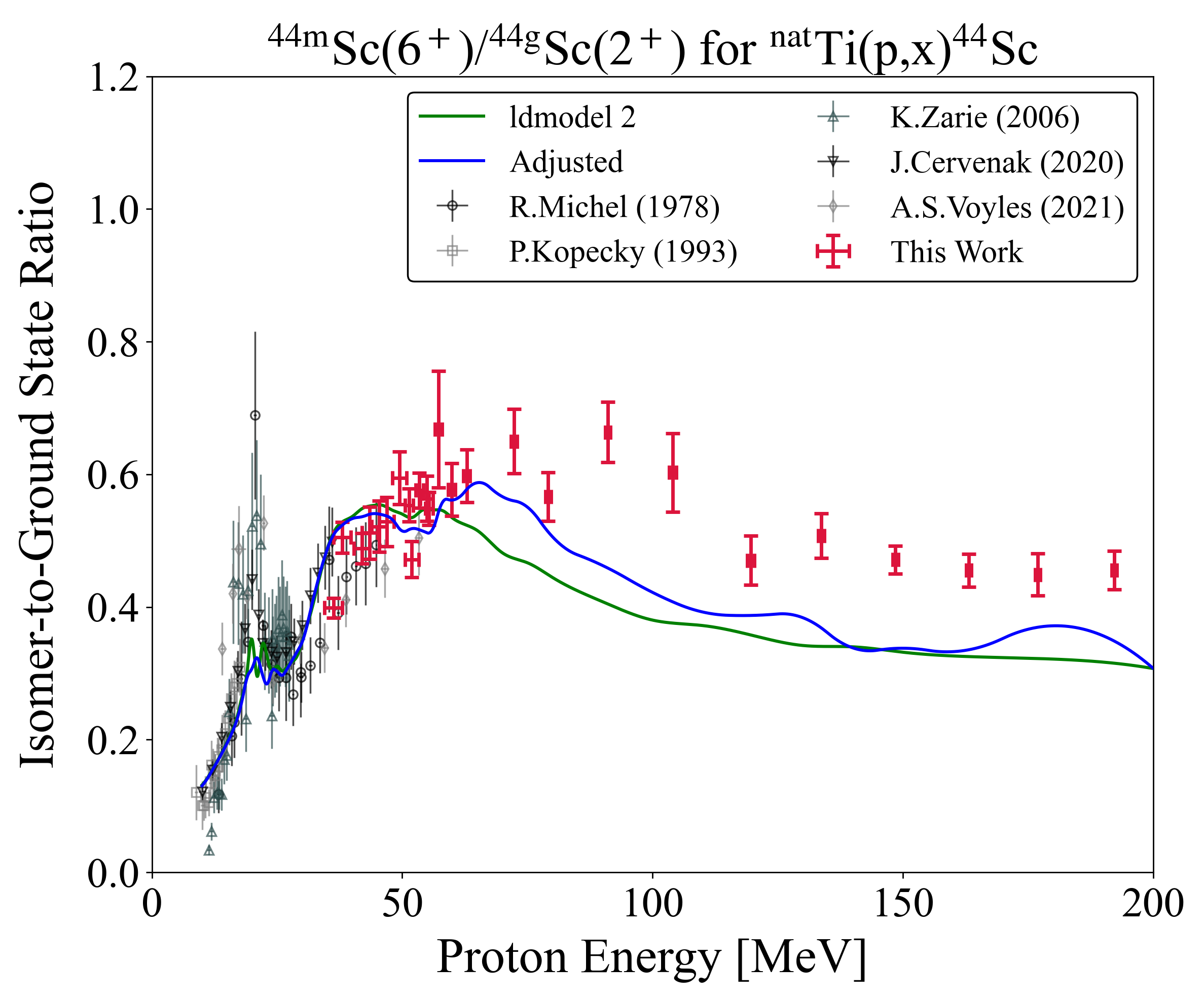}}
\vspace{-0.65cm}
\caption{Comparison of the TALYS \texttt{ldmodel} 2 model prediction for the isomer-to-ground state production ratio for $^{\mathrm{nat}}$Ti(p,x)$^{\mathrm{44m/g}}$Sc with a TALYS fit using adjusted parameters, including a spin cut-off increase.}
\label{Ti_44Sc_mgRatio_Proc}
\end{figure}

A single iteration of the \textcite{Fox2020:NbLa} fitting procedure was additionally applied for $^{\mathrm{nat}}$Ti(p,x) to try and glean more insight on the effect of level density choice for the relevant nuclei. It was found that an overall best fit to the multiple observed residual product channels (see Table \ref{Ti_xsValues} for product list) was still achieved using \texttt{ldmodel} 2 but that an energy-dependent increase in the spin cut-off parameter was also included among the model adjustments. The spin cut-off increase, set by the procedure to begin globally at $E_p=40$\,MeV in this case, broadens the width of the angular momentum distribution of the level densities involved in the $^{\mathrm{nat}}$Ti(p,x) reaction \cite{TalysManual}. This adjusted best fit can be seen versus the unadjusted \texttt{ldmodel} 2 case for the isomer-to-ground state ratio in Figure \ref{Ti_44Sc_mgRatio_Proc}. The mispredictions of the adjusted fit beyond $\approx$60\,MeV are not necessarily unexpected since these calculations are significantly complicated due to a polyisotopic target. However, since special attention was paid to adjusting this $^{\mathrm{44m/g}}$Sc ratio, the lasting mispredictions are likely more attributable to fundamental issues in the base pre-equilibrium model rather than parameter tuning.

It is interesting to observe that beyond $\approx$125\,MeV, the ratio remains relatively constant, thereby indicating a limit to the maximum amount of angular momentum that can be imparted to the system. This is a reflection of the mechanics of the pre-equilibrium process.
% but maybe also is a function of nuclear viscosity \cite{}.

This is evidently only an elementary investigation of the angular momentum in $^{44}$Sc and neighbouring nuclei, and a detailed investigation is outside the intent of this paper. Altogether, this discussion is still presented to inform the value and scarcity of these types of ratio datasets over wide energy regions, and to provide motivation for further analysis.

\vspace{-0.15cm}
\vspace{-0.3cm}

\subsection{\label{}{$^{\mathrm{nat}}$Cu(p,x) Cross Sections}}
The numerous $^{\mathrm{nat}}$Cu(p,x) cross sections measured here are in good agreement with the existing body of literature data and help to populate the more sparse regions of measurements between 100--200\,MeV. Plots of these copper excitation functions are provided in Appendix \ref{Appendix_Plots}. Similar to the $^{73}$Se results (Figure \ref{As_73SE}), the $^{\mathrm{nat}}$Cu(p,x) comparisons with existing data lend credence to our analysis methodology as well as our extensions to energy regions with no prior cross section measurements.

%These data contributions include XXXXXX.... to IAEA-recommended monitor cross sections, maybe useful for ocntinued evaluation to improve data availaibility/monitors in higher energy experiments
%\subsection{\label{}{$^{\mathrm{nat}}$Cu(p,x)$^{61}$Cu Cross Section}}
%Close agreement with the large body of experimental data, save for a slight over-prediction of the compound peak magnitude, though in agreement at the 2-$\sigma$ level. Extends the body of data well beyond 75 MeV proton energy, where only ~3 datasets previously contributed. Because the cross sections in this experiment are measured relative to the 2017 IAEA-recommended monitor cross sections, this measurement may be particularly useful if the natCu(p,x)61Cu reaction were to be included in a future evaluation, which may be unlikely due to the potential for secondary neutron contamination in this channel.

\subsection{Predicted Physical Thick Target Yields}
\begin{figure}[H]
{\includegraphics[width=1.0\columnwidth]{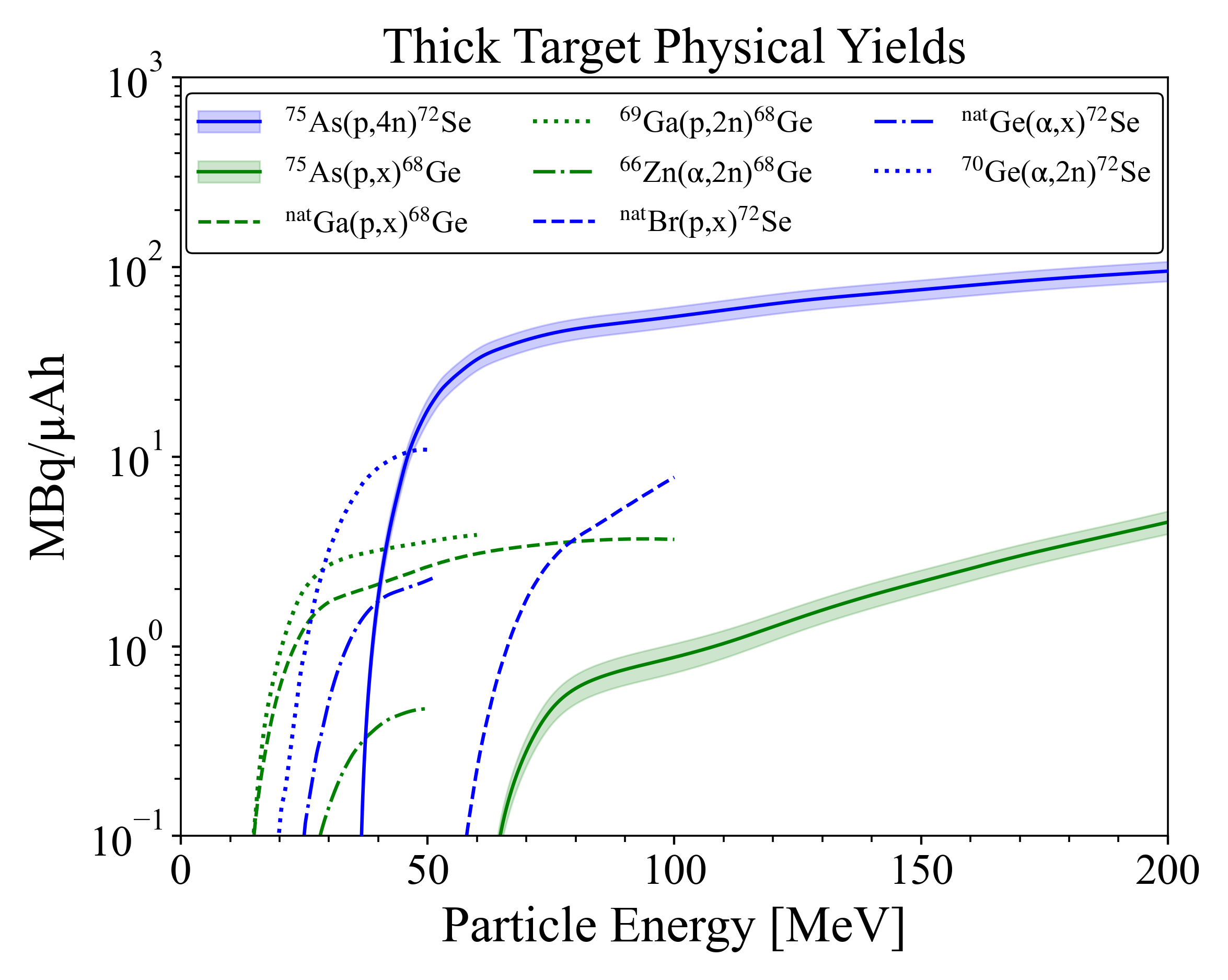}}
\vspace{-0.65cm}
\caption{Yields for the PET generator radionuclides $^{72}$Se and $^{68}$Ge according to established production routes and the new arsenic-based routes measured in this work \cite{Levkovski1991:MiddleMass,Nagame1989,Ruddy1969,Porile1963,Amiel1959,Fassbender2001:Br,Villiers2002,AdamRebeles2013,Hermanne2015,
Takacs2016:GetoSe,Mushtaq1990}.}
\label{As_PhysicalYields}
\end{figure}

Instantaneous thick target yields for $^{75}$As(p,x)$^{72}$Se,\,$^{68}$Ge were calculated from the measured cross section results and are plotted in Figure \ref{As_PhysicalYields}. A comparison to the yields from earlier discussed established production routes for these generator nuclei in Section \ref{Introduction} are also included.

\vspace{-0.15cm}
\vspace{-0.3cm}

The data from TREND suggests that across all relevant incident particle energies beyond reaction threshold, the $^{75}$As(p,4n)$^{72}$Se is the optimal production pathway to the $^{72}$Se/$^{72}$As generator system. The arsenic target route offers an increase in yield of greater than an order of magnitude versus the current methods, while still affording radioisotopically pure production as best as possible. Specifically, no charged-particle production route to the $^{72}$Se/$^{72}$As generator system is uncontaminated from $^{75-73}$Se co-production. However, it is expected that $^{72}$As will be efficiently separated from the parent $^{72}$Se when needed, and that the co-produced $^{75-73}$Se will also follow the chemical separation \cite{Sanders2020:Radioarsenic,Ballard2012:SeAsGenerator}. Of course, any $^{73}$Se contaminant is much shorter lived than $^{72}$Se and can be decayed out to reach a more pure $^{72}$Se starting condition regardless. Further, $^{75}$Se production is energetically unfavorable in the p+$^{75}$As production conditions for $^{72}$Se, meaning any in-grown $^{75}$As prior to separation will be both minimal and stable. The $^{75}$As(p,4n) pathway also avoids any potential long-lived $^{74,73,71}$As contamination present from Ge target routes. In total, arsenic-based production of $^{72}$Se gives the best chance to produce and collect a radiochemically-pure $^{72}$As daughter.

%any $^{73}$Se contaminant is much shorter lived than $^{72}$Se and can be decayed out to reach nearly pure 72Se 
%meaning that after irradiation, decay, and separation,  good ability to collect radiochemically-pure daughter. 

%Conversely, $^{75}$Se is a long-lived nuclide whose decay cannot be waited out but $^{75}$Se production is energetically unfavorable in the production conditions for $^{72}$Se. Therefore any small amount that is produced meaning any in-grown $^{75}$As will be both minimal and stable.

It is seen that at incident proton energies nearing 200\,MeV, the yield from $^{75}$As(p,x)$^{68}$Ge can rival and exceed the production route based on already employed natural gallium targets. Specifically, Figure \ref{As_PhysicalYields} predicts an $\approx$18\% increase for the arsenic-based yield at 200\,MeV (${4.5>3.8}$\,MBq/\mmicro Ah). Nevertheless, a p+$^{75}$As approach is expected to co-produce more stable germanium and $^{71}$Ge$\rightarrow ^{71}$Ga contamination versus the p+$^{\textnormal{nat}}$Ga route, leading to reduced $^{68}$Ge specific activity. Arsenic targets would also introduce a need for additional, potential lossy, separation chemistries due to long-lived selenium and arsenic products not present from p+$^{\textnormal{nat}}$Ga. Therefore, uprooting the successful established gallium route for arsenic is unwarranted. Still, this $^{75}$As(p,x)$^{68}$Ge study gives valuable information in the context of total arsenic reactions, contributes to the knowledge base of the essential $^{68}$Ge/$^{68}$Ga system, and demonstrates the importance of measuring these high-energy reactions, which can very easily produce large yields due to the long range of high-energy protons.
%It is seen that at incident proton energies nearing 200\,MeV, the yield from $^{75}$As(p,x)$^{68}$Ge can rival and exceed production routes based on the currently preferred gallium targets. Nevertheless, even with the potential ease resulting from the use of an arsenic target, the slight increase in yield likely does not counterbalance the onset of radioisotope co-production contamination issues not present with gallium.

%in targetry moving to an arsenic-centric production approach, the slight increase in yield likely does not counterbalance the onset of radioisotope co-production contamination issues not present with gallium.

%Figure \ref{OtherYields} provides physical yield calculations for other measured arsenic and selenium residual products of interest.
%where the measured data exists, with some interpolation/extrapolation, so restricting range to 35--200 MeV
%Comments on the comparisons/range, what gets achieved better from the different formation choices

\section{Charged-Particle Reaction Modeling}
The effort to explore and improve the current nuclear reaction models for charged-particles, and perhaps more specifically charged-particles at high incident energies, is continued in this work. Explicitly, the TALYS residual product based fitting procedure presented by \textcite{Fox2020:NbLa} is applied to $^{75}$As(p,x) given the unique, large body of proton-induced data measured here.

The nine reaction channels $^{75}$As(p,x)$^{75,73,72}$Se, $^{74,73,71}$As, $^{69}$Ge, $^{68,67}$Ga were simultaneously used for the parameter adjustment investigation. $^{73}$Se, $^{73}$As, $^{69}$Ge, and $^{68}$Ga were considered as the most important fitting cases due to a combination of factors such as cross section magnitude, diversification of particle emission types, and impact on production competition with neighbouring nuclei.

\subsection{Deformation Effect of $^{75}$As}
While the cases of $^{93}$Nb(p,x) in \textcite{Fox2020:NbLa} and of $^{75}$As(p,x) here have similar attributes - both utilize data from the same experiments, which cover the same energy range of interest, and both are monoisotopic targets in nearby mass ranges - the documented deformation of $^{75}$As is a notable change from the spherical $^{93}$Nb \cite{Effenberger1982,Verma2013:AsBackBending,Goriely2007:HFB,Li2017:AsStapler}. This potentially introduces a complication to the direct application of the fitting procedure from \textcite{Fox2020:NbLa}. Specifically, it would be necessary to address coupled-channels (CC) calculations or other angular momentum modifications to the typical spherically symmetric Hauser-Feshbach formalism prior to any further parameter changes \cite{Grimes2013:ModHF}.
%, else risking inherently poor predictive power for cross sections due to a flawed foundation \cite{Grimes2013:ModHF}.

The RIPL-3 imported TALYS value for the $^{75}$As quadrupole deformation parameter is -0.25, which suggests a strongly oblate deformation \cite{TalysManual,RIPL3}. In fact, RIPL-3 lists strong oblate deformation for the arsenic isotopes $A=68-76$. While some experimental evidence supports these values for the neutron deficient cases and transitions around $N=Z$, it is quite rare that the neutron rich isotopes would demonstrate oblate rather than prolate deformation \cite{Obertelli2011:AsShape,Bruce2000:AsShape}. An investigation using a Nilsson diagram gives further support that $^{75}$As is actually prolate in nature. Finally, ENSDF and the original datasets incorporated into the structure evaluation provide experimental evidence of the prolate condition for $^{75}$As and actually list a quadrupole deformation parameter of +0.314 (6) \cite{Effenberger1982}. This prolate value appears to be both physically and historically more correct than the RIPL-3 $\beta_2=-0.25$ and is therefore taken as the $^{75}$As deformation in the analysis that follows.
%$^{75}$As is a statically deformed nucleus in its ground state that has been shown to have collective nature by experiment \cite{Effenberger1982,Verma2013:AsBackBending,Goriely2007:HFB,Li2017:AsStapler}. 
%It appears as though an erroneous oblate deformation value has been propagated and reached the current RIPL-3 database.

TALYS, however, does not include any deformation coupling schemes for arsenic isotopes and as a result, a spherical OMP basis is used in the predictive calculations, thereby potentially neglecting a significant physics aspect of the problem. It was therefore necessary to manually create a coupling scheme to see whether this has an effect on final results. Yet, the level scheme of $^{75}$As does not present any ideal vibrational or rotational bands for coupling and its deformation is very likely either soft vibrational or soft rotational \cite{Kawano2015:CoHArsenic,Shibata2012:EvalAs}. 

On further examination, the $3/2^-$ ground state with the $5/2^-$ level at 279.543\,keV and the $7/2^-$ level at 821.620\,keV appear to form a rotational band. The $5/2^-$ level shows the expected strong $\gamma$-ray transition ($I_\gamma=100.0\ (5)\%$) of M1 character to the ground state, while the $7/2^-$ excited level shows both a strong E2 transition to the ground state ($I_\gamma=100.0\ (15)\%$) and weaker M1 transition to the $5/2^-$ level ($I_\gamma=9.6\ (11)\%$), generally in line with behaviour expected from a rotational band. Further, the $7/2^-$ E2 transition is 23.0 (24) Weisskopf units \cite{DataSheetsA75}, providing evidence for its collective behaviour. This three-level rotational band coupling scheme was added to TALYS. 

It was also noticed that the neighbouring nuclei $^{76,74}$Se and $^{76,74}$Ge demonstrate vibrational character \cite{DataSheetsA74,DataSheetsA76} and have vibrational coupling schemes implemented in TALYS for CC calculations ($^{76}$Ge has actually recently been shown as rigid triaxially deformed \cite{Ayangeakaa2019}). These neighbouring properties provide motivation to model the arsenic target as soft vibrational rather than rotational.
%Making parameter adjustments to the default TALYS prediction are therefore not worthwhile before addressing this starting point for the calculations.
%TALYS is only capable of implementing a deformed OMP through the coupled-channels (CC) approach. 

%An examination of the level scheme of natural arsenic certainly shows that it is not a perfect rotor. However, the $3/2^-$ ground state with the $5/2^-$ level at 279.543\,keV and the $7/2^-$ level at 821.620\,keV appear to form a rotational band. The $5/2^-$ level shows the expected strong $\gamma$-ray transition ($I_\gamma=100.0 (5)\%$) of M1 character to the ground state, while the $7/2^-$ excited level shows both a strong E2 transition to the ground state ($I_\gamma=100.0 (15)\%$) and weaker M1 transition to the $5/2^-$ level ($I_\gamma=9.6 (11)\%$), generally in line with behaviour expected from a rotational band. Further, the $7/2^-$ E2 transition is $20-40$ Weisskopf units, providing definitive evidence for its collective behaviour. 
%The creation of a three-level rotational band coupling scheme is in line with multiple other coupling schemes already included in TALYS, such as for $^{151}$Eu, and $^{153}$Eu.

Unfortunately, TALYS's implementation of the ECIS-06 code for optical model and CC calculations is unsuited for a pure vibrational coupling scheme for odd-Z nuclei, and the weak-coupling model has to be used in such cases. Moreover, the only odd-Z nucleus with any sort of vibrational deformation file in TALYS is $^{241}$Am, where vibrational collectivity is built on top of rotational character. Therefore, taking the $^{241}$Am deformation formatting as a guide, a weak vibrational band consisting of the $^{75}$As $9/2^+$ (303.9243\,keV), $5/2^+$ (400.6583\,keV), and $1/2^+$ (860.0\,keV) levels were added to a second created coupling scheme including the prior discussed rotational band. In this suggested vibrational band, the $1/2^+$ level is dominated by transition to $5/2^+$, which then has an E2 transition to the $9/2^+$ of 76.4 (25) Weisskopf units \cite{DataSheetsA75}. The $9/2^+$ de-excitation is dominated by E3 decay to the ground state. This mixed rotational+vibrational coupling scheme was also added to TALYS.

Elsewhere, this treatment for adjusting the global spherical optical model by a CC approach to implement a deformed optical model for $^{75}$As calculations has been used in \textcite{Shibata2012:EvalAs} and \textcite{Kawano2015:CoHArsenic}. The \textcite{Shibata2012:EvalAs} work is an evaluation of neutron nuclear data on $^{75}$As up to 20 MeV for JENDL-4 and uses a similar rotational coupling scheme to the one presented here but substitutes the $5/2^-$ level at 279.543\,keV with a $5/2^-$ level at 572.41\,keV. \textcite{Shibata2012:EvalAs} uses the quadrupole deformation parameter $\beta_2=-0.19$ within a rigid-rotator model. In their evaluation, they found it necessary to additionally tune the matrix element parameter as well as the pickup and knockout contributions for their pre-equilibrium model relevant to the residual product cross sections of (n,$\gamma$), (n,p), (n,2n), and (n,$\alpha$). However, the JENDL-4 evaluation still found limited success in fitting the $^{75}$As(n,p) channel after accounting for both deformation and pre-equilibrium changes. \textcite{Shibata2012:EvalAs} considered other solutions attempts that included level density and optical model parameter changes concerning both $^{75}$As and $^{75}$Ge but could not simultaneously improve the (n,p) channel while maintaining good global behaviour elsewhere.

\textcite{Kawano2015:CoHArsenic} performed their CC calculations using the CoH reaction code and probed the collectivity effects in $^{75}$As for incident neutrons. They explored the total and some close-to-target residual product cross sections up to 20 MeV, similar to \textcite{Shibata2012:EvalAs}. In comparison to ENDF/B-VII.0 results, the \textcite{Kawano2015:CoHArsenic} calculations demonstrated improvement in reproducing the total cross section but did require model parameter adjustments for the individual reaction channels, not always yielding satisfactory results. \textcite{Kawano2015:CoHArsenic} used the RIPL-3 suggested strong oblate deformation of arsenic.

In this p+$^{75}$As modeling work, the CC calculations in TALYS for arsenic, when invoking either the custom rotational+vibrational deformation or the custom pure rotational deformation scheme, together with the ENSDF-accepted prolate deformation parameter $\beta_2=+0.314$ (6), proved to have minimal impact on the predictions for residual product excitation functions. Any alterations that were present were not seen to be consistent improvements versus the default spherical optical model calculations. This is not an entirely unusual result given the higher energies under consideration and the overall expected lower level of collectivity for this target nucleus. It should be noted that this is not an exhaustive investigation of arsenic deformation, CC calculations, or collectivity models, and no structure or theory statements can be made. This result is only a statement of the sensitivity of the modeling under the conditions of this work.
%Perhaps deformation coupling schemes need to be even added for all the relevant arsenic isotopes accessible through high-energy p+$^{75}$As to better explore CC calculations. 

Given the observed unremarkable changes, the inability to disentangle effects of CC calculations from more dominating level density, optical model, and pre-equilibrium parameter adjustments, and the imperfections of previously established deformed fitting approaches, the decision was made to treat $^{75}$As spherically within TALYS and implement the fitting procedure from \textcite{Fox2020:NbLa} identically.
%we feel satisfied to not explore further Hauser-Feshbach modifications. Instead, this work will treat $^{75}$As spherically within TALYS and implement the fitting procedure from \textcite{Fox2020:NbLa} identically.

%In this study, a comparison of the default TALYS calculation and the deformed TALYS prediction from a coupling scheme for $^{75}$As(p,x)$^{73,72}$Se, $^{72}$As, $^{68}$Ge is shown in Figures \ref{XXXXXX}. It is clear that the deformation inclusion is not a full solution for the charged-particle induced reactions, largely in contrast to the neutron OMP results of \textcite{Nobre2014:DeformedOMP} and, to some extent, \textcite{Kawano2015:CoHArsenic} as well. The compound peak magnitude in Figures \ref{XXXXXX} is reduced in the deformed calculation compared to the default case, which is overall similar behaviour to deformed neutron OMP calculations, which act to reduce the total reaction cross section. No experimental $^{75}$As(p,tot) total reaction cross section or elastic scattering data exists to act as a comparison or constraint in this analysis, however. 

\subsection{\label{FitProc}Fitting Procedure Applied to $^{75}$As(p,x)}
Firstly, the application of microscopic level density models proved beneficial as compared to the default phenomenological Gilbert-Cameron constant temperature model or the placement of compound peak centroids. However, it was seen that no one microscopic level density model best reproduced the excitation functions across all the observables. Instead, level density calculations from Goriely’s tables using the Skyrme effective interaction (\texttt{ldmodel} 4) \cite{ldmodel4} proved to be most accurate for the close-to-target residual products, and specifically for $^{72-76}$Se and their competition with close-to-target arsenic products. Yet, applying \texttt{ldmodel} 4 to all nuclei involved in $^{75}$As(p,x) created pre-equilibrium tails biased too high above the experimental data for Ga, Ge, and other $\alpha$-emission residual product excitation functions farther from the target. Conversely, it was observed that the temperature-dependent Hartree-Fock-Bogolyubov level density calculations using the Gogny force (\texttt{ldmodel} 6) \cite{ldmodel6} did not suffer from the magnitude bias problems in the far-from-target channels, but failed to model the close-to-target Se and their competing products unlike \texttt{ldmodel} 4. Therefore, two microscopic level density models were used, where \texttt{ldmodel} 4 was applied to the aforementioned grouping of selenium nuclei and \texttt{ldmodel} 6 was applied for all else. Further details of these level density considerations can be reviewed in Section \ref{ldadjust_analysis}.

The pre-equilibrium parameter adjustments in the next portion of the procedure were indeed found to follow the systematic trend described in \textcite{Fox2020:NbLa}, with \texttt{M2constant}=0.80, \texttt{M2limit}=3.9, and \texttt{M2shift}=0.55. Furthermore, the value for the constant of the proton and neutron single-particle level density parameter used for calculations of the exciton model particle-hole state densities was altered from its default \texttt{Kph}=15 to \texttt{Kph}=15.16. Other pre-equilibrium modeling changes were manipulations of the stripping and knockout reaction contributions for outgoing alpha, deuteron, triton, and $^{3}$He particles. These manipulations were performed using the TALYS \texttt{Cstrip} and \texttt{Cknock} keywords. The precise adjusted values can be viewed in Table \ref{AsAdjustedParams} in Appendix \ref{Appendix_Params}.

Subsequent iterative simultaneous tuning of optical model and individual level density parameters were needed to aid the compound reaction regime and to fix erroneous production competitions between clustered products. 

The need for nuclide-specific level density changes arises from discrepancies between measured and modeled data where global changes to exciton or optical model parameters can not resolve the singular problems. These nuclide-specific adjustments were most evident for $^{73}$As production, where both the adjusted fit to this point and the default calculation were nearly 200\,mb smaller than the observed results. As in \textcite{Fox2020:NbLa}, these level density manipulations per nuclide could be performed with the TALYS \texttt{ctable} and \texttt{ptable} commands when microscopic level density models are implemented.

The effects of \texttt{ctable} and \texttt{ptable} to create an adjusted level density $\rho(E_x,J,\pi)$ are explicitly given by,
\begin{gather}\label{mic_adjust}
\rho(E_x,J,\pi)=\exp(c\sqrt{E_x-\delta})\rho_{mic}(E_x-\delta,J,\pi),
\end{gather}
where \texttt{ctable} is the $c$ constant, \texttt{ptable} is the $\delta$ constant (denoted as the ``pairing shift"), and $\rho_{mic}(E_x-\delta,J,\pi)$ are the tabulated microscopic level density calculations as a function of excitation energy $E_x$, angular momentum $J$, and parity $\pi$. The produced tables in TALYS have not been adjusted to experimental data and have $c=0$ and $\delta=0$ by default. The implementation of \texttt{ctable} and \texttt{ptable} under the definition of Equation (\ref{mic_adjust}) then provides necessary scaling flexibility at both low and high energies \cite{TalysManual}.

Since the production of $^{73}$As is most heavily correlated with the neighbouring exit channels $^{72,73}$Se and $^{74}$As, the \texttt{ctable} and \texttt{ptable} effects on $^{73}$As necessitated corresponding nuclide-specific level density changes in $^{72,73}$Se and $^{74}$As as well.

The most suitable optical model adjustments were found to be \texttt{d1adjust n}=1.75 and \texttt{d1adjust p}=1.55, which multiply the energy-dependent imaginary surface-central potential well depth for neutron and protons, respectively. These multiplicative changes lead to increased particle emission from the surface region of the nucleus, and thus to increased emission of high-energetic particles, particularly at lower incident proton energies. In turn, these alterations create a more pronounced pre-equilibrium spectrum that contributes additional production within the compound regions of residual product excitation functions and some additional production to their tails.

Although these are not unsubstantial multiplication factors from the default 1.0 values, the energy dependence of the surface potential means that the adjustment impact is large in the vicinity of low threshold channels for lower incident proton energies but becomes only a minor change above $\approx$50\,MeV. This is mirrored by the volume potentials that increase and dominate absorption/emission as energies reach $\approx$50\,MeV and beyond. For example, at $E_p=20$\,MeV, the default imaginary surface-central potential well depth for protons on $^{75}$As is 8.4\,MeV while the adjusted well depth is 1.55$\times$ larger at 13.0\,MeV. This 4.6\,MeV difference is a relevant change around low residual product threshold energies but by $E_p=75$\,MeV, this default versus adjusted well depth difference is reduced to just 1.5\,MeV. The difference then falls below 1\,MeV at $E_p=90$\,MeV, and is reduced down to 0.1\,MeV at $E_p=200$\,MeV. Similar behaviour is true for the change to the imaginary surface-central potential well depth for neutrons. Furthermore, at $E_p=20$\,MeV, the imaginary volume potential is $5-7\times$ smaller than the imaginary surface potential for both neutrons and protons in the adjusted case, but by $E_p=75$\,MeV, the imaginary volume potential has grown to be $2-3\times$ larger. The imaginary volume potential becomes increasingly more dominant, growing to be $50-70\times$ larger by $E_p=200$\,MeV.

It is possible that portions of the \texttt{d1adjust} changes should actually be substituted with changes to the imaginary surface diffusivity parameter, but this cannot be unambiguously determined using only residual product cross section data and instead requires angle-differential cross section information \cite{KD2003:OMP}. This limited diversity of high-$E_p$ fit data is a common theme that permeates the limitations of this approach to parameter adjustments as well as prevents much physical meaning to be gleaned from the modeling. These limitations are further explored in Section \ref{Limitations}.

An additional increase to proton absorptivity and emissivity across a wider range of energies, to increase peaks and tails for numerous channels consistently, was still warranted by the experimental data. This was implemented with an increase to the imaginary volume potential well depth for protons by \texttt{w1adjust p}=1.21.
%increase absorption and emissivity across a wider range of energies and helps to necessarily increase peaks and tails for numerous channels consistently, as required by the experimental data.
%increase the cross section magnitudes for the considered channels across all energies while maintaining the underlying shape set by the pre-equilibrium adjustments.
%The magnitude changes for the w1adjustttttt  w2adjusttxxxxx reach maximums of $\approx$3\,MeV difference for protons (adjusted 16.1\,MeV vs.... default 13.4\,MeV) and $\approx$1\,MeV difference for neutrons (adjusted 12.5\,MeV vs.... default 11.7\,MeV) at 200\,MeV.

The default TALYS alpha optical model of \textcite{Avrigeanu2014} was deliberately chosen as it performed best for the considered As and Ge channels. The deuteron optical model of \textcite{Han2006} was applied instead of the default model from standard Watanabe folding \cite{Watanabe1958}. This deuteron adjustment is minor ($\leq 5$\,mb) compared to the alpha model effect but does better match the experimental peak and tail behaviour in observed residual product channels for $A\leq 72$.
%There is no noticeable effect for the deuteron model choice for $A>71$.

%Subsequent tuning of optical model parameters were needed to aid the compound reaction regime and increase compound peak magnitudes for Se and As channels with $A>70$, which threshold at lower energies relative to farther-from-target nuclei. These adjustments included \texttt{d1adjust n}=1.75 and \texttt{d1adjust p}=1.55, which multiply the energy-dependent imaginary surface-central potential well depth for neutron and protons, respectively. These multiplicative increases

%ANd to provide some more context on the size of these change, at 20\,MeV the ratio of the imaginary volume potential to the surface potential for neutrons in the adjusted case is approximately 7\% smaller than the same ratio in the default case. 
%The ratio might not be the best way to compare overall because as the number get smaller at higher energies the changes to ratios become larger, but this at least gives some idea at the peak of the change caused by d1adjust........ Maybe this ratio argument is not super relevant, you don't want to give off the idea that you're not making changes, because you defintiely want to be making changes, and you don't want to get caught in some loop where some one can say oh these changes are tooo big, so maybe just giving straight up the absolute magnitude of the well depth changes is your best bet in all these cases.

Lastly, an additional minor nuclide-specific case for level density adjustments that became relevant as a result of iterating over the above parameter changes was $^{71}$As. This adjustment included corresponding small level density changes to $^{68}$As, $^{69}$Ge, and $^{69}$Ga as a function of correlated production competition.

The lone prominent outstanding modeling discrepancy among the considered channels was an overprediction of $^{67}$Ga production. It is likely that this difference represents a sensitivity limit for this fitting procedure through a manual approach. Moreover, given the massive parameter space for adjustments in TALYS,  it is realistic that the fitting here ends in a local variance minimum, unable to perfectly match all prioritized ($\approx$15\% of total cross section) and minor ($\leq 5$\% of total cross section) residual products. We can correct for this $^{67}$Ga error by reducing the nuclide-specific level density, but this change is likely a compensating correction in this context and does not contribute to any increase in predictive power.

All parameter changes creating this total adjusted fit are given in Table \ref{AsAdjustedParams} in Appendix \ref{Appendix_Params}. Figure \ref{Proc_Fits} presents the adjusted fit compared to the default TALYS calculation for the nine considered reaction channels up to an incident proton energy of 200\,MeV.

%This 67Ga change is an adjustment that could only be made after the fact and after acquiring data in the channels 67Ga and 66Ga, so it is not necessaarily contributing to any predictive power, it is just an example of where sensitivity for this procedure cannot get to for all these channels, we simply could not predict it under this manusal sensitivty study and the unrestricted parameter space, along with the more improtant channels.
%still consistent issues for the chain of Ga products
%using \texttt{ldmodel 5} for $^{69}$Ga puts most things in order
%$^{67}$ was one prediction that remained poor and was not possible to correct under this manual sensitvity study
%Perhaps some local minimum issue and just a sensitivity limit on this manual approach, not able to do everyrthing, it's not a full blwon evluation solution, just a step towards the thought process and noticing larger fixes
%does not come with associated physical meaning and its implications on the global fit confidence are discussed in section ...

Overall, we put forth a large number of level density scalings, either directly or as a correlation consequence, and though this is not unexpected given the prior lack of data and ambiguity for the reactions and energies of interest \cite{ldmodel6,Nobre2020:LD}, it is important to reflect on the intricacies of performing such a number of scalings. This discussion is presented in Section \ref{ldadjust_analysis}.

Additionally, context for our suggested parameter adjustments can be gleaned from the ``best" parameters file for n+$^{75}$As included with TALYS-1.95 \cite{TalysManual}. This ``best" parameterization contains multiple level density scalings (with the back-shifted Fermi gas model as a base) in addition to multiple optical model real potential radii and diffusivity adjustments, some of which reach upwards of 11\% changed from default and are made energy-dependent. Similar stripping and knock-out contributions to our suggested adjustments in this work exist as well in the ``best" file. Overall, our adjustments generally work to avoid potential unphysical changes to geometry parameters and the real potential instead to focus on the imaginary potential. This focus is likely more appropriate for high-energy residual product cross section datasets versus the lower-energy scattering and resonance data important to the development of the ``best" n+$^{75}$As parameters.

\begin{figure*}[!t]
	\subfloat[\label{75Se_Proc}]
		{\includegraphics[width=0.315\textwidth]{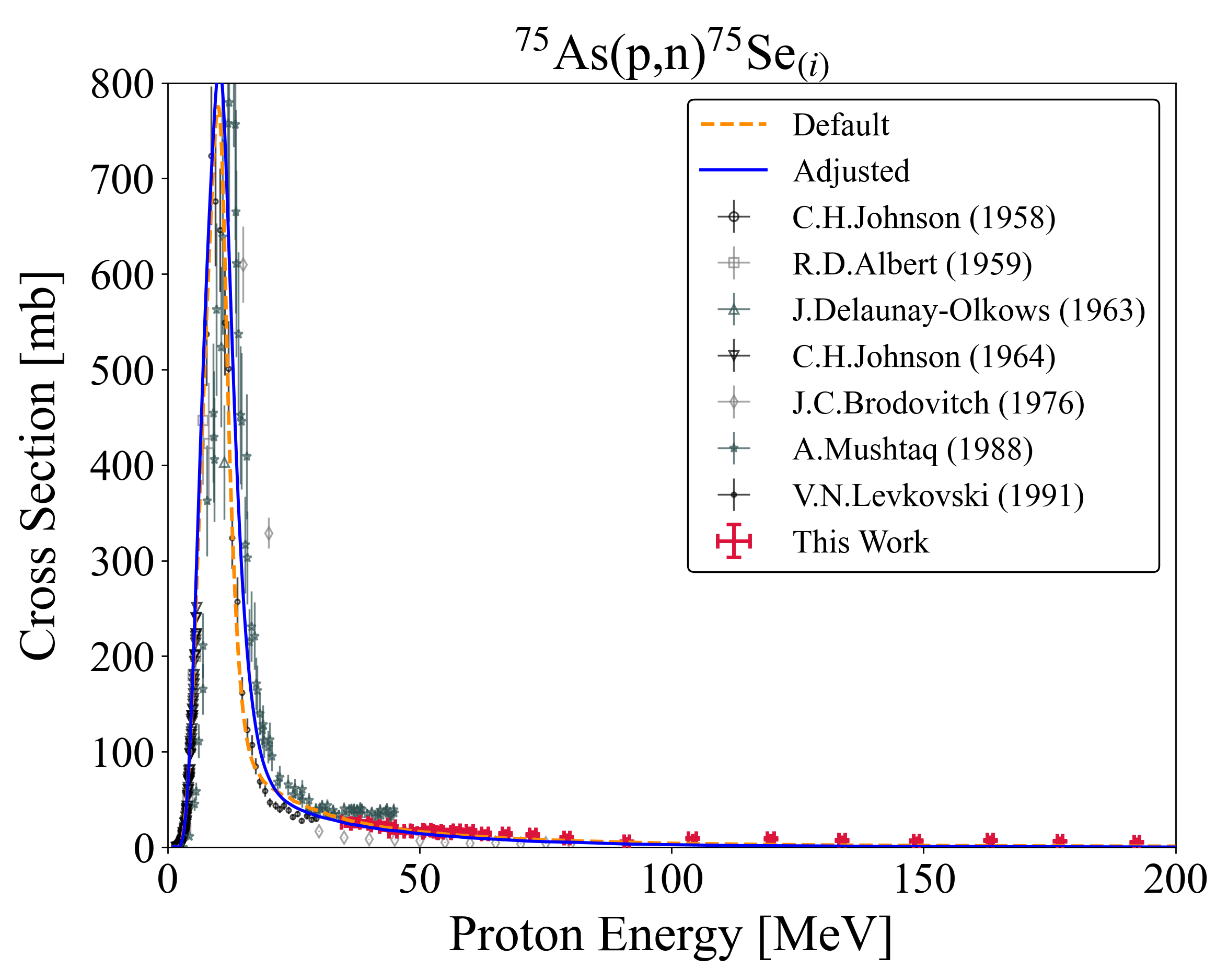}}
			\subfloat[\label{74As_Proc}]
		{\includegraphics[width=0.315\textwidth]{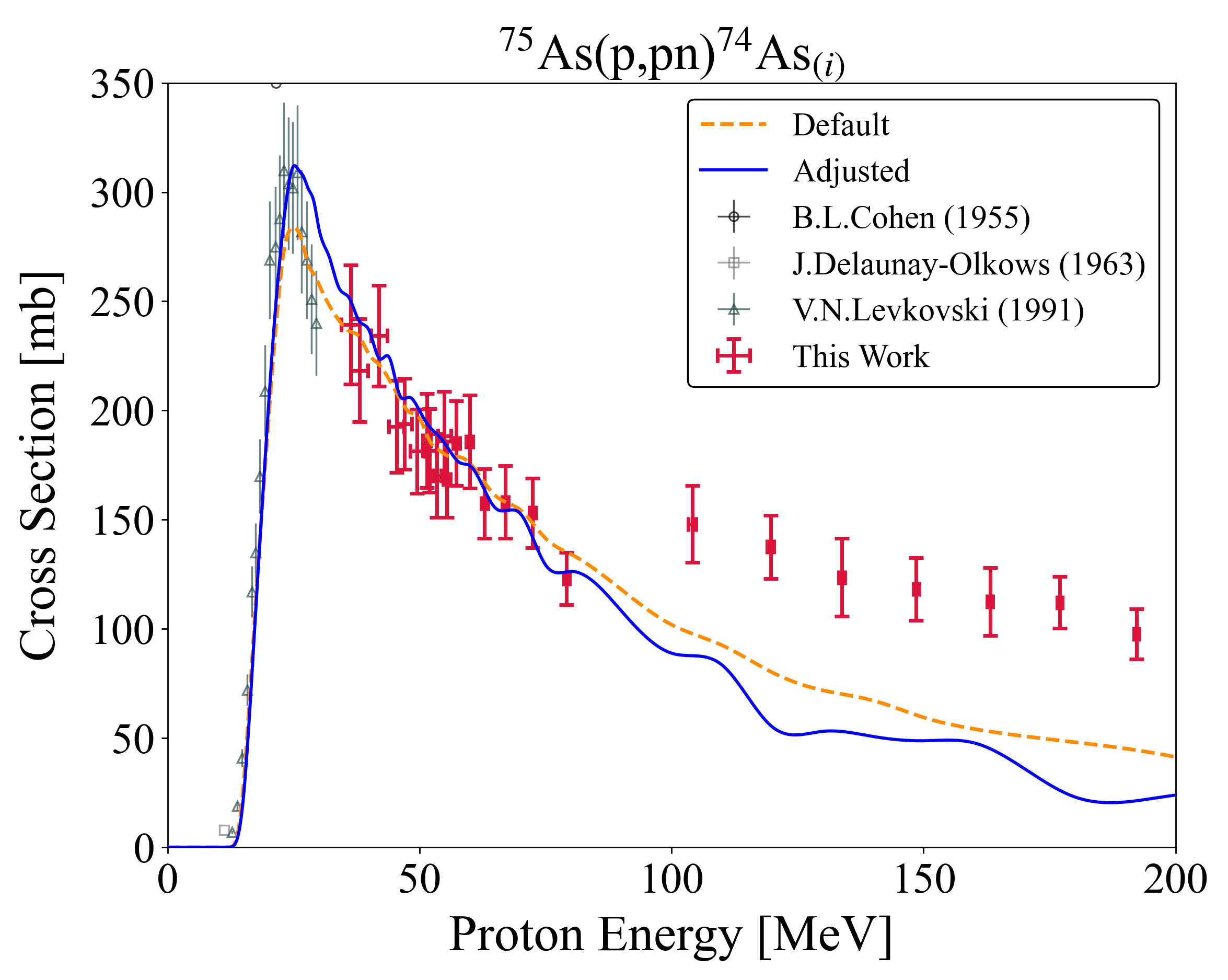}}
	\subfloat[\label{73Se_Proc}]
		{\includegraphics[width=0.315\textwidth]{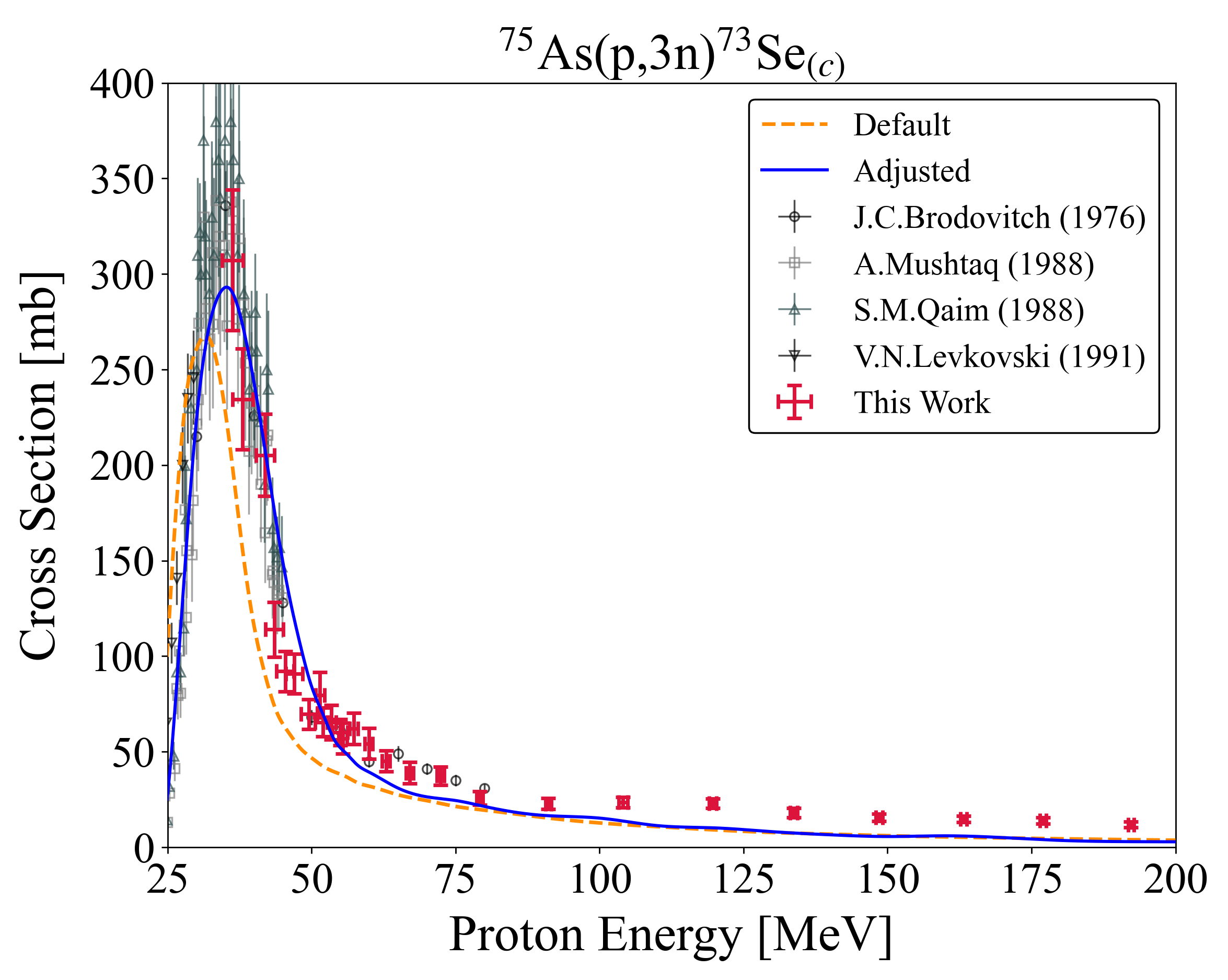}}\\
	\subfloat[\label{73As_Proc}]
		{\includegraphics[width=0.315\textwidth]{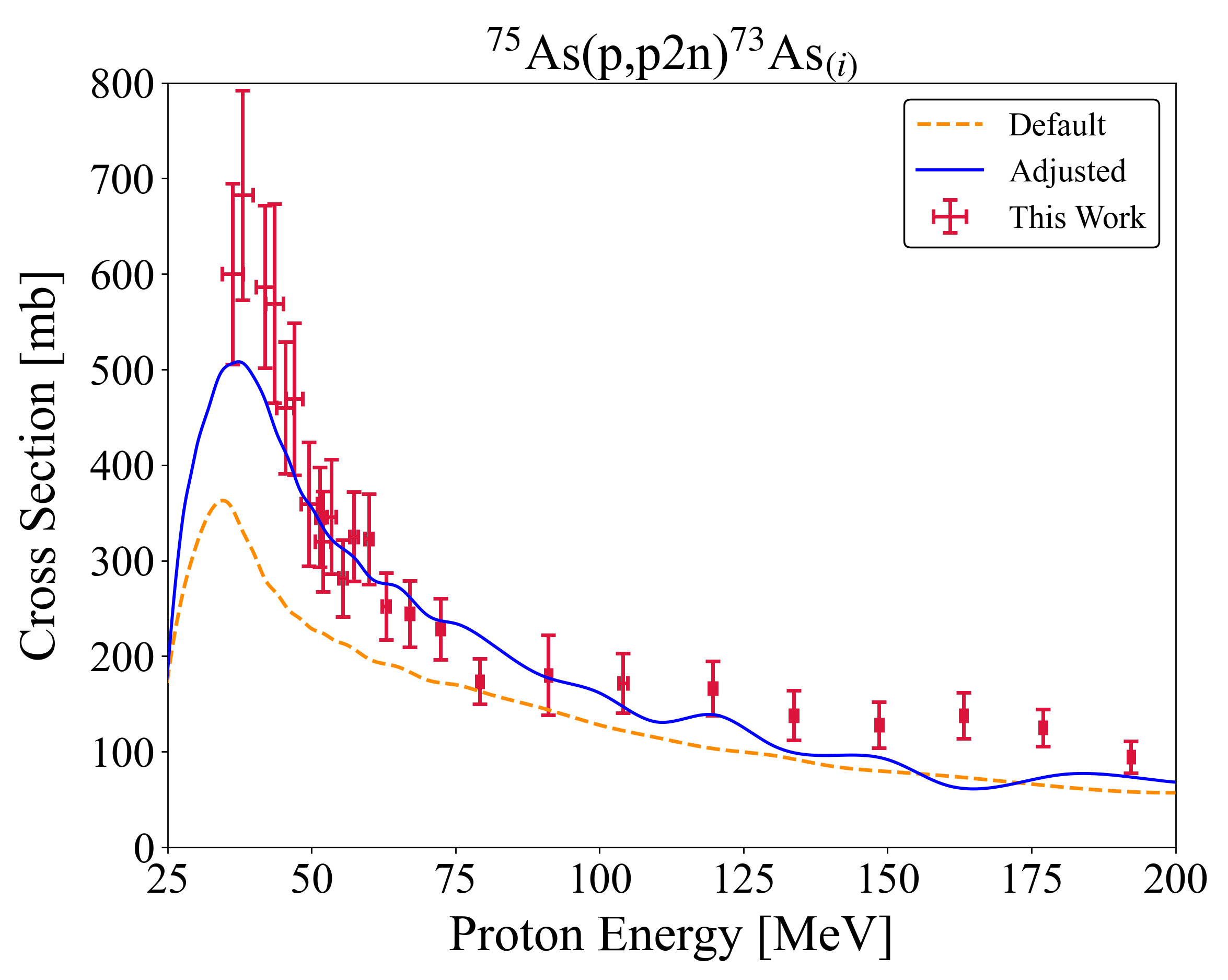}}
	\subfloat[\label{72Se_Proc}]
		{\includegraphics[width=0.315\textwidth]{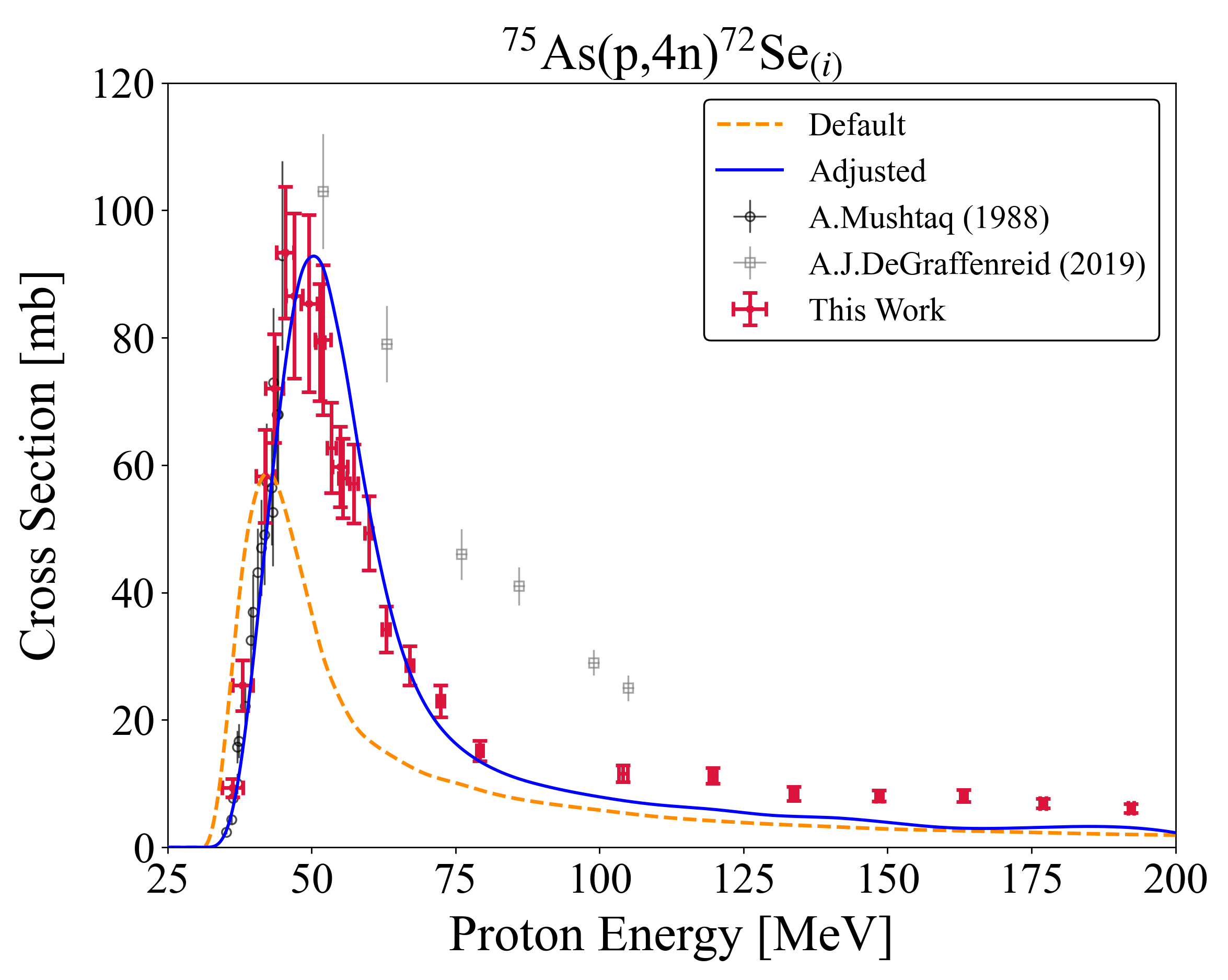}}
	\subfloat[\label{71As_Proc}]
		{\includegraphics[width=0.315\textwidth]{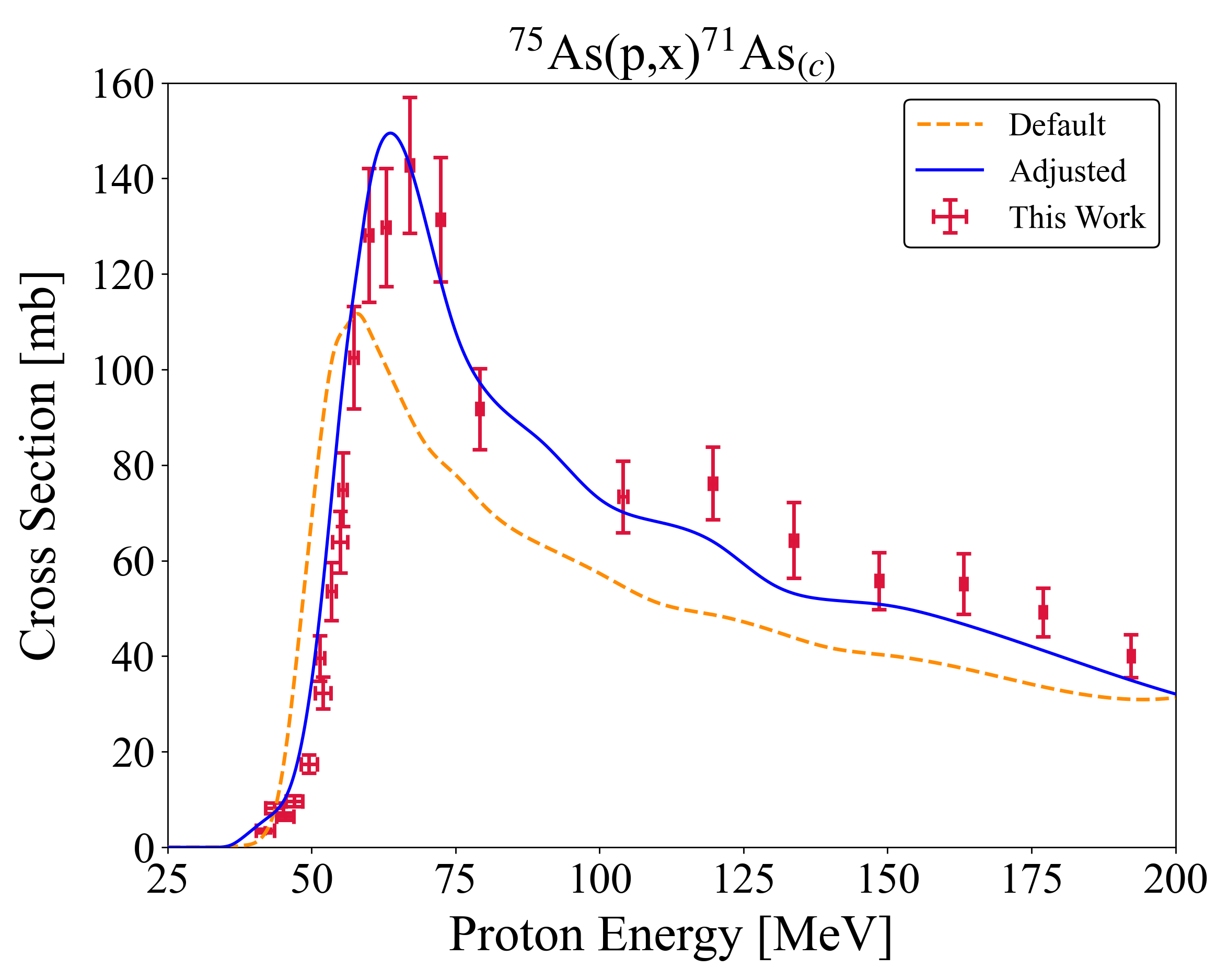}}\\
	\subfloat[\label{69Ge_Proc}]
		{\includegraphics[width=0.315\textwidth]{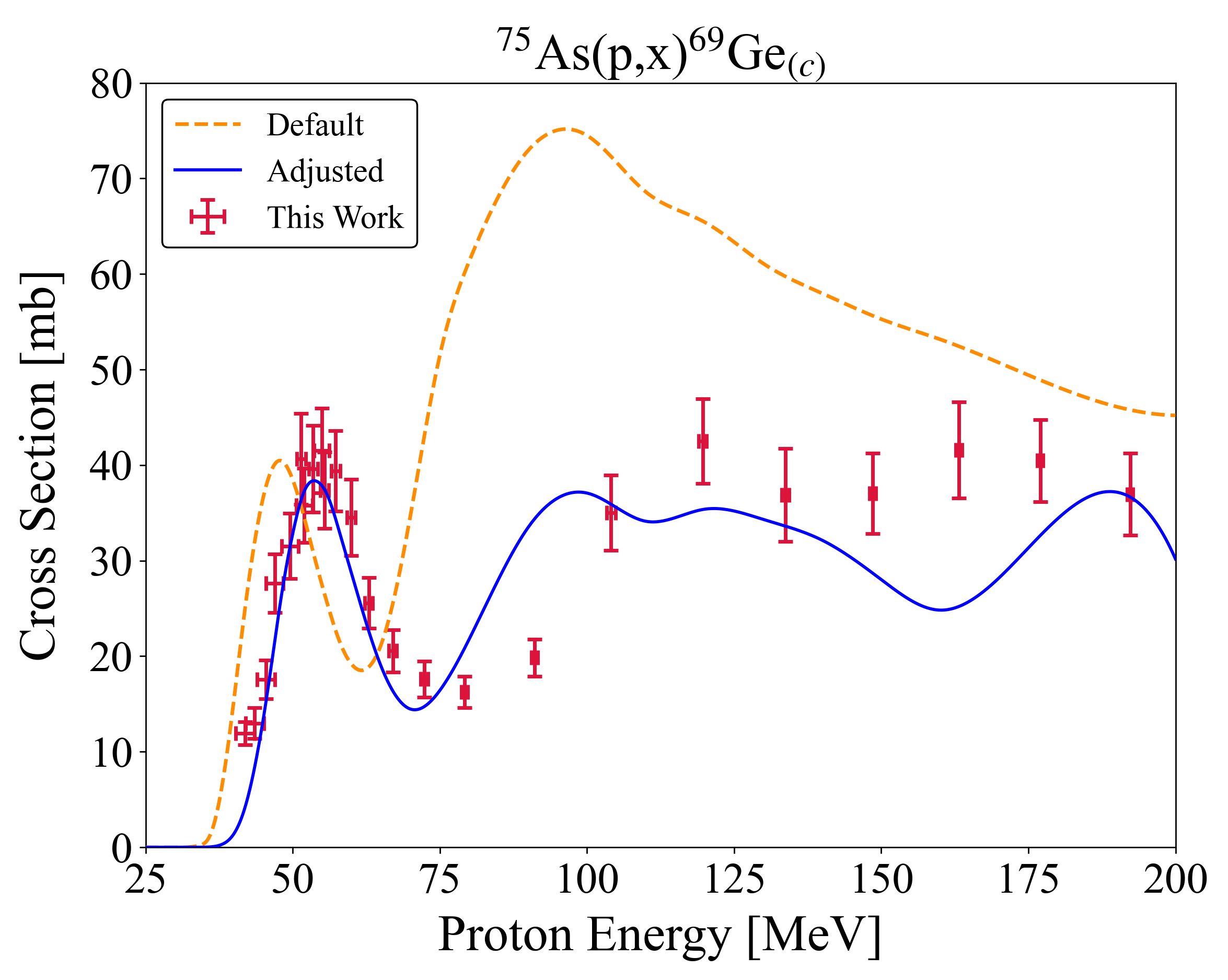}}
	\subfloat[\label{68Ga_Proc}]
		{\includegraphics[width=0.315\textwidth]{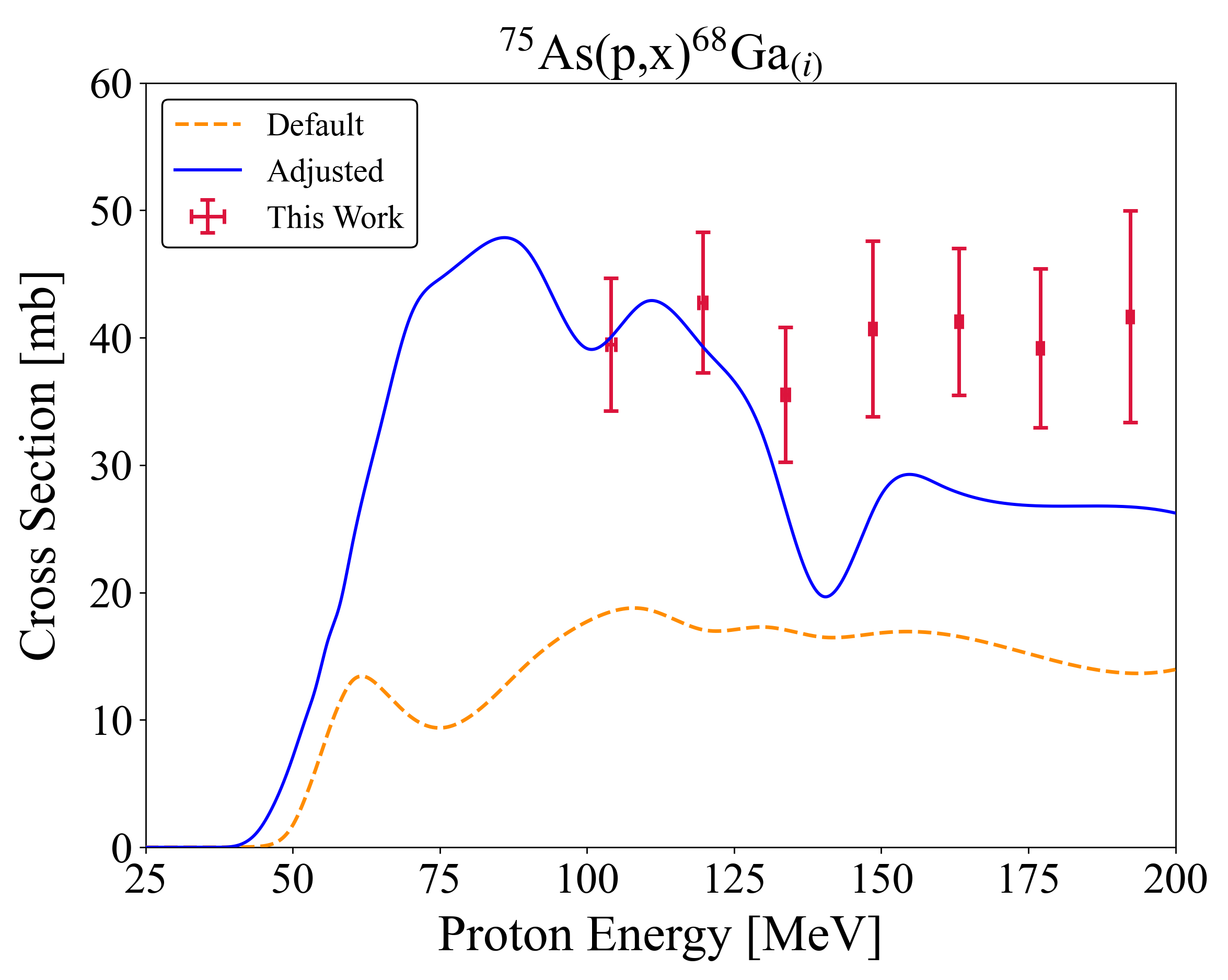}}
	\subfloat[\label{67Ga_Proc}]
		{\includegraphics[width=0.315\textwidth]{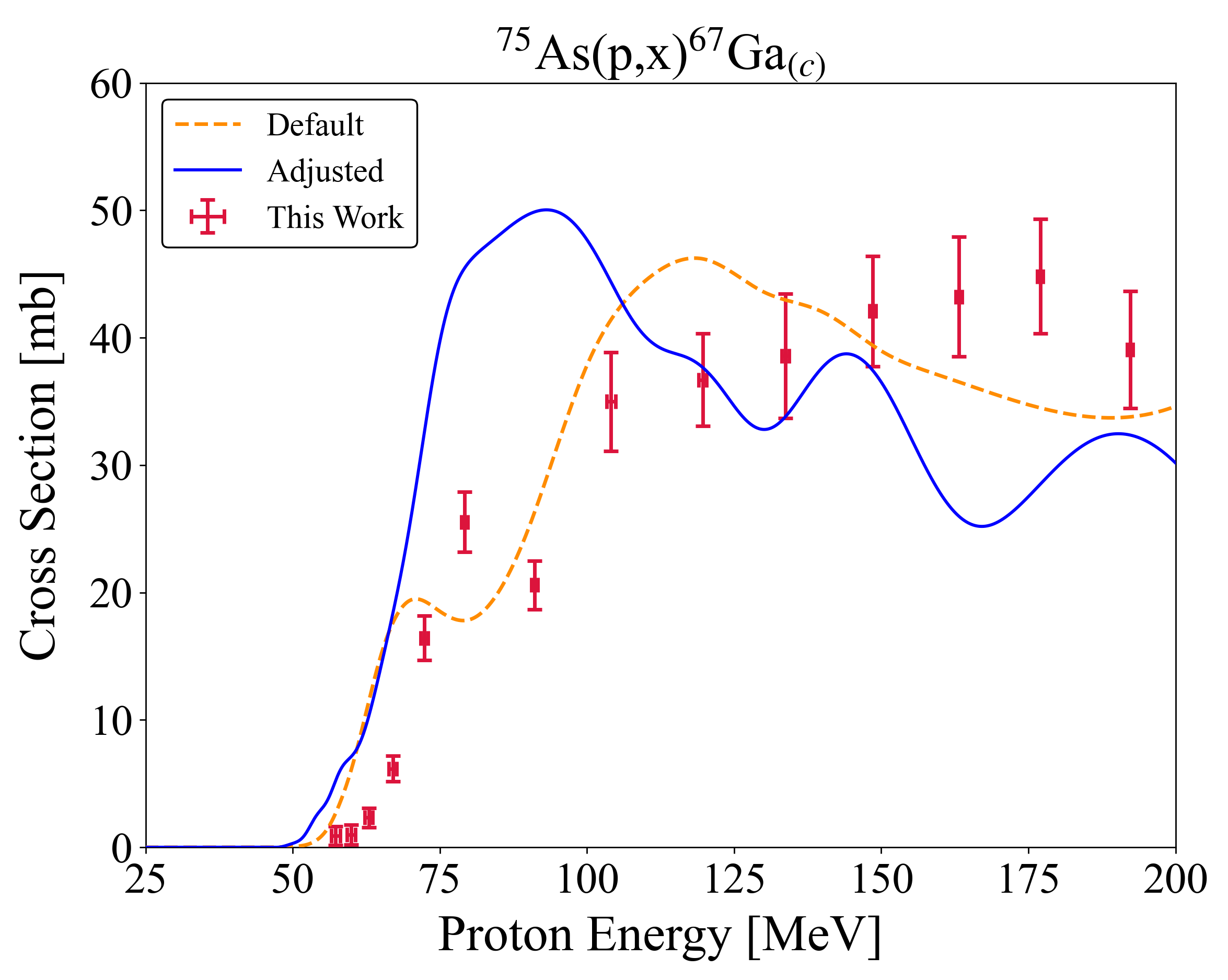}}
	\caption{TALYS default and adjusted calculations for residual products of proton-induced reactions on arsenic up to 200\,MeV.}\label{Proc_Fits}
\end{figure*}

\subsubsection{\label{ldadjust_analysis}Level Density Adjustments}
\begin{figure*}[!t]
	\subfloat[\label{ld_74As}]
		{\includegraphics[width=0.315\textwidth]{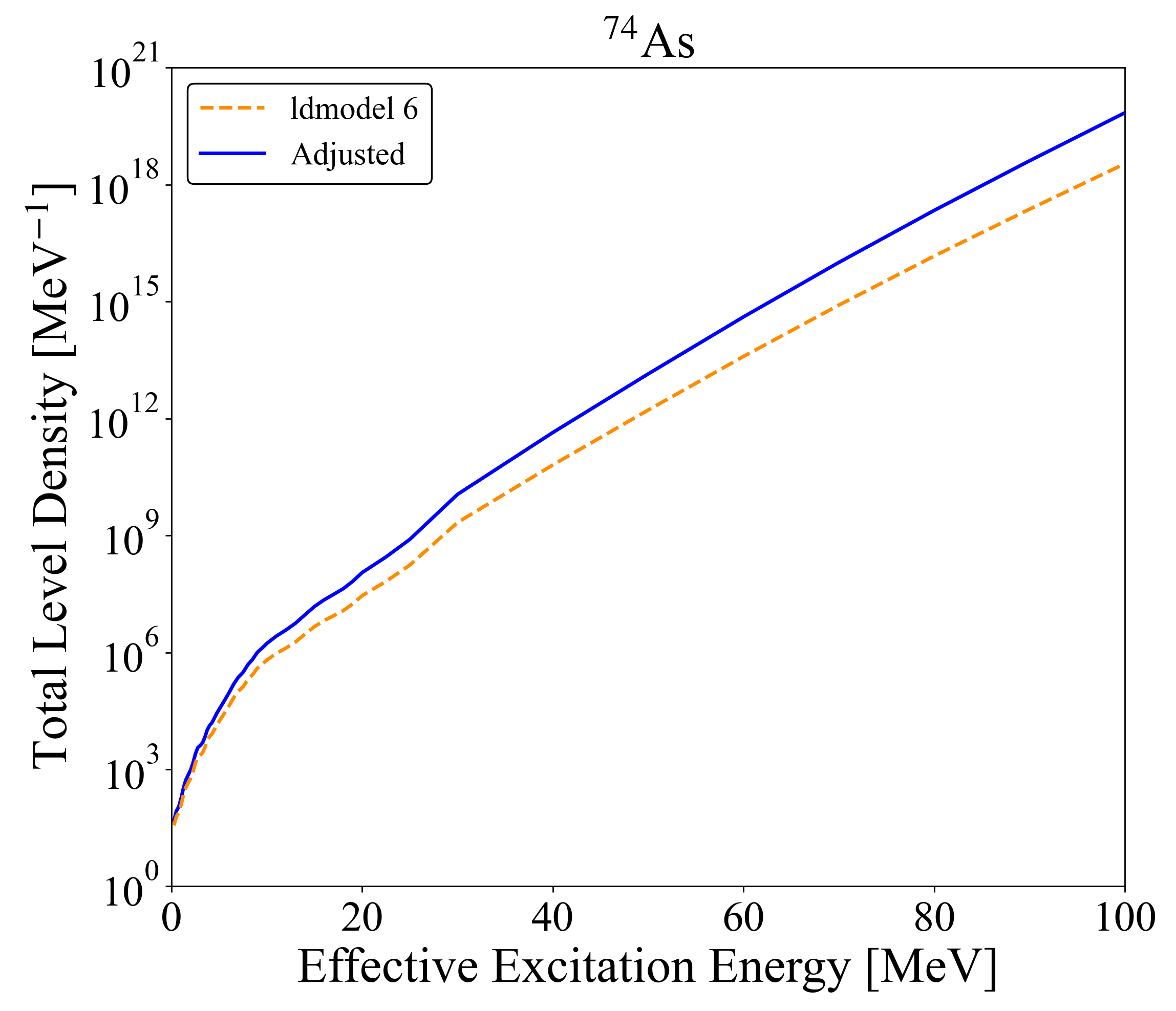}}
	\subfloat[\label{ld_73Se}]
		{\includegraphics[width=0.315\textwidth]{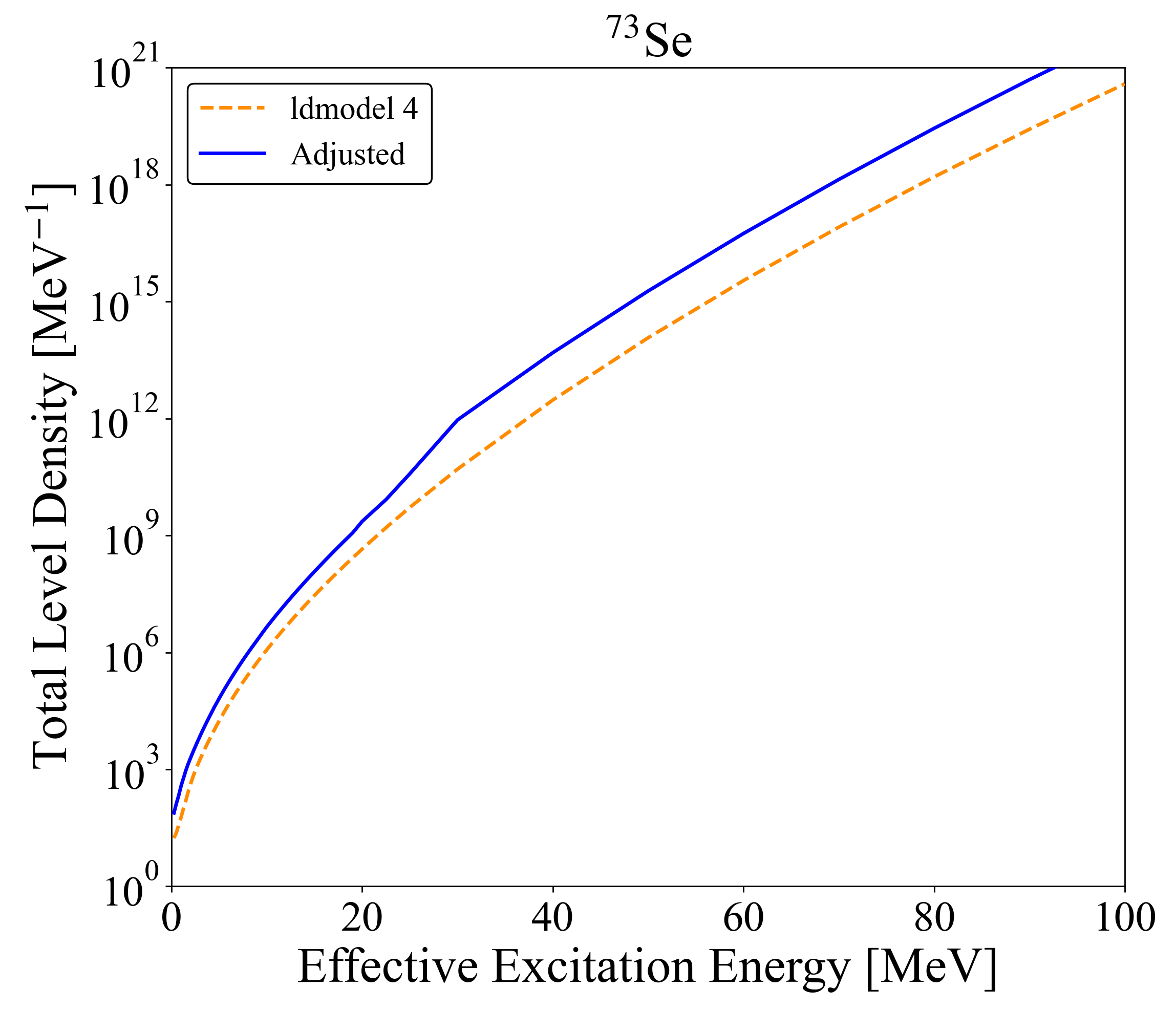}}
	\subfloat[\label{ld_73As}]
		{\includegraphics[width=0.315\textwidth]{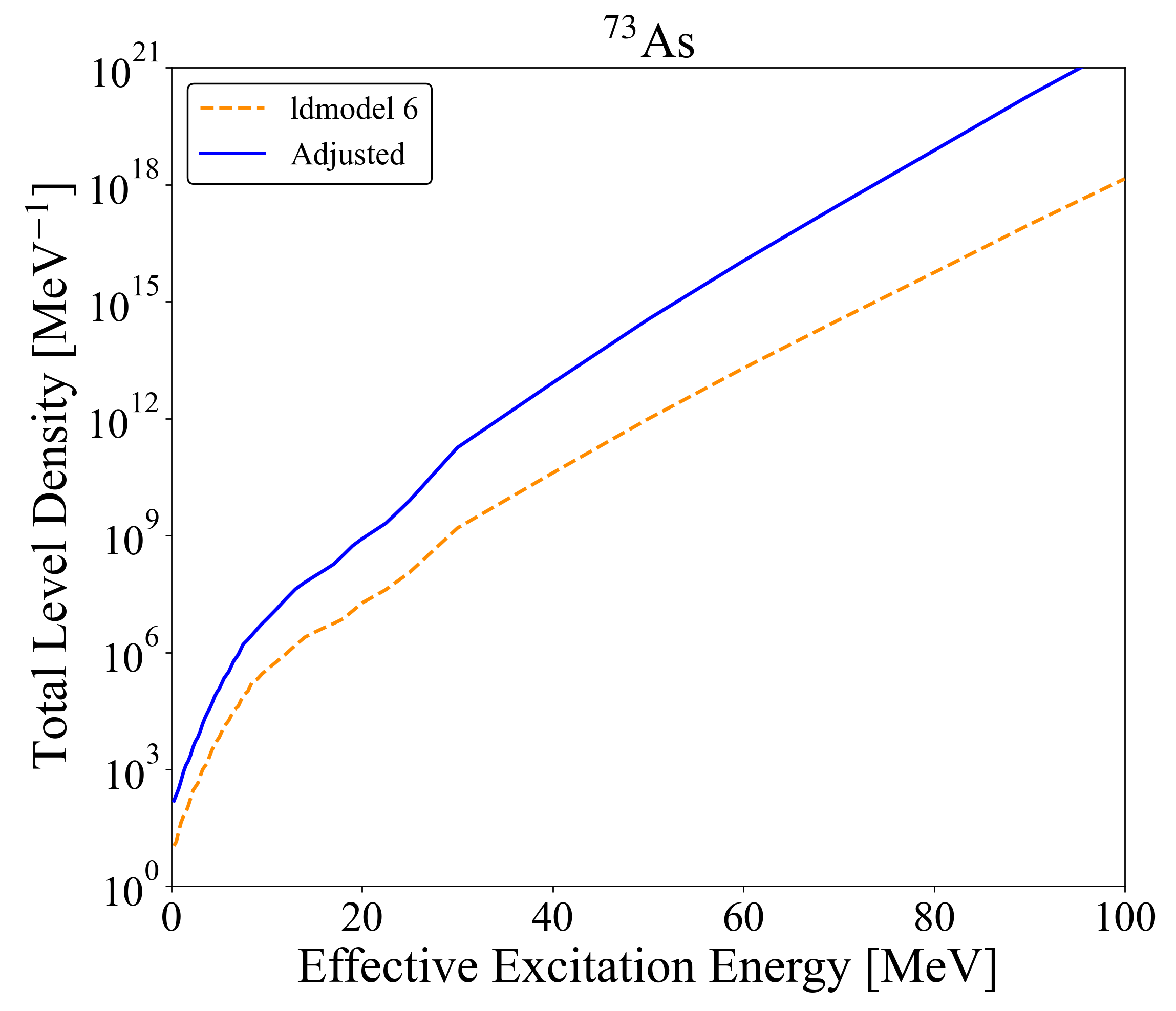}}\\
	\subfloat[\label{ld_72Se}]
		{\includegraphics[width=0.315\textwidth]{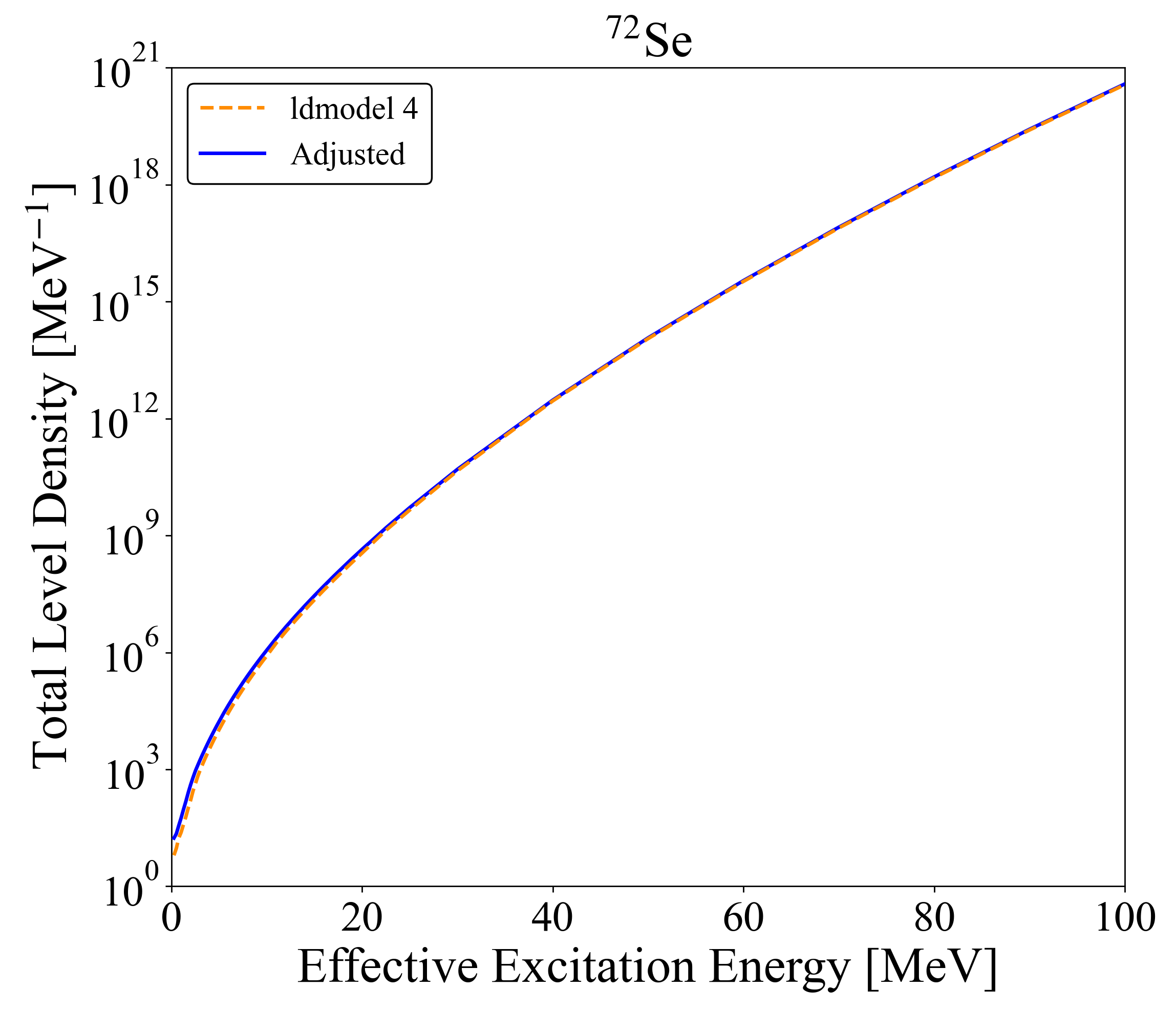}}
	\subfloat[\label{ld_71As}]
		{\includegraphics[width=0.315\textwidth]{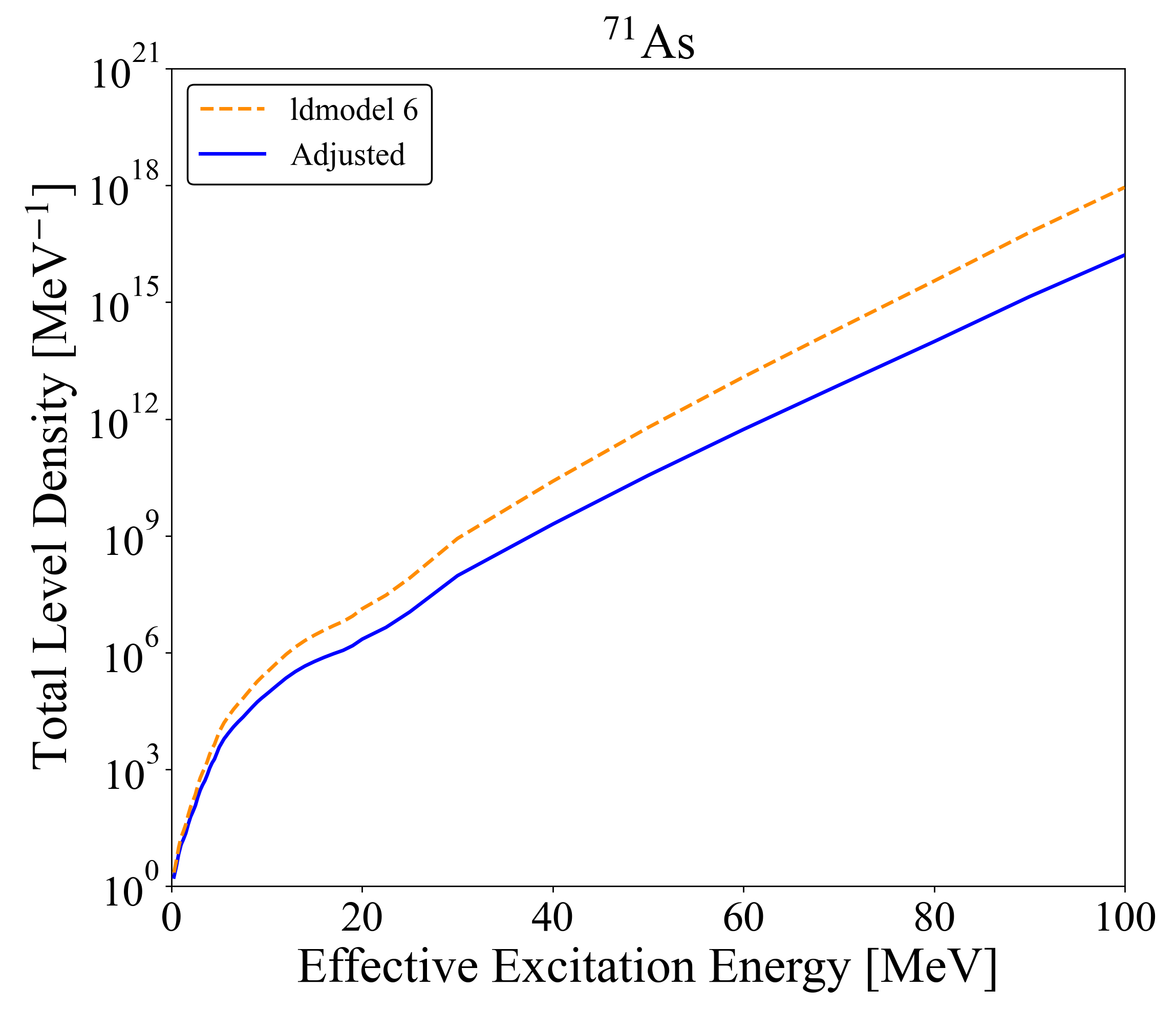}}
	\subfloat[\label{ld_69Ge}]
		{\includegraphics[width=0.315\textwidth]{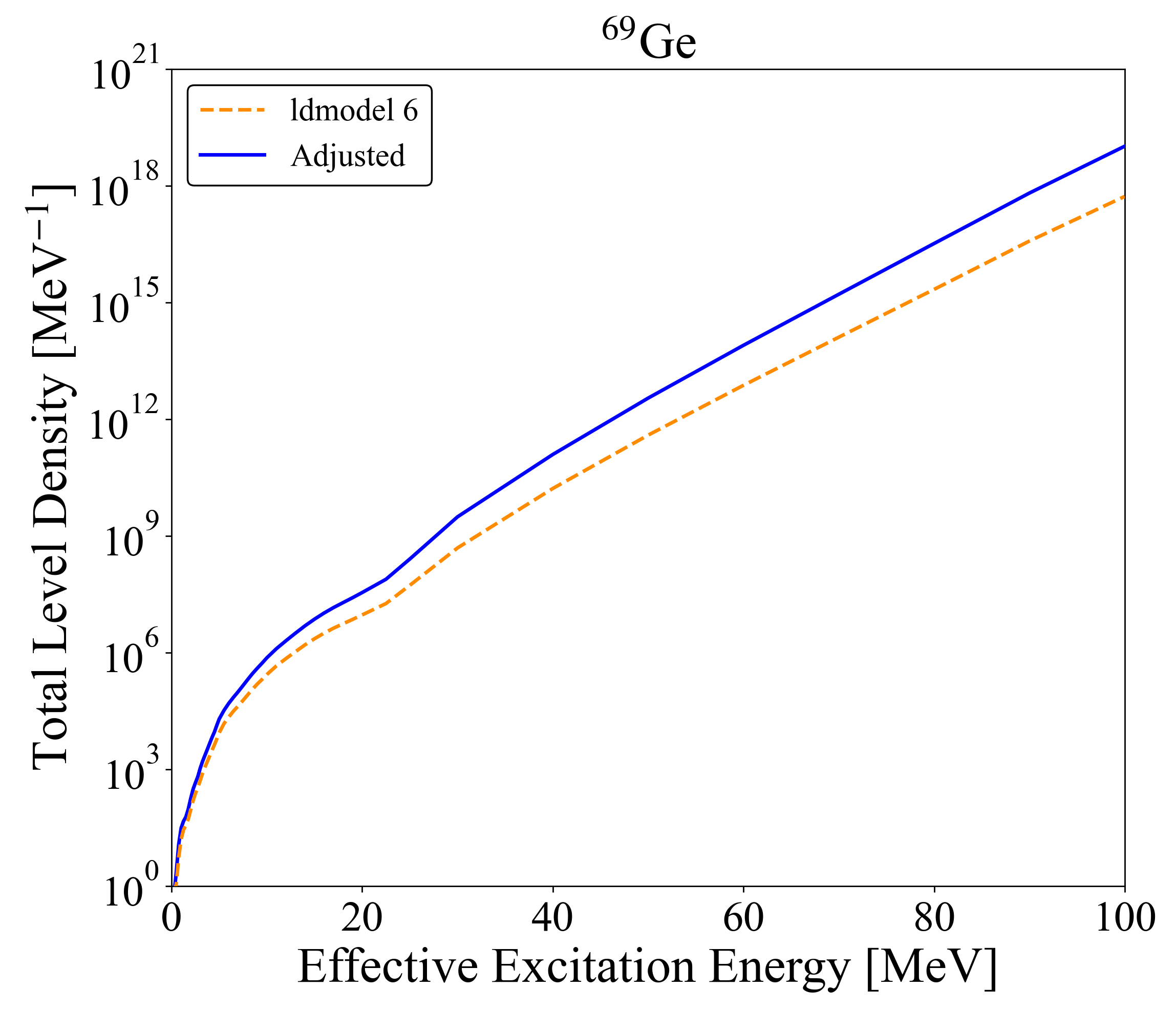}}\\
	\subfloat[\label{ld_69Ga}]
		{\includegraphics[width=0.315\textwidth]{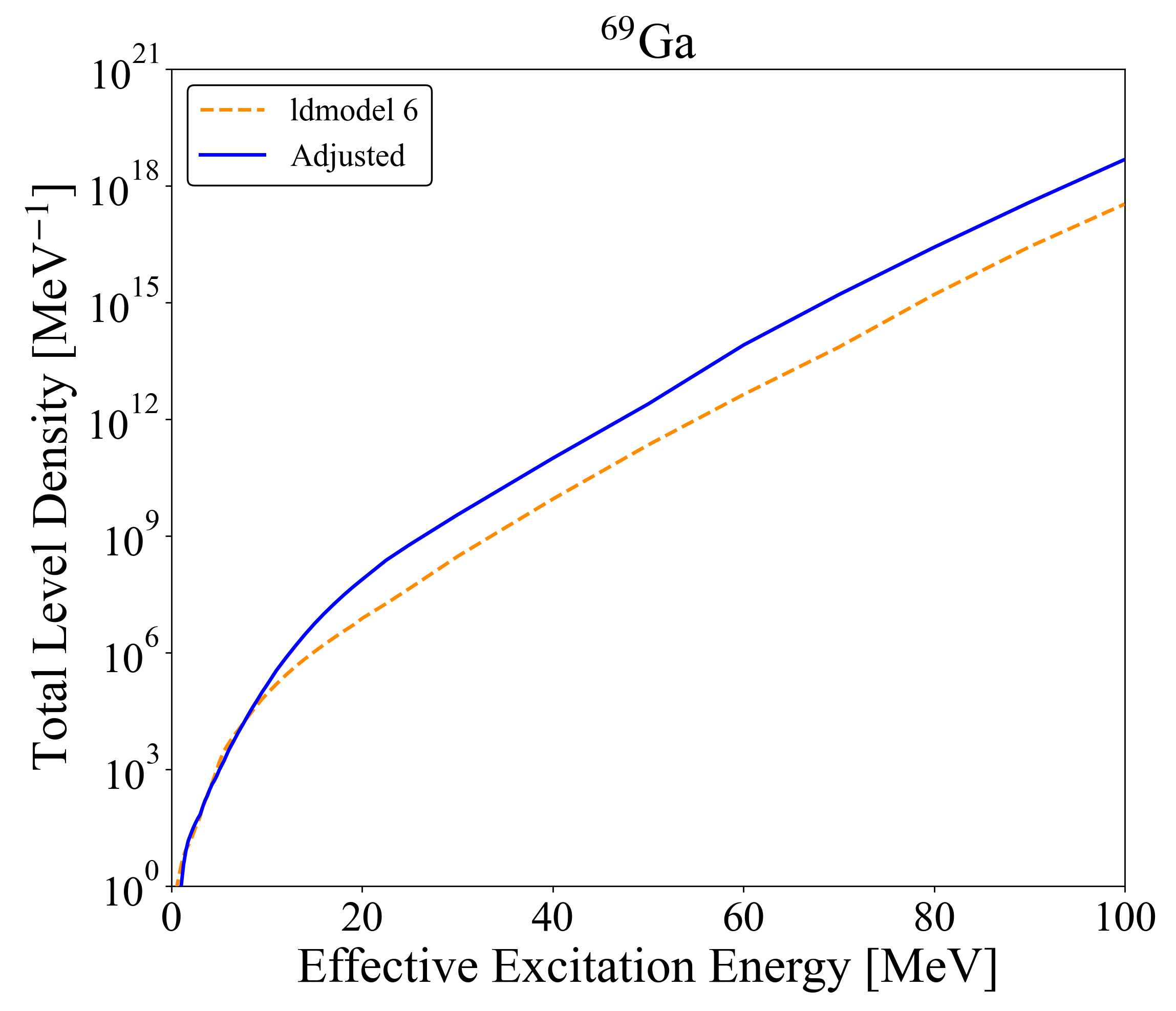}}
	\subfloat[\label{ld_68As}]
		{\includegraphics[width=0.315\textwidth]{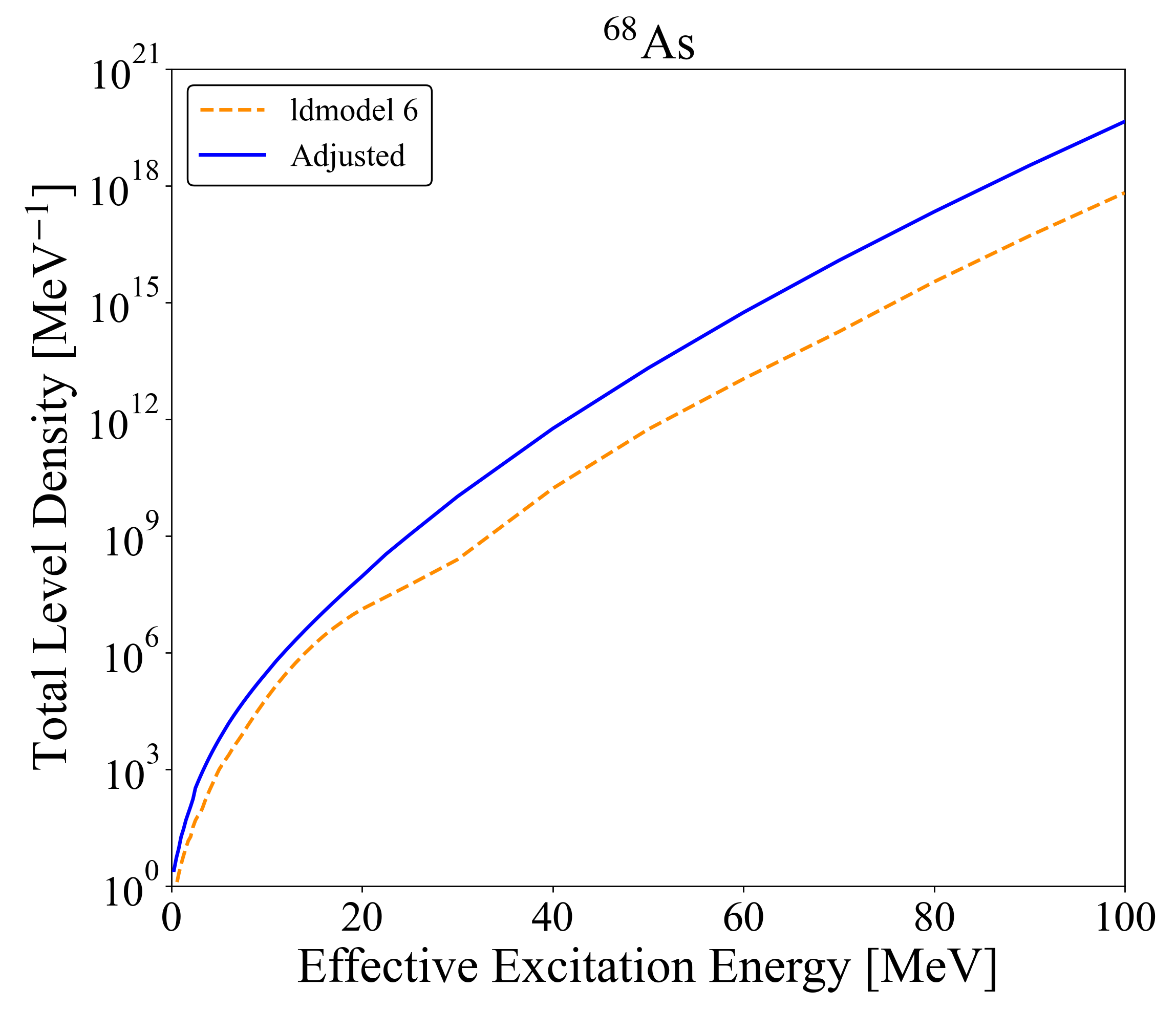}}
	\subfloat[\label{ld_67Ga}]
		{\includegraphics[width=0.315\textwidth]{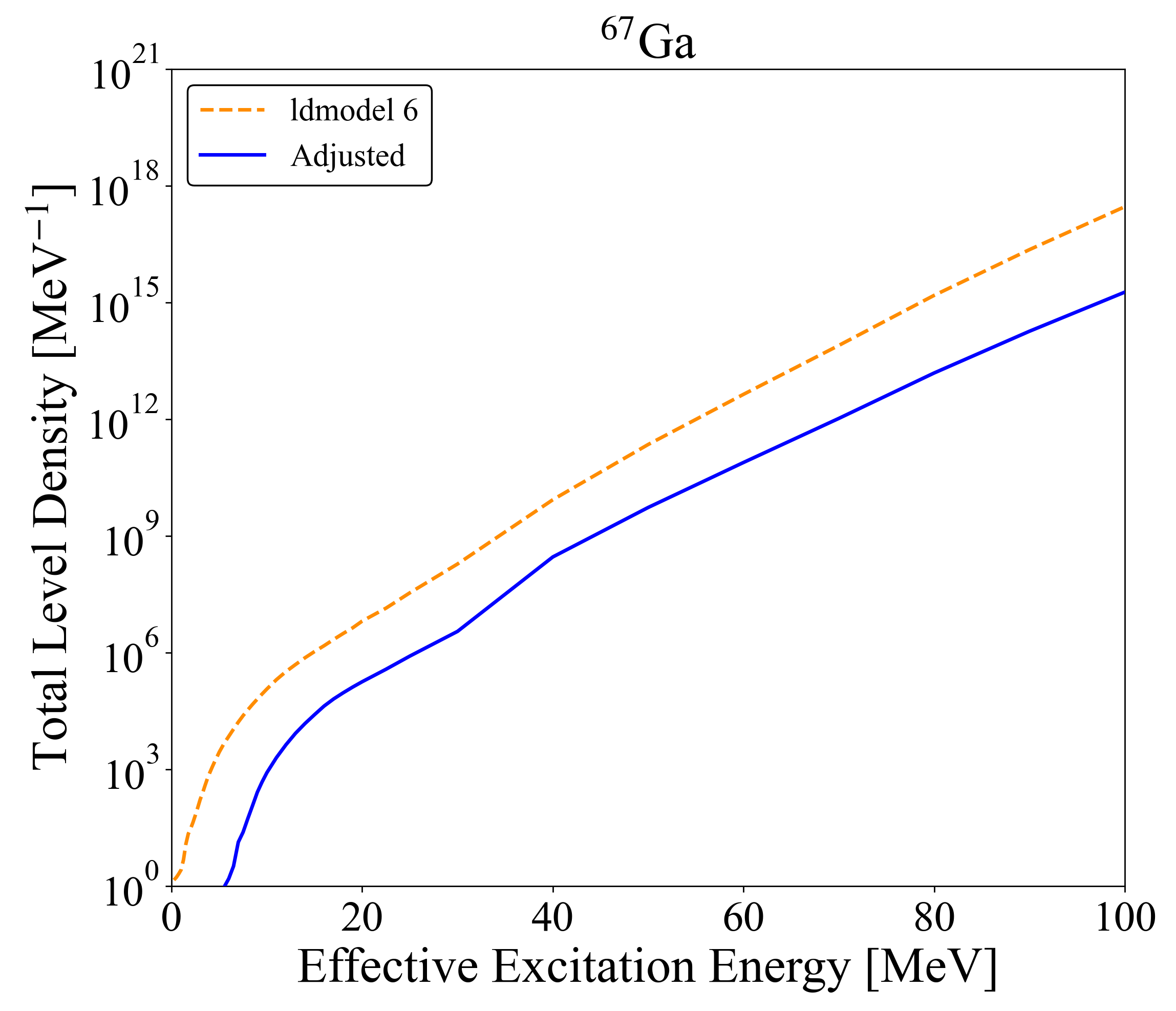}}
	\caption{Magnitude of all level density scalings implemented as part of the global fitting procedure for residual products of proton-induced reactions on arsenic up to 200\,MeV.}\label{ldadjust_figures}
\end{figure*}

Figure \ref{ldadjust_figures} directly shows the magnitude of all manually adjusted level density cases with reference to the base \texttt{ldmodel} choice. The total level density of $^{73}$As has been significantly increased (Figure \ref{ld_73As}), as warranted by the experimental data, while a significant decrease is seen in $^{67}$Ga although for less direct reasons (Figure \ref{ld_67Ga}).

The recommendation of these many level density changes is substantiated by the global fit success seen in Figure \ref{Proc_Fits} and described in Section \ref{Validation} but still requires more scrutiny. Furthermore, it is necessary to, at minimum, consider the impact of these level density changes on the residual product channels for which there were no experimental data and were unaccounted for either independently or cumulatively throughout the fitting procedure. Specifically, due to limitations of the activation technique to measure stable or some very-short lived nuclei production, the $^{75}$As(p,x)$^{74}$Se, $^{74-70}$Ge, and $^{73,71-69}$Ga channels, for $A>65$ and $Z>30$, were hidden from the fitting observations. Accordingly, it is essential to have a ``performance check" for these hidden channels, where the TALYS default and adjusted fits can be compared to monitor for any egregious shape or magnitude changes brought on by the level density adjustments. 

\begin{figure}[t]
{\includegraphics[width=1.0\columnwidth]{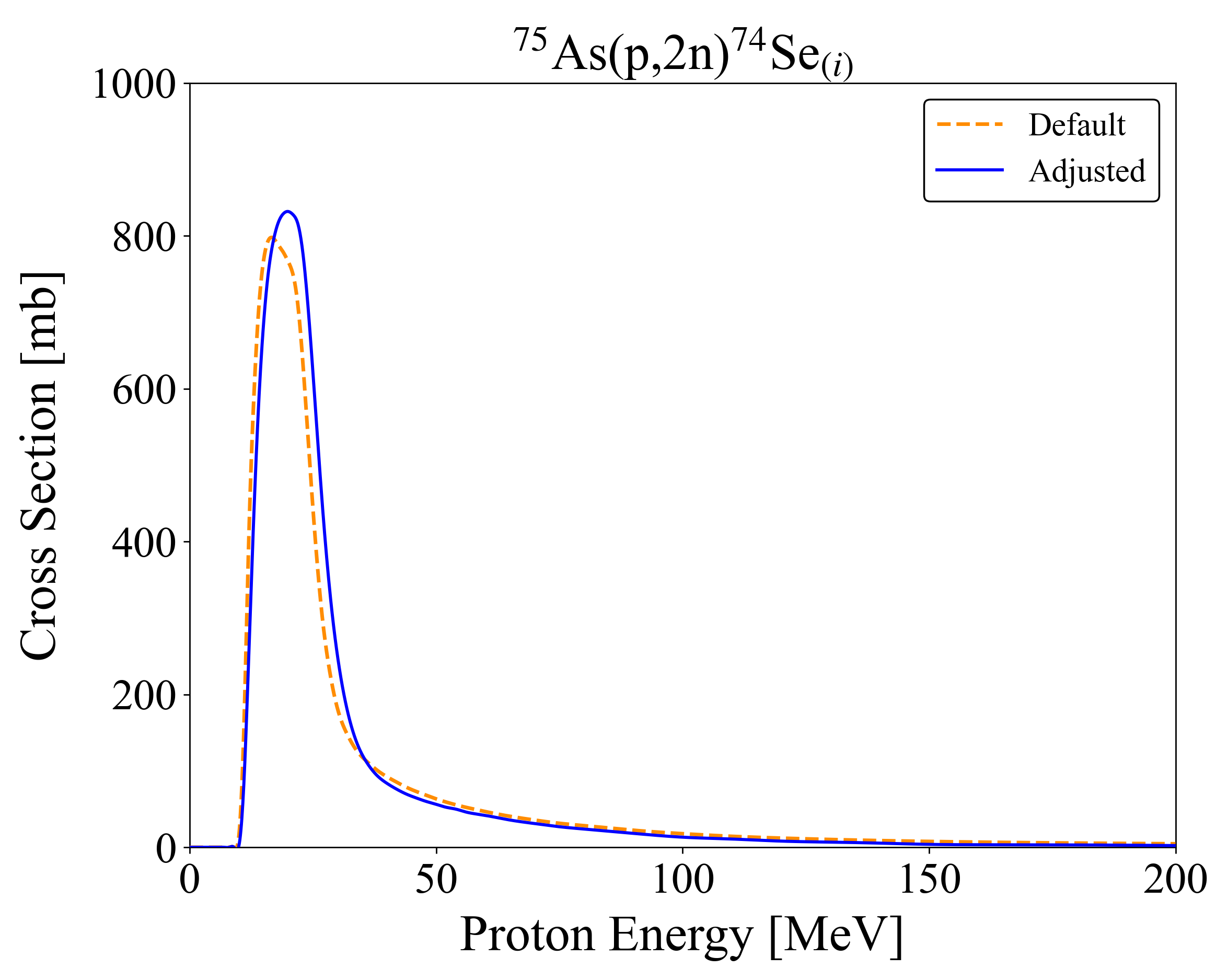}}
\vspace{-0.65cm}
\caption{Performance check for behaviour of the adjusted fit in $^{74}$Se, the largest unobserved channel from the fitting procedure.}
\vspace{-0.3cm}
\label{74Se_Proc}
\end{figure}

The fit performance for stable $^{74}$Se is of particular interest since had there been experimental data, the channel would have been one of the prominent excitation functions for the fitting procedure. This $^{74}$Se performance check is given in Figure \ref{74Se_Proc} and the difference between the default and adjusted is certainly acceptable.

In the unobserved Ge and Ga channels, there continues to be no obviously incorrect changes from the default to adjusted cases. Magnitude differences for most of these products reach $\approx$5-7\,mb and excitation function shape continuity is maintained within expectations. The adjusted $^{70}$Ge production is the most significantly changed hidden channel from default, with a maximum difference of $\approx$40\,mb in the compound peak region. Therefore, when confined to residual product datasets, there are no obvious indications that the bulk of the level density adjustments made here are not viable. Even if new experimental data were to be collected for these ``hidden" channels, which disagreed with the adjusted fit, it is likely that since no drastic changes have been made, the parameters can be properly updated to include the new information.

It is also worth remarking that using multiple level density models in this work is not a qualification or statement that one model more accurately reflects physical behaviour. Instead, we can only conclude that multiple level density models, and nuclide-specific changes, were simply scalings needed to best match the available experimental data, which has been seen in other work as well \cite{Nobre2020:LD,Kawano2006:ld}. There is likely no clear physical insight about the models that can be taken from these fits alone.
% performs better than another or is more correct than another within some set of circumstances

Perhaps some of the need for scaling is due to inconsistent or lacking discrete level data that feeds into the level density models. The residual products of interest generally exist off the line of stability and resonance parameters are unknown \cite{Kawano2006:ld,Grimes2008:ld,Roy2020:ld}. In $^{70}$As and $^{72}$As, only 68 and 65 experimental discrete levels, respectively, as stored in the RIPL-3 database, inform the level density calculations \cite{TalysManual}. This is compared to isotopes such as $^{71,73,75}$As where over 120 experimental levels each are used.
% and $^{74}$As where only 7 levels are theoretically calculated.
%, meaning that 30+ levels are generated from theoretical structure calculations to reach the TALYS 100 level threshold \cite{TalysManual}.

A similar pattern exists for $^{72}$Se where only 52 experimental discrete levels inform calculations, as compared to much more well-characterized $^{73-76}$Se isotopes. If there are missing levels relatively low in the level scheme, then the level density model may be adjusted to the wrong number of assumed complete levels. This lack of structure data exists for $^{66,67}$Ge and $^{66,68}$Ga within their respective isotope chains as well. Ultimately, it is conceivable that incomplete structure data leads to numerous compensating level density effects in this mass region, which may themselves be a key contributor to the adjusted scalings as opposed to any inherent issues with the models \cite{Kawano2006:ld,Grimes2008:ld,Roy2020:ld}.

It is also possible that a disregard of isospin effects in the current TALYS calculations, missing collective enhancement effects for nuclei far from stability, or deterioration of the microscopic level density models altogether at the high excitation energies relevant to this work, have prompted the need for the corrective scalings \cite{Roy2020:ld,Grimes1972:Isospin}. A future experiment examining $\alpha$+$^{72}$Ge, which populates the same $^{76}$Se compound system as p+$^{75}$As but with different isospin, could provide some additional information. Relevant future research may also include examinations of smooth transitions between different level density models as a function of mass difference from the target nucleus or separate structure-based level density model-mixing schemes. The implementation of different level density models in this manner merits specific interest because it has shown to be needed for p+$^{75}$As here and p+$^{93}$Nb in \textcite{Fox2020:NbLa}.
%or collective effects which impact models far from stability
%N=Z discussion here?, something about alpha preformation or collectivity increases that gets modeled better by 6 rather than 4??
%These nuclide-specific cases include ... As an interative consequence to these changes, competitions with correlated production channels were altered, and required compensating nuclide-speciific level density adjustements in ...

The overall viability of the level density adjustments in this modeling work in combination with the other modeling parameter changes are further justified in Section \ref{Validation}.

\newpage
\vspace{-0.15cm}

\subsection{\label{Validation}Parameter Adjustment Validation}
\begin{figure*}[!t]
	\subfloat[\label{72As_Proc}]
		{\includegraphics[width=0.315\textwidth]{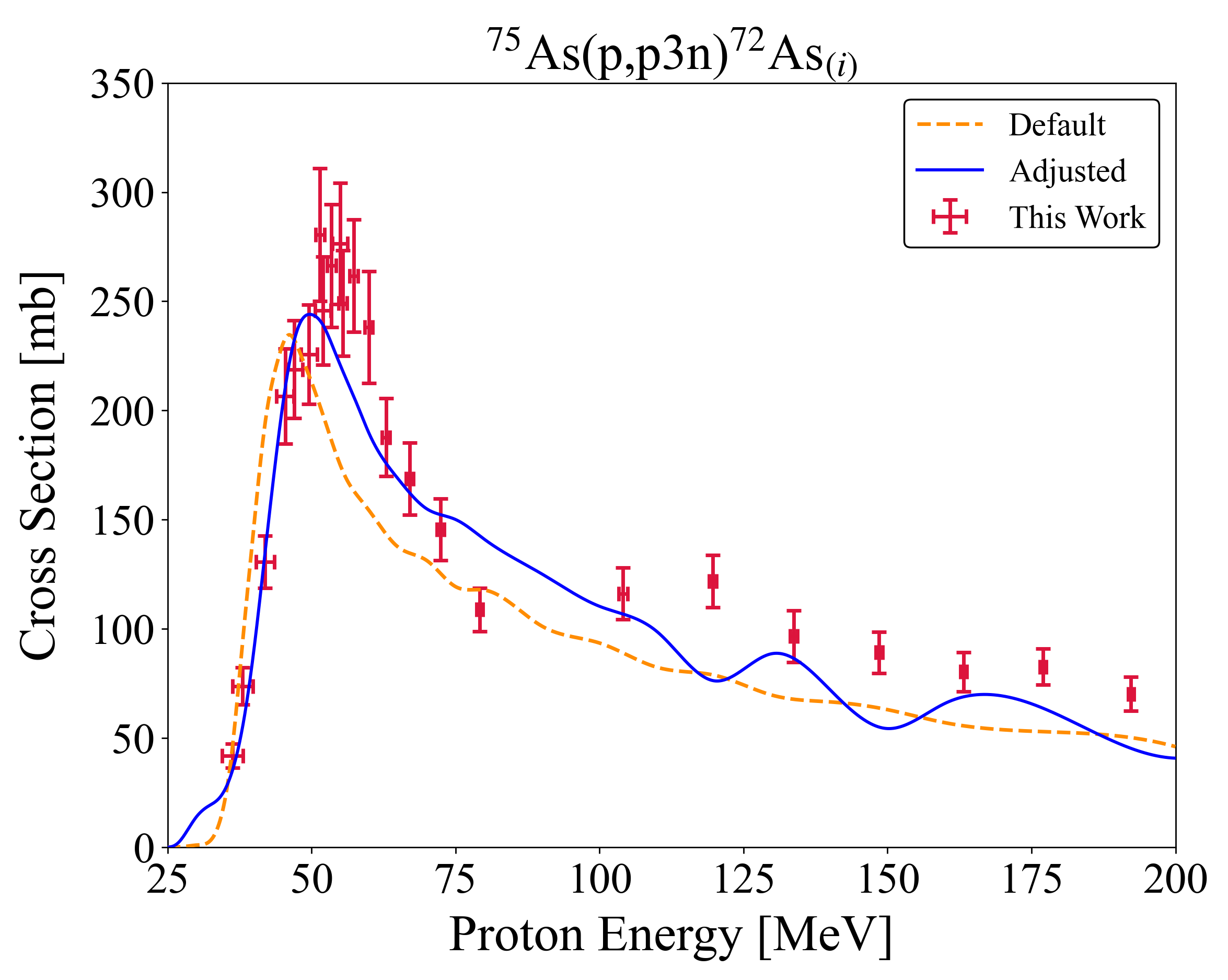}}
			\subfloat[\label{72Ga_Proc}]
		{\includegraphics[width=0.315\textwidth]{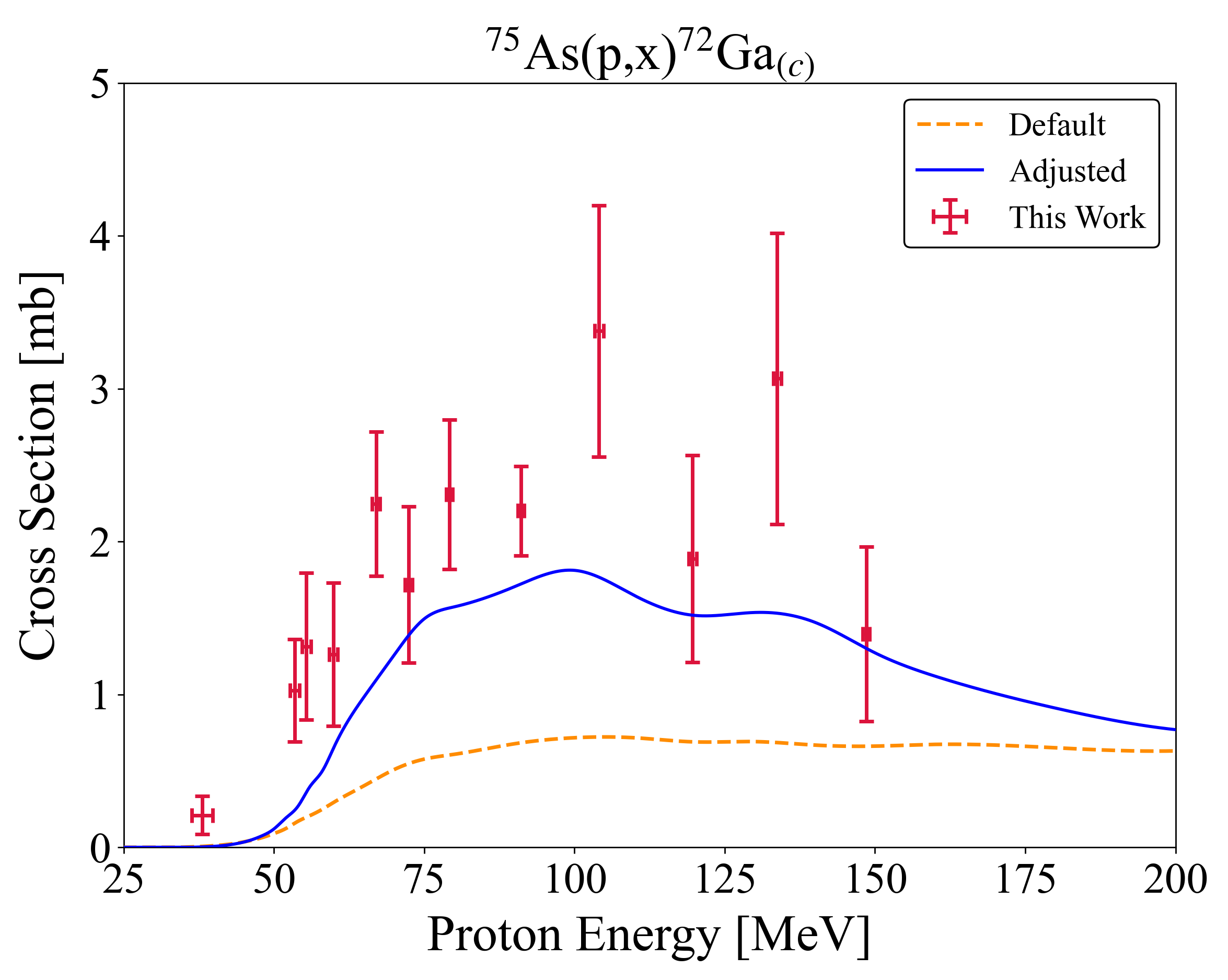}}
	\subfloat[\label{70As_Proc}]
		{\includegraphics[width=0.315\textwidth]{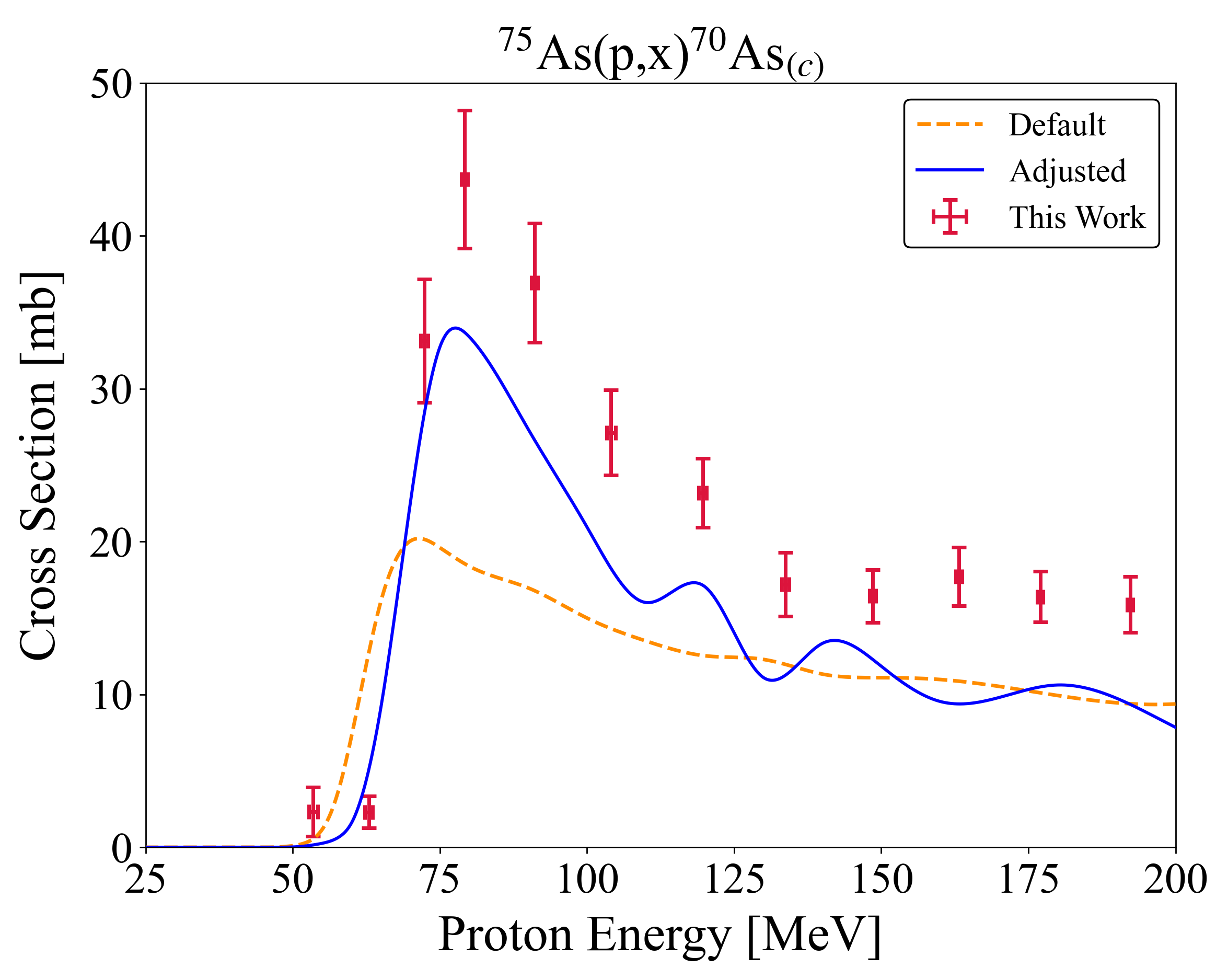}}\\
			\subfloat[\label{69mZn_Proc}]
		{\includegraphics[width=0.315\textwidth]{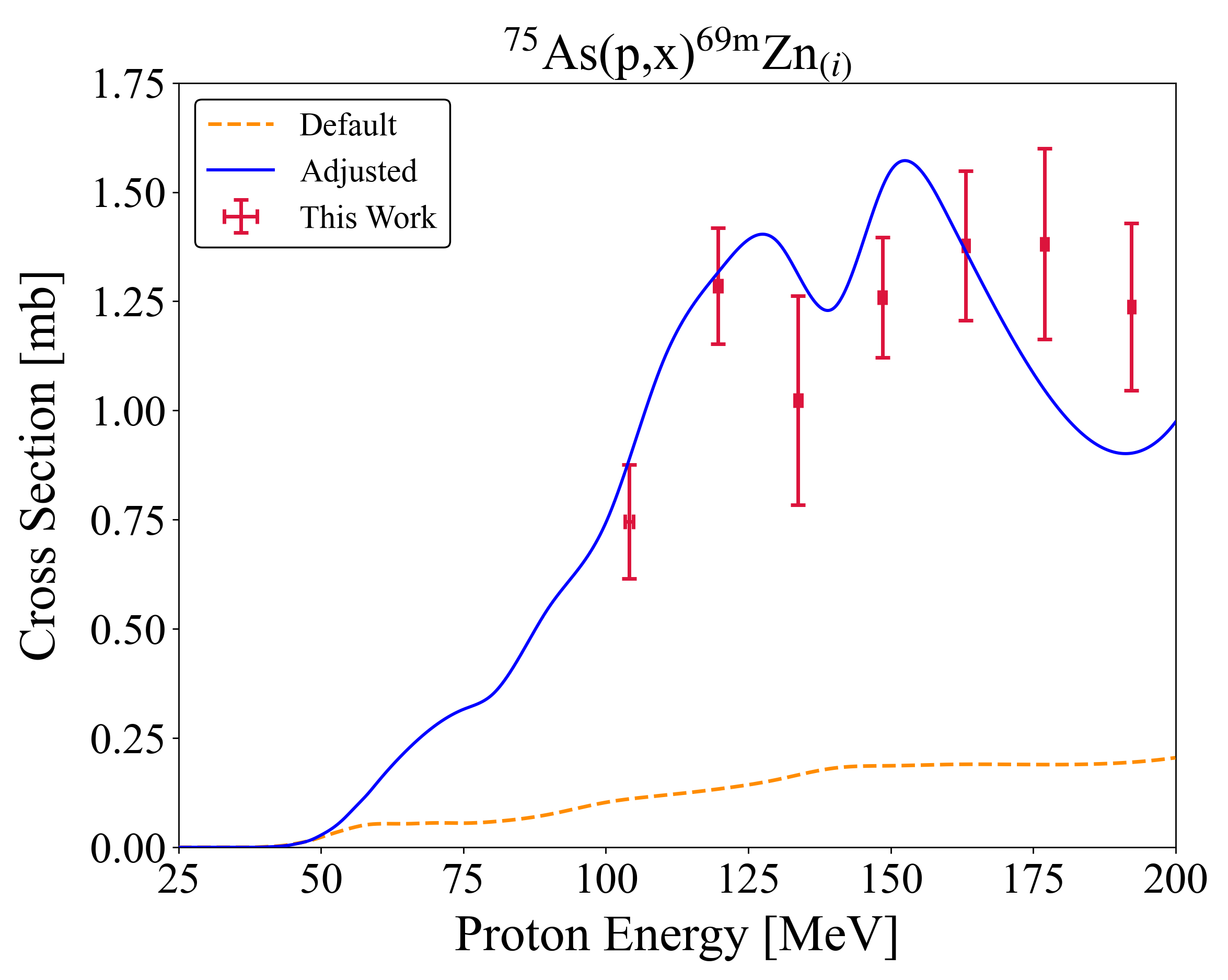}}
	\subfloat[\label{68Ge_Proc}]
		{\includegraphics[width=0.315\textwidth]{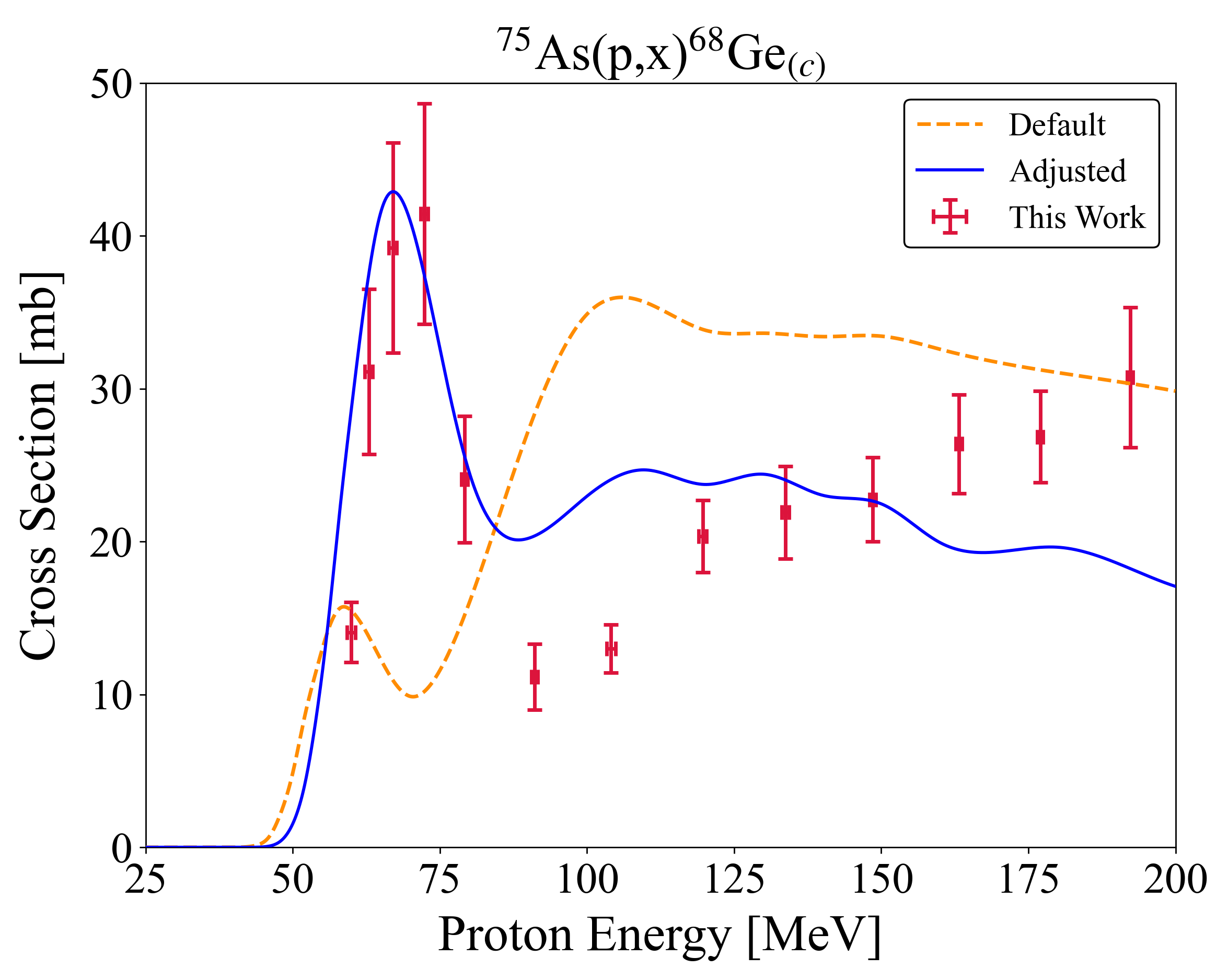}}
	\subfloat[\label{66Ge_Proc}]
		{\includegraphics[width=0.315\textwidth]{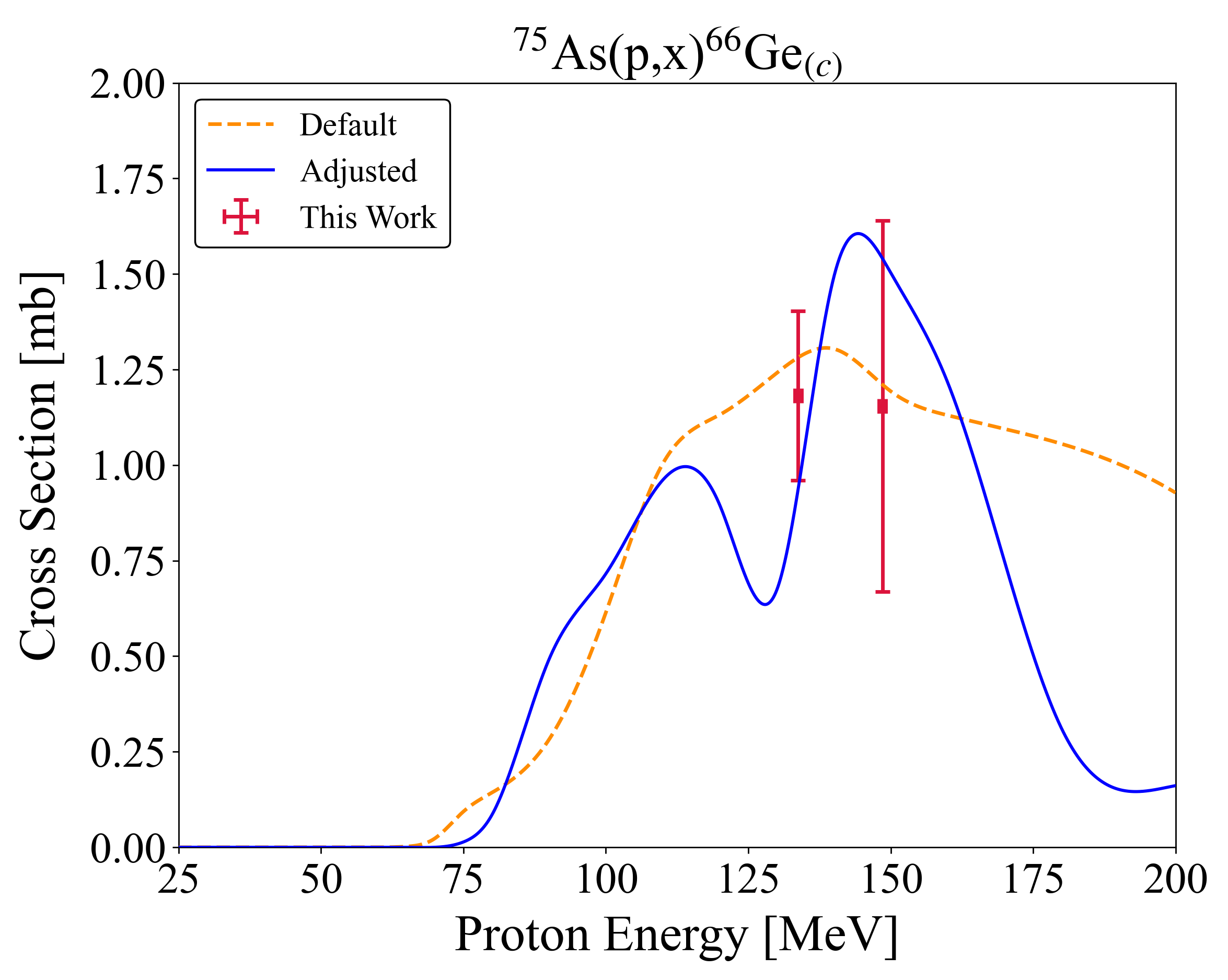}}\\
	\subfloat[\label{66Ga_Proc}]
		{\includegraphics[width=0.315\textwidth]{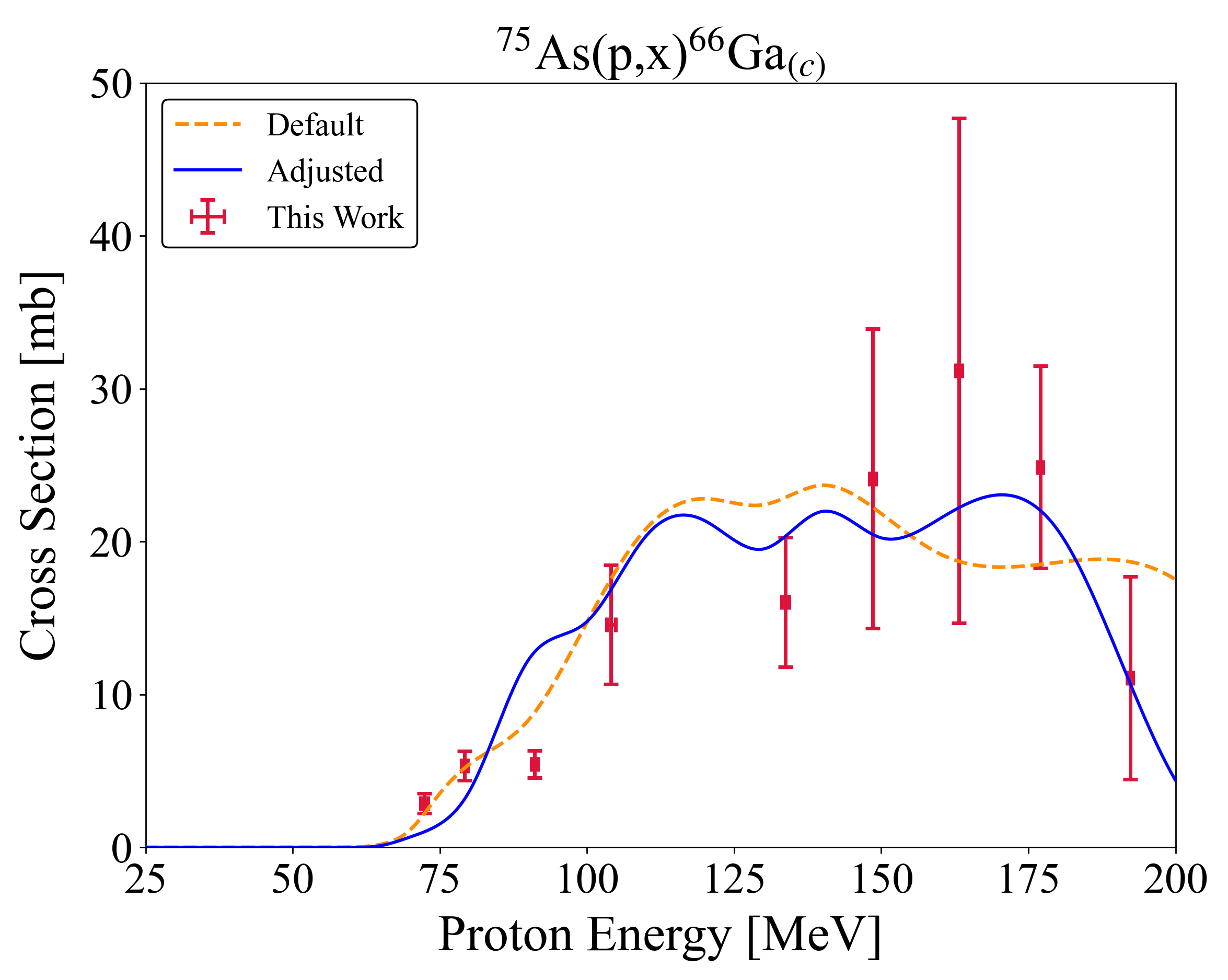}}
	\subfloat[\label{65Zn_Proc}]
		{\includegraphics[width=0.315\textwidth]{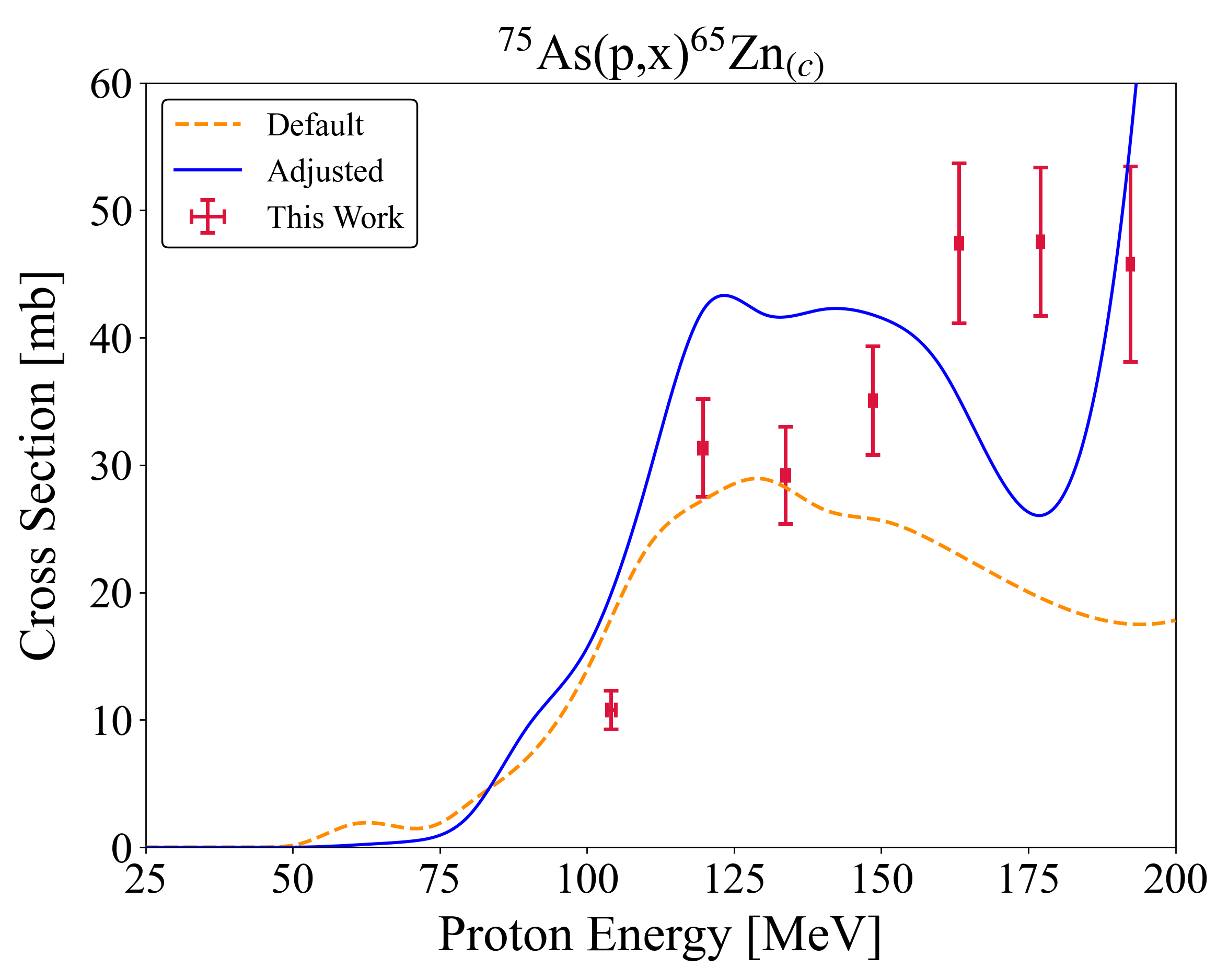}}
	\subfloat[\label{60Co_Proc}]
		{\includegraphics[width=0.315\textwidth]{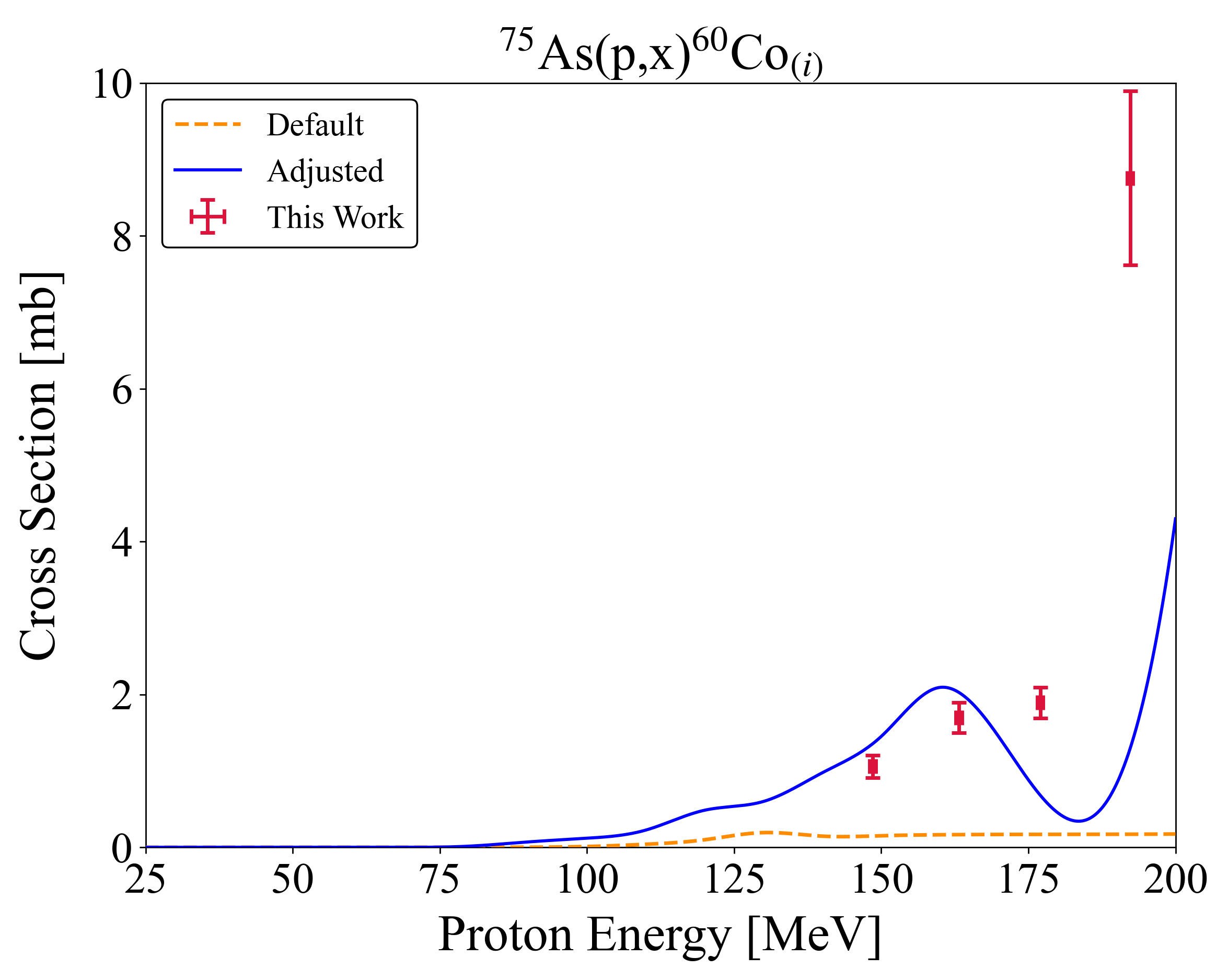}}\\
	\subfloat[\label{58Co_Proc}]
		{\includegraphics[width=0.315\textwidth]{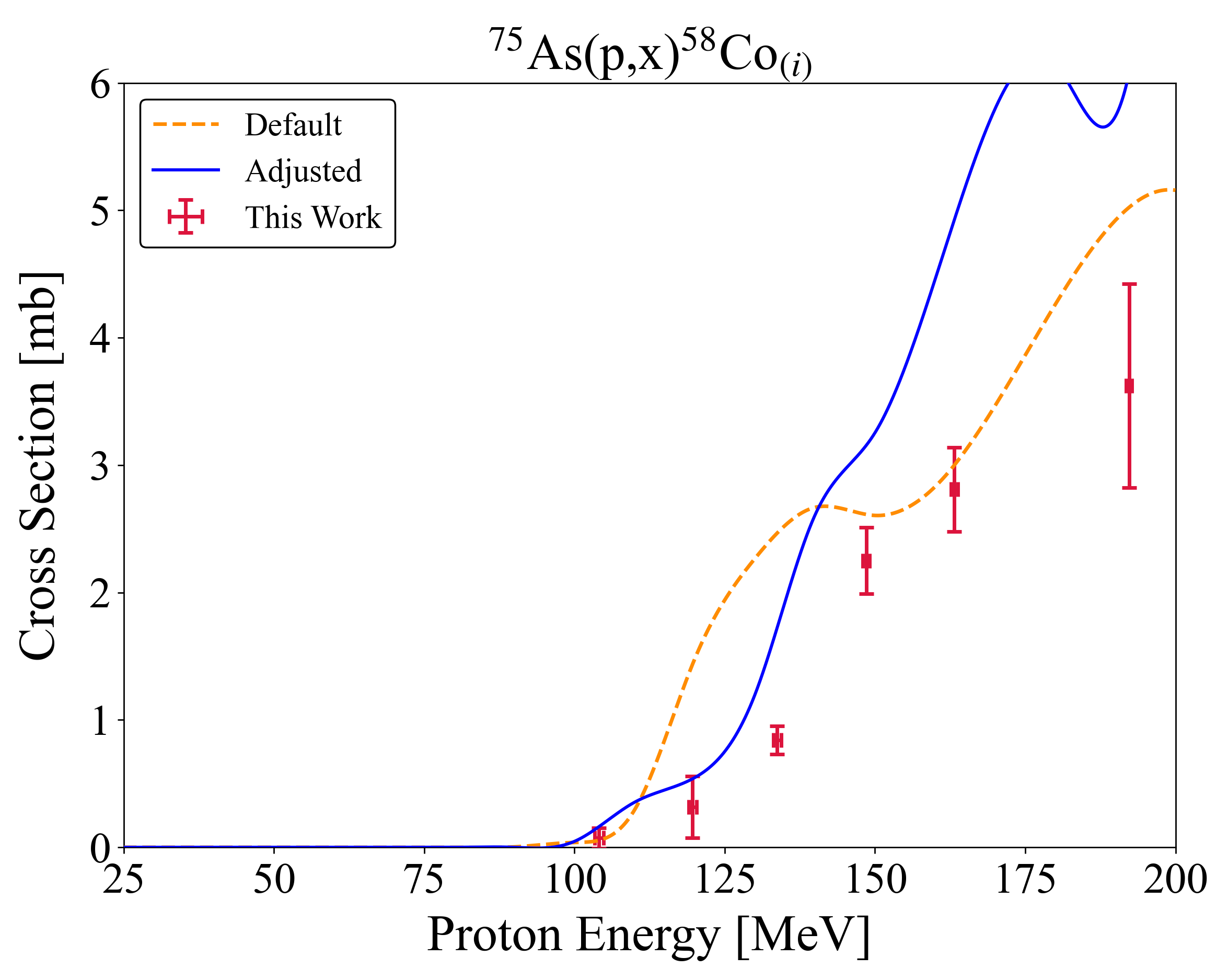}}
	\subfloat[\label{57Co_Proc}]
		{\includegraphics[width=0.315\textwidth]{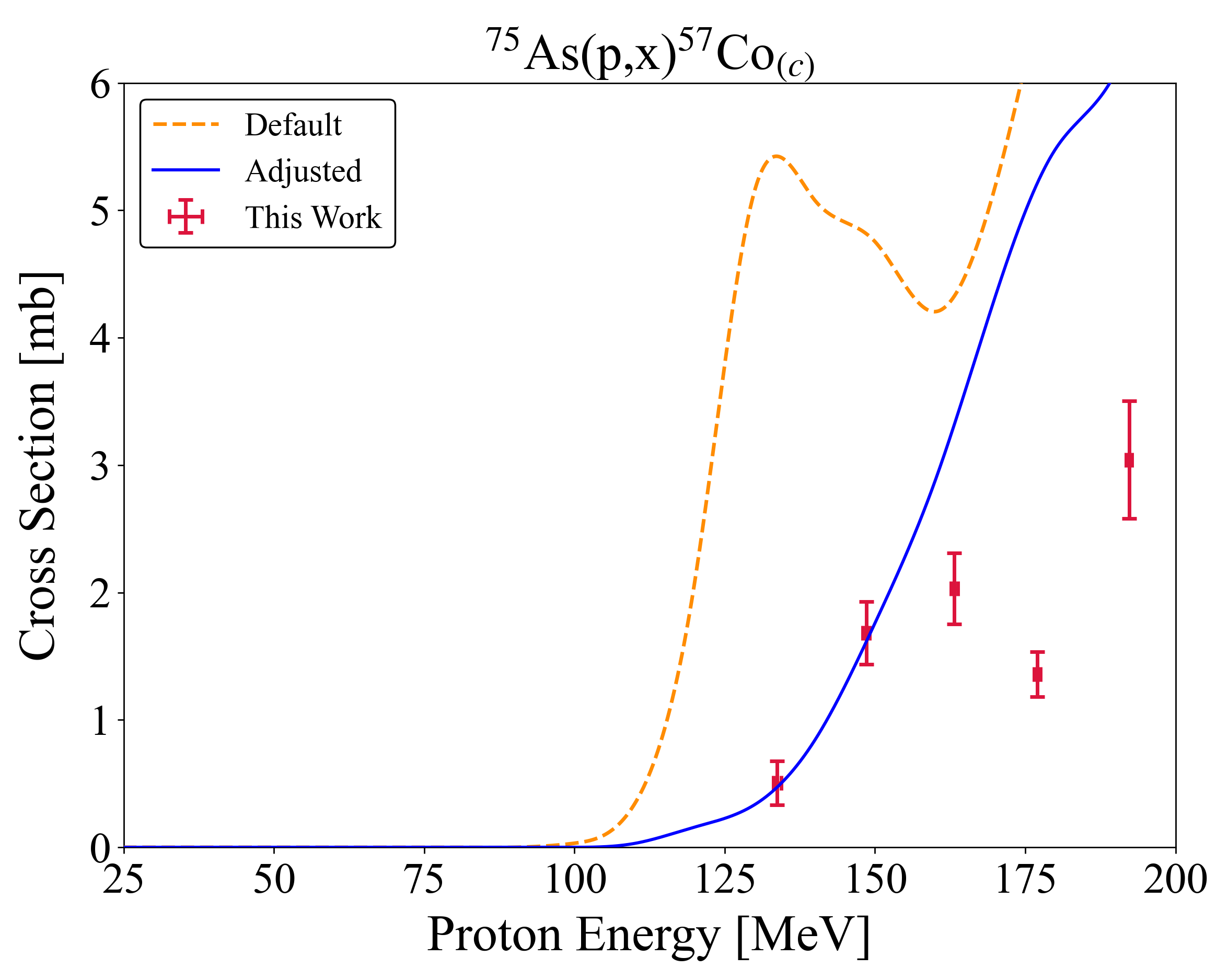}}
	\subfloat[\label{56Co_Proc}]
		{\includegraphics[width=0.315\textwidth]{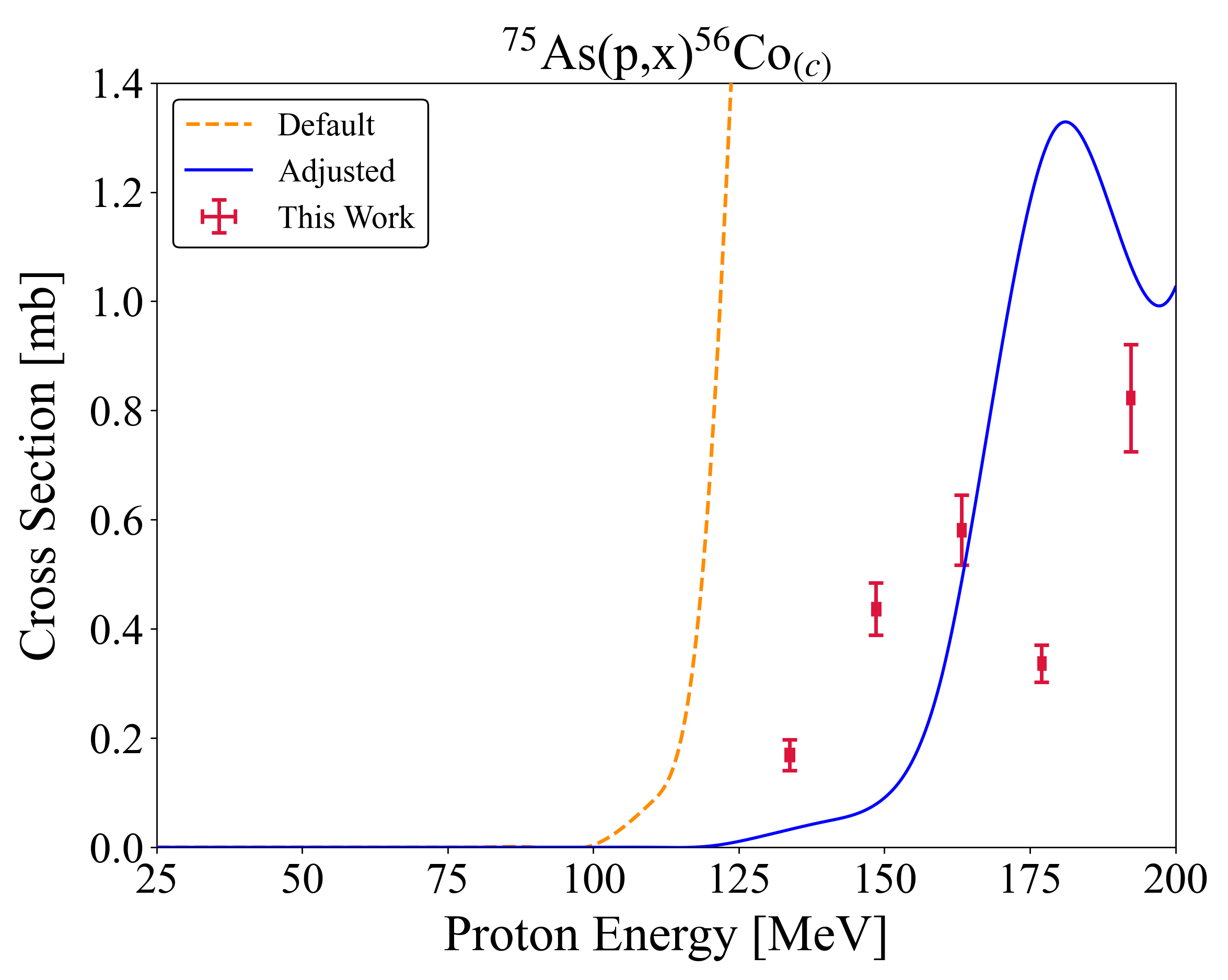}}
	\caption{TALYS default and adjusted calculations extended to residual products not used in the parameter adjustment sensitivity studies.}\label{Proc_FitsValidation}
\end{figure*}

As proposed in \textcite{Fox2020:NbLa}, validation for the suggested parameter changes can be performed by applying the adjusted fit to reaction channels not included in the initial adjustment sensitivity studies in Section \ref{FitProc}. In this work, the validation channels $^{75}$As(p,x)$^{72,70}$As, $^{68,66}$Ge, $^{72,66}$Ga, $^{\mathrm{69m},65}$Zn, $^{60,58,57,56}$Co help test for cumulative cross section effects and far-from-target modeling stability. $^{75}$As(p,p3n)$^{72}$As is in fact independently measured here and meets many of the \textcite{Fox2020:NbLa} criteria to be included as a residual product channel in the initial parameter optimization. However, since numerous neighboring competing Se and As channels to $^{72}$As were already incorporated in Section \ref{FitProc}, it became more worthwhile to save $^{72}$As production as a significant channel for validation. Figure \ref{Proc_FitsValidation} demonstrates the adjusted fit behaviour in these validation channels, where consistently improved predictive power is seen.

\begin{figure}[H]
{\includegraphics[width=1.0\columnwidth]{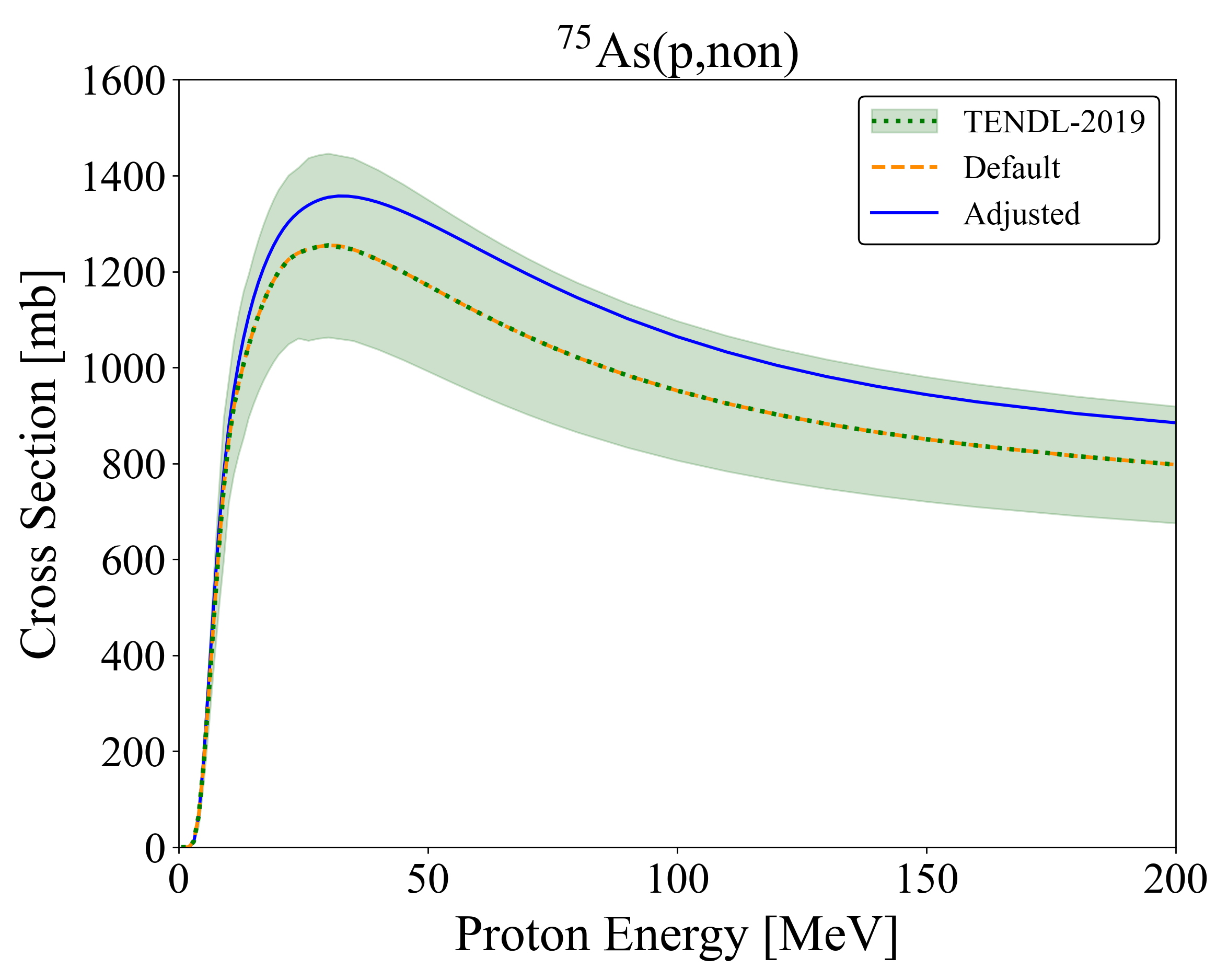}}
\vspace{-0.65cm}
\caption{Comparison of evaluated and theoretical non-elastic cross sections. The filled error bands are associated with the TENDL data.}
\label{75AsTot_Proc}
\end{figure}

It is also possible to further analyze the total non-elastic cross section predictions of the default and adjusted TALYS models, together with the TENDL evaluation (Figure \ref{75AsTot_Proc}). No experimental data points guide the $^{75}$As(p,non) predictions, commensurate with the little prior published data for the residual product excitation functions as a whole. Even with the new data results of this paper, due to the unseen reaction products described in Section \ref{ldadjust_analysis}, it is not viable to derive any $^{75}$As(p,non) data points from summing the measured cross sections. The adjusted (p,non) remains within the TENDL uncertainty band and its increase versus the default is defensible. Specifically, the adjusted (p,non) shares the same shape as the TENDL evaluation and the default prediction, which are based on global fits to other targets, and the increase in magnitude is validated based on changes seen in residual product channels such as $^{73}$As in Figure \ref{73As_Proc}.
%, where the TALYS default underpredicted the experimental data by $\approx$200\,mb.

A $\chi^2_{tot}$ descriptive metric for comparing the default and adjusted TALYS fits across all presented excitation functions, following the formalism described in \textcite{Fox2020:NbLa}, is given in Table \ref{globalchi}. Both weighting methodologies yield similar results and the adjusted fit is seen to outperform the default prediction. The $\chi^2_{tot}$ values are partially deflated relative to the $^{93}$Nb(p,x) and $^{139}$La(p,x) results in \textcite{Fox2020:NbLa} on account of the heavy dependence on the arsenic cross section measurements provided in this work and their associated larger uncertainties (9.0--15\%) stemming from the electroplating process. Consequently, the $\chi^2_{tot}$ results are more usefully viewed as a relative measure between fits rather than as absolute measure of goodness.

\vspace{-0.2cm}
\begin{table}[H]
\caption{Global $\chi^2$ metric describing goodness-of-fit for the default and adjusted TALYS calculations of $^{75}$As(p,x). Low $\chi^2_{tot}$ values, and a case of $\chi^2_{tot}<1.0$, are seen as a function of large weights associated with the measured arsenic data.}
\label{globalchi}
\begin{ruledtabular}
\begin{tabular}{lcc}
Weighting Method&Default $\chi^2_{tot}$ &Adjusted $\chi^2_{tot}$ \\[0.1cm]
\hline\\[-0.25cm]
Cumulative $\sigma$     & 2.55 & 0.58\\[0.1cm]

Maximum $\sigma$    & 3.58 & 1.25\\
\end{tabular}
\end{ruledtabular}
\end{table}
\vspace{-0.5cm}

\subsection{\label{Limitations}Alternative Solutions and Limitations of the Fitting Procedure}
The \texttt{M2constant}=0.80, \texttt{M2limit}=3.9, and \texttt{M2shift}=0.55 exciton model adjustments suggested in this paper match the trend of \texttt{M2constant}$<$1.0, \texttt{M2limit}$>$1.0, and \texttt{M2shift}$<$1.0 changes from the $^{93}$Nb(p,x) and $^{139}$La(p,x) fitting cases in \textcite{Fox2020:NbLa}. As a result, the same systematic behaviour of a relative decrease for internal transition rates at intermediate proton energies ($E_p=20-60$\,MeV) in the exciton model as derived from the Nb and La cases is seen in the As as well. These determined \texttt{M2} pre-equilibrium adjustments therefore continue to be indicative of a needed global enhancement to the two-component exciton model.

However, due to the mathematical formulation of the exciton model in TALYS, which can be reviewed in detail in \textcite{Koning2004:GlobalPEMexcitonOMP}, it has been found that in fact \texttt{M2constant}$<$1.0, \texttt{M2limit}$>$1.0, and \texttt{M2shift}$<$1.0 is not a required condition to generate the systematic behaviour. Instead, numerous sets of (\texttt{M2constant,M2limit,M2shift}) will reproduce the same decrease for internal transition rates and replicate the residual product cross section predictions of Section \ref{FitProc}. For example, both (2.45, 0.7, 1.2) and (1.1, 2.85, 0.7) satisfy these conditions for the $^{75}$As(p,x) fitting. Thus, the transition rate trend result from \textcite{Fox2020:NbLa} is corroborated in this work but the \texttt{M2} adjustment requirements to create this trend are revised. Moreover, since multiple triplets all predict the expected systematic behaviour for the reaction phase space transitioning between the Hauser-Feshbach and exciton models for nuclear reactions, it is not possible to conclude which triplet is more accurate without more diversified datasets such as particle emission spectra or prompt gamma yields by $^{75}$As(p,x$\gamma$) \cite{Nobre2020:LD}.

\clearpage

%Note that in context, chi squred in ok for this type of validation, but a move to an autoimated procedure, where there are few experimental data points, relatvie to something like low energyt enutron evaluations, provides mathmetatical artifacts wehre objectively poort parameters sets can perform well for the fitting of resiudal products in terms of chi squared results. ANd in the case of the asrenic here at the moment, it is heavily dependent only on our dataset, which has fairly large weights (in the 11-14\% range), which contributes to probably a smaller chi squared than represnetative of the actuality of the fit
%\clearpage
\vspace{-0.15cm}
\vspace{-0.3cm}

Indeed, this lack of diversified datasets is the overall limiting factor of the fitting procedure in its current state. The TALYS parameter space is extremely large and the effects of many parameters are hidden from high-energy residual product modeling. Furthermore, the secondary effects from pre-equilibrium, optical model, level density, and coupled-channels changes that are made cannot be deduced without other data types, which detracts from physical insights that can be made about the modeling physics in this work \cite{Vagena2021}. Prompt gamma data or emission spectra could act to concretely identify certain parameters as well as greatly reduce the remaining parameter space, all creating a more suitable and physical fit solution. Of course, these additional data types would themselves only be able to inform small portions of the incident energy range explored through stacked-target activation and would not be as useful without the abundance of residual product data. Clearly, continued high-energy reaction measurements of all types are both needed and complementary.

The size of the parameter space is a further limiting element since it leads to local minimum results for the fitting procedure, as was discussed for $^{67}$Ga in this work. The implementation of automated searching and/or machine learning could likely mitigate this problem and would be in line with the sentiment of evaluators in the nuclear data community \cite{WANDA2020,WANDA2021}.

Overall, these shortcomings emphasize that the thought process of the \citet{Fox2020:NbLa} fitting procedure is most relevant, and not every individual result -- at this stage, because it principally builds evaluation considerations into nuclear data measurements. This is an important introductory step for an area where no formalism or data existed, as the evolution of this type of thought process better aligns data work and evaluations as a necessary path forward.

\section{Conclusions}
\vspace{-2mm}
This work furthers the Tri-laboratory Effort in Nuclear Data by reporting 55 sets of measured $^{75}$As(p,x), $^{\textnormal{nat}}$Cu(p,x), and $^{\textnormal{nat}}$Ti(p,x) residual product cross sections between 35 and 200 MeV. The measured data most notably include the first cross section results for $^{75}$As(p,x)$^{68}$Ge and the best characterized excitation function of $^{75}$As(p,x)$^{72}$Se to-date, which are important for the production of the $^{68}$Ge/$^{68}$Ga and $^{72}$Se/$^{72}$As PET generator systems.

We have additionally continued to develop the \textcite{Fox2020:NbLa} formalism for high-energy reaction modeling using the newly available measured $^{75}$As(p,x) data. The modeling study in this paper corroborated the pre-equilibrium exciton model findings presented in \textcite{Fox2020:NbLa} surrounding the transition between the compound and pre-equilibrium regions in TALYS. Furthermore, we provided an in-depth discussion on the limitations to modeling predictive power caused by the lack of level density knowledge for nuclei off of stability.

This paper merges experimental work and evaluation techniques for high-energy charged-particle isotope production in a continuance of the initial analysis of this kind. The consideration of these different aspects of the nuclear data pipeline together is a priority moving forwards that will benefit future data compilation, evaluation, and application.

%And continue with goal for bringing evaluation thoughts/techniques for high-energy chargped-aprticle data as a priority and advocationg for further work, measurments, and incorporations into the nuclear data pipeline.
%no one is doing this type of work for high-energies. so here's a good starting point, residual product data, a lot of it, covering a broad energy range. This is how we start using it and what we can build on top of

%%%%%%%%%%%%%%%%%%
%\section*{Data Availability Statement}
\vspace{2mm}
The $\gamma$-ray spectra and all other raw data created during this research are openly available at \url{http://doi.org/10.5281/zenodo.4648950} \cite{ZenodoTREND}. On publication, the experimentally determined cross sections will be uploaded to the EXFOR database.

\vspace{-4mm}
\section*{Acknowledgments}
\vspace{-4mm}
This research was supported by the U.S. Department of Energy Isotope Program, managed by the Office of Science for Isotope R\&D and Production, and was carried out under Lawrence Berkeley National Laboratory (Contract No. DE-AC02-05CH11231), Los Alamos National Laboratory (Contract No. 89233218CNA000001), and Brookhaven National Laboratory (Contract No. DEAC02-98CH10886). The authors acknowledge the assistance and support of Brien Ninemire, Scott Small, Nick Brickner, Devin Thatcher, and all the rest of the operations, research, and facilities staff of the LBNL 88-Inch Cyclotron. We also thank David Reass and Mike Connors at LANSCE-IPF, the LANL C-NR Countroom operators, and the LANSCE Accelerator Operations staff.  The authors acknowledge Deepak Raparia, head of the Pre-Injector Systems group at CAD-BNL, for LINAC beam tuning for the experiment and all members of the BNL Medical Isotope Research and Production group for their assistance. We are grateful to Patrick Sullivan and John Aloi of the BNL Radiological Control Division for the Health Physics support. Sumanta Nayak is acknowledged for the engineering and Frank Naase for the IT support.
%%%%%%%%%%%%%%%%%%
\newpage
\appendix
%%%%%%%%%%%%%%%%%%
\section{\label{Appendix_OtherStacks}Additional LANL and BNL Target Stack Information}
Details of the stacked-targets irradiated at LANL and BNL are given in Tables \ref{LANLStack} and \ref{BNLStack}, taken directly from \textcite{Fox2020:NbLa}.

\begin{longtable}{lP{1.3cm}P{1.8cm}P{1.8cm}}
\caption{Target stack design for irradiation at IPF. The proton beam initially hits the stainless steel plate (SS-SN1) after passing through the upstream Inconel beam entrance window, a water cooling channel, and the target box aluminum window. The thickness and areal density measurements are prior to any application of the variance minimization techniques described in this work.}\label{LANLStack}\\ %The beam is subsequently transported through the rest of the shown stack order.

\hline\hline\\[-0.27cm]
\multirow{4}{*}{Target Layer}&\multirow{4}{=}{\centering Thickness [$\mmicro$m]}&\multirow{4}{=}{\centering Areal Density [mg/cm$^2$]}&\multirow{4}{=}{\centering Areal Density Uncertainty [\%]}\\ \\ \\ \\[0.1cm]
\hline\\[-0.22cm]
\endfirsthead

%\multicolumn{4}{c}%
{{\tablename\ \thetable{} -- cont.}} \\
\hline\hline\\[-0.27cm]
\multirow{4}{*}{Target Layer}&\multirow{4}{=}{\centering Thickness [$\mmicro$m]}&\multirow{4}{=}{\centering Areal Density [mg/cm$^2$]}&\multirow{4}{=}{\centering Areal Density Uncertainty [\%]}\\ \\ \\ \\[0.1cm]
\hline\\[-0.22cm]
\endhead

%\hline \multicolumn{3}{r}{{Continued on next page}}
\hline Continued on next page&&&
\endfoot

\hline\hline
\endlastfoot

SS-SN1 Profile Monitor    &130.0 &100.12&0.07 \\[0.1cm]

Al-SN1                             &27.33 &7.51  &0.21  \\[0.1cm]

Nb-SN1                           &25.75  &23.08 &0.12  \\[0.1cm]

As-SN1                            &4.27 &2.45 &8.2 \\[0.1cm]

Ti-SN1                               &25.00 &11.265 &1.0 \\[0.1cm]

Cu-SN1                             &24.33 &19.04 &0.13 \\[0.1cm]

Al Degrader 01                 &6307.0 &1702.89 &0.001 \\[0.1cm]

Al-SN2                              &26.67 &7.58  &0.32  \\[0.1cm]

Nb-SN2                            &24.75  &22.67 &0.08  \\[0.1cm]

As-SN2                            &4.30 &2.46 &8.3 \\[0.1cm]

Ti-SN2                             &25.00 &11.265 &1.0 \\[0.1cm]

Cu-SN2                           &24.00 &18.90 &0.36 \\[0.1cm]

Al Degrader 02                 &3185.5 &860.09 &0.02 \\[0.1cm]

Al-SN3                             &26.67 &7.38  &0.22  \\[0.1cm]

Nb-SN3                          &24.50  &22.83 &0.03  \\[0.1cm]

As-SN3                          &3.62 &2.07 &9.0 \\[0.1cm]

Ti-SN3                            &25.00 &11.265 &1.0 \\[0.1cm]

Cu-SN3                           &23.33 &19.38 &0.11 \\[0.1cm]

Al Degrader 03                 &2304.5  &622.22 &0.06 \\[0.1cm]

Al-SN4                            &28.00 &7.34  &0.18  \\[0.1cm]

Nb-SN4                           &25.50  &22.57 &0.16  \\[0.1cm]

As-SN4                           &3.54 &2.03 &9.2 \\[0.1cm]

Ti-SN4                             &25.00 &11.265 &1.0 \\[0.1cm]

Cu-SN4                             &24.67 &19.24 &0.11 \\[0.1cm]

Al Degrader 04                  &1581.3  &426.94&0.04 \\[0.1cm]

Al-SN5                               &27.00 &7.48  &0.44  \\[0.1cm]

Nb-SN5                              &24.75  &22.78 &0.12  \\[0.1cm]

As-SN5                               &3.90 &2.23 &8.7 \\[0.1cm]

Ti-SN5                               &25.00 &11.265 &1.0 \\[0.1cm]

Cu-SN5                             &25.00 &19.09 &0.17 \\[0.1cm]

Al Degrader 05                  &1033.8 &279.11 &0.06 \\[0.1cm]

Al-SN6                               &28.67 &7.44  &0.25  \\[0.1cm]

Nb-SN6                              &25.25  &22.80 &0.08  \\[0.1cm]

As-SN6                               &3.11 &1.78 &10 \\[0.1cm]

Ti-SN6                                 &25.00 &11.265 &1.0 \\[0.1cm]

Cu-SN6                               &24.33 &19.50 &0.16 \\[0.1cm]

Al Degrader 06                    &834.8  &225.38 &0.22 \\[0.1cm]

Al-SN7                                  &28.33 &7.56  &0.15  \\[0.1cm]

Nb-SN7                              &25.50  &22.62 &0.06  \\[0.1cm]

As-SN7                               &2.79 &1.59 &9.2 \\[0.1cm]

Ti-SN7                                 &25.00 &11.265 &1.0 \\[0.1cm]

Cu-SN7                               &23.67 &18.79 &0.04 \\[0.1cm]

Al Degrader 07                     &513.5  &138.65 &0.10 \\[0.1cm]

Al-SN8                                  &27.67 &7.56  &0.10  \\[0.1cm]

Nb-SN8                               &25.50  &22.95 &0.45  \\[0.1cm]

As-SN8                               &2.20 &1.26 &9.0 \\[0.1cm]

Ti-SN8                                 &25.00 &11.265 &1.0 \\[0.1cm]

Cu-SN8                               &24.00 &19.06 &0.23 \\[0.1cm]
                 
Al Degrader 08                   &517.3  &139.66 &0.43 \\[0.1cm]

Al-SN9                                &27.00 &7.47  &0.36  \\[0.1cm]

Nb-SN9                               &25.00  &22.53 &0.24  \\[0.1cm]

As-SN9                               &2.57 &1.47 &9.9 \\[0.1cm]

Ti-SN9                                  &25.00 &11.265 &1.0 \\[0.1cm]

Cu-SN9                                 &26.33 &19.19 &0.12 \\[0.1cm]

Al Degrader 09                     &517.8  &139.79 &0.09 \\[0.1cm]

Al-SN10                                &28.00 &7.41  &0.17  \\[0.1cm]

Nb-SN10                               &24.75  &22.82 &0.02  \\[0.1cm]

As-SN10                               &1.94 &1.11 &10 \\[0.1cm]

Ti-SN10                                  &25.00 &11.265 &1.0 \\[0.1cm]

Cu-SN10                                 &25.67 &18.87 &0.18 \\[0.1cm]

SS-SN10 Profile Monitor          &130.0 &100.12&0.07 \\[0.1cm]
\end{longtable}
%\clearpage
\newpage

\newpage
\begin{table}[!t]
\caption{Target stack design for irradiation at BLIP. The proton beam initially hits the stainless steel plate after passing through the upstream beam windows, water cooling channels, and target box aluminum window. The thickness and areal density measurements are prior to any application of the variance minimization techniques described in this work.}
\label{BNLStack}
\begin{ruledtabular}
\begin{tabular}{lP{1.3cm}P{1.8cm}P{1.8cm}}
%\begin{tabular}{lccc}
\multirow{4}{*}{Target Layer}&\multirow{4}{=}{\centering Thickness [$\mmicro$m]}&\multirow{4}{=}{\centering Areal Density [mg/cm$^2$]}&\multirow{4}{=}{\centering Areal Density Uncertainty [\%]}\\ \\ \\ \\[0.1cm]
\hline\\[-0.22cm]
SS Profile Monitor               &120.2 &95.16&0.58 \\[0.1cm]

Cu-SN1                              &26.00 &22.34  &0.10  \\[0.1cm]

Nb-SN1                             &25.75  &22.75 &0.25  \\[0.1cm]

As-SN1                             &1.89 &1.08 &9.9 \\[0.1cm]

Ti-SN1                                &25.00 &11.265 &1.0 \\[0.1cm]

Cu Degrader 01               &5261.1 &4708.07 &0.02 \\[0.1cm]

Cu-SN2                           &26.75 &22.41  &0.11  \\[0.1cm]

Nb-SN2                          &24.75  &22.91 &0.19  \\[0.1cm]

As-SN2                            &2.94 &1.68 &9.0 \\[0.1cm]

Ti-SN2                            &25.00 &11.265 &1.0 \\[0.1cm]

Cu Degrader 02              &4490.7 &4018.99 &0.04 \\[0.1cm]

Cu-SN3                            &26.50 &22.26  &0.05  \\[0.1cm]

Nb-SN3                           &24.00  &22.67 &0.31  \\[0.1cm]

As-SN3                           &3.06 &1.75 &10 \\[0.1cm]

Ti-SN3                            &25.00 &11.265 &1.0 \\[0.1cm]

Cu Degrader 03             &4501.8  &4028.84 &0.03 \\[0.1cm]

Cu-SN4                          &26.00 &22.29  &0.15  \\[0.1cm]

Nb-SN4                          &24.75  &22.70 &0.23  \\[0.1cm]

As-SN4                           &4.85 &2.78 &9.9 \\[0.1cm]

Ti-SN4                              &25.00 &11.265 &1.0 \\[0.1cm]

Cu Degrader 04               &4243.9  &3797.96&0.03 \\[0.1cm]

Cu-SN5                            &25.50 &22.35  &0.04  \\[0.1cm]

Nb-SN5                            &25.00  &22.54 &0.12  \\[0.1cm]

As-SN5                            &7.26 &4.15 &12 \\[0.1cm]

Ti-SN5                             &25.00 &11.265 &1.0 \\[0.1cm]

Cu Degrader 05               &3733.8 &3341.56 &0.03 \\[0.1cm]

Cu-SN6                           &26.25 &22.34  &0.08  \\[0.1cm]

Nb-SN6                           &25.00  &22.36 &0.24  \\[0.1cm]

As-SN6                           &4.93 &2.82 &9.0 \\[0.1cm]

Ti-SN6                             &25.00 &11.265 &1.0 \\[0.1cm]

Cu Degrader 06               &3783.0  &3385.41 &0.04 \\[0.1cm]

Cu-SN7                            &25.75 &22.26  &0.09  \\[0.1cm]

Nb-SN7                            &25.75  &22.62 &0.10  \\[0.1cm]

As-SN7                            &12.62 &7.22&9.3 \\[0.1cm]

Ti-SN7                            &25.00 &11.265 &1.0 \\[0.1cm]
\end{tabular}
\end{ruledtabular}
\end{table}
\newpage

\section{\label{Appendix_VarMin}Proton Current Variance Minimization}
\begin{figure}[H]
	{\includegraphics[width=1.0\columnwidth]{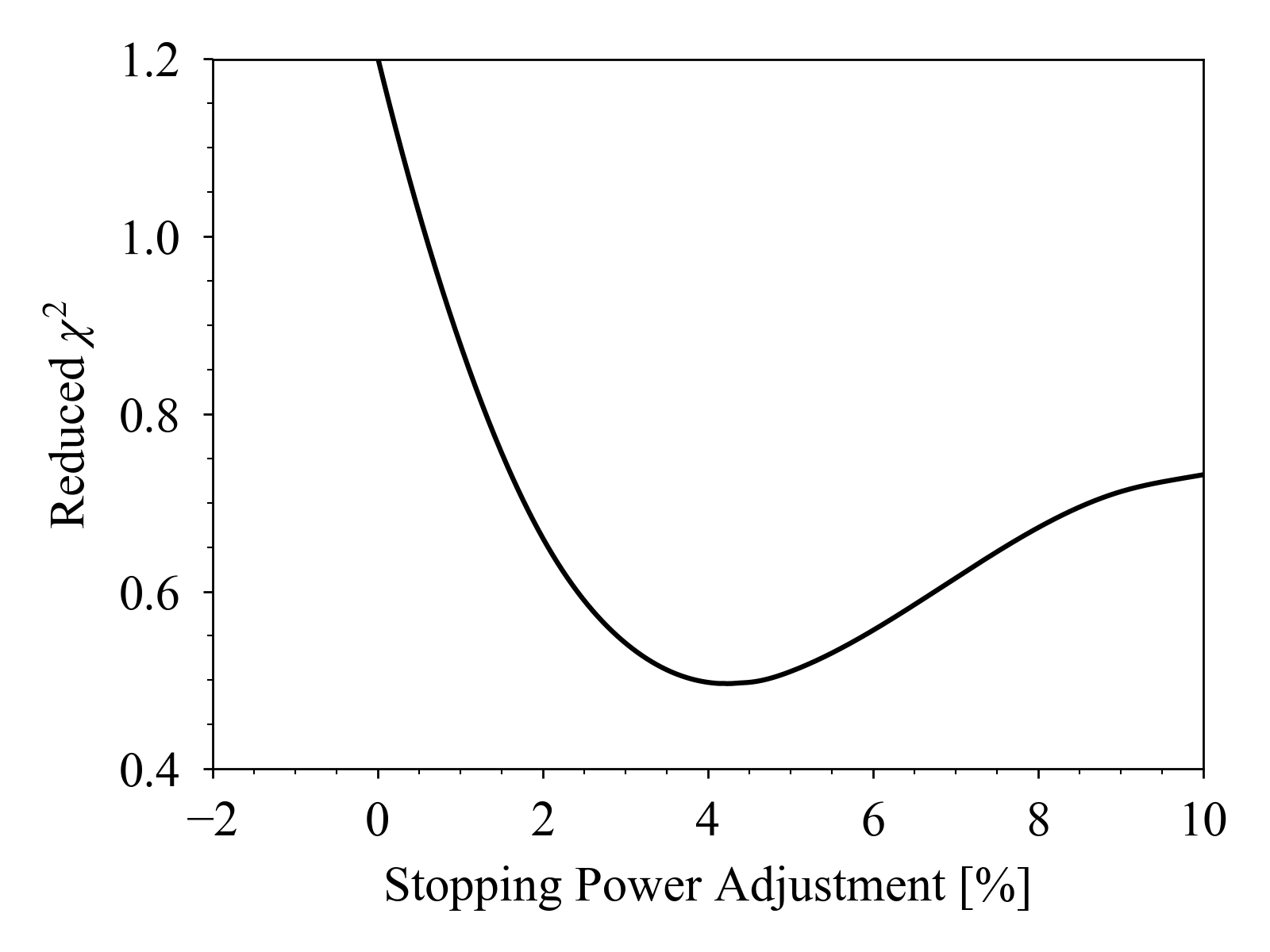}}
\vspace{-0.65cm}
\caption{Result of $\chi^2$ optimization used in the variance minimization of the global linear fit to the monitor fluence data, indicating a required increase to stopping power in transport simulations.}\label{VarMinChi}
	{\includegraphics[width=1.0\columnwidth]{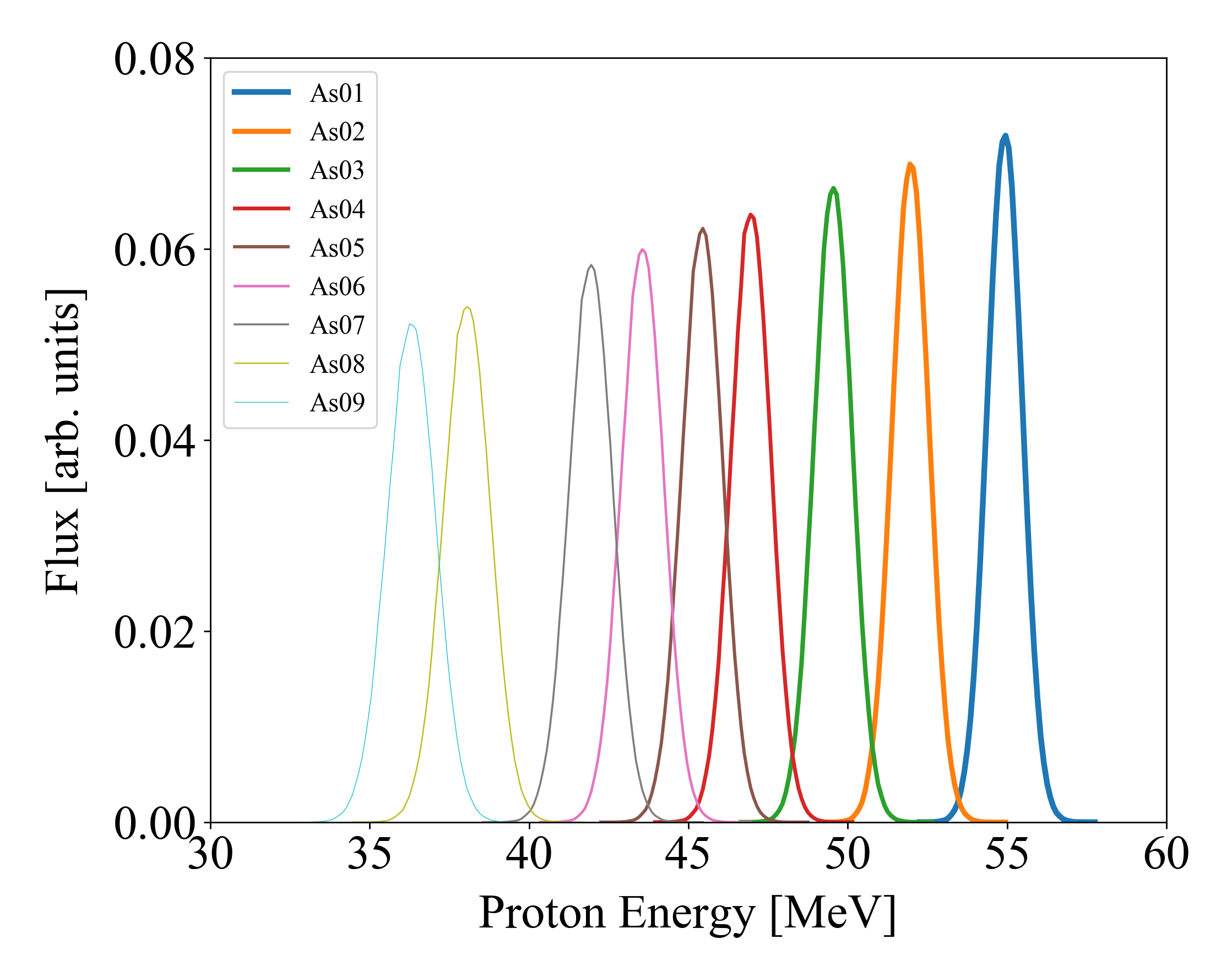}}
\vspace{-0.65cm}
\caption{Visualization of the calculated proton energy spectrum for each arsenic target in the LBNL stack, following variance minimization.}\label{AZ_Flux_Output}
\end{figure}

The applied variance minimization technique for the LBNL irradiation is summarized in Figure \ref{VarMinChi}. A 4.23\% increase to stopping power in simulations, implemented through an equivalent increase to degraders' effective density in the stack, best reduced proton fluence measurement disagreements between different monitor channels in each energy compartment. This is in general agreement with results of past stacked-target work that have shown a needed modest positive enhancement to the stopping power of +2--5\% \cite{Voyles2018:Nb,Graves2016:StackTarget,Morrell2020}.

%The disagreement was most noticeable near the rear of each stack where contribu- tions of poor stopping power characterization, straggling, and systematic uncertainties from upstream components became most compounded.

The associated proton flux spectrum propagating through the stack, after variance minimization, is provided in Figure \ref{AZ_Flux_Output}. The energy assignments for each foil in a stack are then the flux-averaged energies from the spectrum with uncertainties per foil taken as the full width at half maximum.

This same calculation methodology can be reviewed in detail for the LANL and BNL stacks in \textcite{Fox2020:NbLa}
\vspace{-0.3cm}

\section{\label{Appendix_Plots}Measured Excitation Functions}
\begin{figure}[H]
	{\includegraphics[width=1.0\columnwidth]{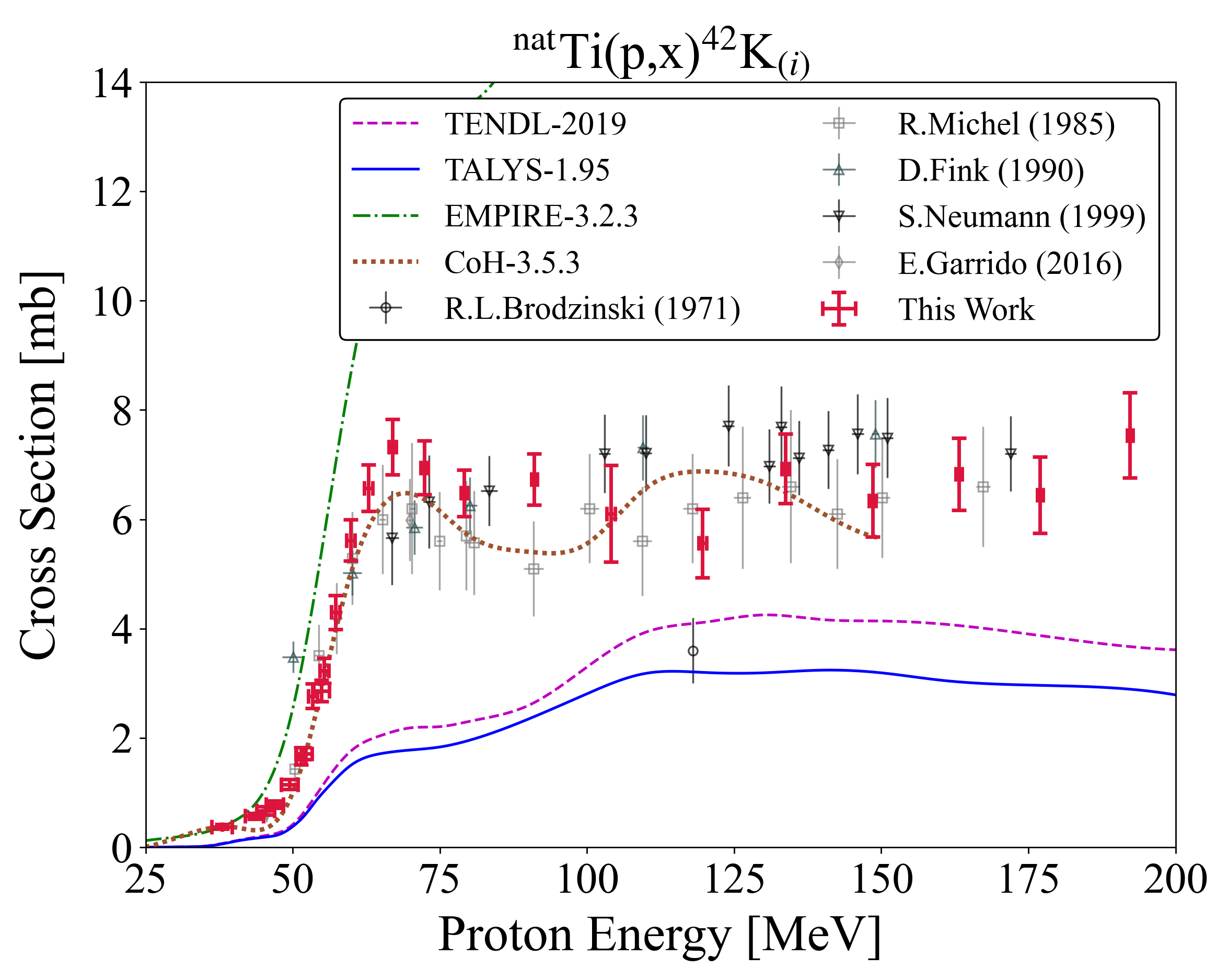}}
\vspace{-0.65cm}
\caption{Experimental and theoretical cross sections for $^{42}$K production.}\label{Ti_42K}
	{\includegraphics[width=1.0\columnwidth]{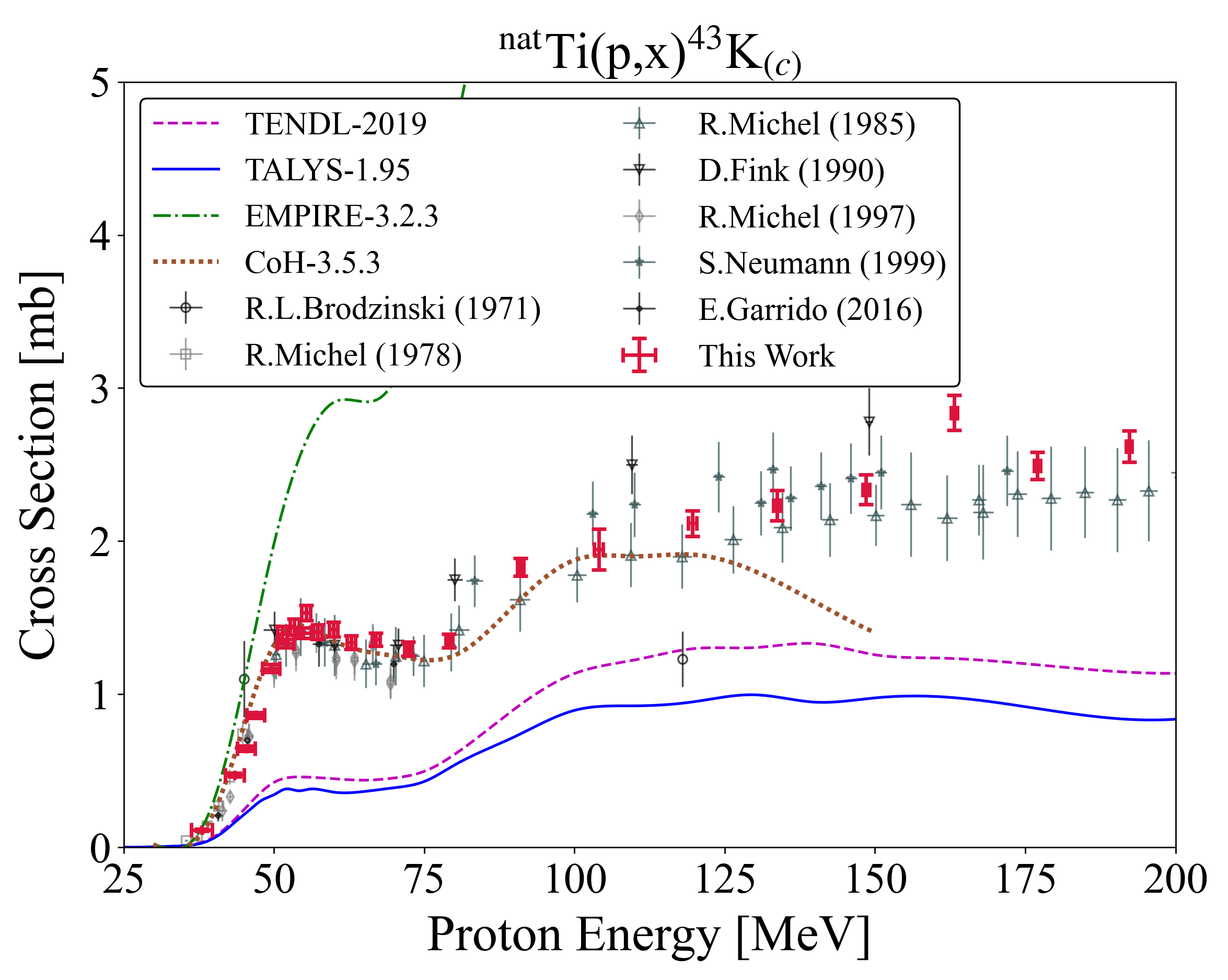}}
\vspace{-0.65cm}
\caption{Experimental and theoretical cross sections for $^{43}$K production.}\label{Ti_43K}
	{\includegraphics[width=1.0\columnwidth]{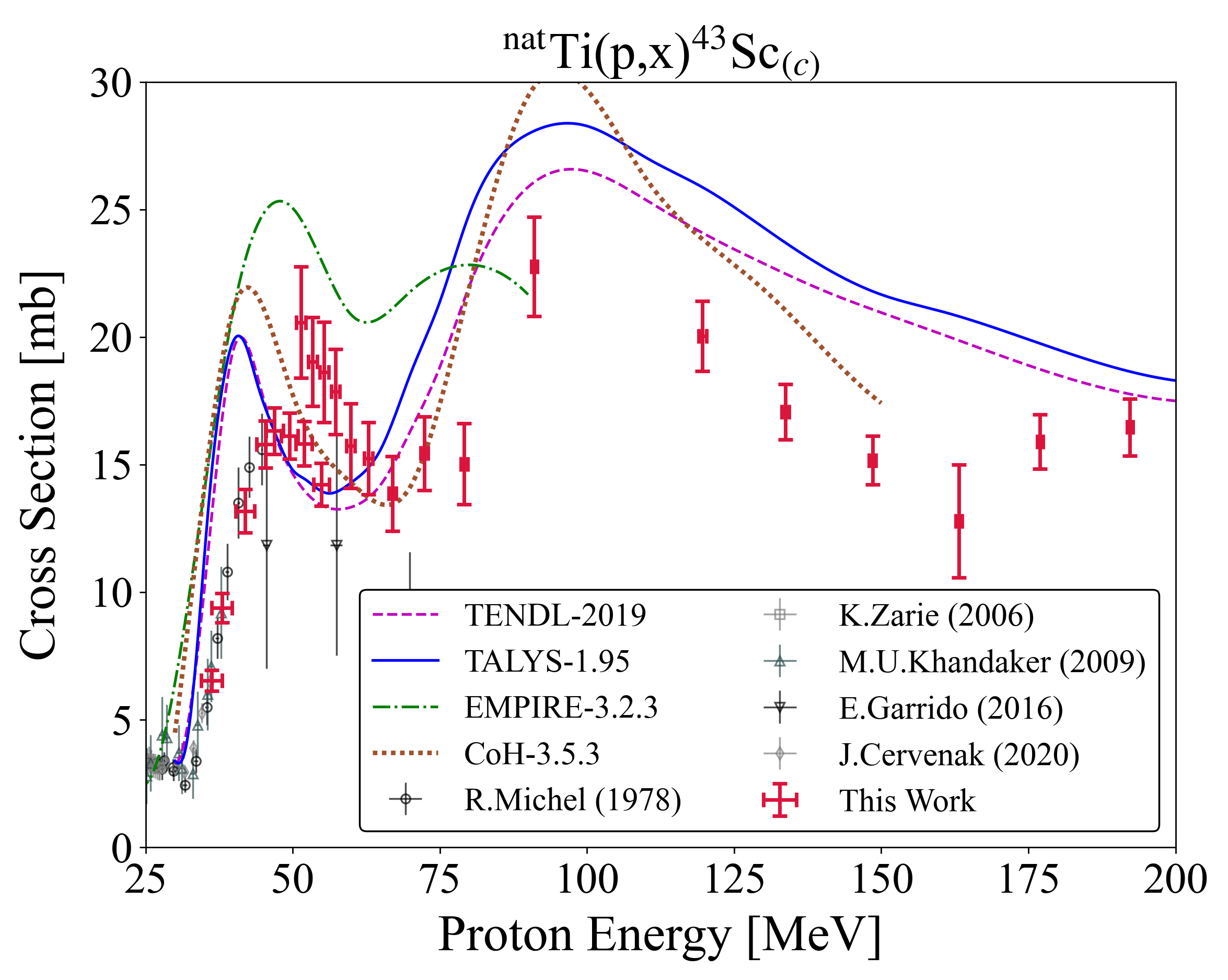}}
\vspace{-0.65cm}
\caption{Experimental and theoretical cross sections for $^{43}$Sc production.}\label{Ti_43SC}
\end{figure}

\begin{figure}[H]
	{\includegraphics[width=1.0\columnwidth]{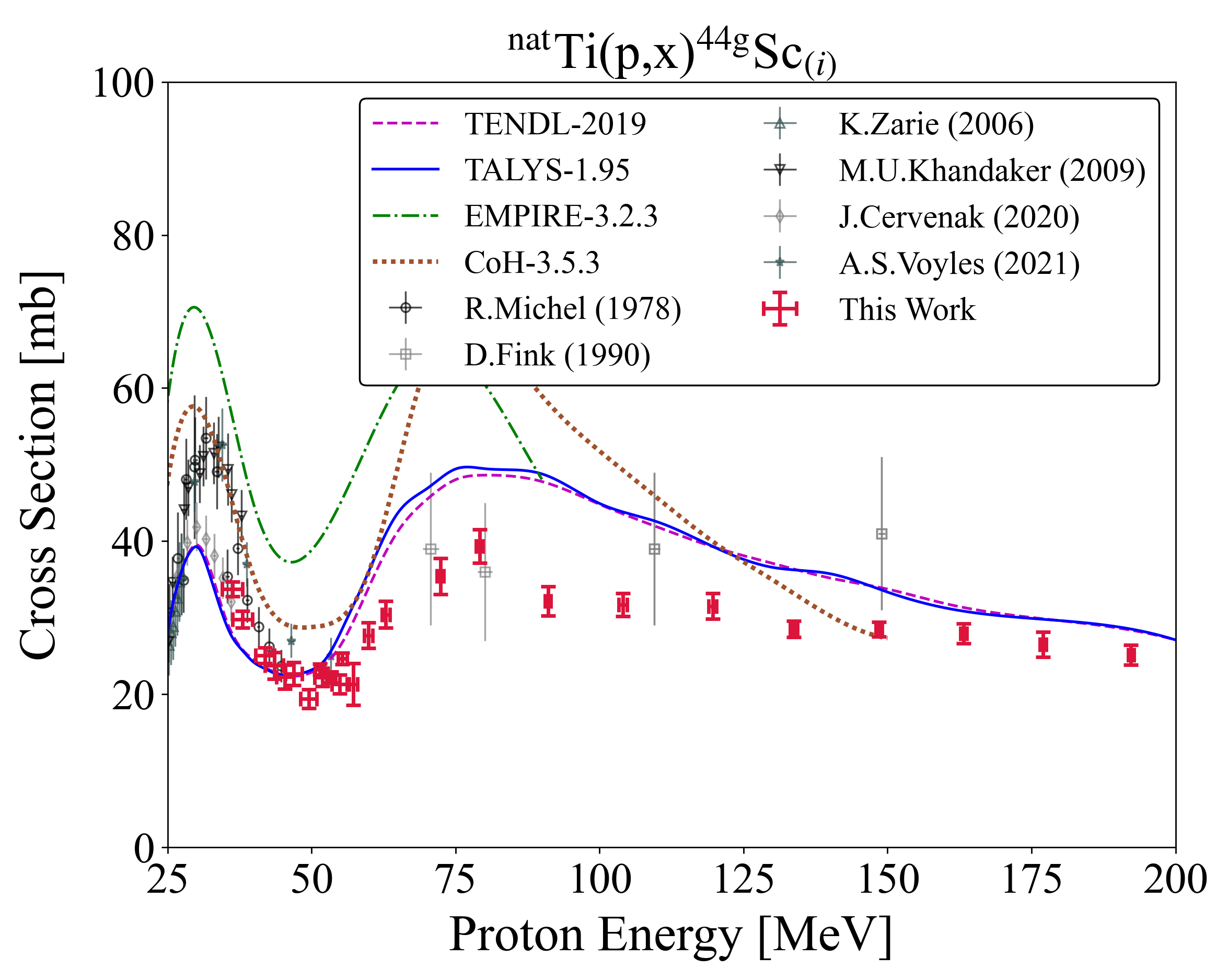}}
\vspace{-0.65cm}
\caption{Experimental and theoretical cross sections for $^{\textnormal{44g}}$Sc production.}\label{Ti_44gSC}
	{\includegraphics[width=1.0\columnwidth]{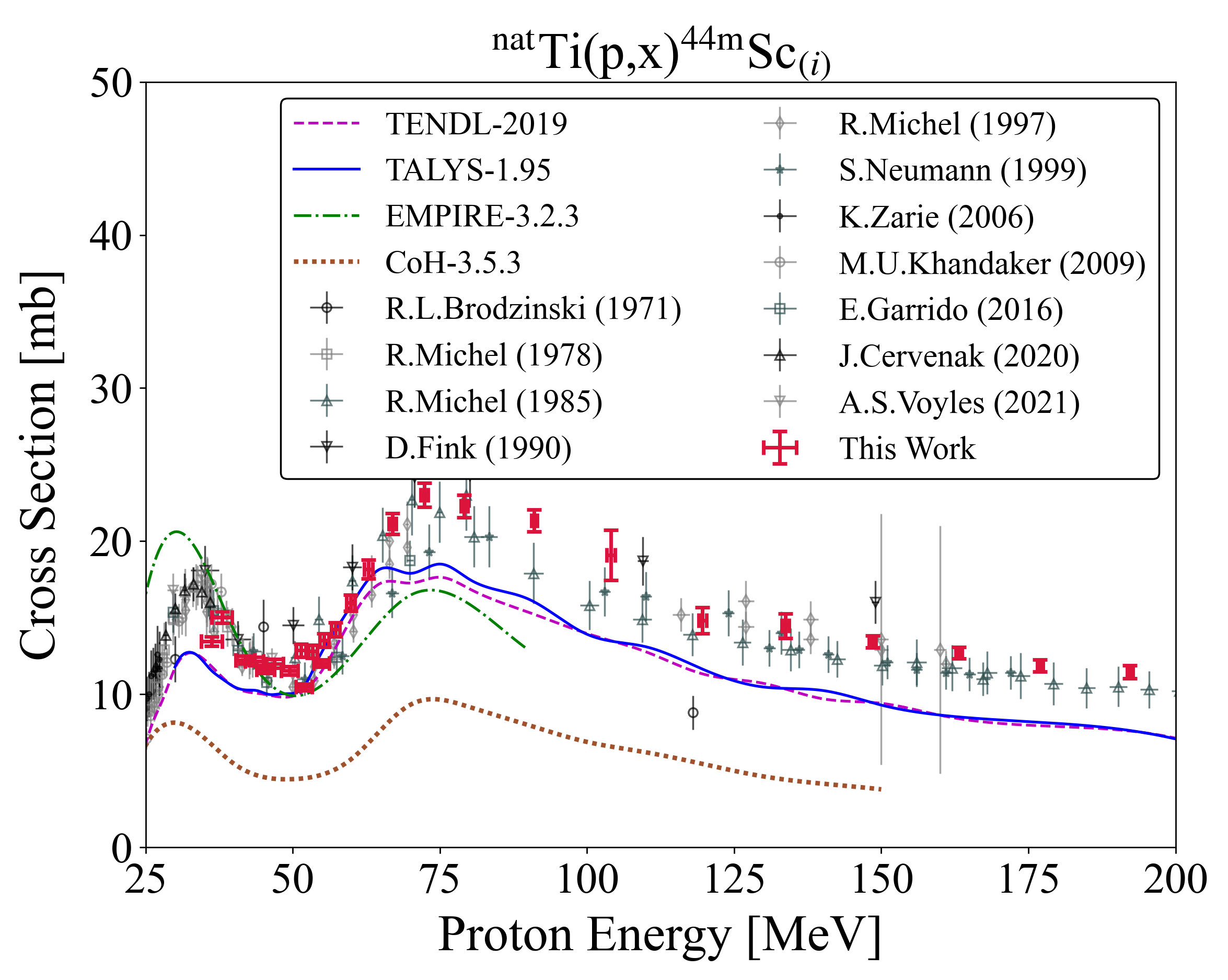}}
\vspace{-0.65cm}
\caption{Experimental and theoretical cross sections for $^{\textnormal{44m}}$Sc production.}\label{Ti_44mSC}
	{\includegraphics[width=1.0\columnwidth]{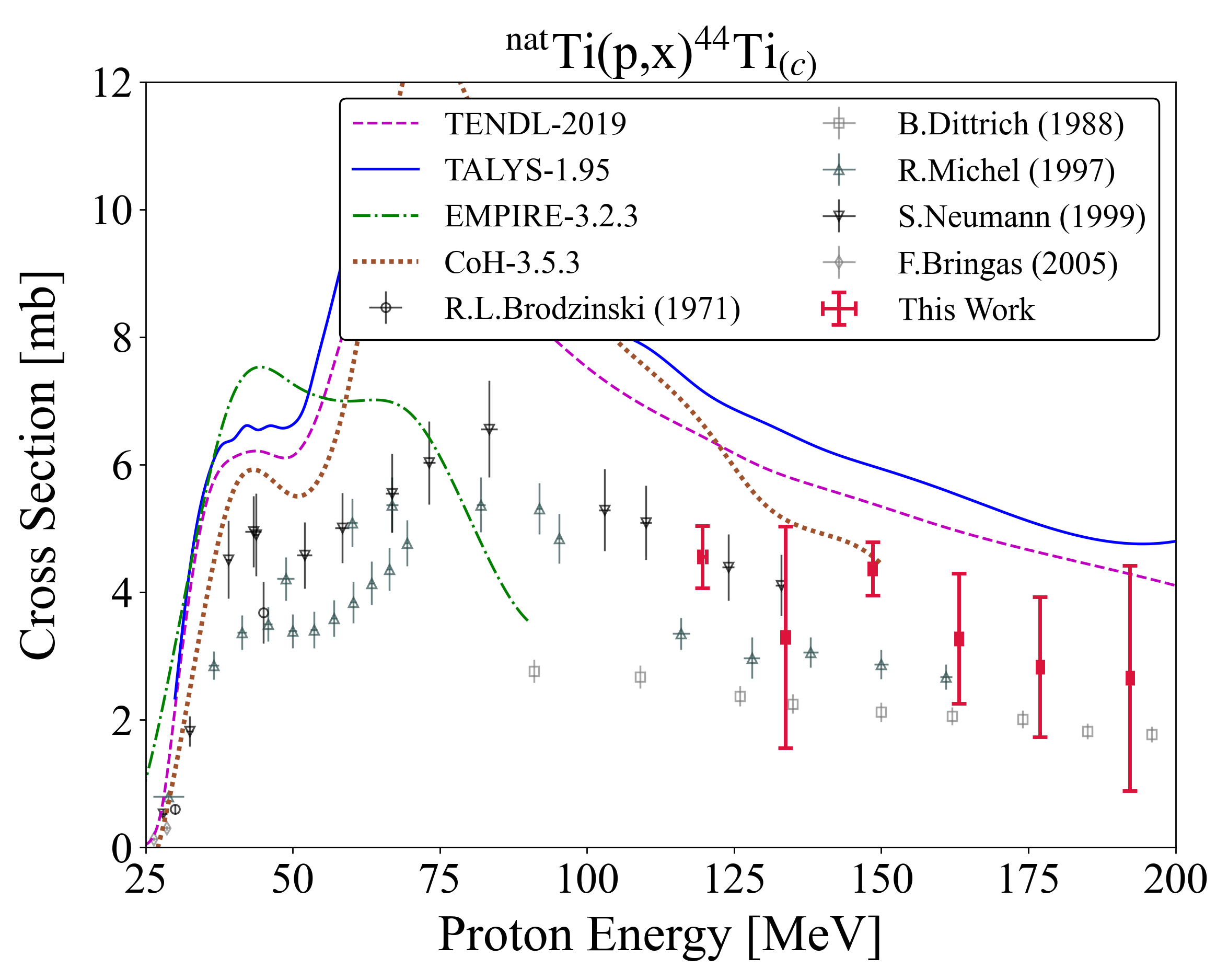}}
\vspace{-0.65cm}
\caption{Experimental and theoretical cross sections for $^{44}$Ti production.}\label{Ti_44TI}
\end{figure}

\begin{figure}[H]
	{\includegraphics[width=1.0\columnwidth]{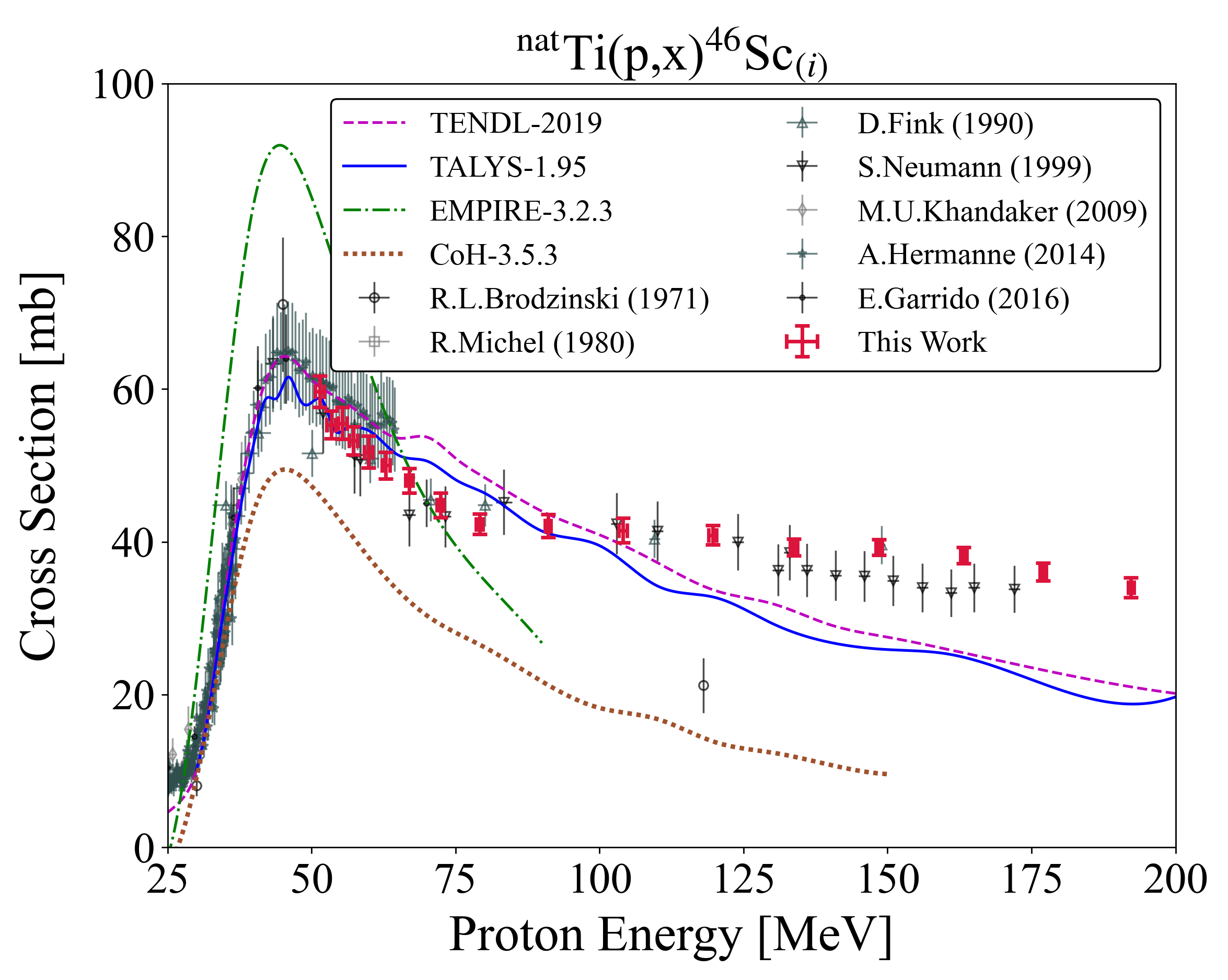}}
\vspace{-0.65cm}
\caption{Experimental and theoretical cross sections for $^{46}$Sc production.}\label{Ti_46SC}
	{\includegraphics[width=1.0\columnwidth]{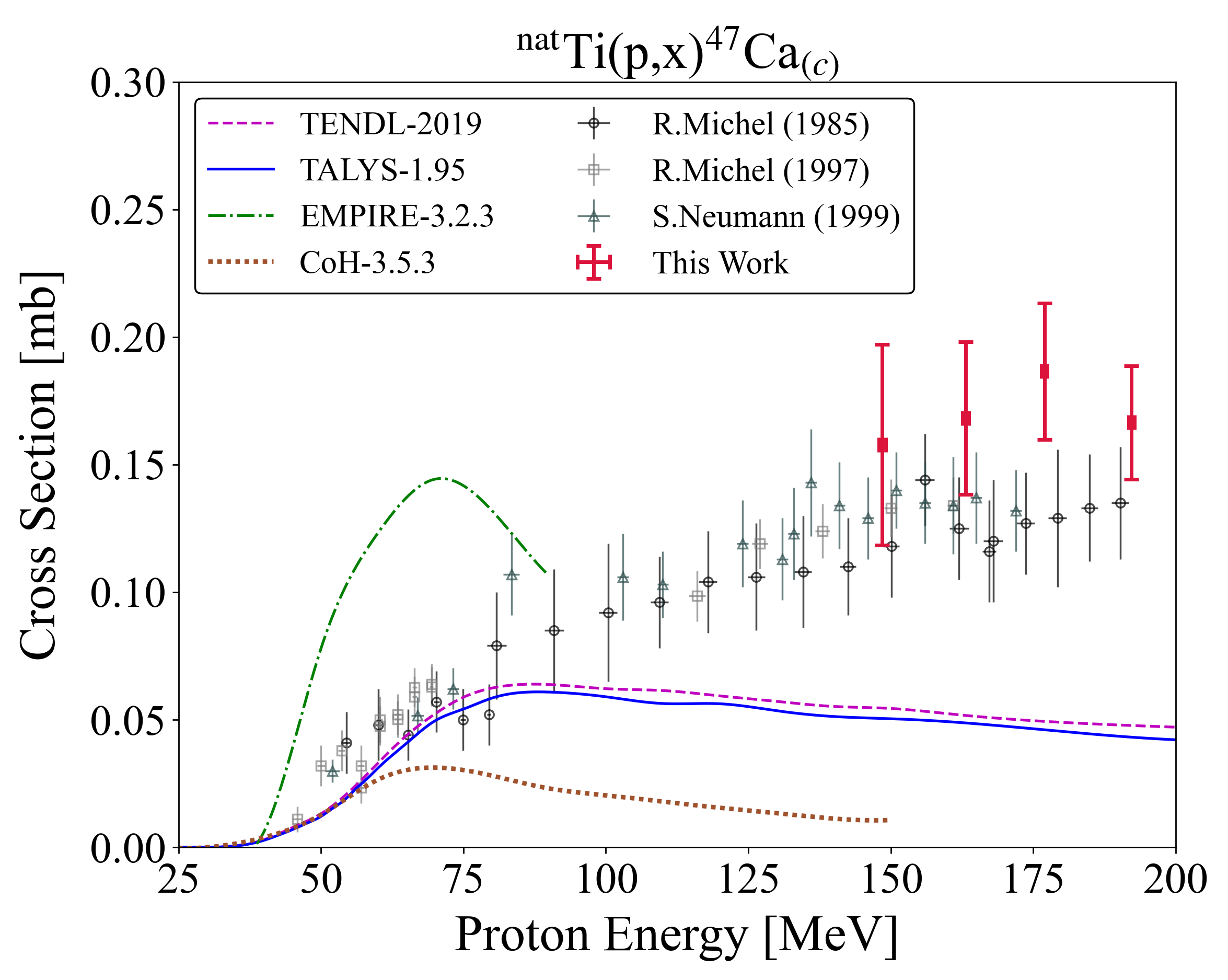}}
\vspace{-0.65cm}
\caption{Experimental and theoretical cross sections for $^{47}$Ca production.}\label{Ti_47CA}
	{\includegraphics[width=1.0\columnwidth]{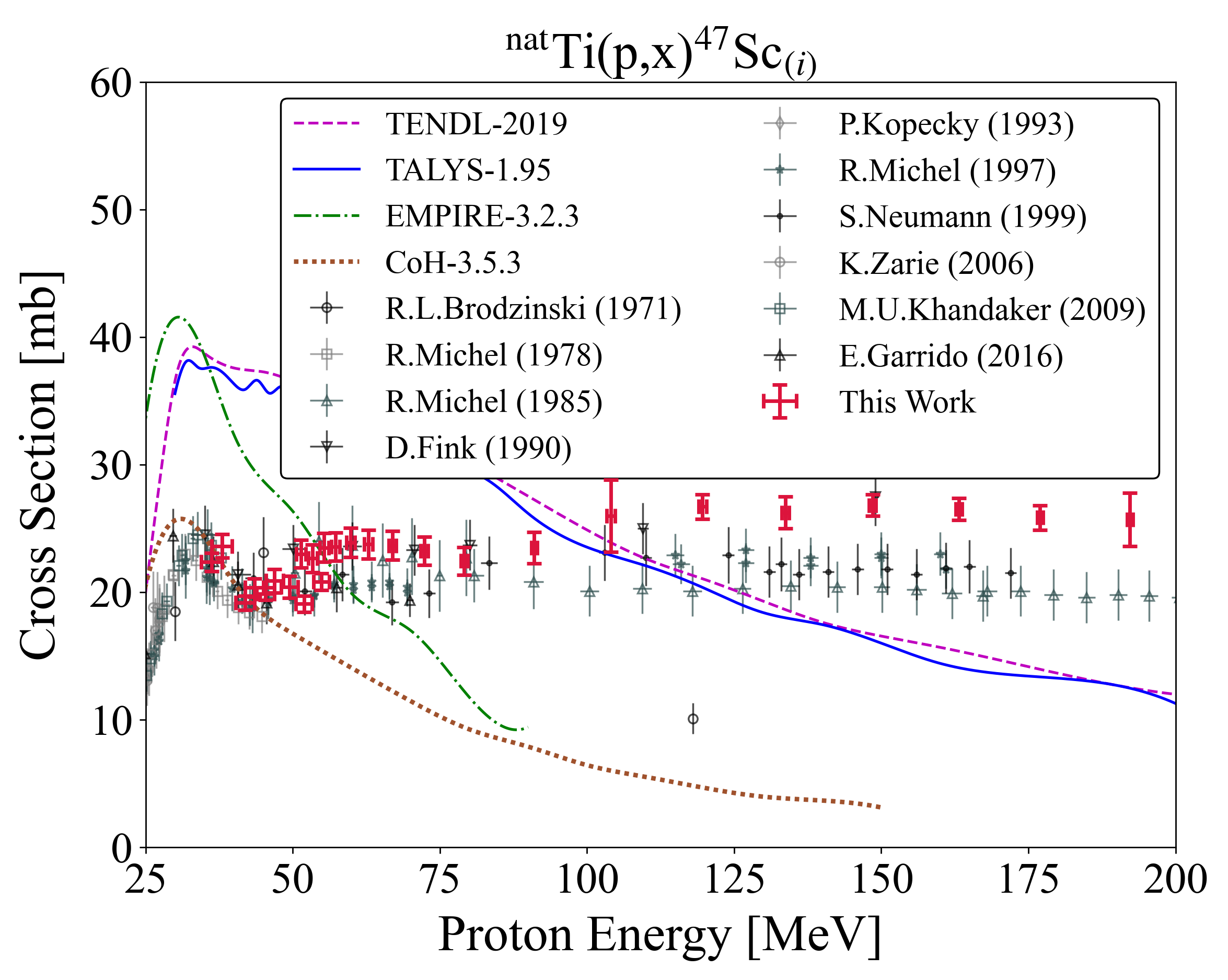}}
\vspace{-0.65cm}
\caption{Experimental and theoretical cross sections for $^{47}$Sc production.}\label{Ti_47SC}
\end{figure}

\begin{figure}[H]
	{\includegraphics[width=1.0\columnwidth]{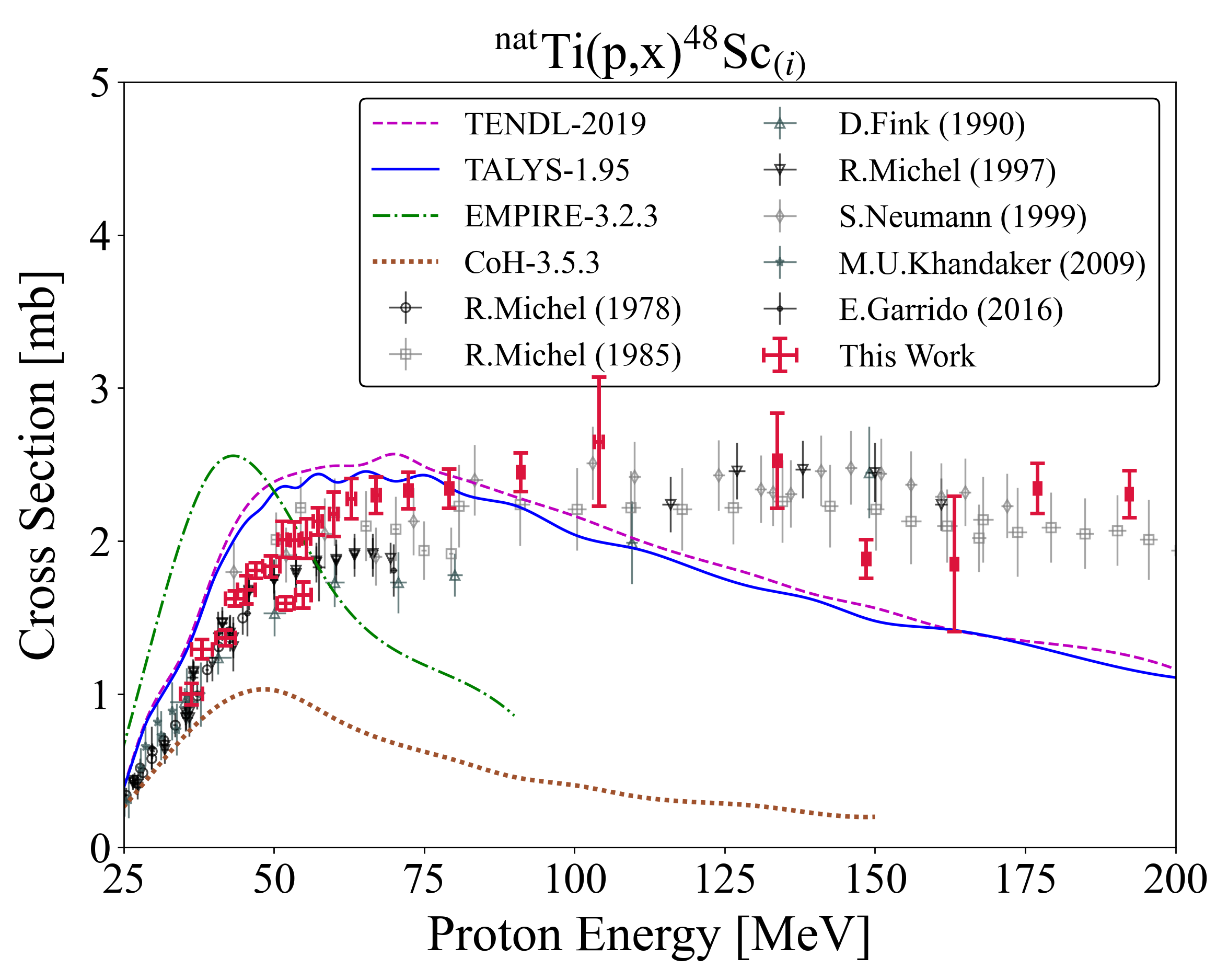}}
\vspace{-0.65cm}
\caption{Experimental and theoretical cross sections for $^{48}$Sc production.}\label{Ti_48SC}
	{\includegraphics[width=1.0\columnwidth]{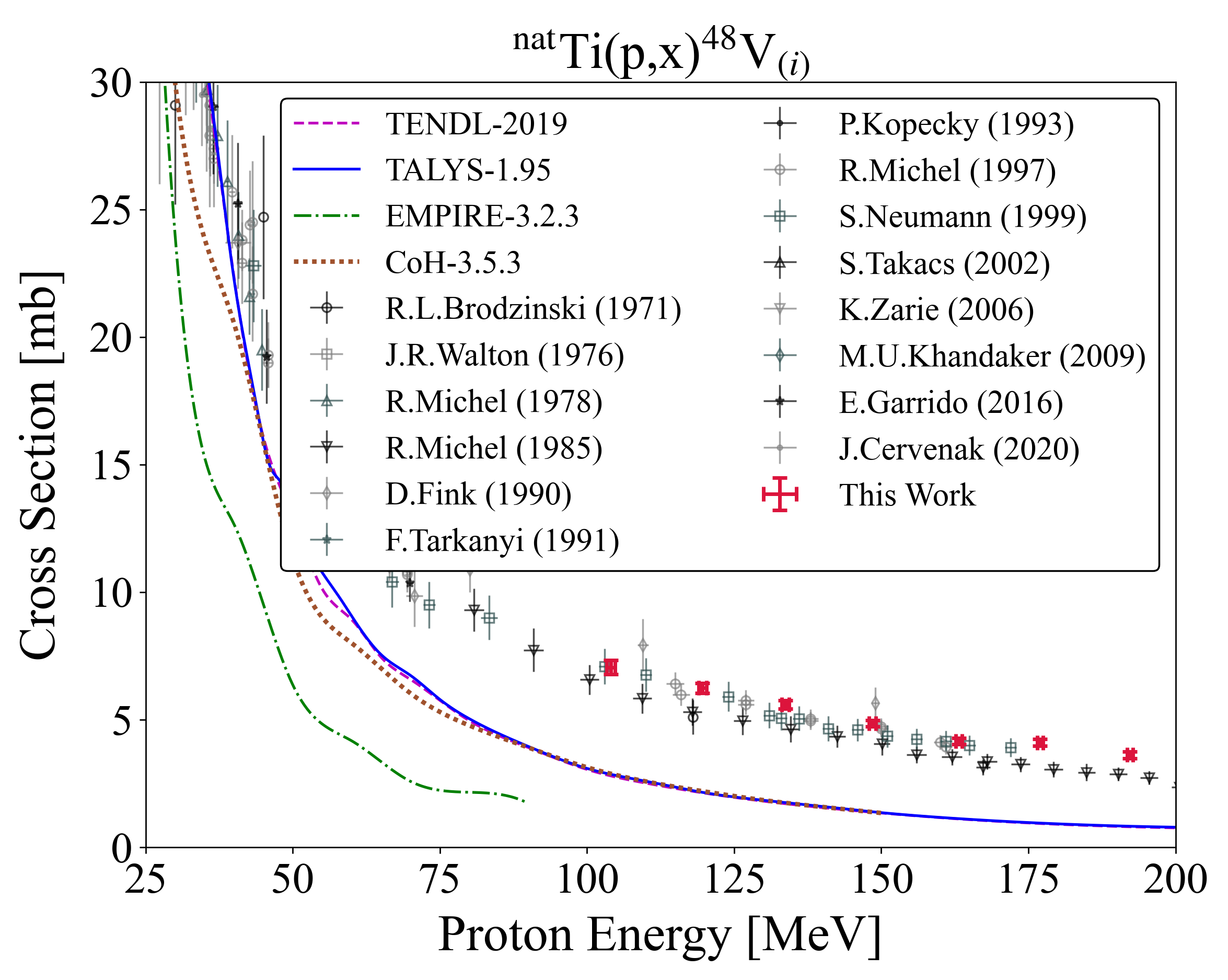}}
\vspace{-0.65cm}
\caption{Experimental and theoretical cross sections for $^{48}$V production.}\label{Ti_48V}
	{\includegraphics[width=1.0\columnwidth]{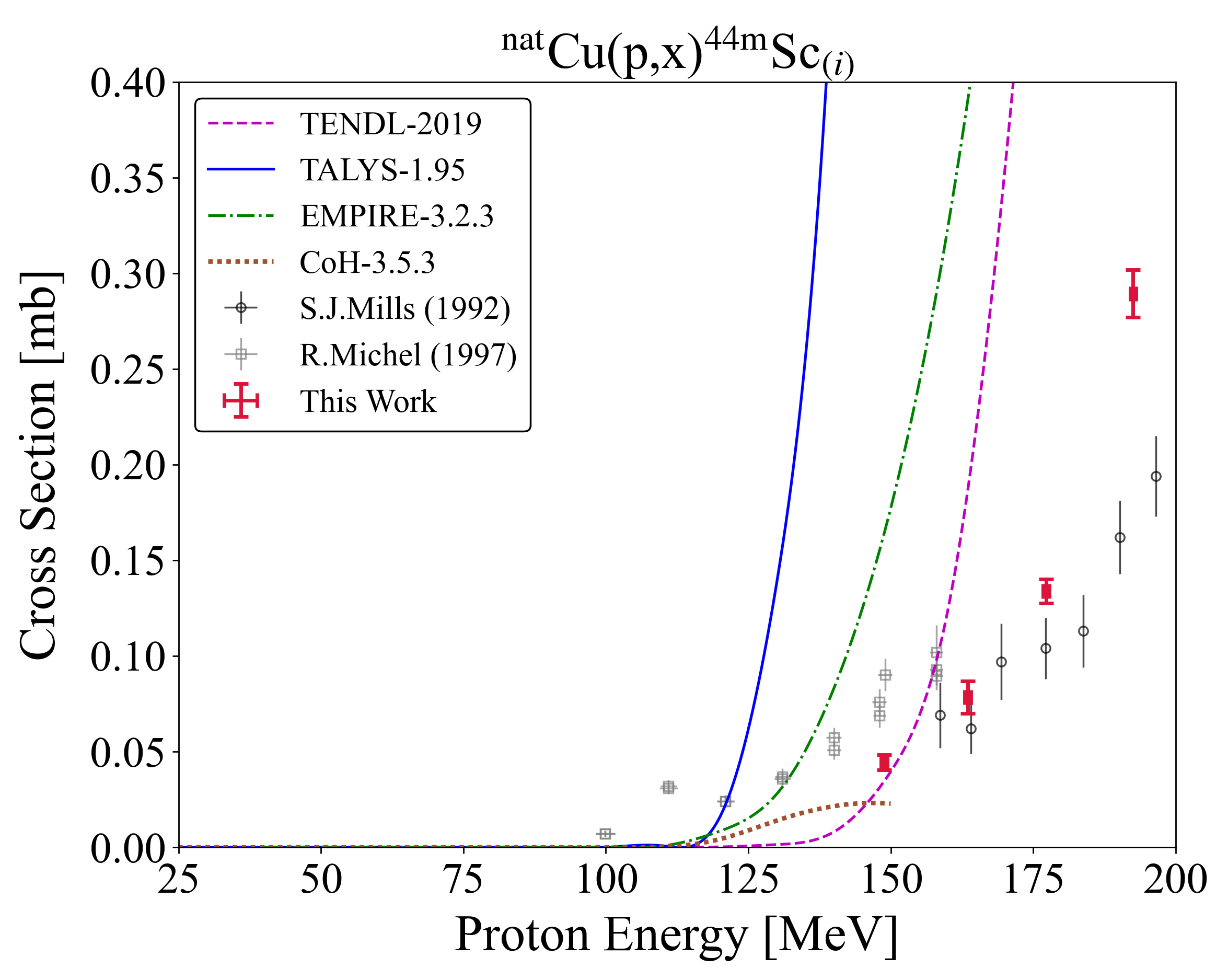}}
\vspace{-0.65cm}
\caption{Experimental and theoretical cross sections for $^{\textnormal{44m}}$Sc production.}\label{Cu_44mSC}
\end{figure}

\begin{figure}[H]
	{\includegraphics[width=1.0\columnwidth]{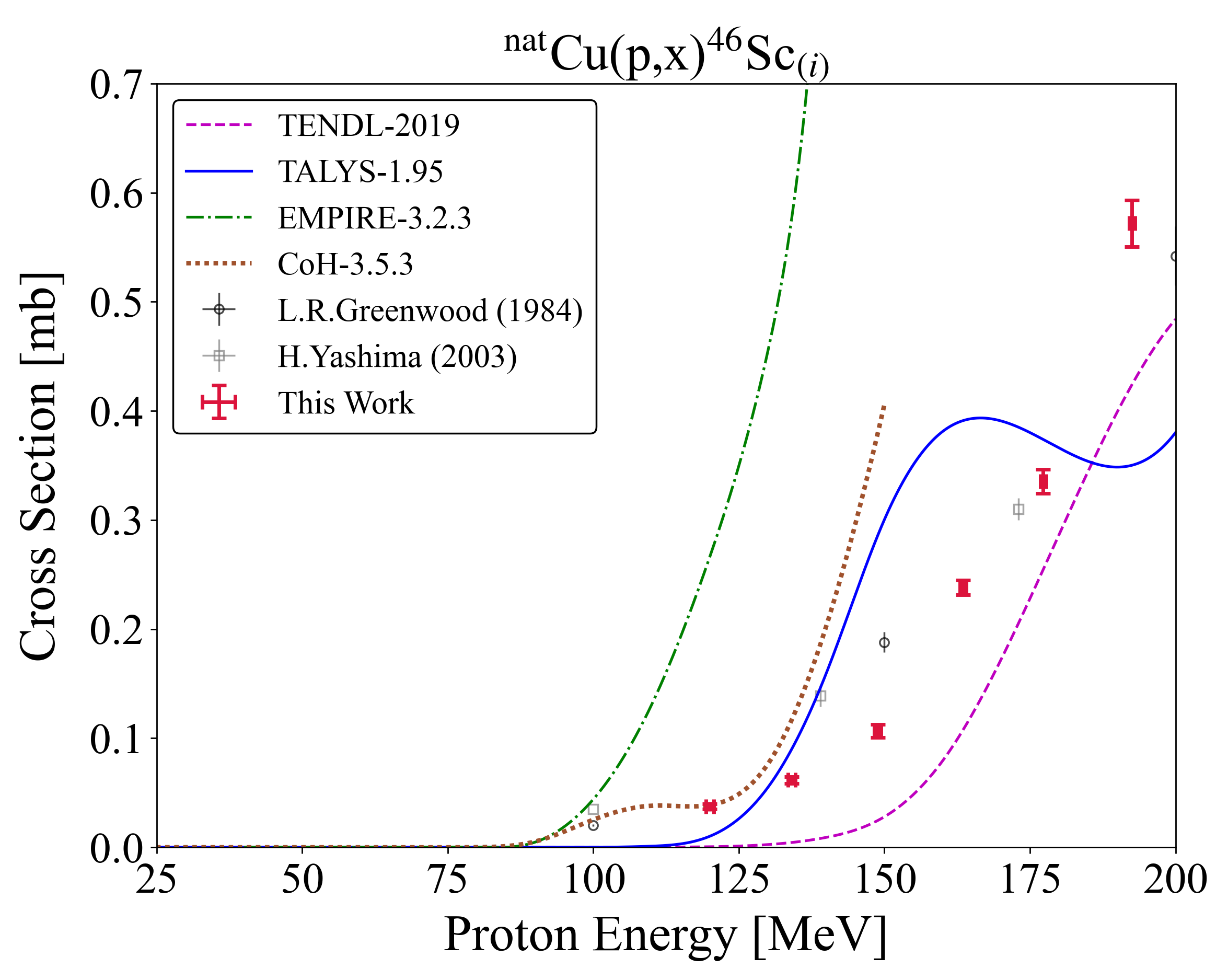}}
\vspace{-0.65cm}
\caption{Experimental and theoretical cross sections for $^{46}$Sc production.}\label{Cu_46SC}
	{\includegraphics[width=1.0\columnwidth]{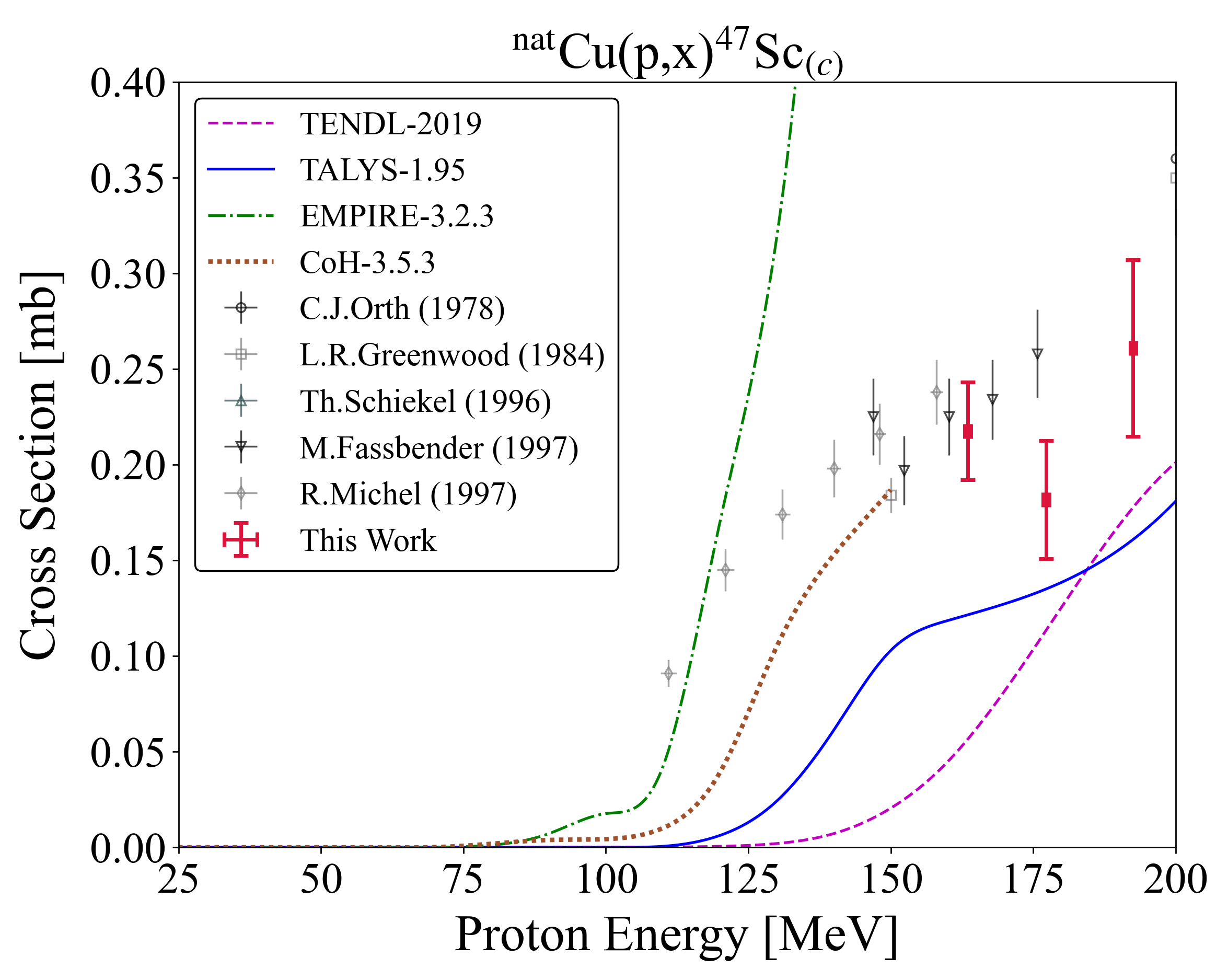}}
\vspace{-0.65cm}
\caption{Experimental and theoretical cross sections for $^{47}$Sc production.}\label{Cu_47SC}
	{\includegraphics[width=1.0\columnwidth]{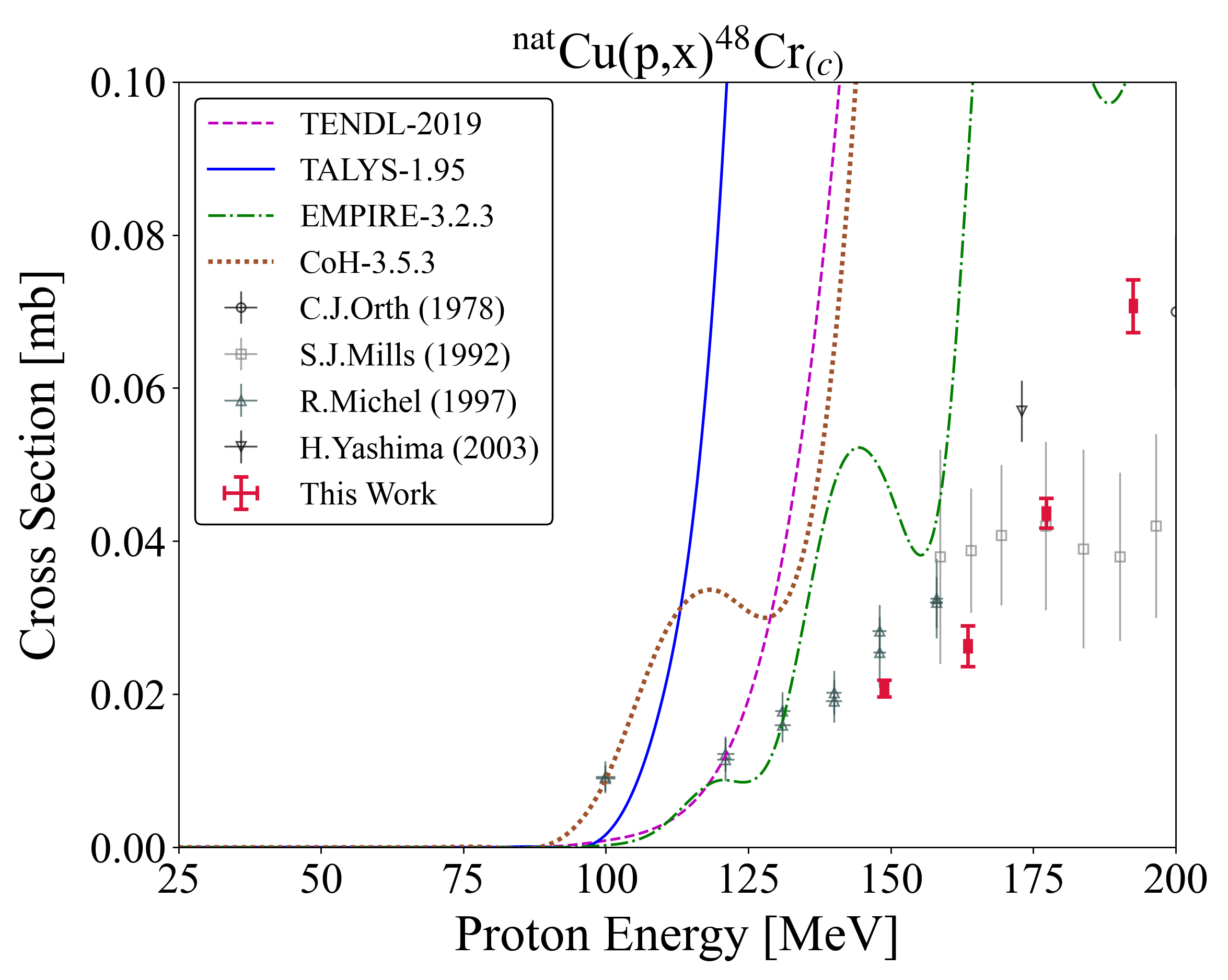}}
\vspace{-0.65cm}
\caption{Experimental and theoretical cross sections for $^{48}$Cr production.}\label{Cu_48CR}
\end{figure}

\begin{figure}[H]
	{\includegraphics[width=1.0\columnwidth]{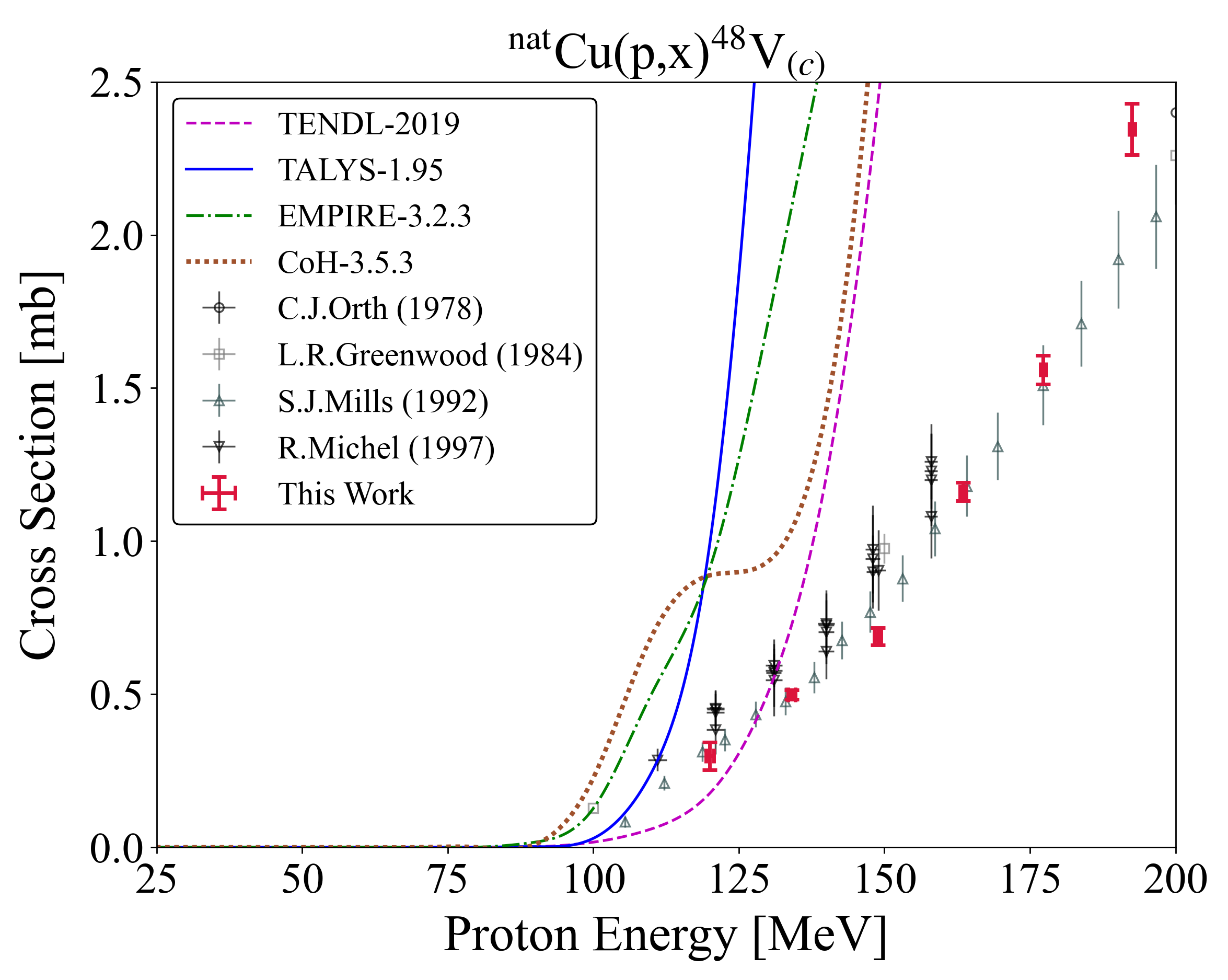}}
\vspace{-0.65cm}
\caption{Experimental and theoretical cross sections for $^{48}$V production.}\label{Cu_48V}
	{\includegraphics[width=1.0\columnwidth]{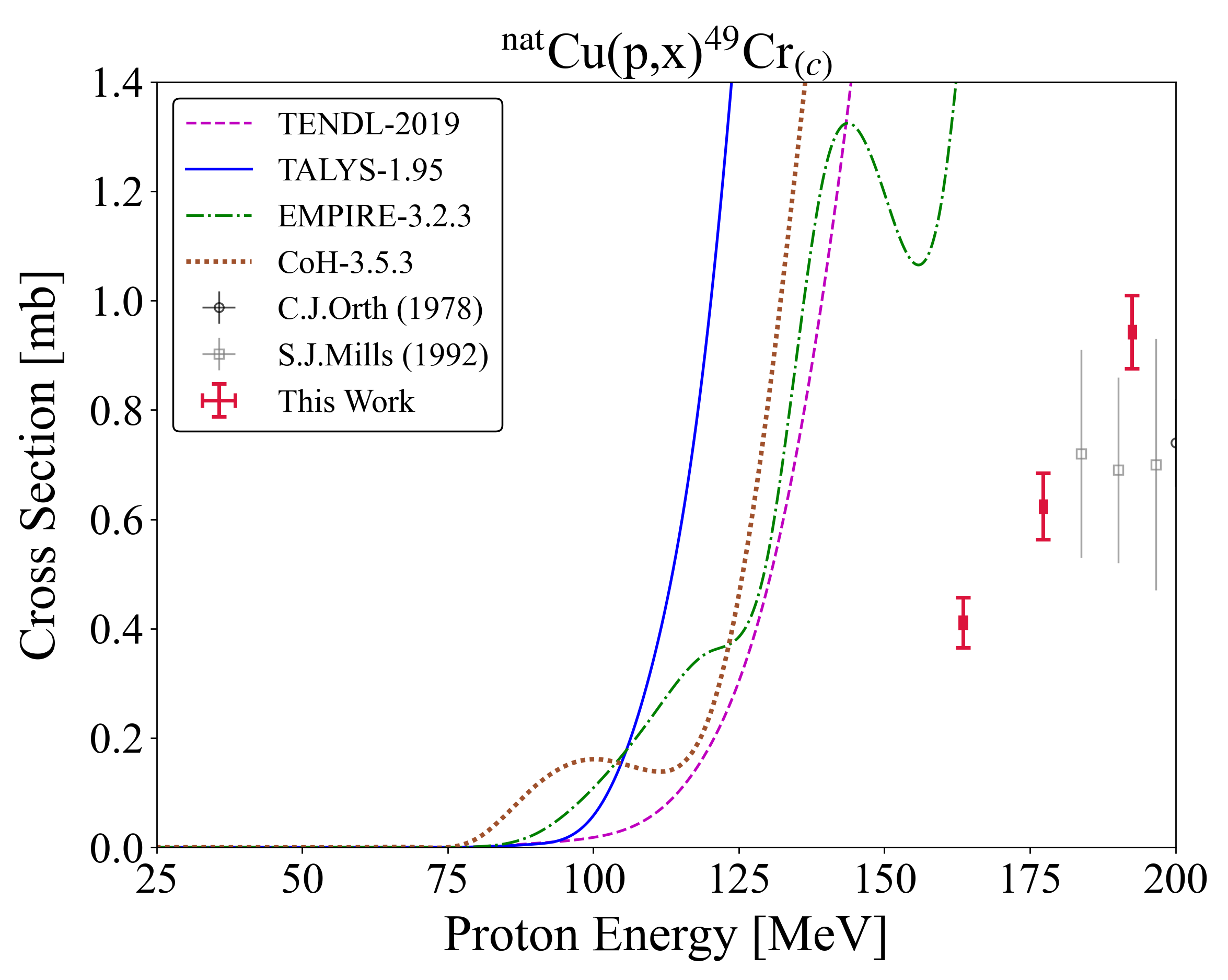}}
\vspace{-0.65cm}
\caption{Experimental and theoretical cross sections for $^{49}$Cr production.}\label{Cu_49CR}
	{\includegraphics[width=1.0\columnwidth]{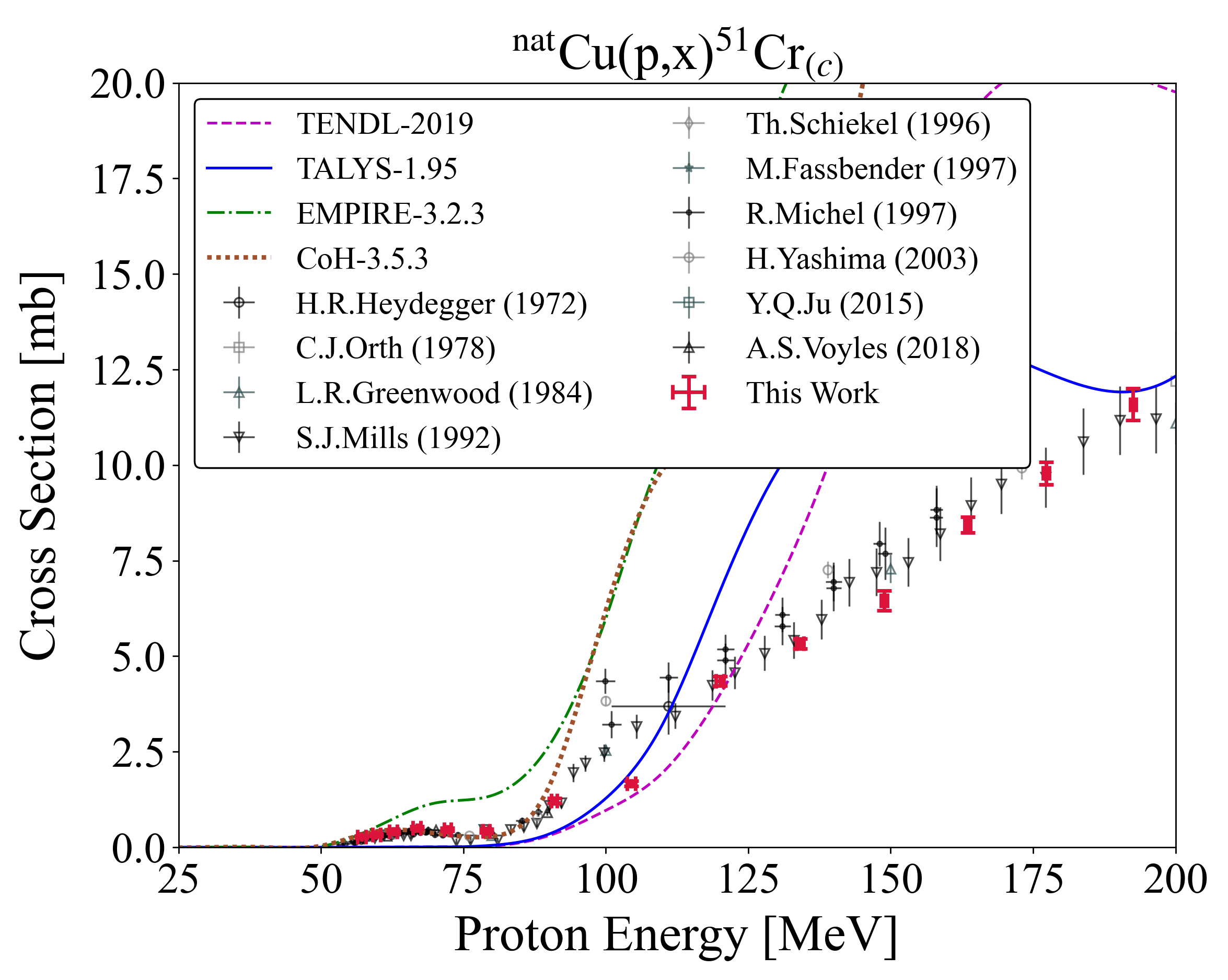}}
\vspace{-0.65cm}
\caption{Experimental and theoretical cross sections for $^{51}$Cr production.}\label{Cu_51CR}
\end{figure}

\begin{figure}[H]
	{\includegraphics[width=1.0\columnwidth]{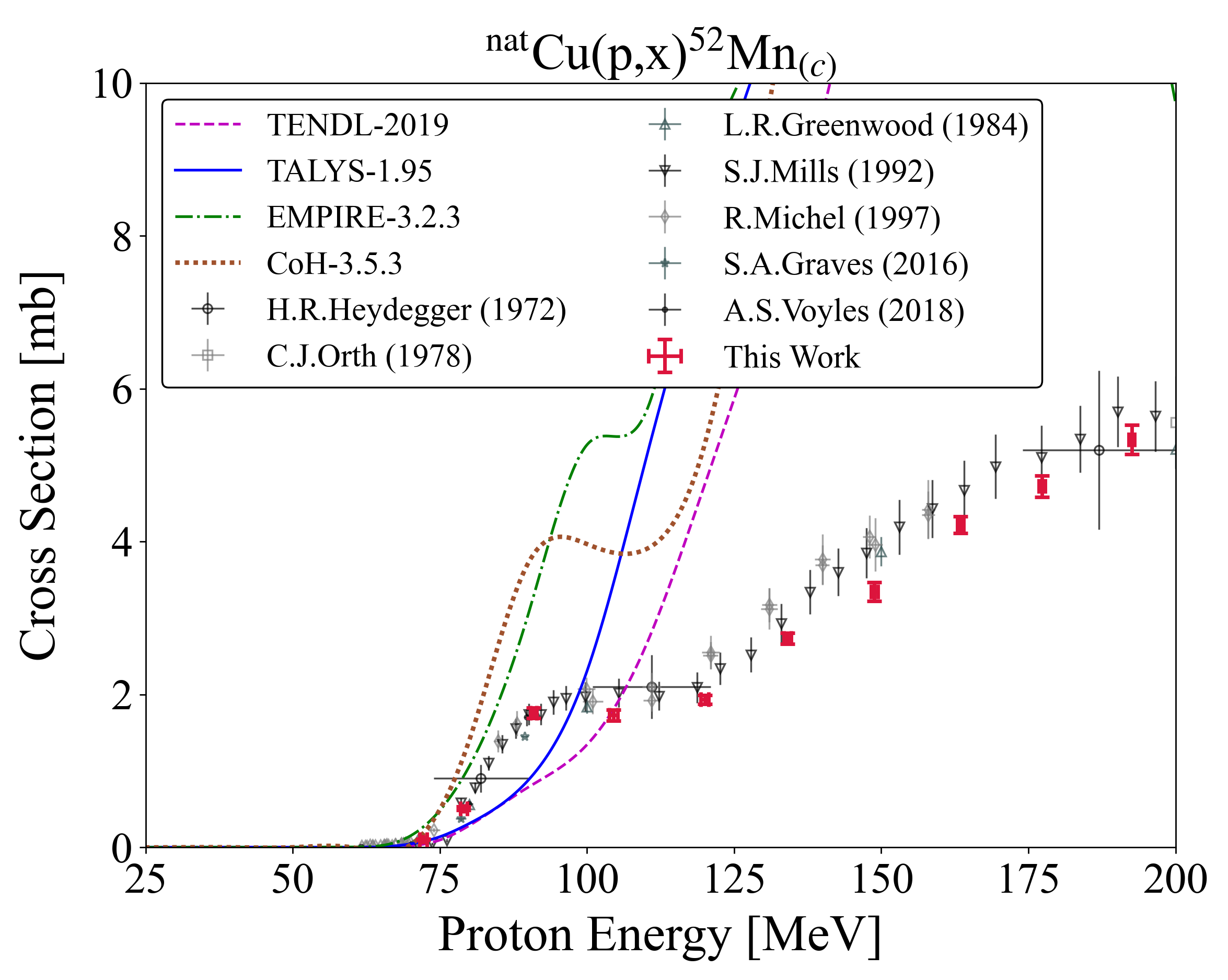}}
\vspace{-0.65cm}
\caption{Experimental and theoretical cross sections for $^{52}$Mn production.}\label{Cu_52MN}
	{\includegraphics[width=1.0\columnwidth]{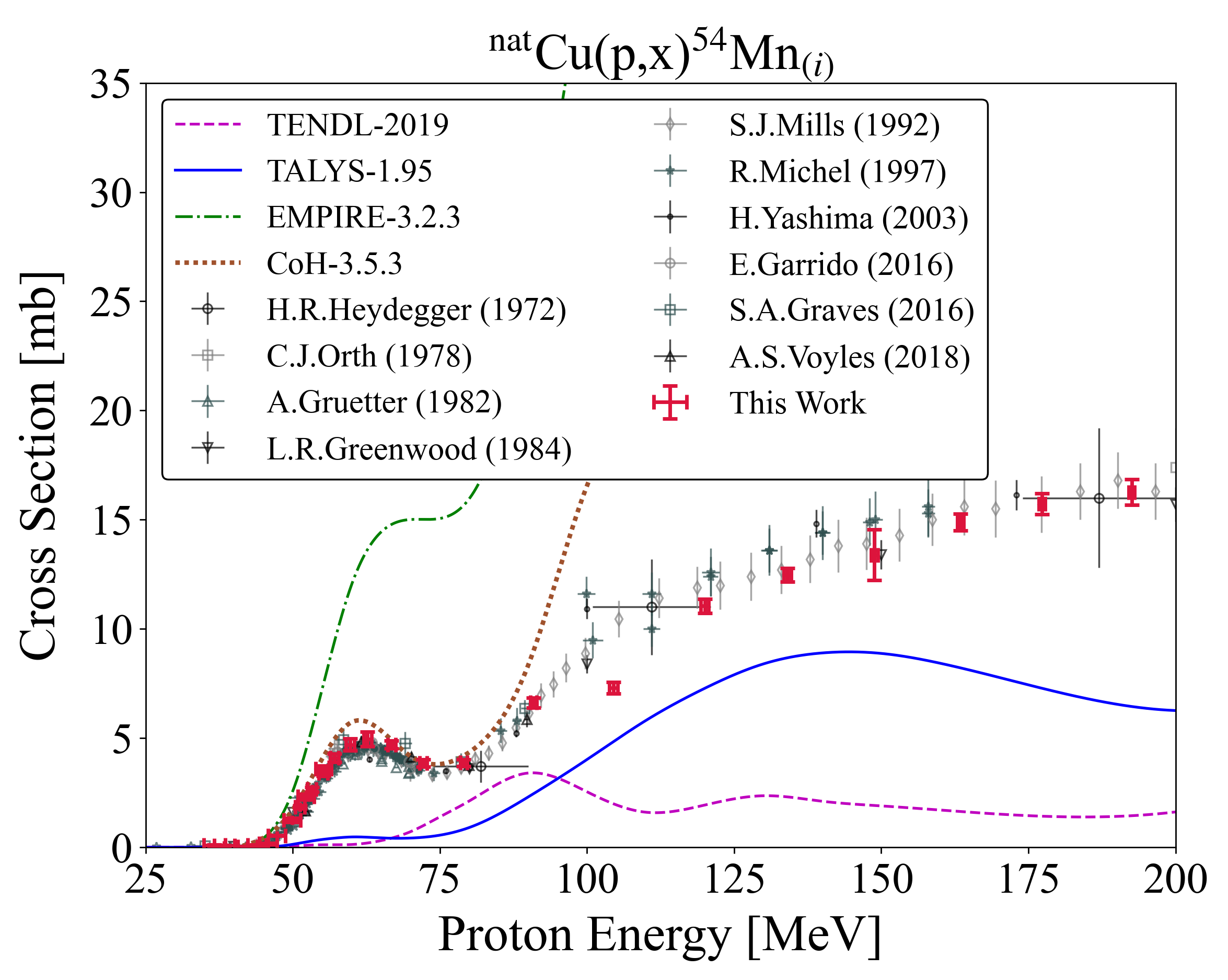}}
\vspace{-0.65cm}
\caption{Experimental and theoretical cross sections for $^{54}$Mn production.}\label{Cu_54MN}
	{\includegraphics[width=1.0\columnwidth]{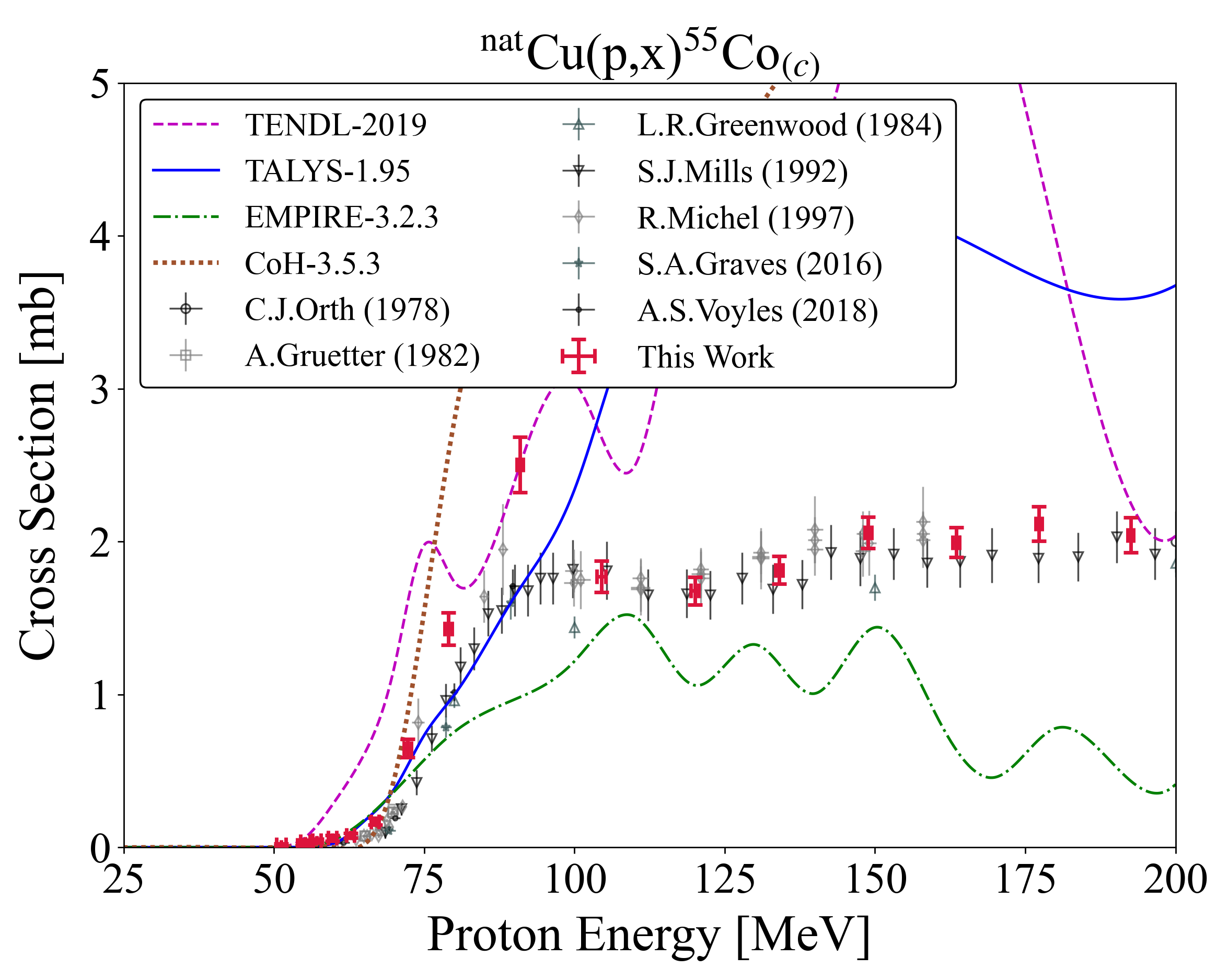}}
\vspace{-0.65cm}
\caption{Experimental and theoretical cross sections for $^{55}$Co production.}\label{Cu_55CO}
\end{figure}

\begin{figure}[H]
	{\includegraphics[width=1.0\columnwidth]{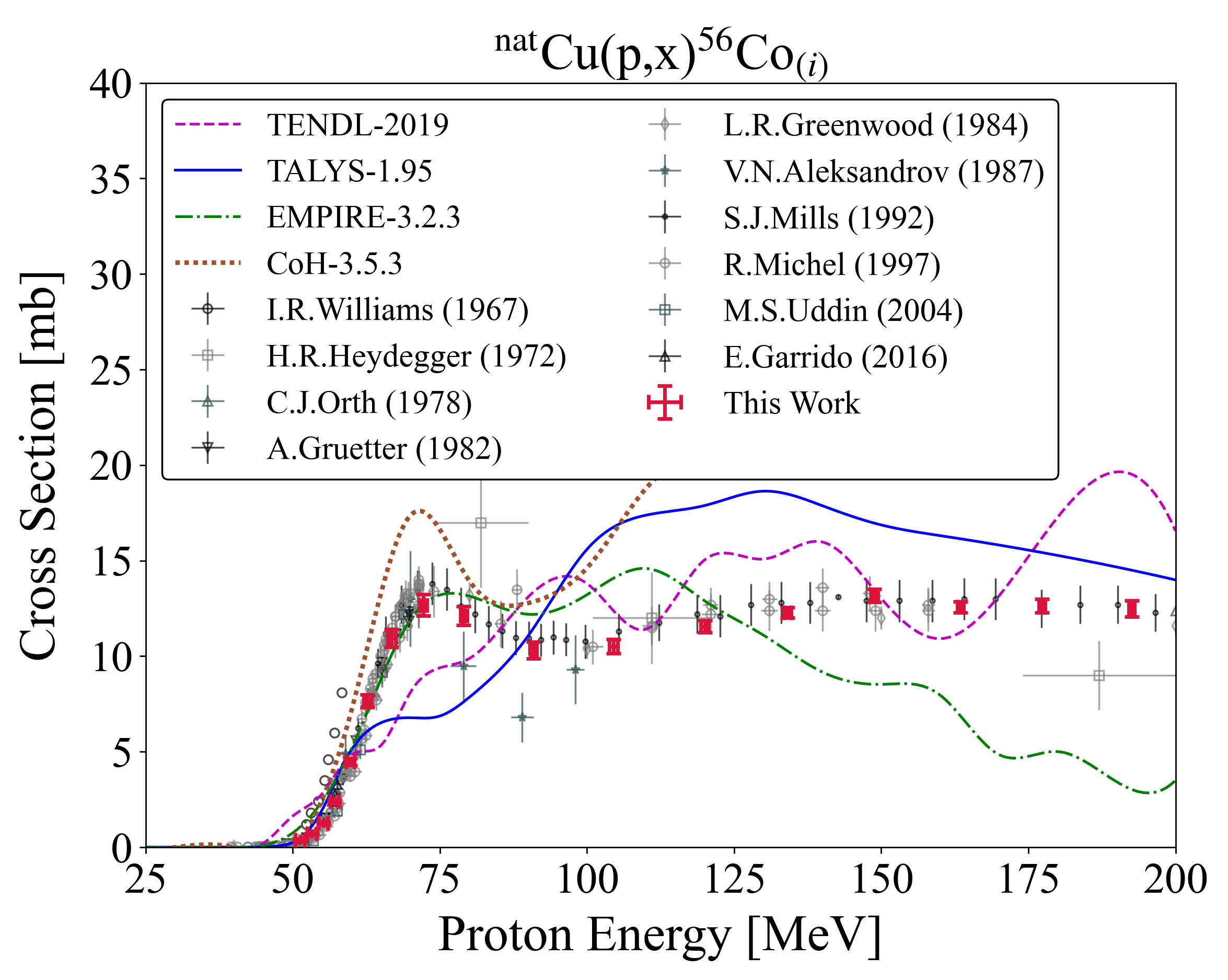}}
\vspace{-0.65cm}
\caption{Experimental and theoretical cross sections for $^{56}$Co production.}\label{Cu_56CO}
	{\includegraphics[width=1.0\columnwidth]{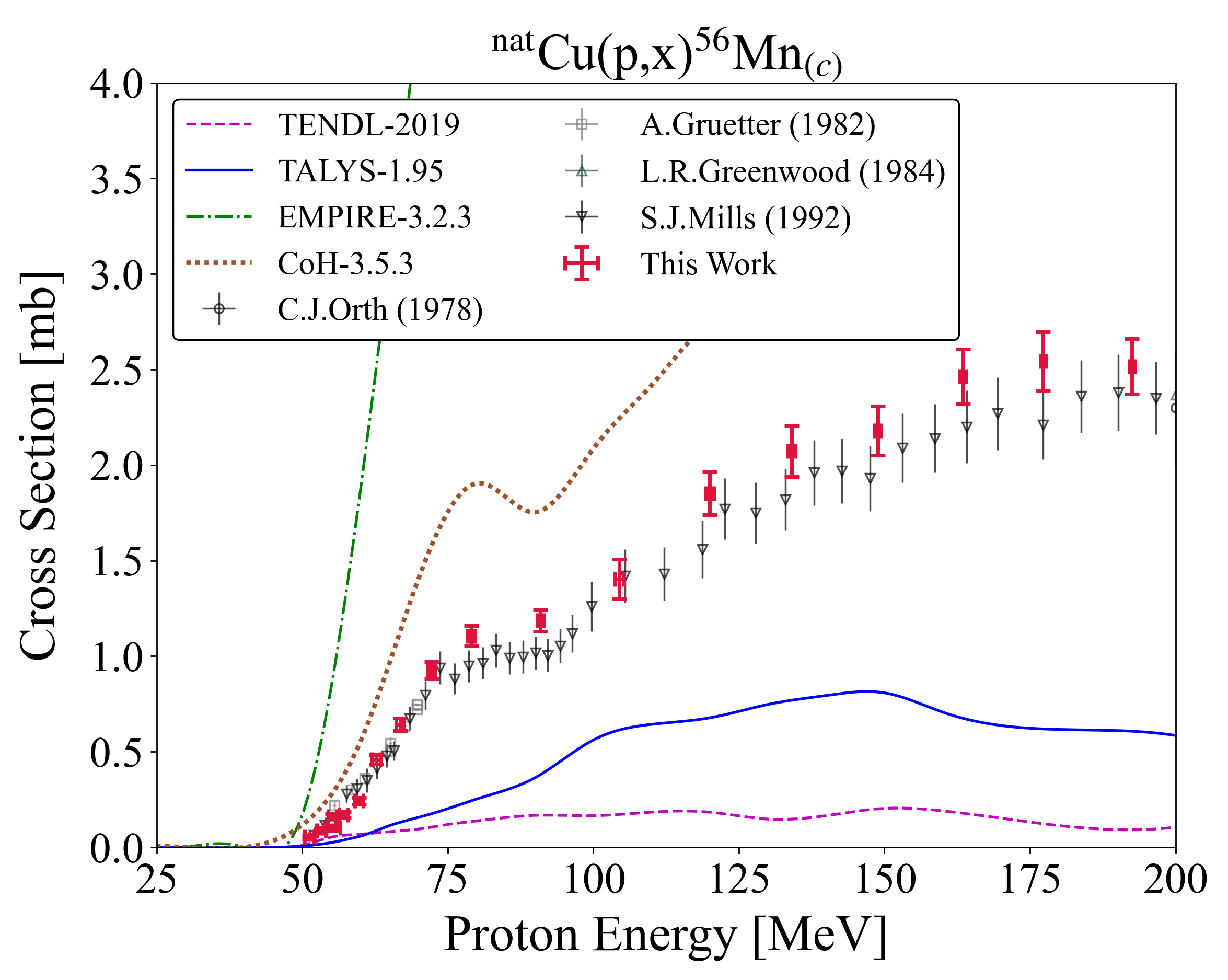}}
\vspace{-0.65cm}
\caption{Experimental and theoretical cross sections for $^{56}$Mn production.}\label{Cu_56MN}
	{\includegraphics[width=1.0\columnwidth]{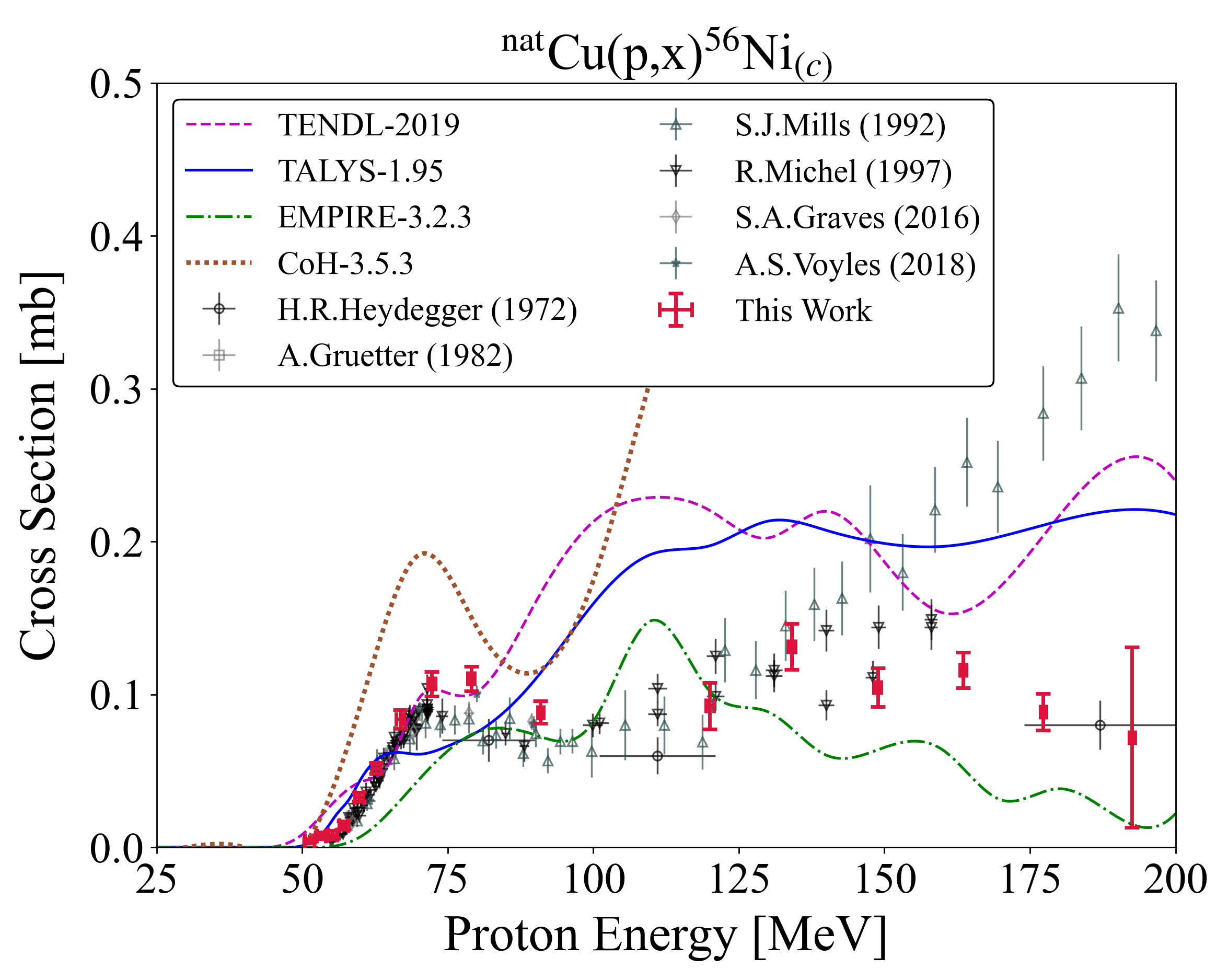}}
\vspace{-0.65cm}
\caption{Experimental and theoretical cross sections for $^{56}$Ni production.}\label{Cu_56NI}
\end{figure}

\begin{figure}[H]
	{\includegraphics[width=1.0\columnwidth]{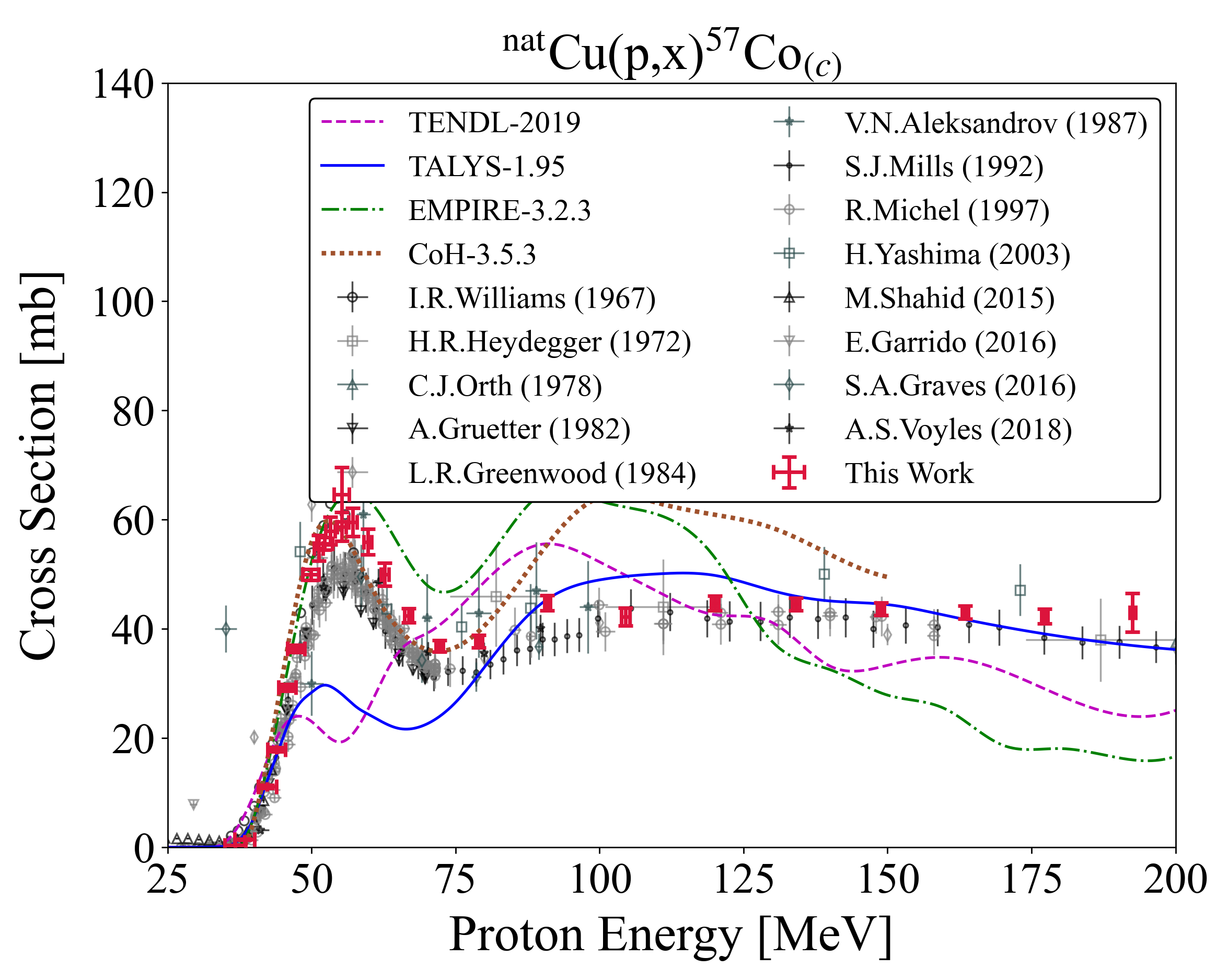}}
\vspace{-0.65cm}
\caption{Experimental and theoretical cross sections for $^{57}$Co production.}\label{Cu_57CO}
	{\includegraphics[width=1.0\columnwidth]{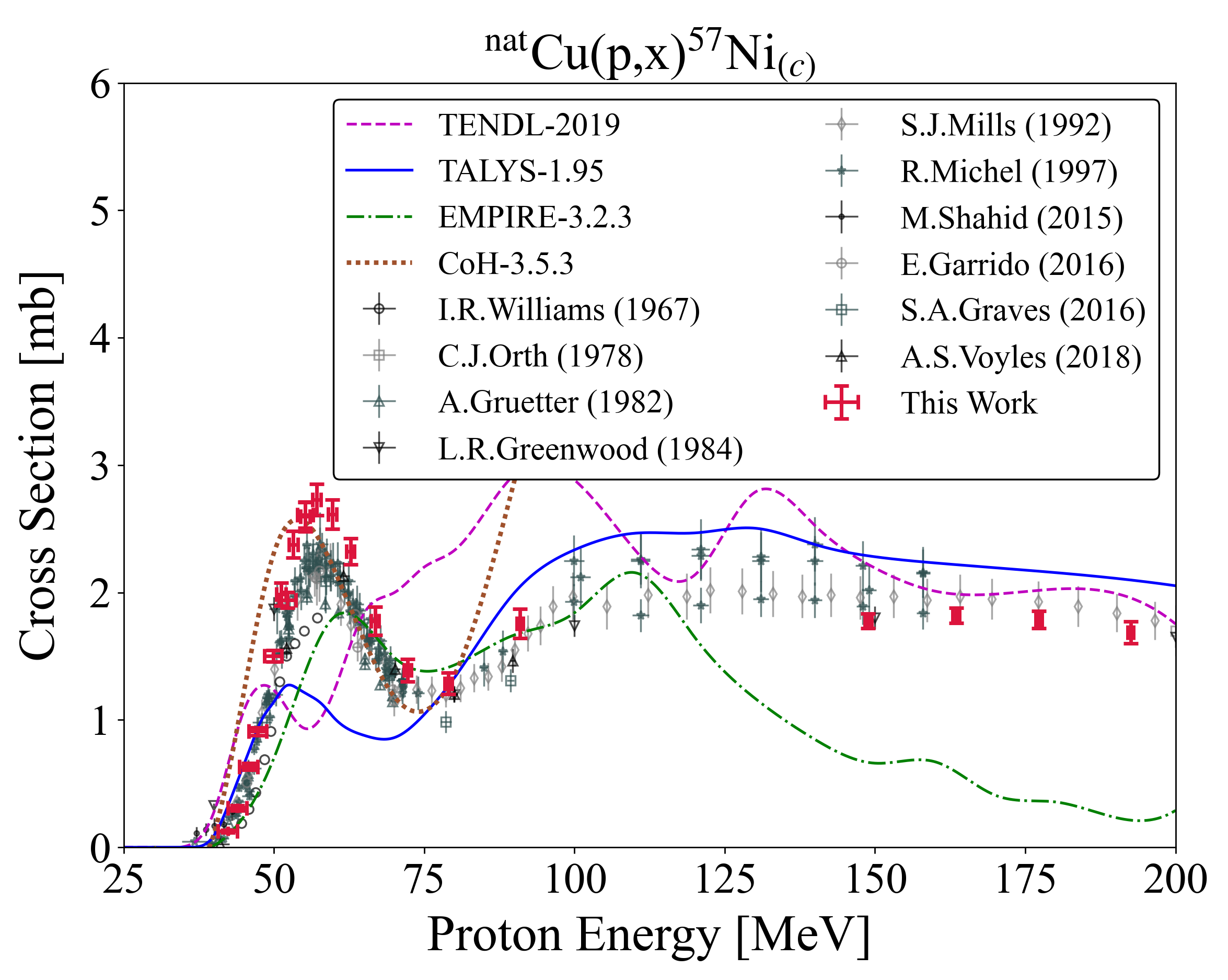}}
\vspace{-0.65cm}
\caption{Experimental and theoretical cross sections for $^{57}$Ni production.}\label{Cu_57NI}
	{\includegraphics[width=1.0\columnwidth]{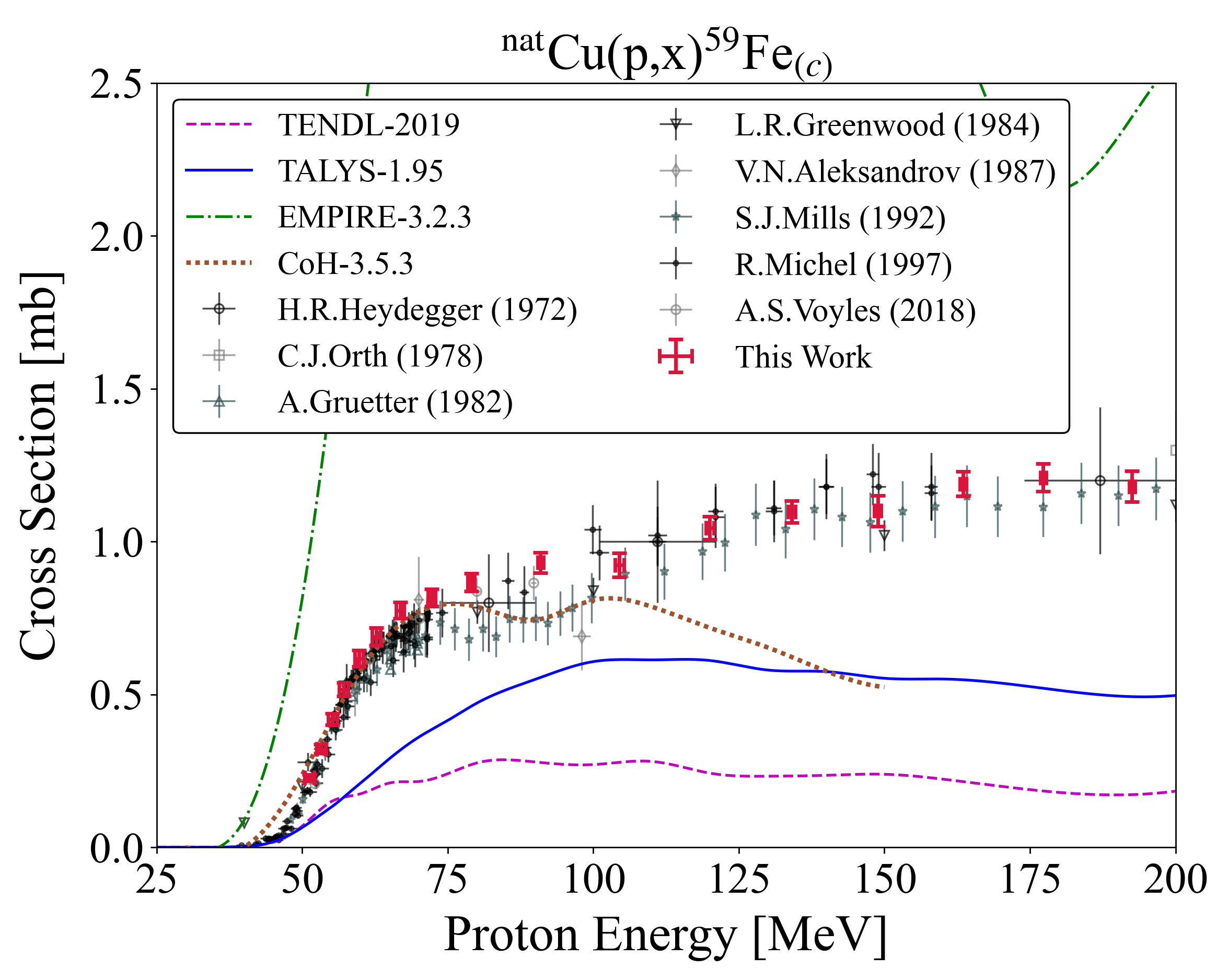}}
\vspace{-0.65cm}
\caption{Experimental and theoretical cross sections for $^{59}$Fe production.}\label{Cu_59FE}
\end{figure}

\begin{figure}[H]
	{\includegraphics[width=1.0\columnwidth]{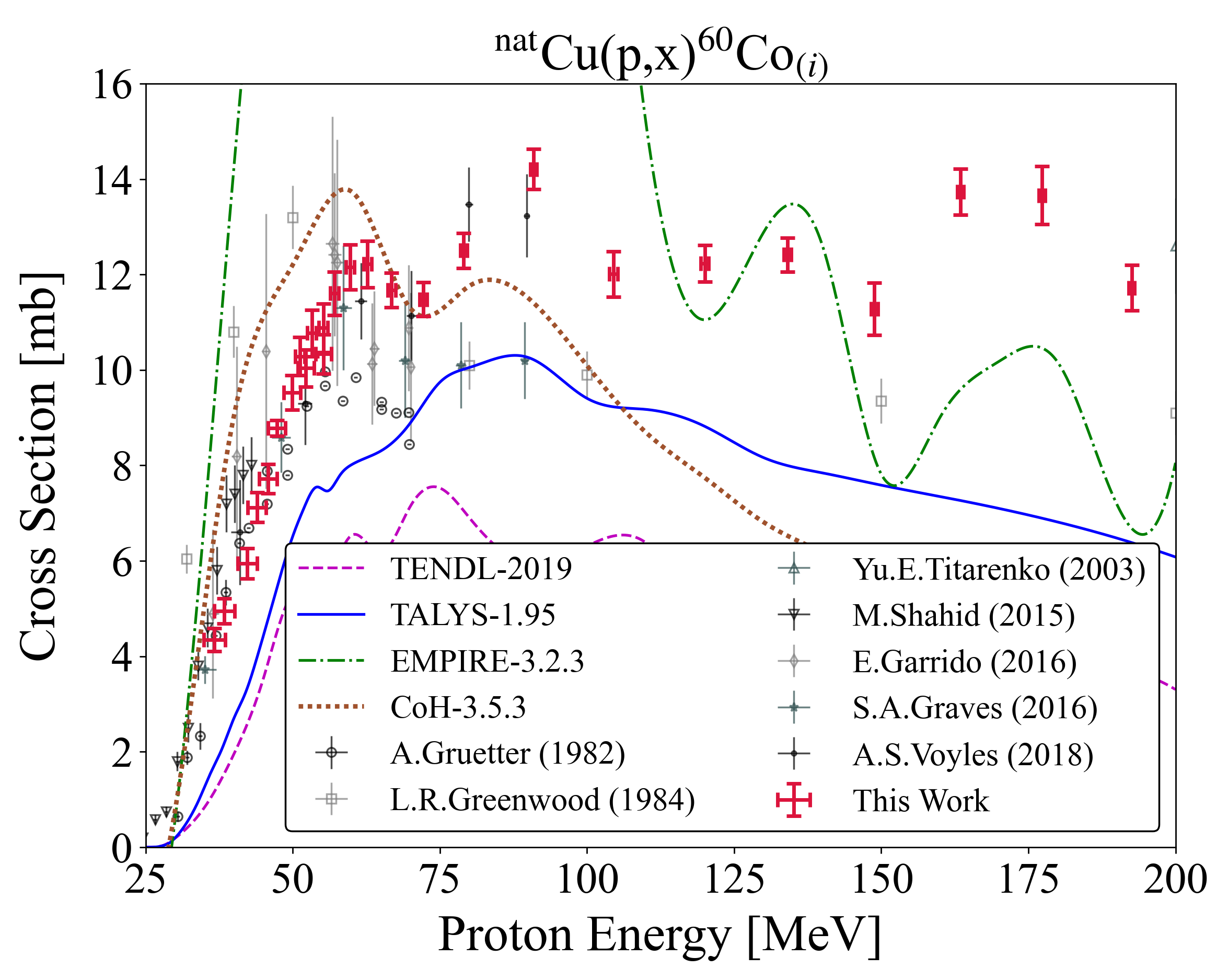}}
\vspace{-0.65cm}
\caption{Experimental and theoretical cross sections for $^{60}$Co production.}\label{Cu_60CO}
	{\includegraphics[width=1.0\columnwidth]{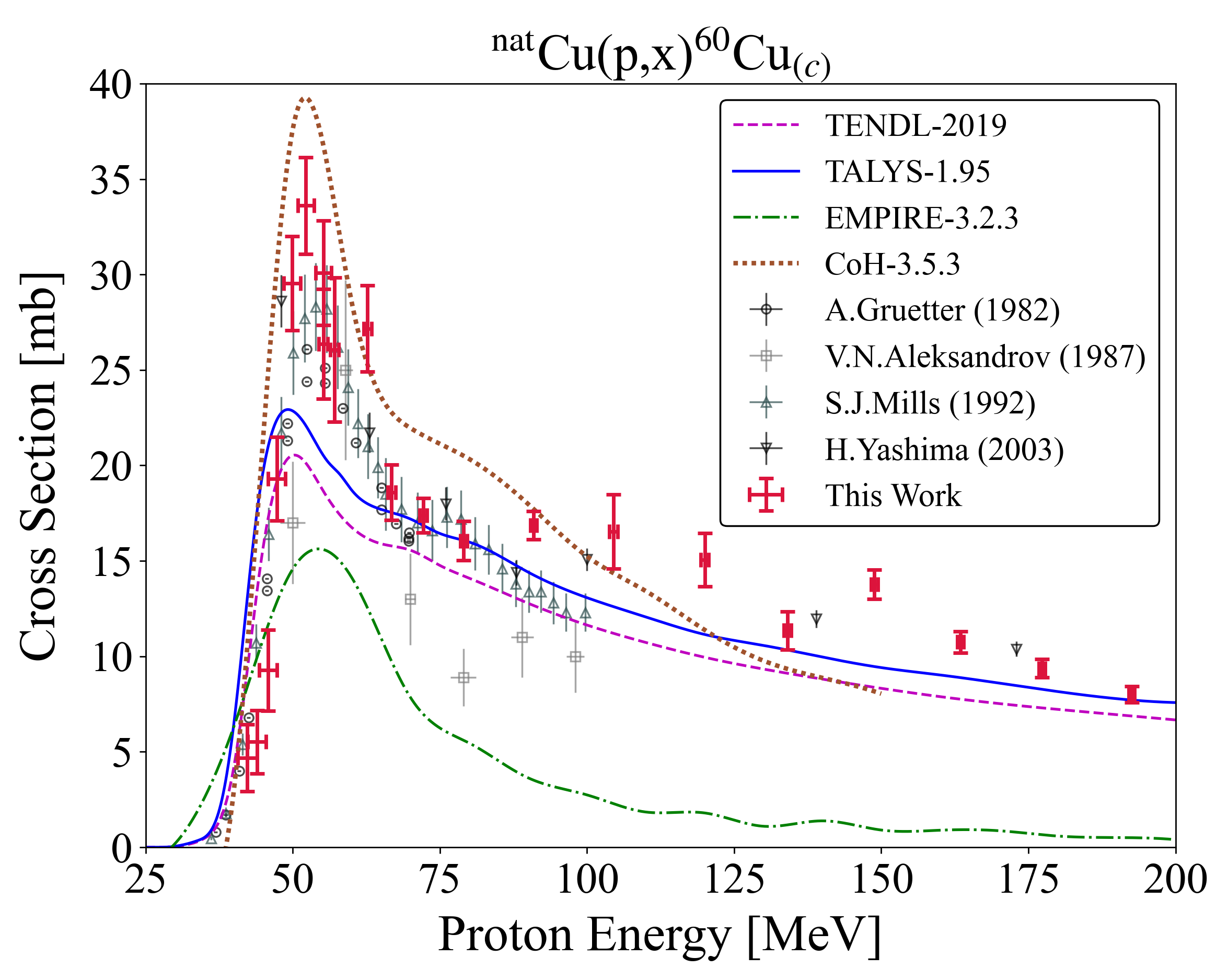}}
\vspace{-0.65cm}
\caption{Experimental and theoretical cross sections for $^{60}$Cu production.}\label{Cu_60CU}
	{\includegraphics[width=1.0\columnwidth]{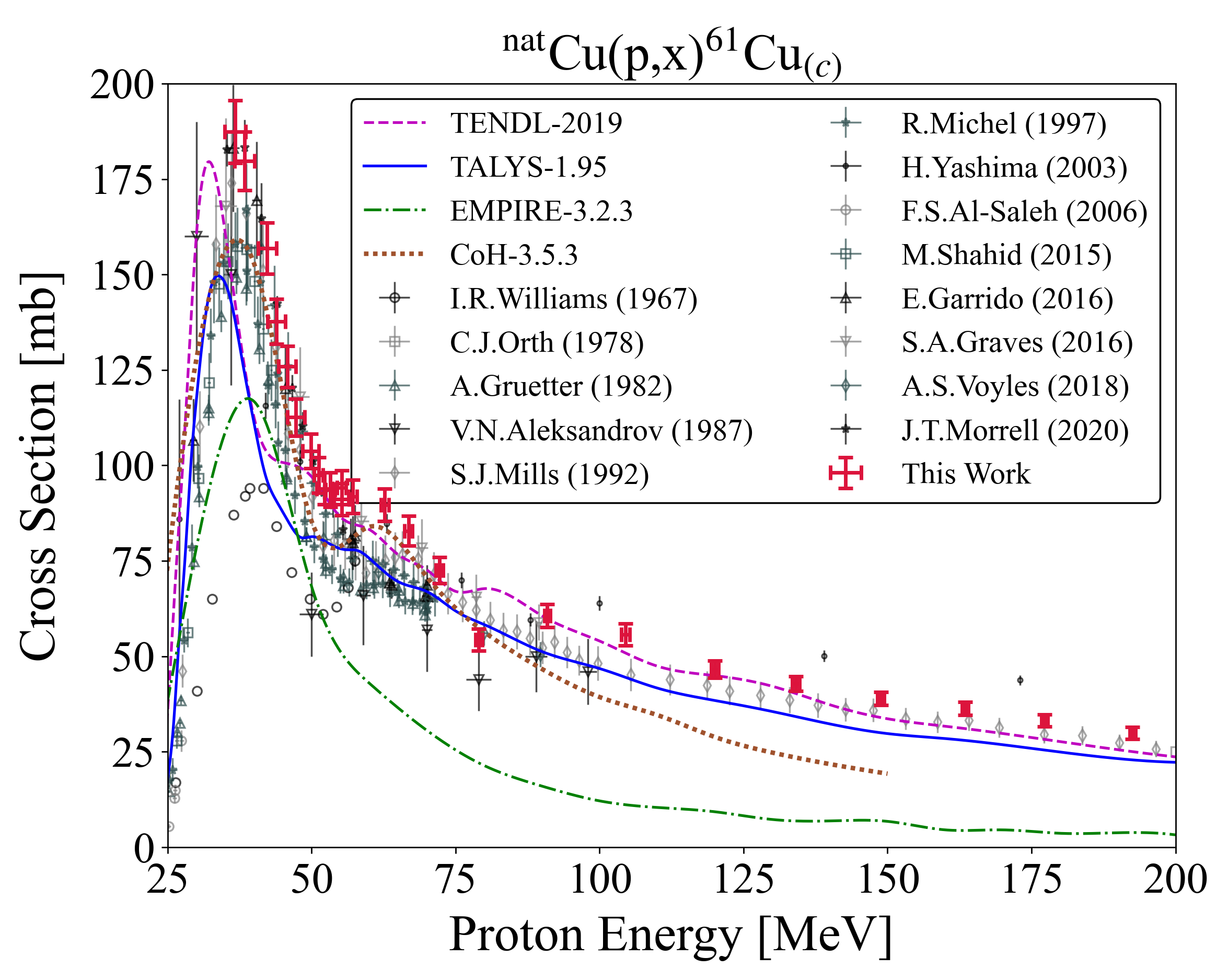}}
\vspace{-0.65cm}
\caption{Experimental and theoretical cross sections for $^{61}$Cu production.}\label{Cu_61CU}
\end{figure}

\begin{figure}[H]
	{\includegraphics[width=1.0\columnwidth]{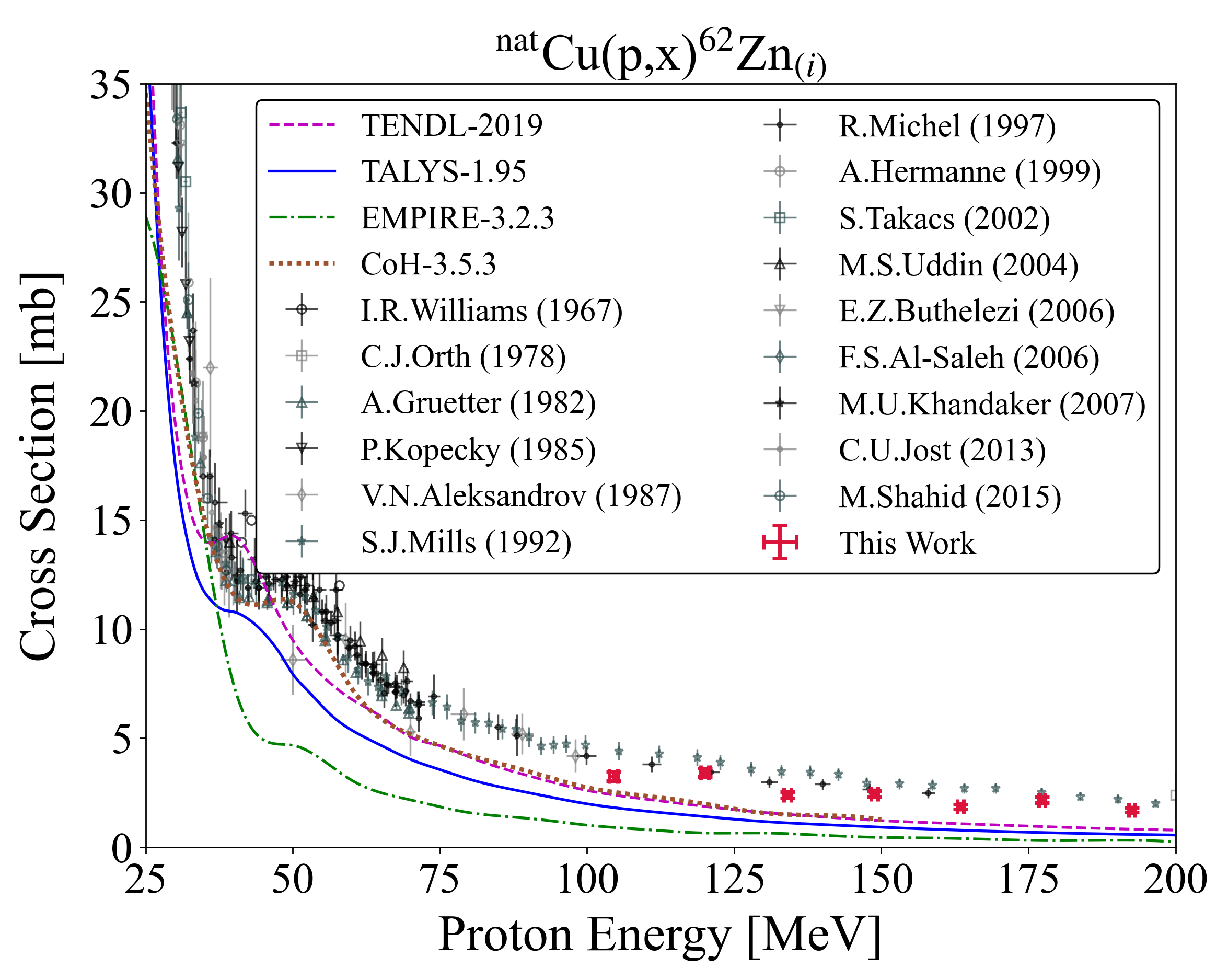}}
\vspace{-0.65cm}
\caption{Experimental and theoretical cross sections for $^{62}$Zn production.}\label{Cu_62ZN}
	{\includegraphics[width=1.0\columnwidth]{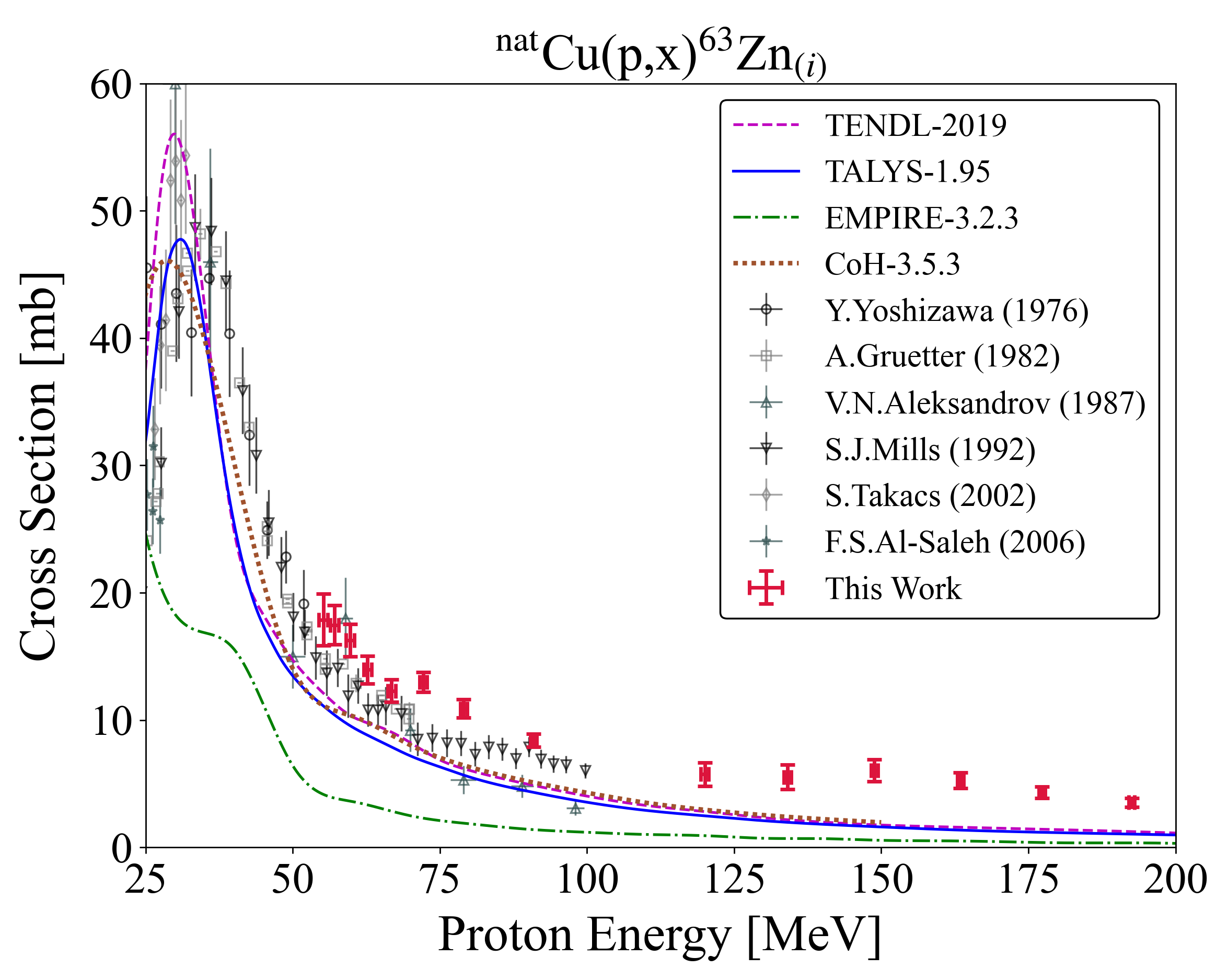}}
\vspace{-0.65cm}
\caption{Experimental and theoretical cross sections for $^{63}$Zn production.}\label{Cu_63ZN}
	{\includegraphics[width=1.0\columnwidth]{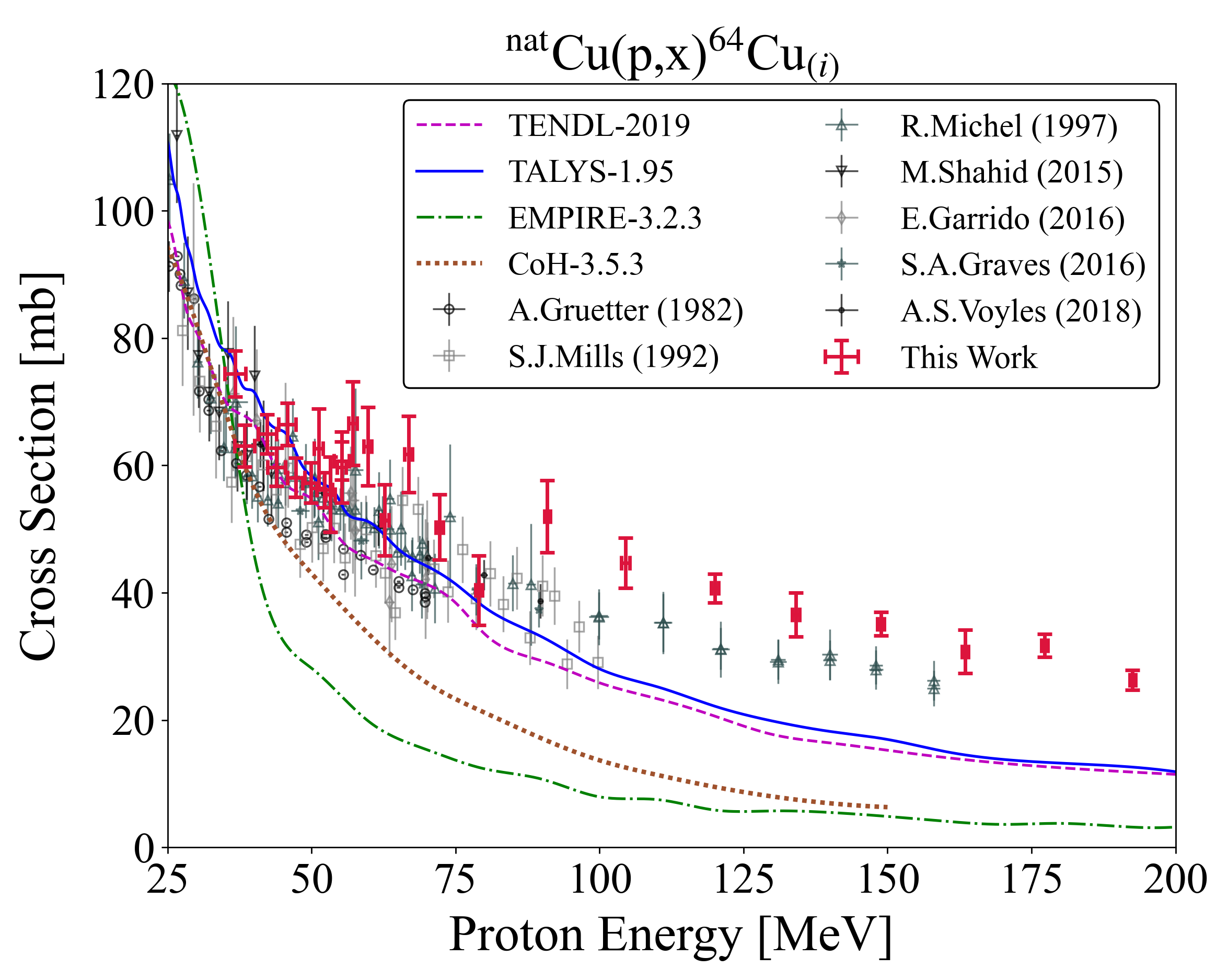}}
\vspace{-0.65cm}
\caption{Experimental and theoretical cross sections for $^{64}$Cu production.}\label{Cu_64CU}
\end{figure}

\begin{figure}[H]
	{\includegraphics[width=1.0\columnwidth]{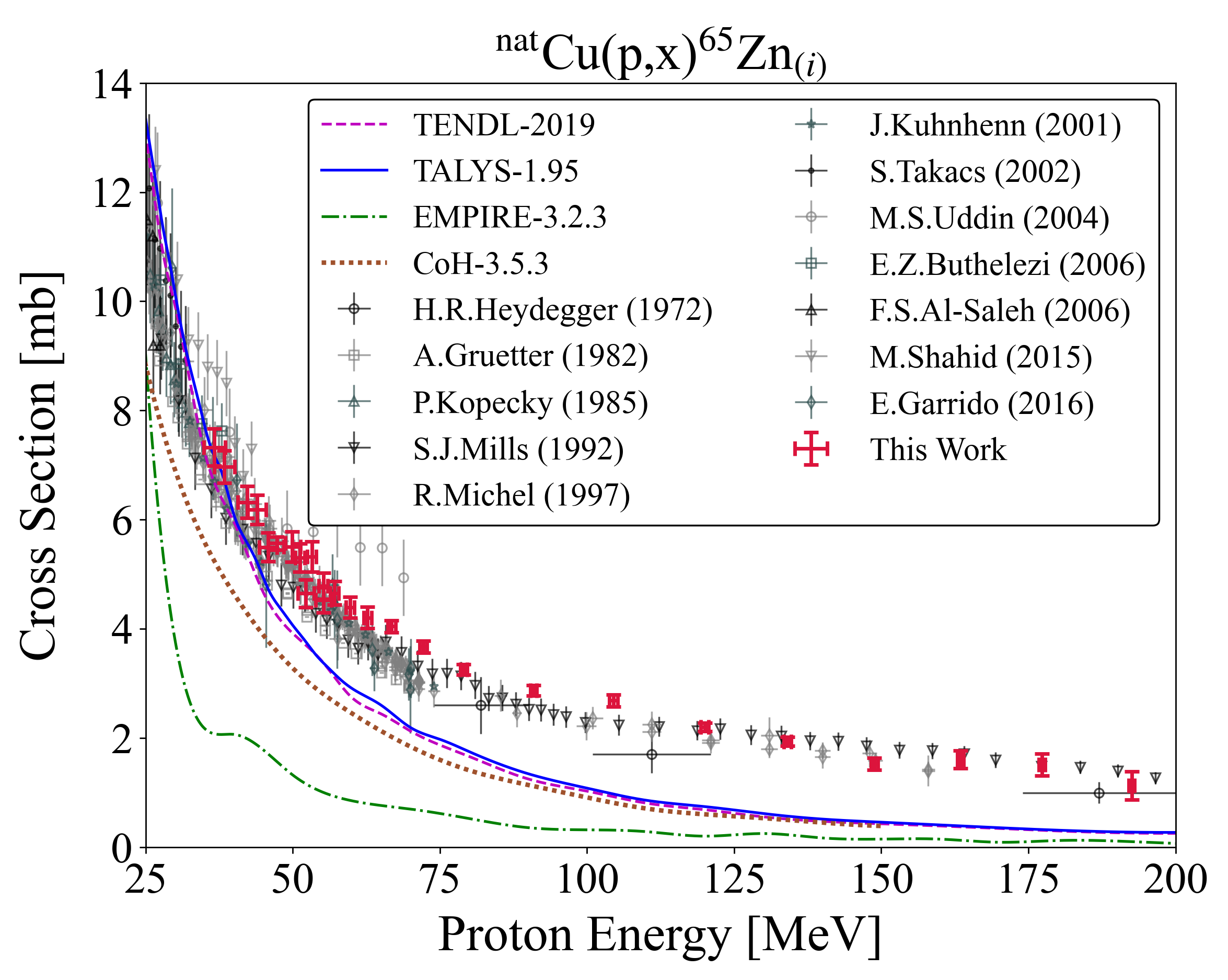}}
\vspace{-0.65cm}
\caption{Experimental and theoretical cross sections for $^{65}$Zn production.}\label{Cu_65ZN}
	{\includegraphics[width=1.0\columnwidth]{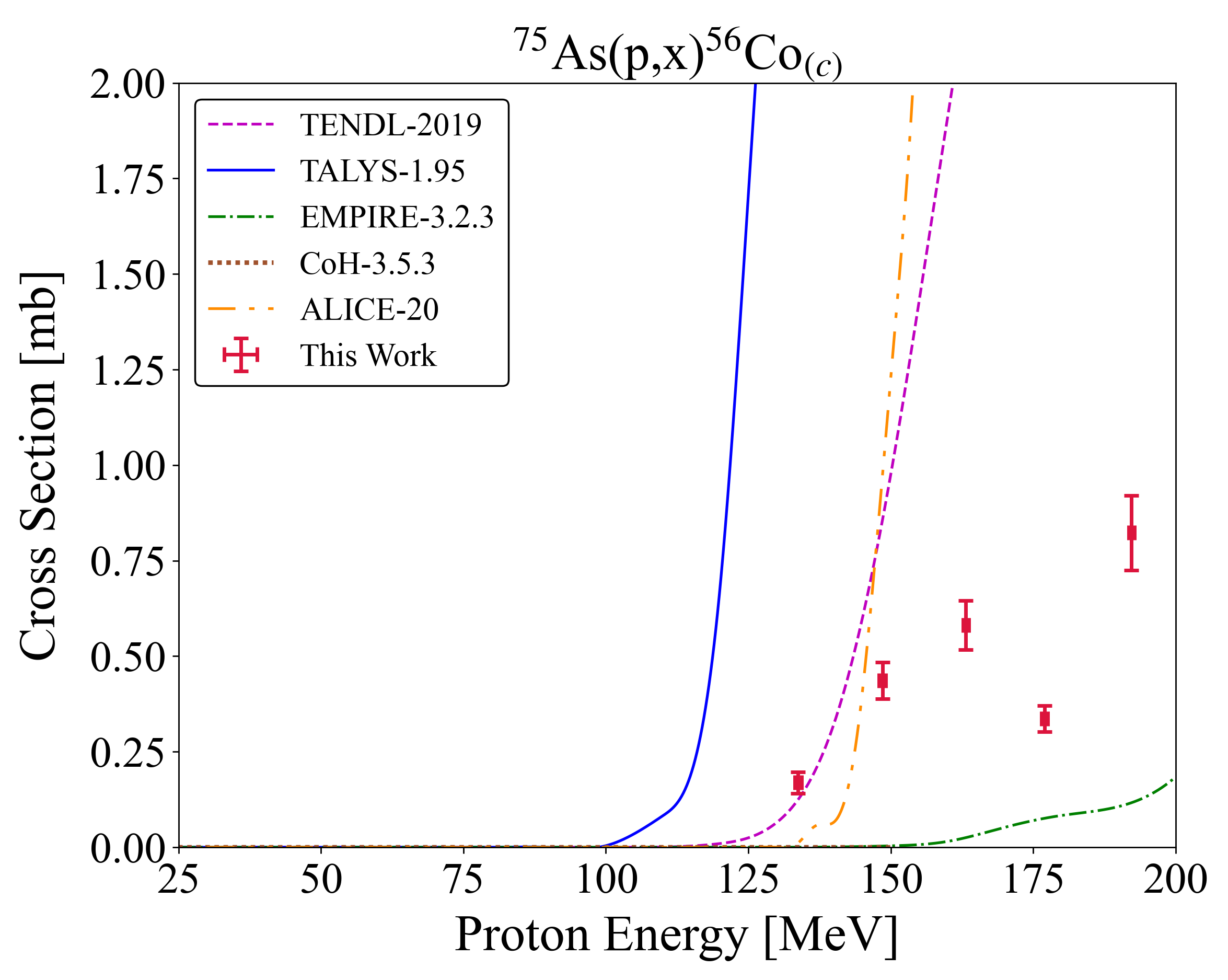}}
\vspace{-0.65cm}
\caption{Experimental and theoretical cross sections for $^{56}$Co production.}\label{As_56CO}
	{\includegraphics[width=1.0\columnwidth]{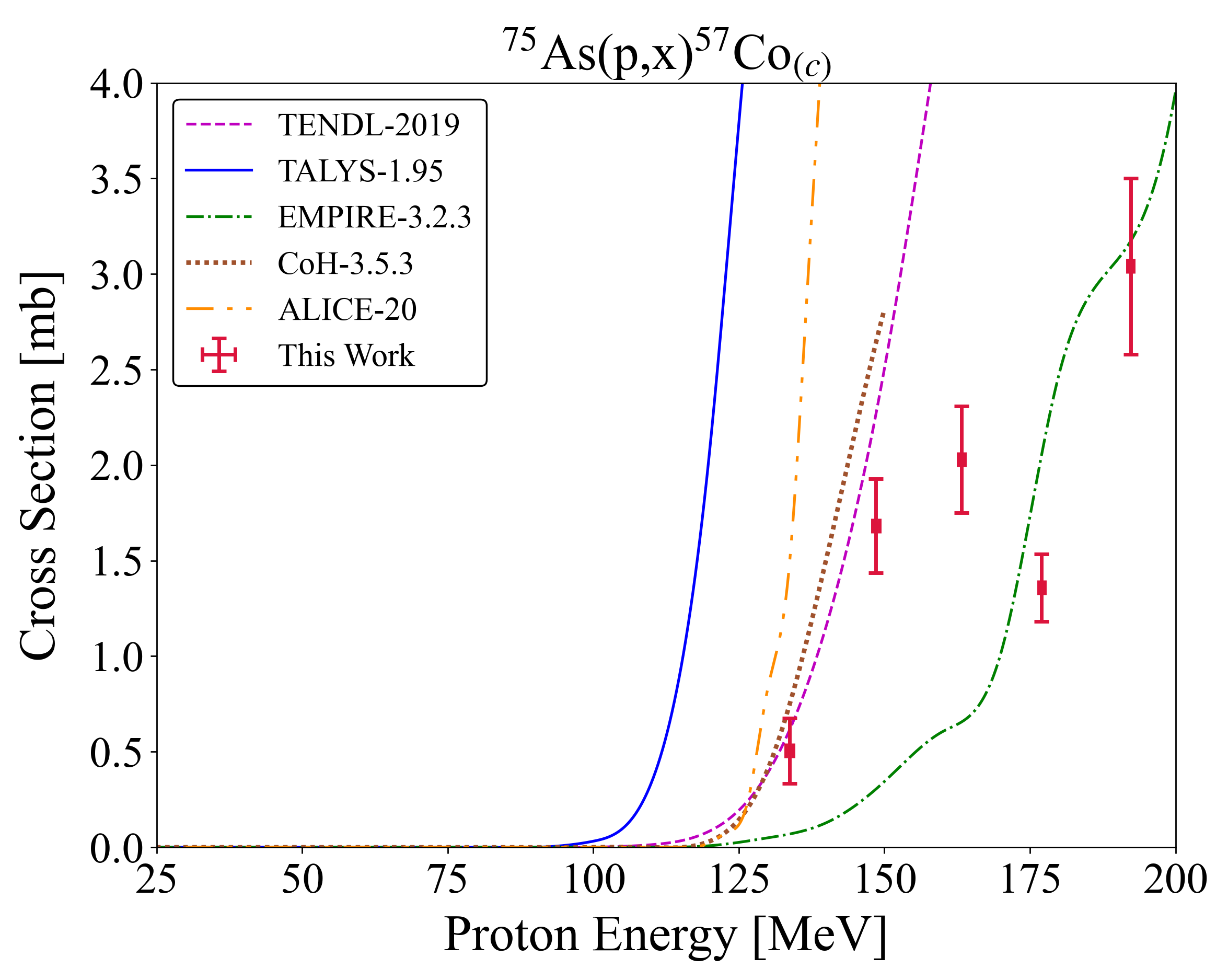}}
\vspace{-0.65cm}
\caption{Experimental and theoretical cross sections for $^{57}$Co production.}\label{As_57CO}
\end{figure}

\begin{figure}[H]
	{\includegraphics[width=1.0\columnwidth]{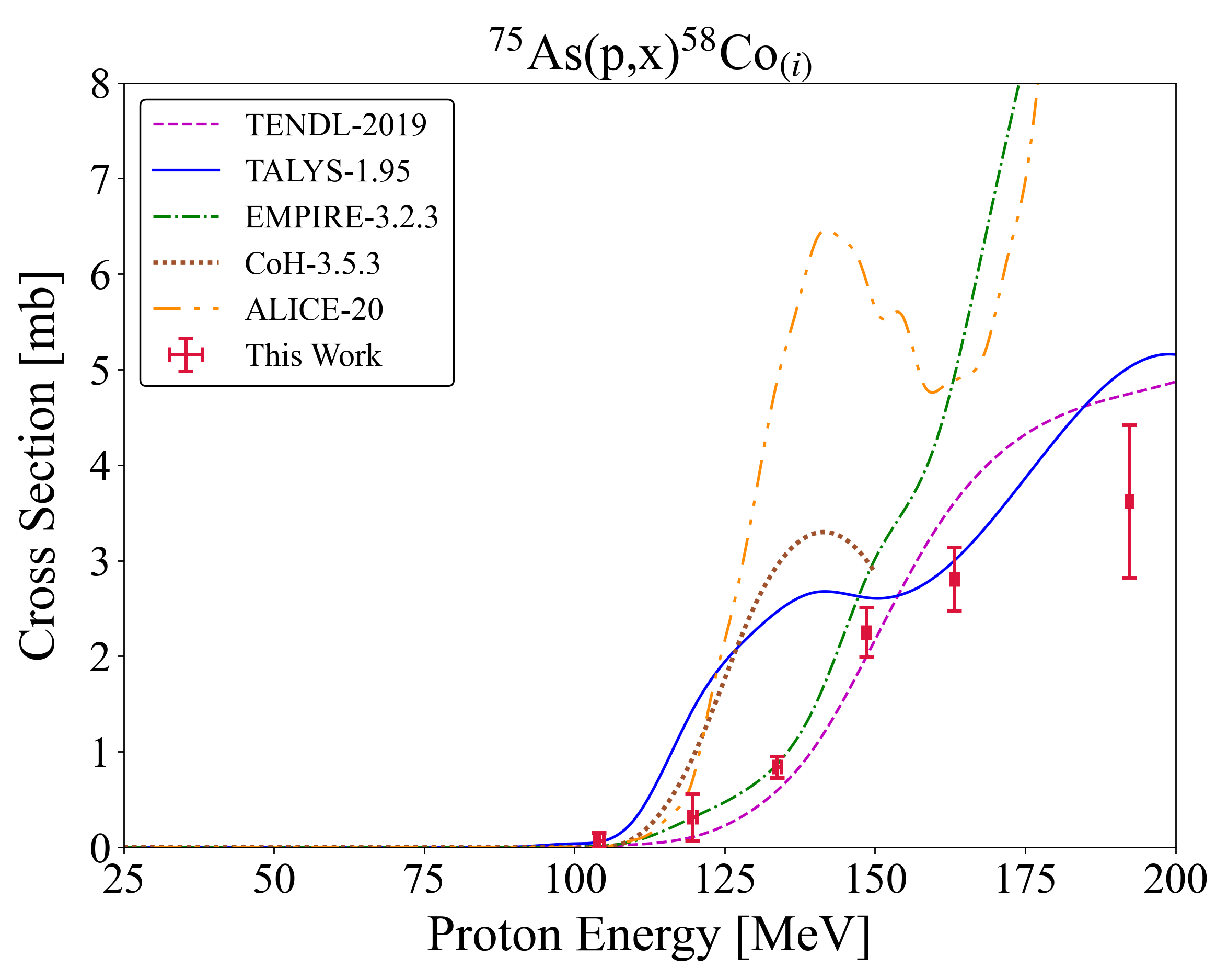}}
\vspace{-0.65cm}
\caption{Experimental and theoretical cross sections for $^{58}$Co production.}\label{As_58CO}
	{\includegraphics[width=1.0\columnwidth]{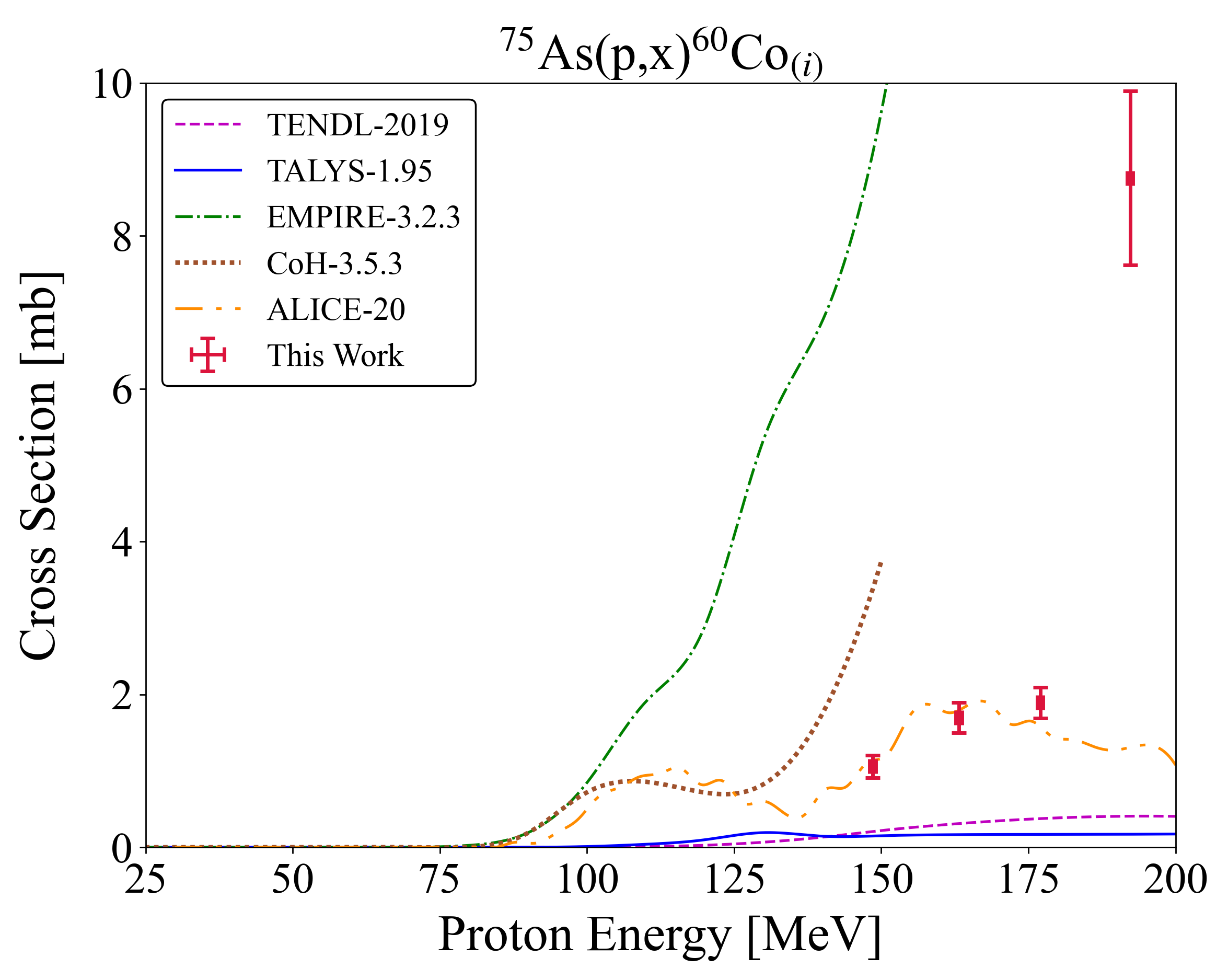}}
\vspace{-0.65cm}
\caption{Experimental and theoretical cross sections for $^{60}$Co production.}\label{As_60CO}
	{\includegraphics[width=1.0\columnwidth]{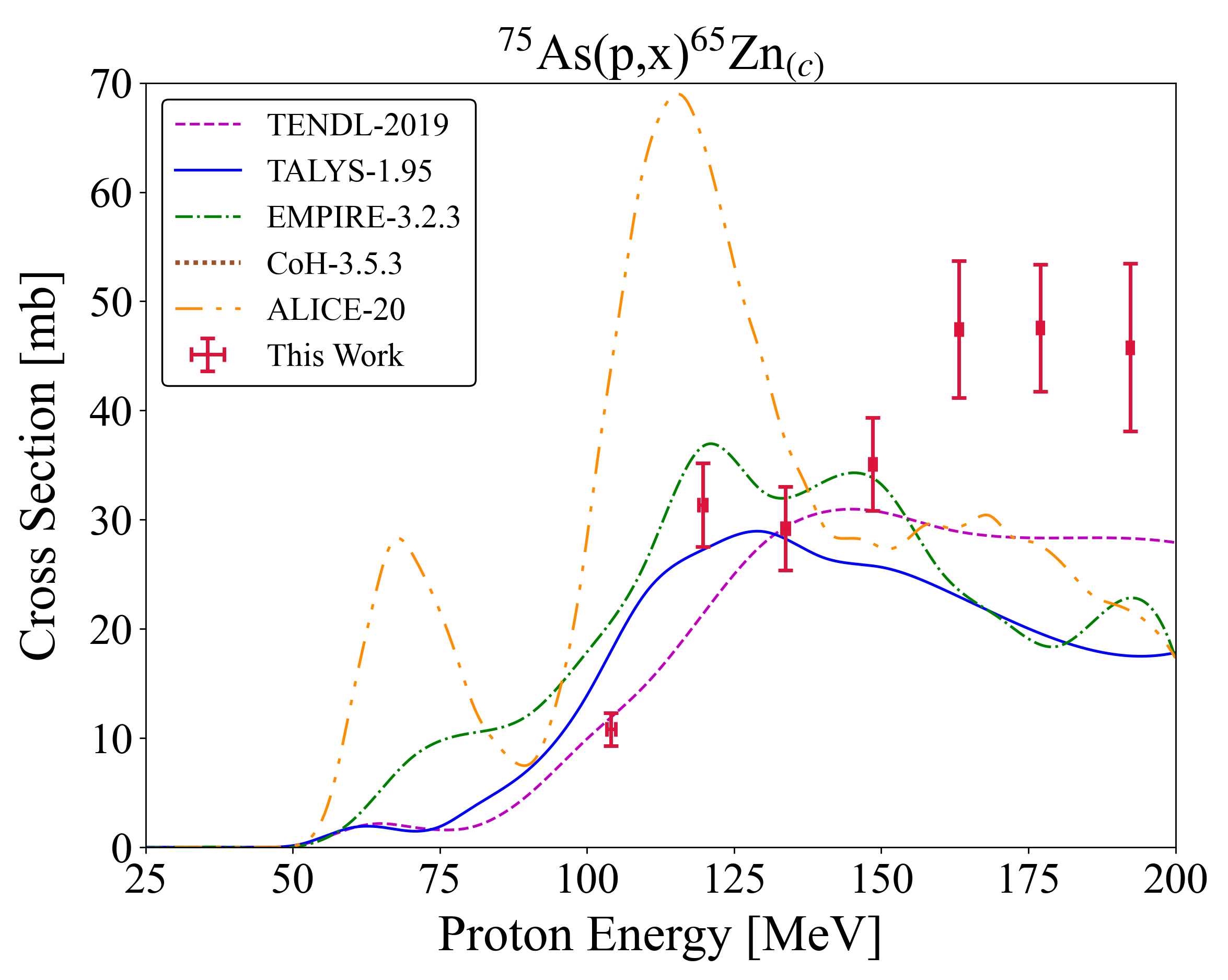}}
\vspace{-0.65cm}
\caption{Experimental and theoretical cross sections for $^{65}$Zn production.}\label{As_65ZN}
\end{figure}

\begin{figure}[H]
	{\includegraphics[width=1.0\columnwidth]{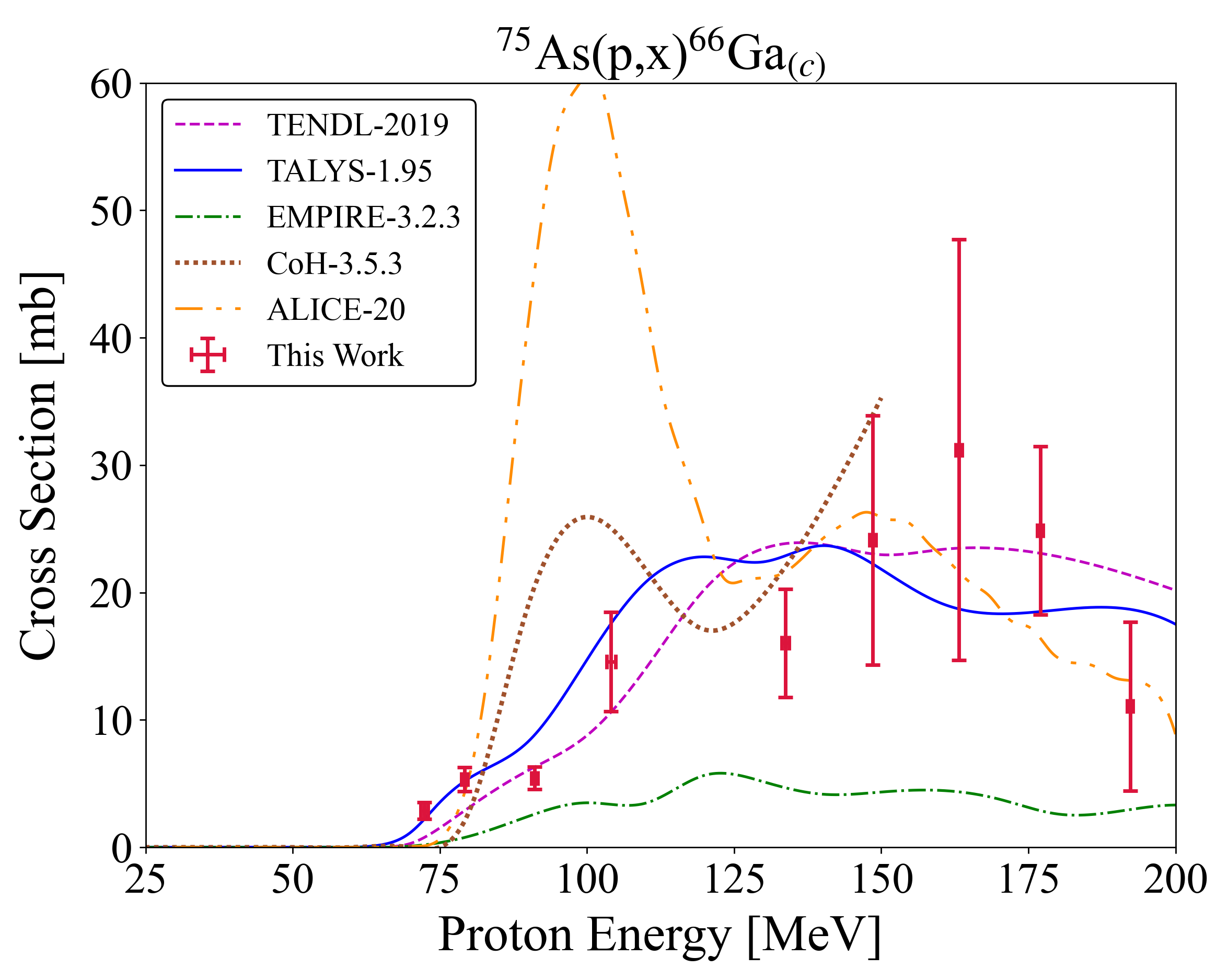}}
\vspace{-0.65cm}
\caption{Experimental and theoretical cross sections for $^{66}$Ga production.}\label{As_66GA}
	{\includegraphics[width=1.0\columnwidth]{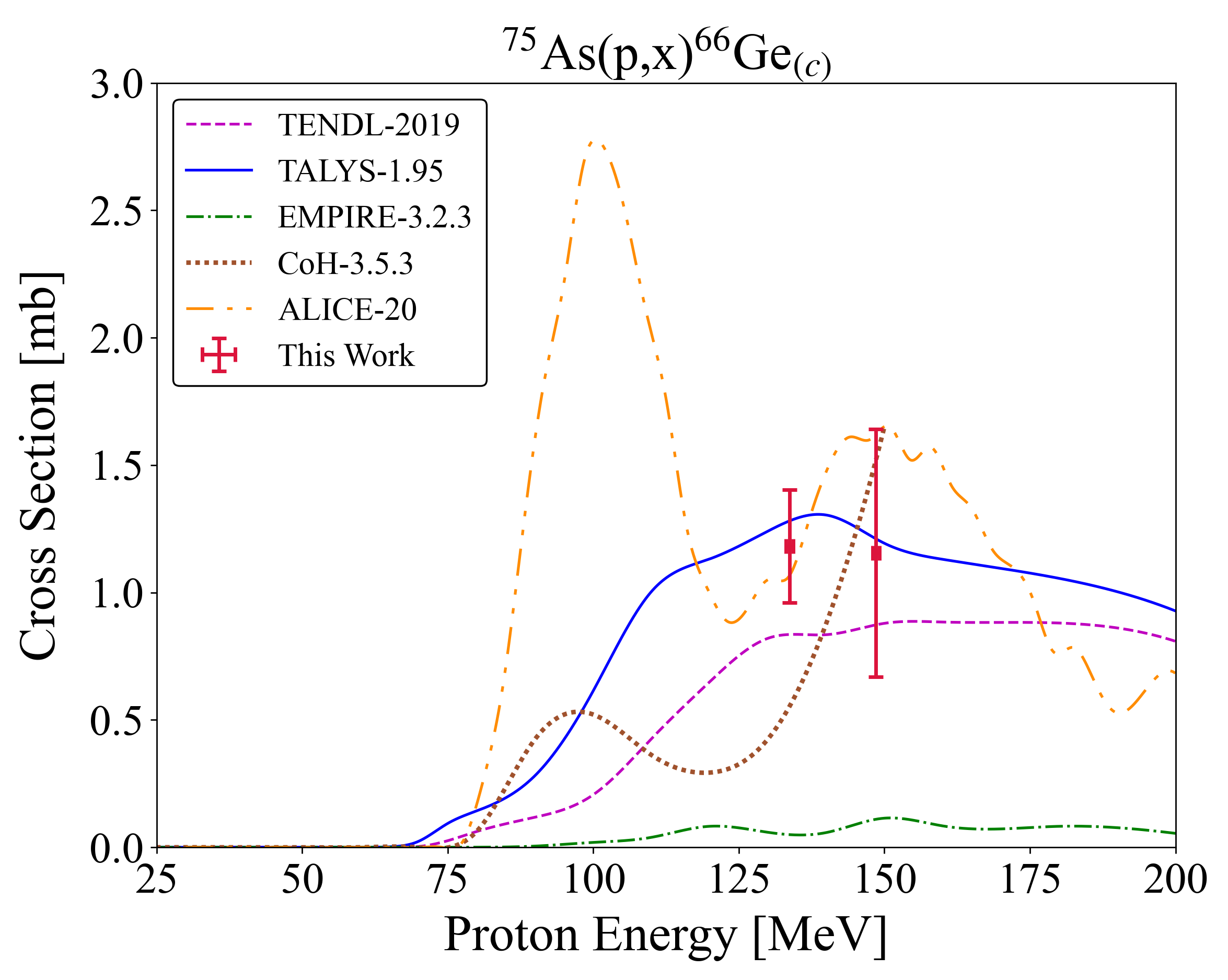}}
\vspace{-0.65cm}
\caption{Experimental and theoretical cross sections for $^{66}$Ge production.}\label{As_66GE}
	{\includegraphics[width=1.0\columnwidth]{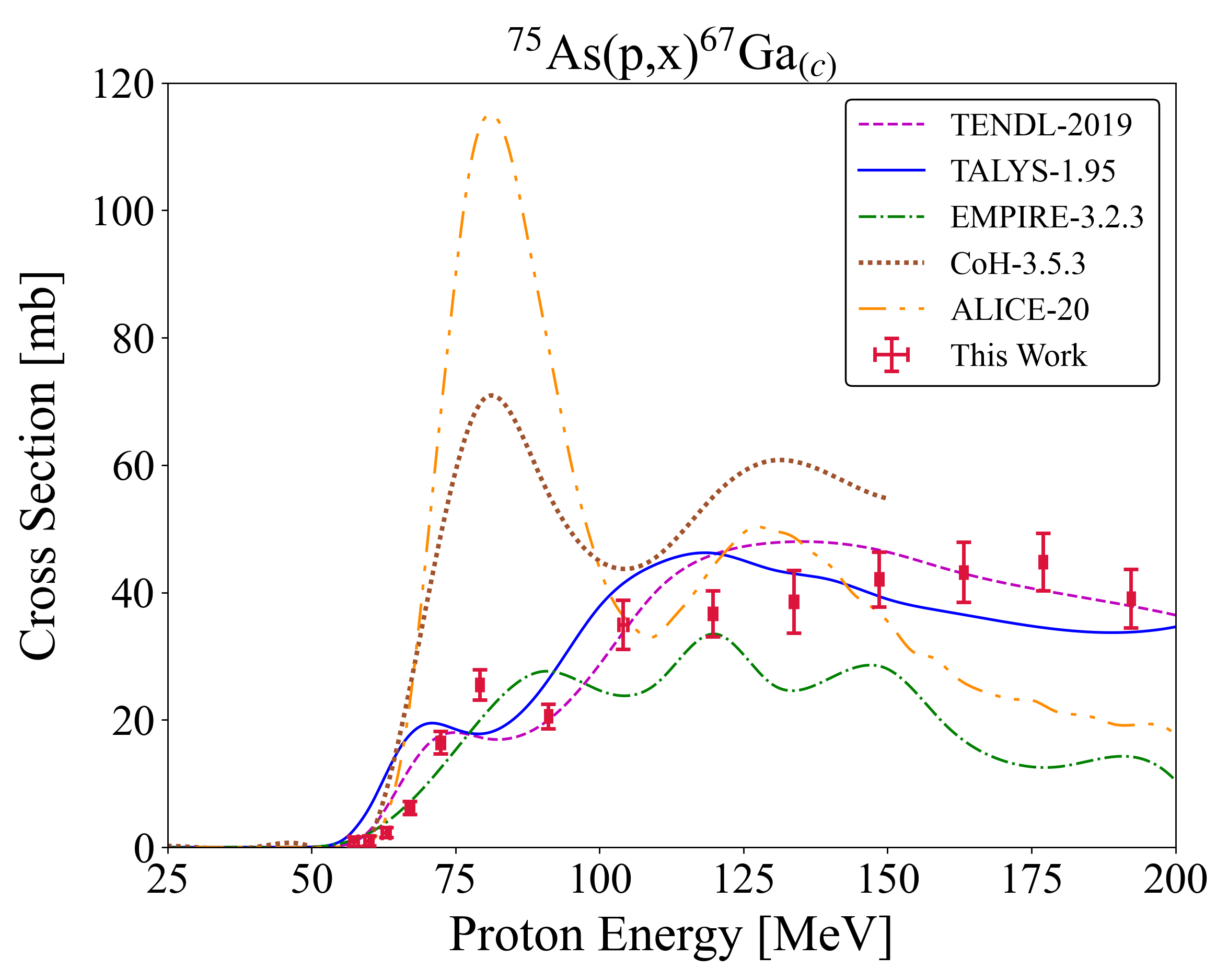}}
\vspace{-0.65cm}
\caption{Experimental and theoretical cross sections for $^{67}$Ga production.}\label{As_67GA}
\end{figure}

\begin{figure}[H]
	{\includegraphics[width=1.0\columnwidth]{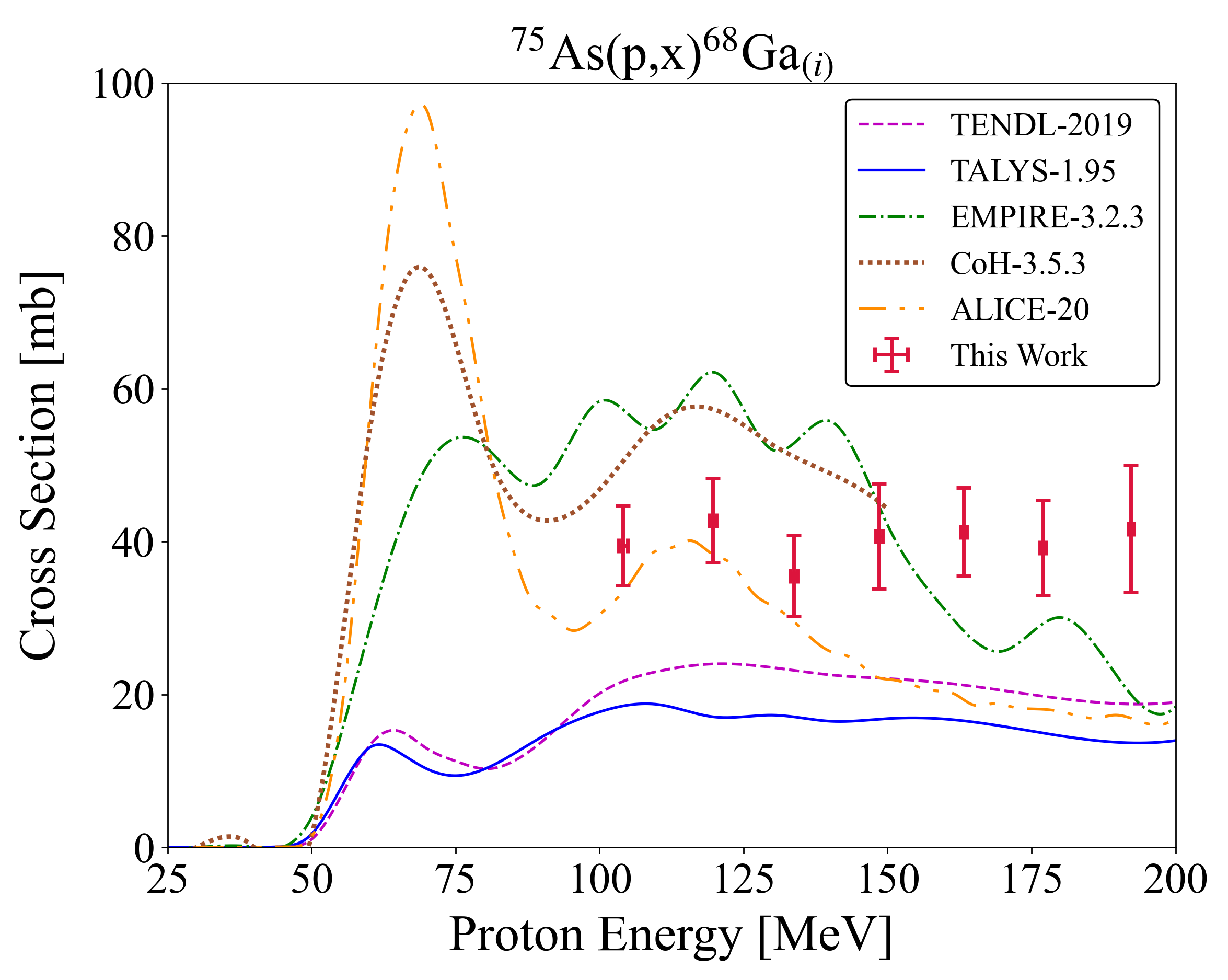}}
\vspace{-0.65cm}
\caption{Experimental and theoretical cross sections for $^{68}$Ga production.}\label{As_68GA}
	{\includegraphics[width=1.0\columnwidth]{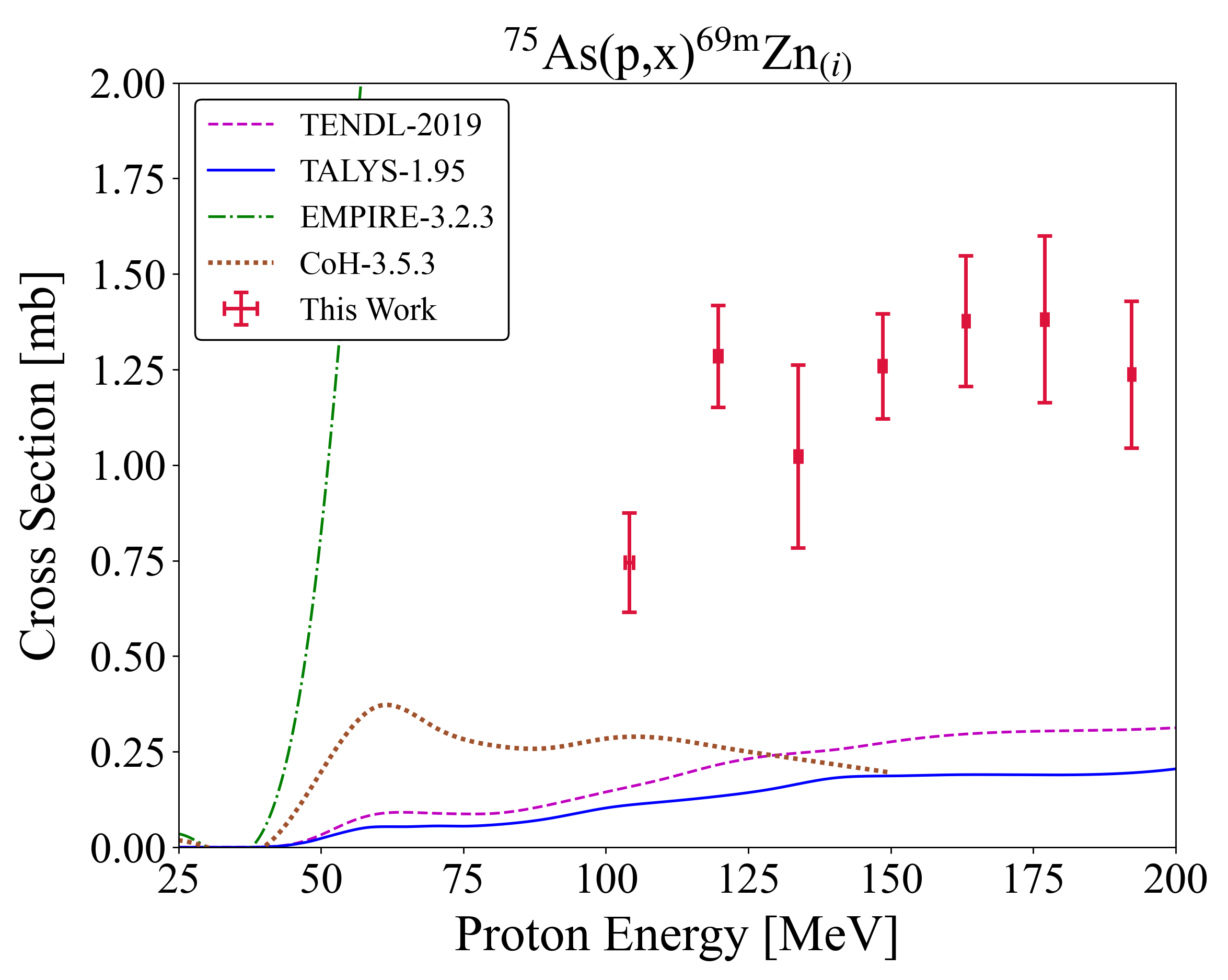}}
\vspace{-0.65cm}
\caption{Experimental and theoretical cross sections for $^{\textnormal{69m}}$Zn production.}\label{As_69ZNm}
	{\includegraphics[width=1.0\columnwidth]{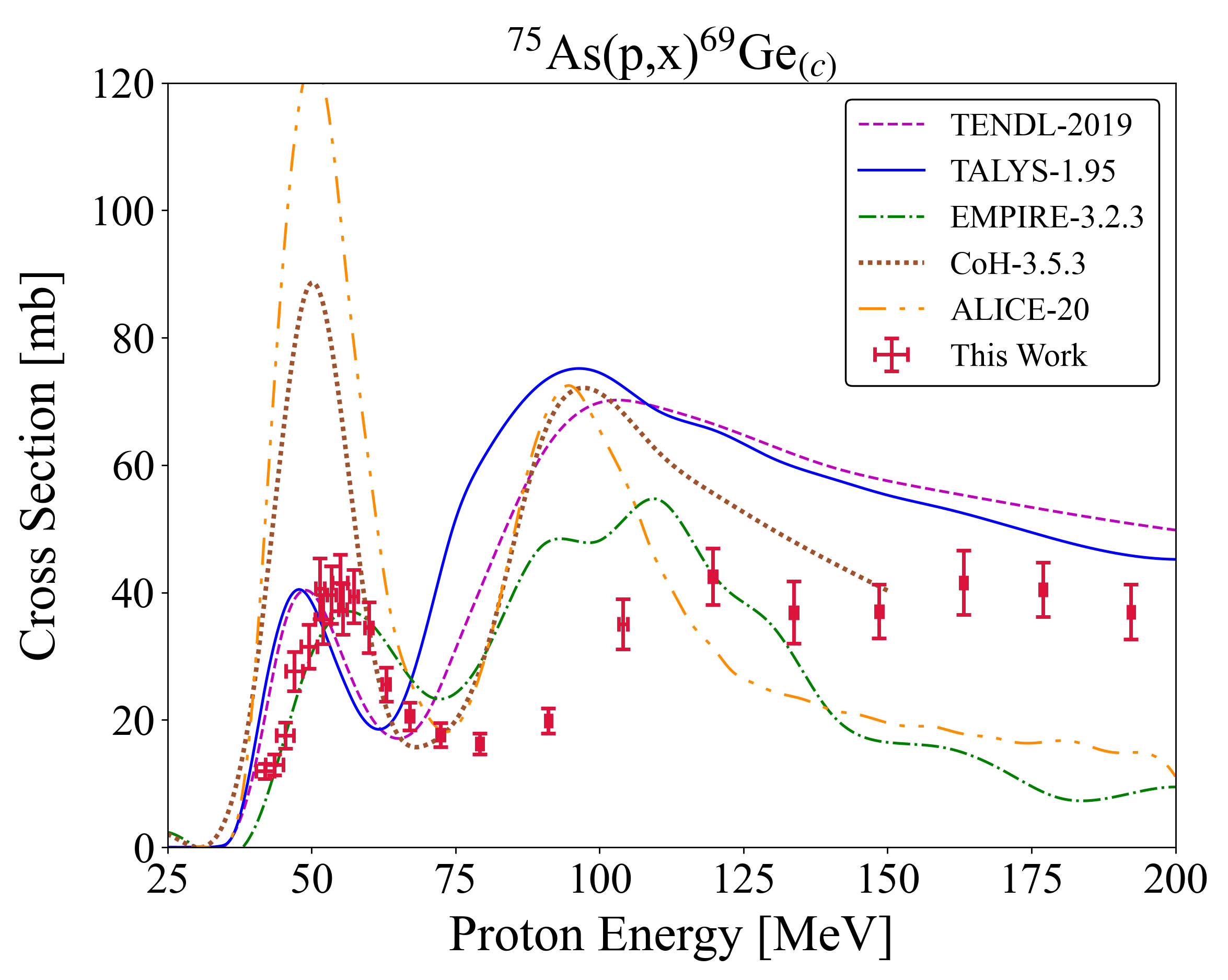}}
\vspace{-0.65cm}
\caption{Experimental and theoretical cross sections for $^{69}$Ge production.}\label{As_69GE}
\end{figure}

\begin{figure}[H]
	{\includegraphics[width=1.0\columnwidth]{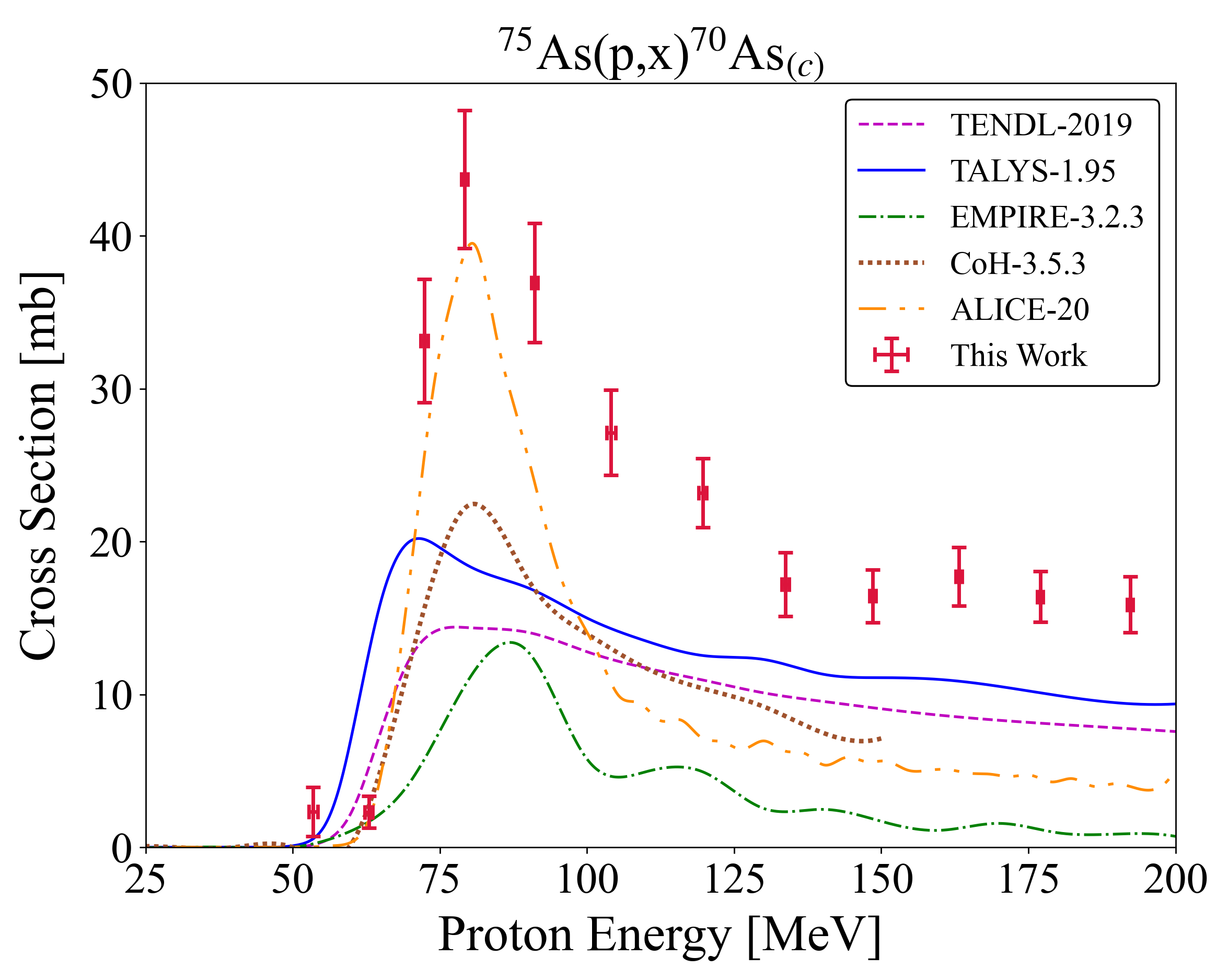}}
\vspace{-0.65cm}
\caption{Experimental and theoretical cross sections for $^{70}$As production.}\label{As_70AS}
	{\includegraphics[width=1.0\columnwidth]{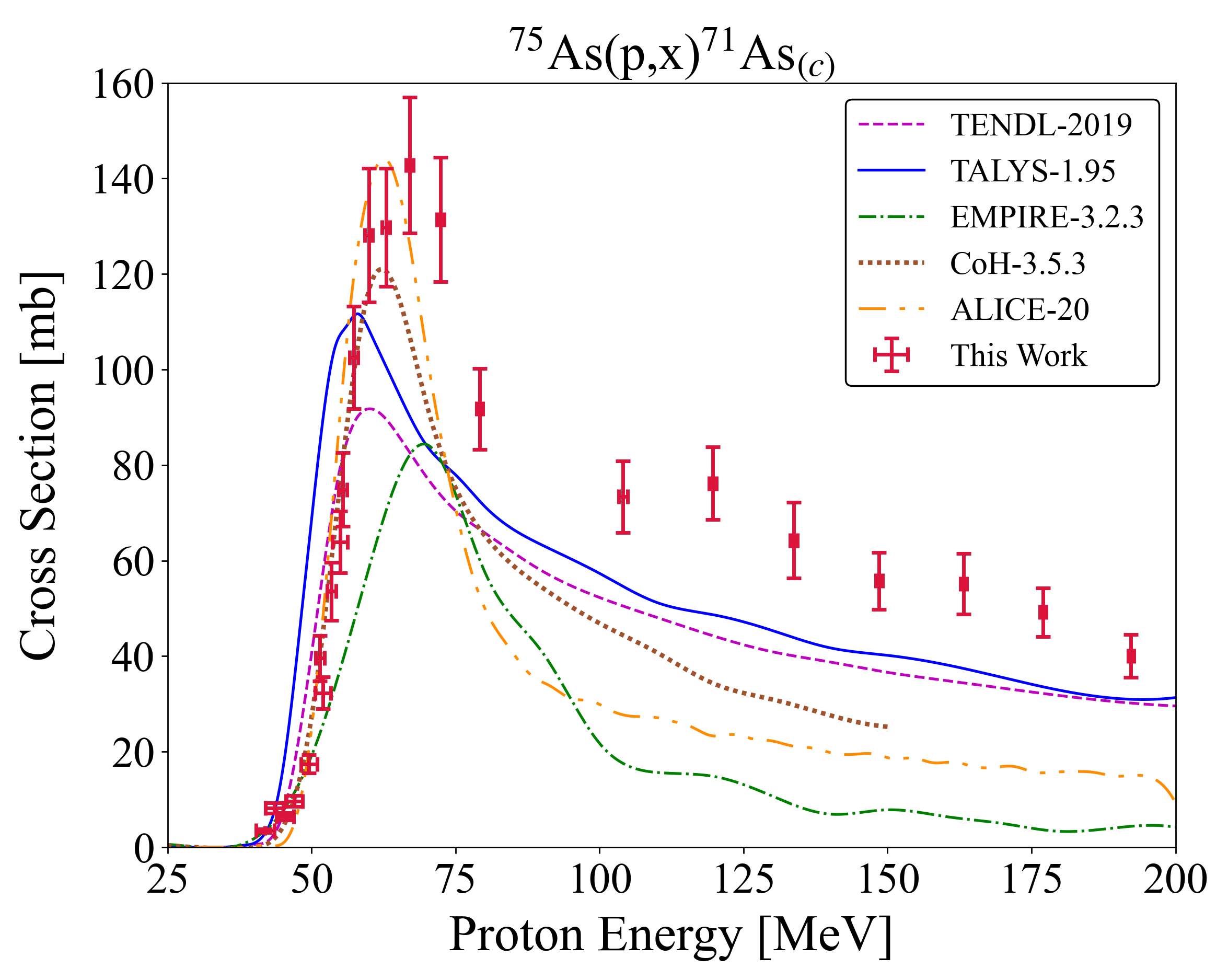}}
\vspace{-0.65cm}
\caption{Experimental and theoretical cross sections for $^{71}$As production.}\label{As_71AS}
	{\includegraphics[width=1.0\columnwidth]{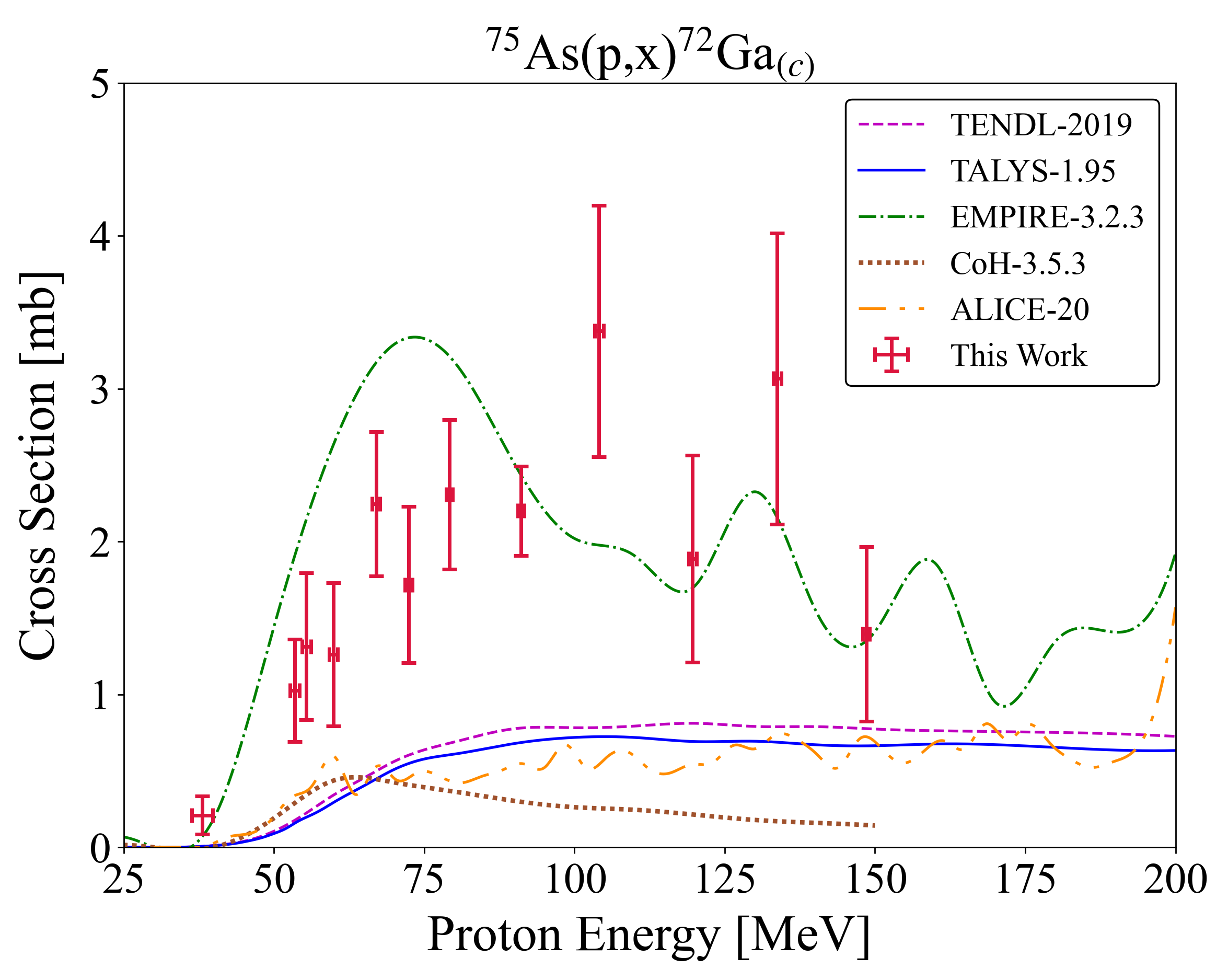}}
\vspace{-0.65cm}
\caption{Experimental and theoretical cross sections for $^{72}$Ga production.}\label{As_72GA}
\end{figure}

\begin{figure}[H]
	{\includegraphics[width=1.0\columnwidth]{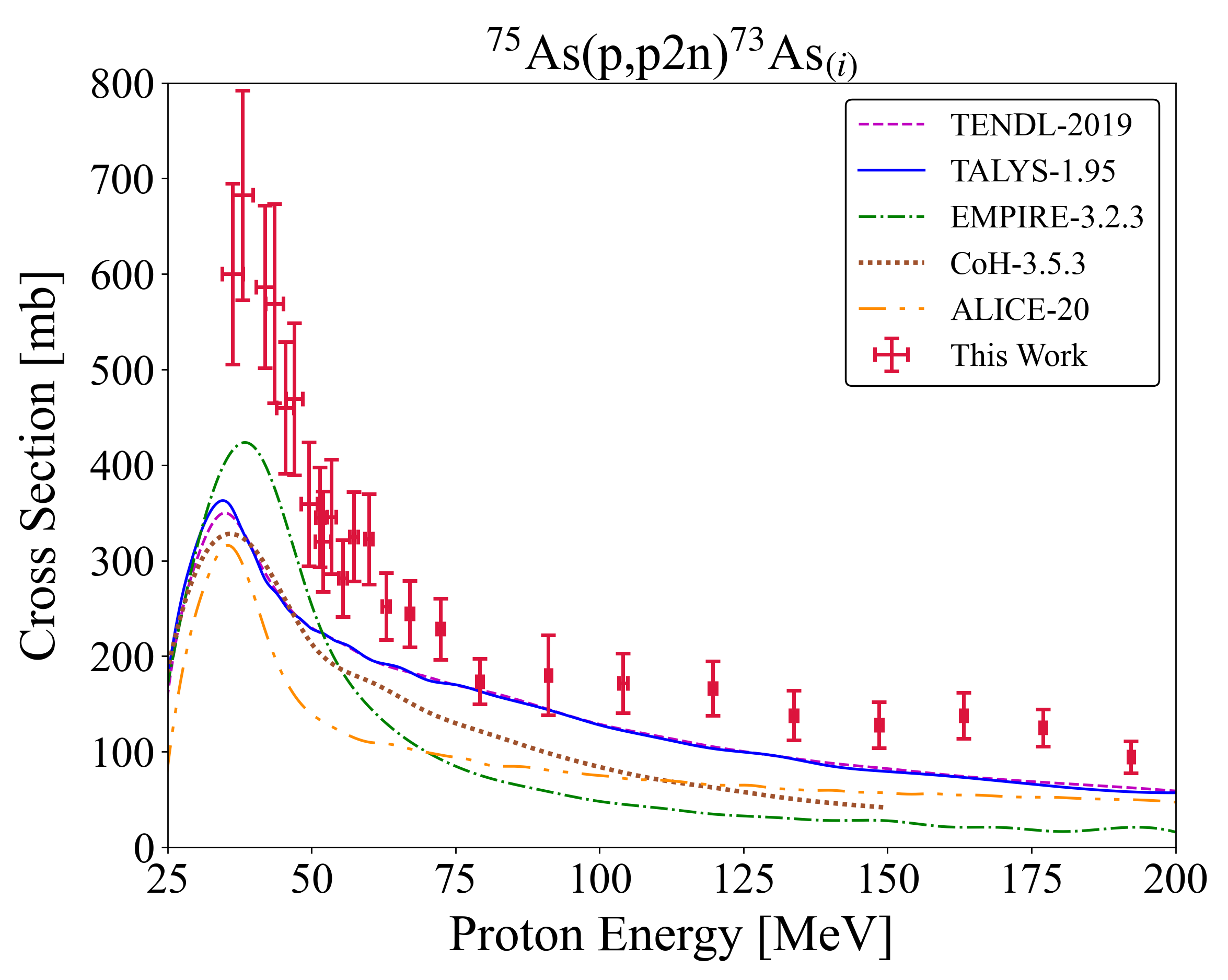}}
\vspace{-0.65cm}
\caption{Experimental and theoretical cross sections for $^{73}$As production.}\label{As_73AS}
	{\includegraphics[width=1.0\columnwidth]{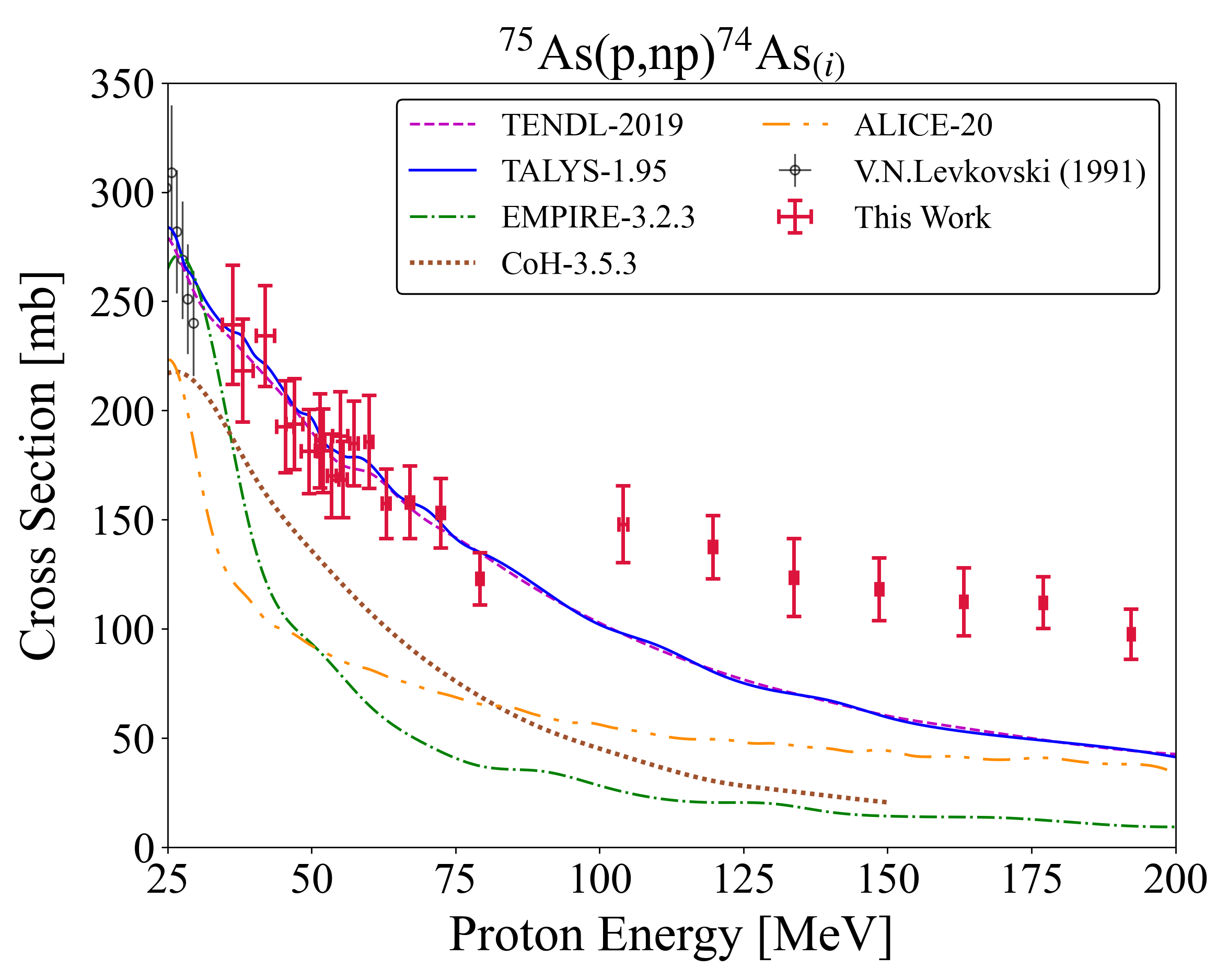}}
\vspace{-0.65cm}
\caption{Experimental and theoretical cross sections for $^{74}$As production.}\label{As_74AS}
	{\includegraphics[width=1.0\columnwidth]{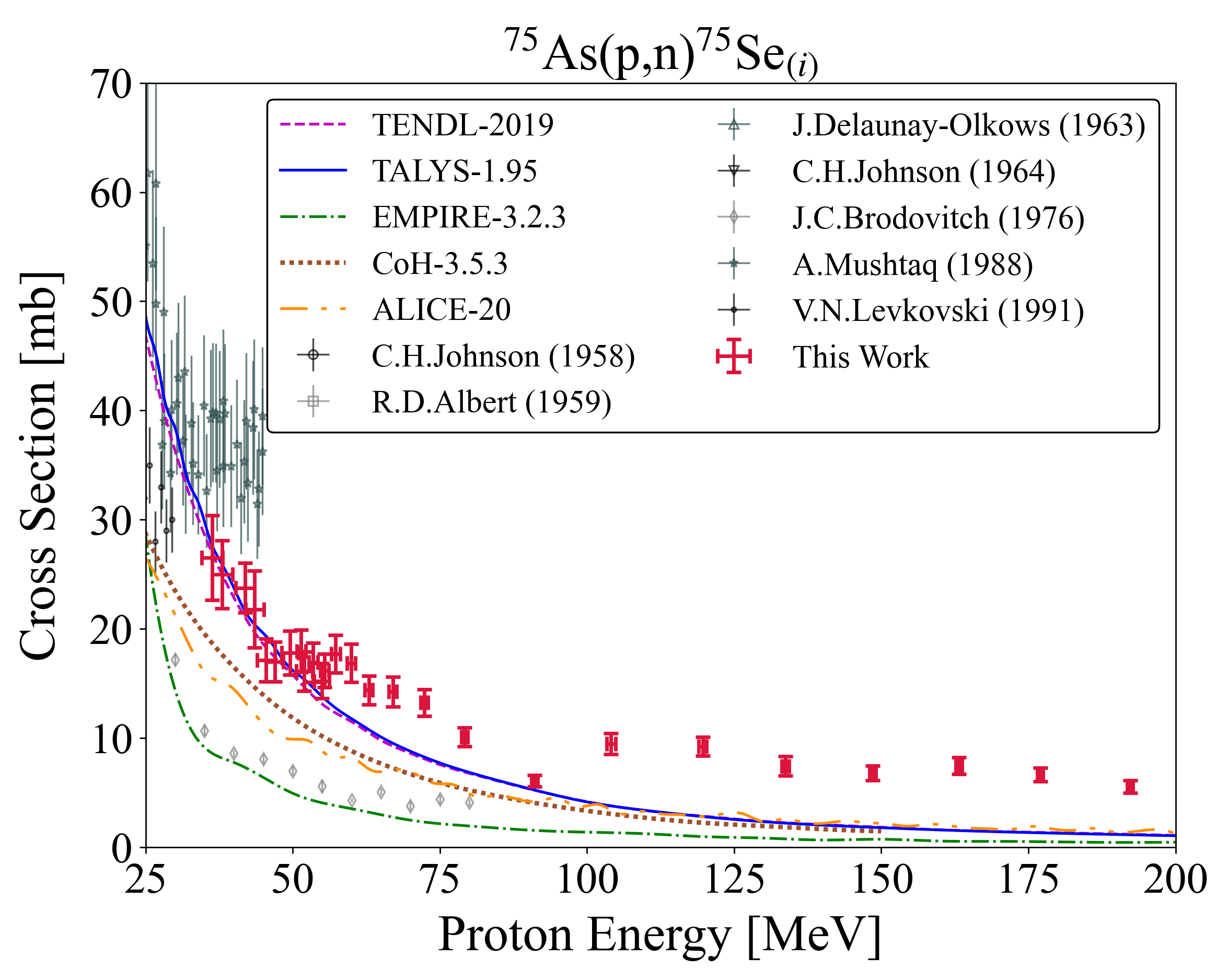}}
\vspace{-0.65cm}
\caption{Experimental and theoretical cross sections for $^{75}$Se production.}\label{As_75SE}
\end{figure}

Plots of extracted cross sections in this work are given (Figures \ref{Ti_42K}--\ref{As_75SE}) with reference to existing literature data, TENDL-2019, and reaction modeling codes TALYS-1.9, EMPIRE-3.2.3, CoH-3.5.3, and ALICE-20 using default parameters \cite{Degraffenreid2019:BNLAs,Mushtaq1988:ProtonsAs,Qaim1988:73SeIsomer,Brodovitch1976:ProtonsPEM,Levkovski1991:MiddleMass,
Cohen1953,Neumann1999:Thesis,Garrido2016:ProtonsTiNiCu,Brodzinki1971:Ti,Michel1985:Ti,Fink1990:Ti,Michel1978:ProtonsTi,
Michel1997:ProtonsTiCuNb,Khandaker2009:ProtonsTi,Zarie2006:Ti,Cervenak2020:Ti,Kopecky1993:Ti,Bringas2005:Ti,Hermanne2014:Ti,
Dittrich1988,Tarkanyi1991,Walton1976,Takacs2002:TiCu,Mills1992:ProtonsCu,Graves2016:StackTarget,Voyles2018:Nb,
Grutter1982:ProtonsCuAl,Heydegger1972,Orth1978,Yashima2003:Cu,Greenwood1984,Williams1967:ProtonsAlFeCu,Shahid2015:ProtonsCu,
Morrell2020,Uddin2004,Hermanne1999,Kopecky1985,Buthelezi2006,Khandaker2007,Yoshizawa1976,AlSaleh2006,Titarenko2003,
Aleksandrov1987,Kuhnhenn2001,Jost2013,Voyles2021:Fe}. Subscripts $(i)$ and $(c)$ in figure titles indicate independent and cumulative cross sections, respectively.

\section{\label{Appendix_Params}TALYS Parameter Adjustments From Fitting Procedure}
The derived parameter adjustments from the fitting procedure applied to the $^{75}$As(p,x) data are listed in Table \ref{AsAdjustedParams}.

\vspace{-0.35cm}
\begin{table}[h]
\vspace{-0.1cm}
\caption{$^{75}$As(p,x) best fit parameter adjustments derived from \textcite{Fox2020:NbLa} procedure. The \texttt{equidistant} keyword adjusts the width of excitation energy binning and will be a default in updated TALYS versions. The \texttt{strength} keyword selects the gamma-ray strength model and has only a small impact in this charged-particle investigation. \texttt{strength} 8 performed comparably or slightly better than the other available models in TALYS.}
\label{AsAdjustedParams}
\begin{ruledtabular}
\begin{tabular}{ll}
\multirow{1}{*}{Parameter}&\multirow{1}{*}Value\\[0.1cm]
\hline\\[-0.22cm]
\multirow{3}{*}{ldmodel}& 6\\& 4 $^{76-72}$Se, $^{68}$As\\& 5 $^{69}$Ga\\[0.1cm]
strength    &8\\[0.1cm]
equidistant    &y\\[0.1cm]
\hline\\[-0.25cm]
M2constant    &0.80\\[0.1cm]
M2limit    &3.9\\[0.1cm]
M2shift   &0.55\\[0.1cm]
Kph   &15.16\\[0.1cm]
\hline\\[-0.25cm]
d1adjust\ p    &1.55\\[0.1cm]
d1adjust\ n    &1.75\\[0.1cm]
w1adjust\ p    &1.21\\[0.1cm]
alphaomp   &6\\[0.1cm]
deuteronomp   &4\\[0.1cm]
\hline\\[-0.25cm]
\multirow{4}{*}{Cstrip}&a 0.85\\& d 2.4\\& h 0.55\\& t 0.55\\[0.1cm]
\multirow{4}{*}{Cknock}&a 0.85\\& d 2.4\\& h 0.55\\& t 0.55\\[0.1cm]
\multirow{6}{*}{ctable}& 34 73 0.24\\& 33 74 0.3\\& 33 73 0.75\\& 33 71 -0.4\\& 32 69 0.285\\& 31 67 -0.45\\[0.1cm]
\multirow{5}{*}{ptable}& 34 73 -0.65\\& 34 72 0.14\\& 33 73 -1.85\\& 32 69 -0.25\\& 31 67 5.5\\[0.1cm]
\end{tabular}
\end{ruledtabular}
\end{table}
\clearpage
%# BEST TALYS FILE n+75As
%# General
%#
%ldmodel 2
%#
%# (n,tot), (n,el), (n,inl)
%#
%rvadjust n 1. 1. 6. 3.5 1.02
%rvadjust n 1. 6. 20. 13. 0.99 
%#
%# (n,p), (n,2n), (n,np)
%#
%rvadjust p 0.89 0.02 #t#
%avadjust p 0.89 0.02 #t#
%gnadjust 33 76 1.11 0.05 #t#
%gpadjust 33 76 1.11 0.05 #t#
%gnadjust 32 75 0.92
%gpadjust 32 75 0.92
%aadjust 33 76 0.95
%aadjust 32 75 1.12
%#
%# (n,a)
%#
%rvadjust a 0.93
%avadjust a 0.93
%Cknock a 0.55
%Cstrip a 0.55
%aadjust       31  72 1.20
%#
%# (n,g)
%#
%gamgamadjust 33 76 1.60 0.10 #t#
%Rgamma 4.
%egradjust       33  76  0.8 E1
%#
%# Other: Isomers, (n,d), (n,t), (n,h) etc.
%#
%Cknock t 1.15
%Cstrip t 1.15
%Cknock h 3.00
%Cstrip h 3.00
%branch 32 75 10 1 0 1.
%branch 32 75 16 1 0 1.
%branch 32 75 3 1 0 1.

%%%%%%%%%%%%%%%%%%%%%%%%%%%%%%%%%%%%%%%%%%%%%%%%%%%%

%%%%%%%%%%%%%%%%%%%%%%%%%%%%%%%%%%%%%%%%%%%%%%%%%%%%
%\bibliography{../TriLab_Exp_Model}

%%%%%%%%%%%%%%%%%%%%%%%%%%%%%%%%%%%%%%%%%%%%%%%%%%%%
\end{document}